\documentclass[reprint]{revtex4-2}

\usepackage{physics,bm,amssymb,amsfonts,amsmath,mathtools}
\usepackage{graphicx}

\begin{document}

\title{Enhanced Electro-Optic Sampling with Quantum Probes}
\author{St\'{e}phane Virally}
\email{stephane.virally@polymtl.ca}
\affiliation{\emph{femtoQ} Lab, Department of Engineering Physics, Polytechnique Montr\'{e}al, Montr\'{e}al, QC H3T 1JK, Canada}
\author{Patrick Cusson}
\affiliation{\emph{femtoQ} Lab, Department of Engineering Physics, Polytechnique Montr\'{e}al, Montr\'{e}al, QC H3T 1JK, Canada}
\author{Denis V. Seletskiy}
\email{denis.seletskiy@polymtl.ca}
\affiliation{\emph{femtoQ} Lab, Department of Engineering Physics, Polytechnique Montr\'{e}al, Montr\'{e}al, QC H3T 1JK, Canada}
\date{\today}

\begin{abstract}
Employing electro-optic sampling (EOS) with ultrashort probe pulses, recent experiments showed direct measurements of quantum vacuum fields and their correlations on subcycle timescales.
Here, we propose a quantum-enhanced EOS where photon-number entangled twin beams are used to derive conditioned non-classical probes.
In the case of the quantum vacuum, this leads to a six-fold improvement in the signal-to-noise ratio over the classically-probed EOS.
In addition, engineering of the conditioning protocol yields a reliable way to extract higher-order moments of the quantum noise distribution and robust discrimination of the input quantum states, for instance a vacuum and a few-photon cat state.
These improvements open a viable route towards robust tomography of quantum fields in space-time, an equivalent of homodyne detection in energy-momentum space, and the possibility of precise experiments in real-space quantum electrodynamics.
\end{abstract}

\maketitle
The underlying relativistic invariance of quantum electrodynamics is best manifested when non-classical fields are expressed directly in space-time coordinates~\cite{Feynman1949}.
Yet, most measurement methods of quantum fields are intrinsically rooted in the reciprocal (energy-momentum) space, following a first quantization step with the Hamiltonian method.
Such is the case of homodyne detection (HD)~\cite{Collett1987} in quantum optics, where information on the quantum state is obtained by registering in a square-law detector its linear superposition with a narrow-band classical field, also termed ``local oscillator'' (LO).
In these measurements, non-vanishing signals only arise where both inputs are commensurate in angular frequency and wavevector ($\Delta\omega\simeq0,\Delta k\simeq0$).

For direct space-time measurement of electromagnetic fields, it is necessary to: 1) replace the narrow-band LO by a short, wideband probe and 2) augment the optical superposition of HD by a wave mixing of the probe and signal. In addition, for quantum signals, one needs to access all moments of the statistics of the measured field amplitude, in the same way that HD with a sufficiently strong LO enables full tomography of quantum states in the frequency domain~\cite{Lvovsky2009}.

The first two requirements are fulfilled in a scenario where a low-frequency signal [$\Omega$, e.g.\ in the teraherz (THz) range] is nonlinearly mixed with a short probe of duration $\Delta t$ at a higher carrier frequency [$\omega$, e.g.\ in the near-infrared, NIR].
The subcycle structure of the THz field can be probed when $\Omega\,\Delta t < 1/2$ and both fields are overlapped within a small space-time volume ($\Delta t\simeq0,\Delta r\simeq0$) ~\cite{Riek2015}.
This setup is at the heart of the electro-optic sampling (EOS) scheme, where the measurement of the instantaneous electric amplitude $\mathcal{E}_\textrm{THz}(t)$ of a classical field is made possible by superposing the weak nonlinear mixing product between this field and the probe, with the strong non-interacted part of the probe~\cite{Gallot1999}.
Thanks to advances in femtosecond (fs) laser technologies~\cite{Brabec2000,Krauss2010,Hassan2016}, shortening of the probe duration has enabled progress of EOS from sub-terahertz (THz) frequencies~\cite{Valdmanis1983} to detection of fields with $\Omega$ components extending to 100 THz and above~\cite{Wu1995, Leitenstorfer1999, Kubler2004, Gaal2007, Sell2008, Tomasino2013, Keiber2016, Riek2017a}, in turn empowering field-resolved spectroscopies~\cite{Huber2001, Jepsen2011}.

\begin{figure*}
    \centering
    \includegraphics[width=\textwidth]{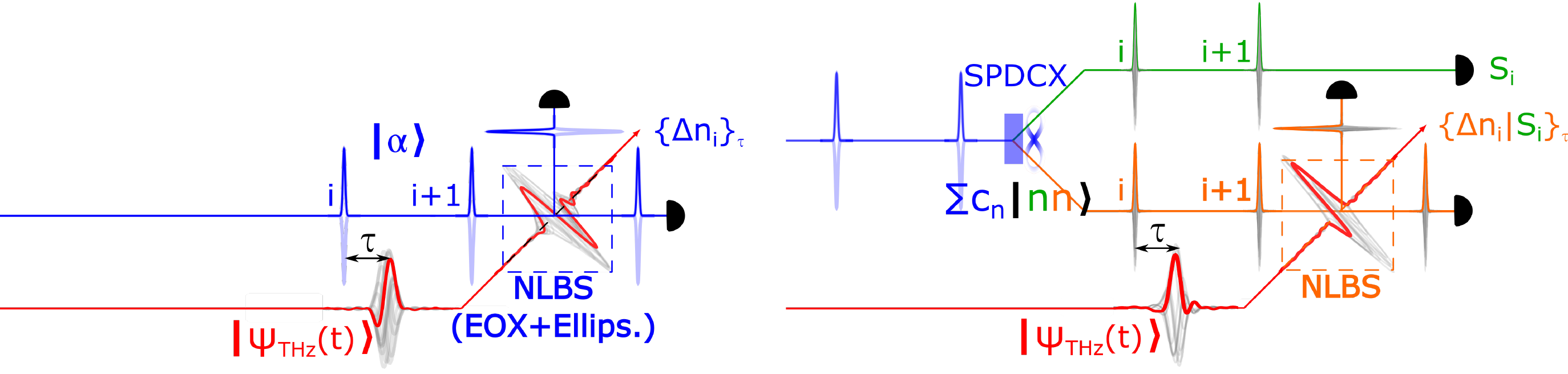}
    \put(-500,30){(a)}
    \put(-245,30){(b)}\\
    \hspace{-0.028\linewidth}
    \includegraphics[width=0.245\linewidth, trim=5 15 55 65, clip]{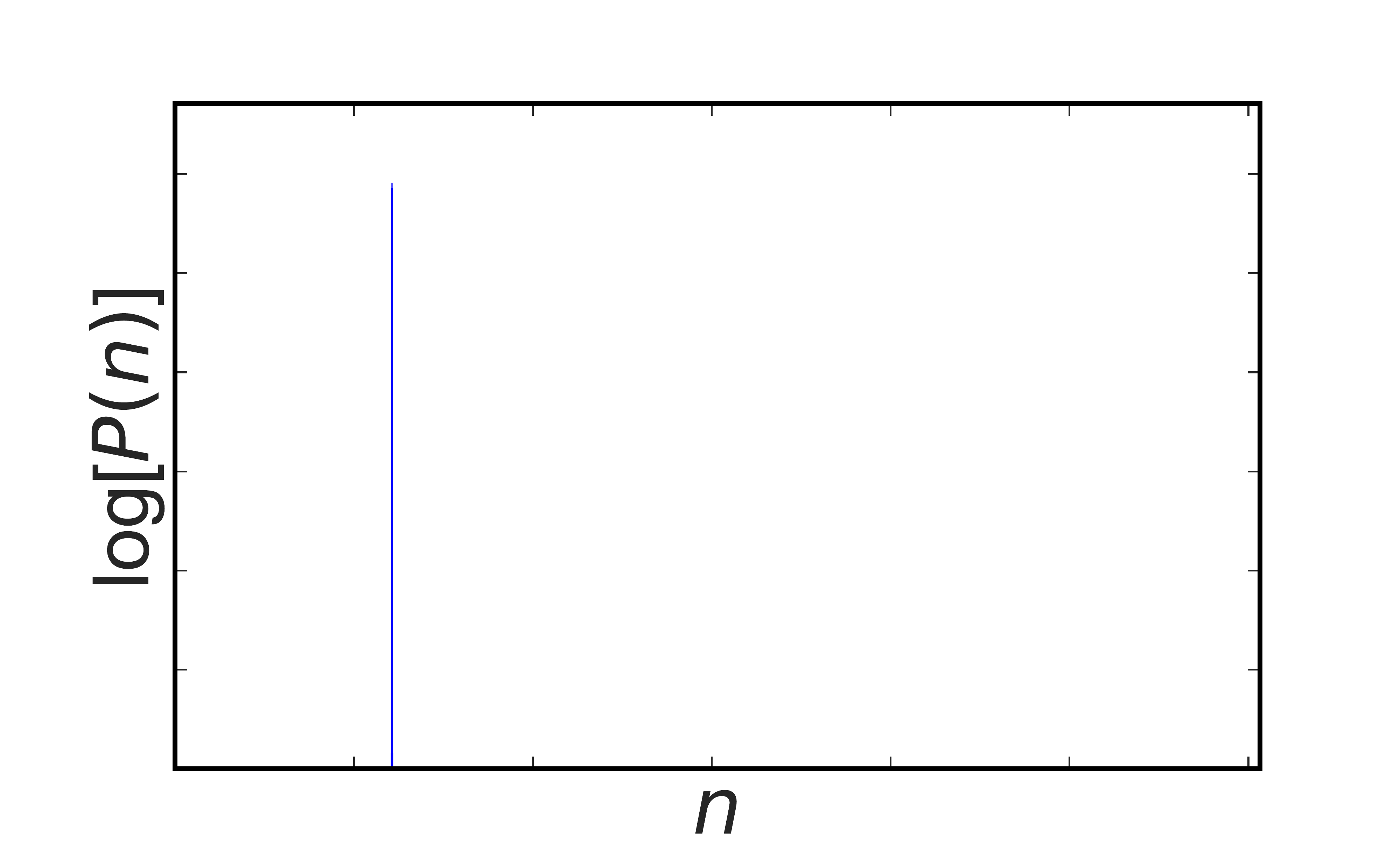}
    \put(-22,25){(c)}
    \hspace{0.005\linewidth}
    \includegraphics[width=0.245\linewidth, trim=5 15 55 65, clip]{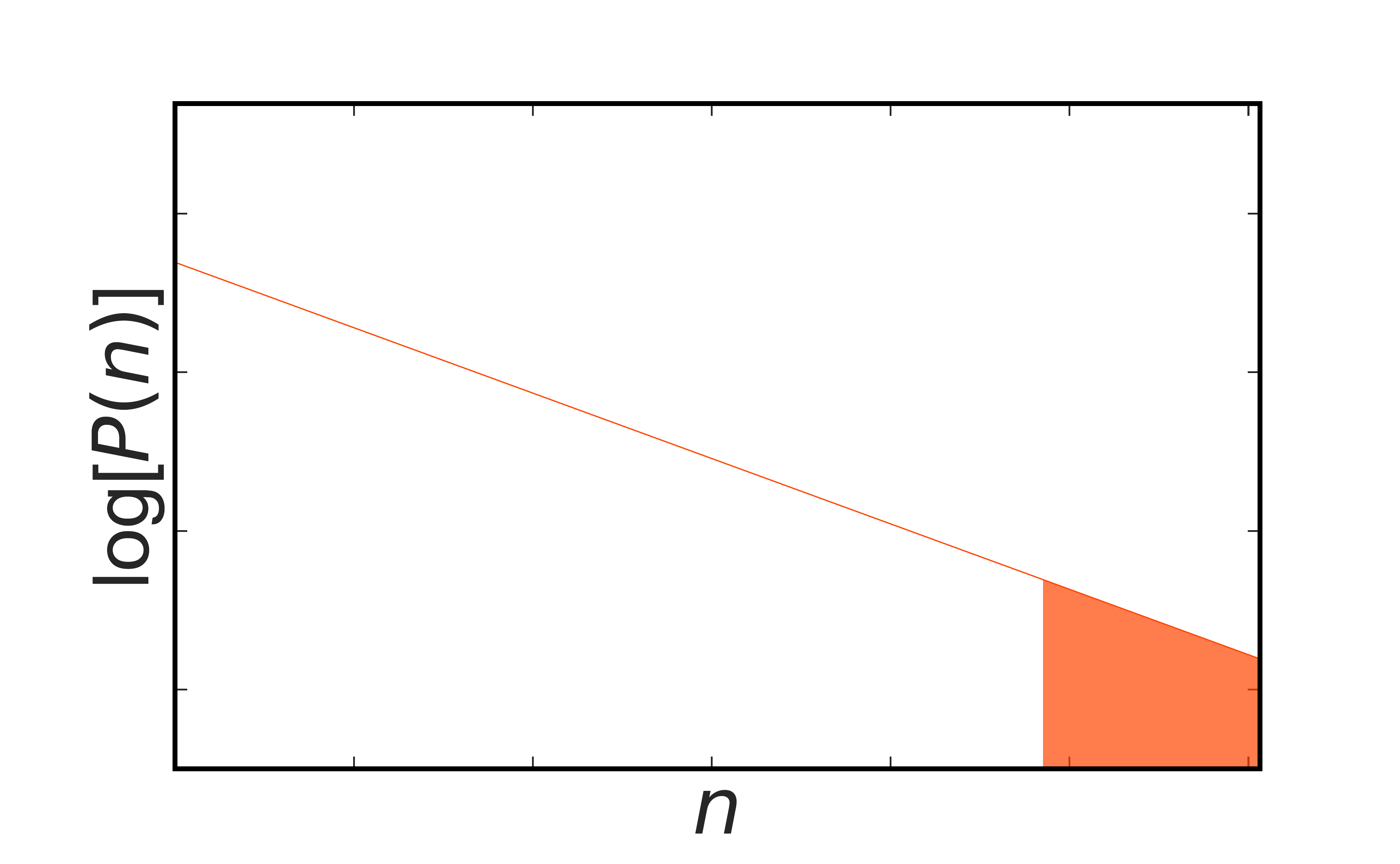}
    \put(-22,25){(d)}
    \hspace{0.005\linewidth}
    \includegraphics[width=0.245\linewidth, trim=5 15 55 65, clip]{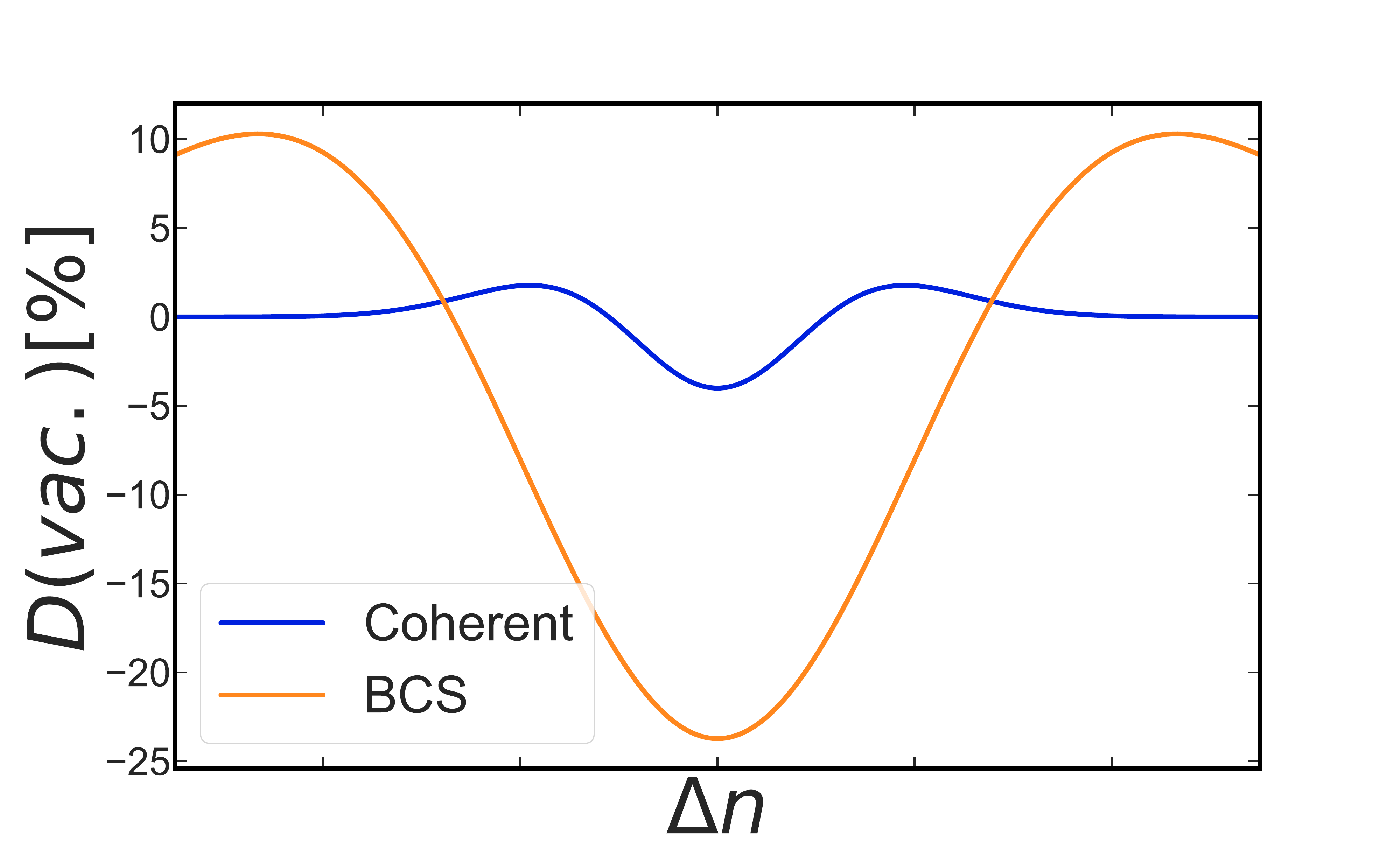}
    \put(-22,25){(e)}
    \hspace{0.005\linewidth}
    \includegraphics[width=0.245\linewidth, trim=5 15 55 65, clip]{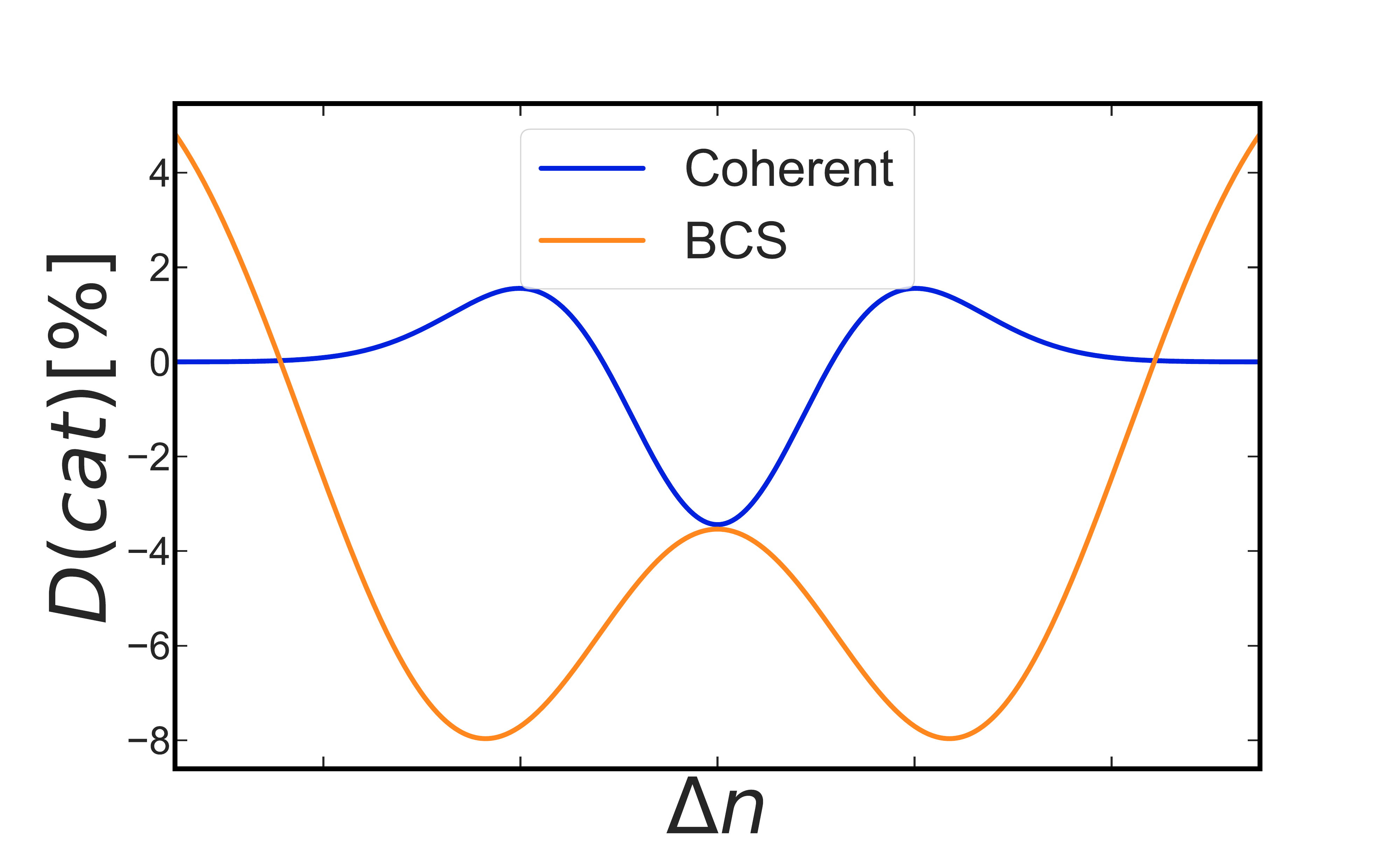}
    \put(-22,25){(f)}
    \vspace{-10pt}
    \caption{EOS setups for statistical sampling of quantum states of THz fields. Top row: (a) typical EOS setup, used for instance to successfully probe the vacuum state of a confined THz mode~\cite{Riek2015}. Detection includes an EOS crystal (EOX) followed by an ellipsometry setup (Ellips.). It can be seen as a nonlinear beam-splitter (NLBS) for the probe. The newly proposed scheme (b) replaces the classical probe of (a) by a non-classical probe (BCS) conditioned on the intensities measured in a twin branch (in green). Bottom row: (c,d) logarithms of the probability distributions of the number of photons in the incoming probe, for the coherent state (c) of classical EOS (a), and for a specific BCS (d) in quantum-enhanced EOS (b); (e,f) relative differential noise amplitude $D$ [Eq.~\eqref{eq:D}] for both cases when the signal is the vacuum state of a confined THz mode such as that of Ref.~\cite{Riek2015} (e). The amplitude is improved by a little less than one order of magnitude. Same comparison for a cat state (f). There is remarkably little difference between the curves for the coherent probe, while the BCS probe clearly differentiates between vacuum and cat.}
    \label{fig:Setup}
\end{figure*}

In contrast to HD, it is very difficult to measure a weak signal such as a quantum state $\lvert\psi_\textrm{THz}\rangle$ with EOS.
In HD, a strong LO can amplify the signal, but in EOS the nonlinear mixing products must remain intrinsically weak to avoid noise induced by higher-order processes.
Nonetheless, recent experiments have ported EOS with classical probes to the quantum regime, measuring the variance $\langle\psi\rvert\hat{\mathcal{E}}^2(t)\lvert\psi\rangle$~\cite{Riek2015,Riek2017} and two-time correlations $\langle\psi\rvert\hat{\mathcal{E}}(t_1)\hat{\mathcal{E}}(t_2)\lvert\psi\rangle$~\cite{Benea-Chelmus2016, Benea-Chelmus2019} of broadband THz vacuum fields.
Such measurements are limited by the shot noise of the probe~\cite{Moskalenko2015}, and the contribution of the quantum signal $\lvert\psi\rangle$ amounts only to a few percent of the total detected noise~\cite{Riek2015,Benea-Chelmus2019}.
This impedes the determination of high statistical moments of the quantum distribution, which is further exacerbated by the Gaussian statistics of the classical probe pulses. 

In this Letter, we show that EOS of quantum fields can be dramatically improved by employing quantum probe pulses. Specifically, entangled bright few-cycle pulses~\cite{Chekhova2015} originating from spontaneous parametric down conversion (SPDC) can be used to carve out non-classical statistics~\cite{Iskhakov2016} in a probe beam by conditioning it on intensity values of its twin. Because the conditioning only selects events in certain bands of intensities, we refer to theses non-classical states as band-conditioned states (BCS).
Compared to classical EOS probes, BCS not only provide a six-fold improvement in the detection of the variance, but also extract additional information on higher statistical moments of the quantum distribution by simple modifications of the conditioning protocol.
This enables robust sampling of non-Gaussian subcycle states of quantum light, for instance a few-photon cat state.
Thanks to these features, BCS are well positioned to fulfill the needs of characterization of time-domain quantum optical beams~\cite{Virally2019} and emerging applications in time-domain quantum spectroscopy~\cite{Kira2006,Dorfman2016,Mukamel2020}.

In order to compare classical and quantum-enhanced detection, we start with the operational principle of EOS using coherent state $\lvert\alpha\rangle$ probes.
As sketched in Fig.~\ref{fig:Setup}(a), a train of NIR probe pulses (\textellipsis, $i$, $i+1$, \textellipsis, in blue) is mixed with a train of identically-prepared THz quantum states $\ket{\Psi_\mathrm{THz}}$ (in red) through second-order nonlinear interaction inside an electro-optic crystal (EOX).
THz-induced birefringence changes the polarization state of the pulsed probe, that is directly proportional to either the amplitude or the Hilbert transform of the THz electric field signal~\cite{Virally2019,Sulzer2020}.
For the amplitude measurement, polarization change is converted by an ellipsometer [a quarter wave plate (QWP) followed by a polarizing beasmplitter (PBS)] into an imbalance of photocounts $\Delta n_i$ and registered by a balanced photodiode pair (BPD)~\cite{Gallot1999}.
The temporal delay $\tau$ introduced between the probe and the signal pulses allows for sampling of field statistics at controlled time slices.
In terms of the probe's polarization state, the effect of EOS (EOX, QWP, PBS) can be mapped onto an effective ``nonlinear beam splitter'' (NLBS)~\ref{appendix}, where the THz amplitude induces an imbalance in an otherwise equal splitting of the probe [Fig.~\ref{fig:Setup}(a)]. 

For an incident signal $\psi$, the BPD pair after the NLBS registers $\Delta n$ with a probability $P(\Delta n; \psi)$, which takes the form
\begin{equation}
    \label{eq:P}
    P(\Delta n;\psi)=\sum_{n=0}^{+\infty}\,P(n)\,\sum_{k=0}^{+\infty} \alpha_k(n,\Delta n)\langle\psi\rvert\hat{\mathcal{E}}^k\lvert\psi\rangle,
\end{equation}
where $P(n)$ is the probability of having $n$ photons in the probe, $\langle\psi\rvert\hat{\mathcal{E}}^k\lvert\psi\rangle$ is the $k^\textrm{th}$ moment of the electric field distribution in the signal, and the $\alpha_k$ are coefficients dependent on the particulars of the setup~\cite{Note1}. 
Eq.~\eqref{eq:P} highlights two main contributions to $P(\Delta n; \psi)$: (i) the second-quantization photocount distribution of the probe and (ii) the statistical distribution of the THz electric field, as a parameter in the binomial distribution characteristic of the NLBS.
Isolation of the contribution of the quantum signal can be made by decoupling it from the probe through a dilation of its space-time volume~\cite{Riek2015, Moskalenko2015}.
In this case, the distribution of EOS photocount difference $P(\Delta n;0)=\sum_nP(n)\,\alpha_0(n,\Delta n)$ depends only on the statistics of the probe, e.~g.~ yielding a Gaussian-distributed shot noise [see Fig.~\ref{fig:Setup}(c)] for a classical probe~\cite{Riek2015}. 
It is thus judicious to define the relative differential noise amplitude
\begin{equation}
    \label{eq:D}
    D(\psi)=\frac{P(\Delta n;\psi)-P(\Delta n;0)}{\max[P(\Delta n;0)]},
\end{equation}
inspired by Refs.~~\cite{Riek2015, Riek2017}, where $\max[P]$ is the maximum of the distribution.
The magnitude of $D$ gives a quantitative understanding of the contribution of the quantum signal to the overall measurement and is therefore a good optimization target for the EOS scheme.

For an incoming broadband THz vacuum state $\lvert\textrm{vac.}\rangle$, experiments have shown~\cite{Riek2015} that mixing in the NLBS results in a $D(\textrm{vac.})$ signal with a peak-to-peak deviation of about 6~\%, as shown in Fig.~\ref{fig:Setup}(e).
It is perhaps not surprising that the weakness of the coupling together with the Gaussian nature of the shot-noise distribution can quickly limit the determination of the full quantum statistics of the signal.
As seen from Eq.~\eqref{eq:P}, the photocount distribution in the probe, and hence its second-quantized state, play a significant role in the $D$ measure.
Indeed, for classical EOS detection, the best signal-to-noise is achieved when the pulsed probe is devoid of technical noise and thus operates at the shot-noise limit, reached with a coherent state $\lvert\alpha\rangle$ characterized by a Poissonian photocount distribution $P(n)$ centered at $\bar{n}=\lvert\alpha\rvert^2$.
Lead by this reasoning, one might conclude that the only strategy for improving the value of $D$ is to increase the power of the coherent probe, thus reducing the relative standard deviation in its photocounts.
This strategy is already employed in the most optimized experiments so far~\cite{Riek2015,Benea-Chelmus2016,Riek2017}. 

As we show below, engineering of the second-quantized state of the probe can dramatically enhance the value of $D$.
Perhaps paradoxically, this shows that it can be beneficial to \emph{add} noise to the EOS probe, so long as its quantum correlations are exploited as a resource for metrology~\cite{Iskhakov2016}.
In our proposed scheme, this can be accomplished with surprisingly minor modifications of the classical EOS setup.
To this end, quantum-enhanced EOS starts with the generation of spatially distinct single-mode photon-number entangled beams [green and orange lines in Fig.~\ref{fig:Setup}(b)] through non-linear mixing of a strong classical pump (blue) with the vacuum in a spontaneous parametric down-conversion crystal (SPDCX).
One train of the entangled pulses (green) produces photo-counts $S_i$ in a square-law detector and the second (orange) is used as a probe in the EOS setup, identical to that of Fig.~\ref{fig:Setup}(a).
For a given temporal delay $\tau$, the statistical readout of the BPD output $\Delta n_i$ can now be conditioned on any distribution of the $S_i$ values, yielding various measurement series $\{\Delta n_i \lvert S_i\}_{\tau}$. 

\begin{figure}
    \centering
    \includegraphics[width=0.96\linewidth, trim=20 15 200 25, clip]{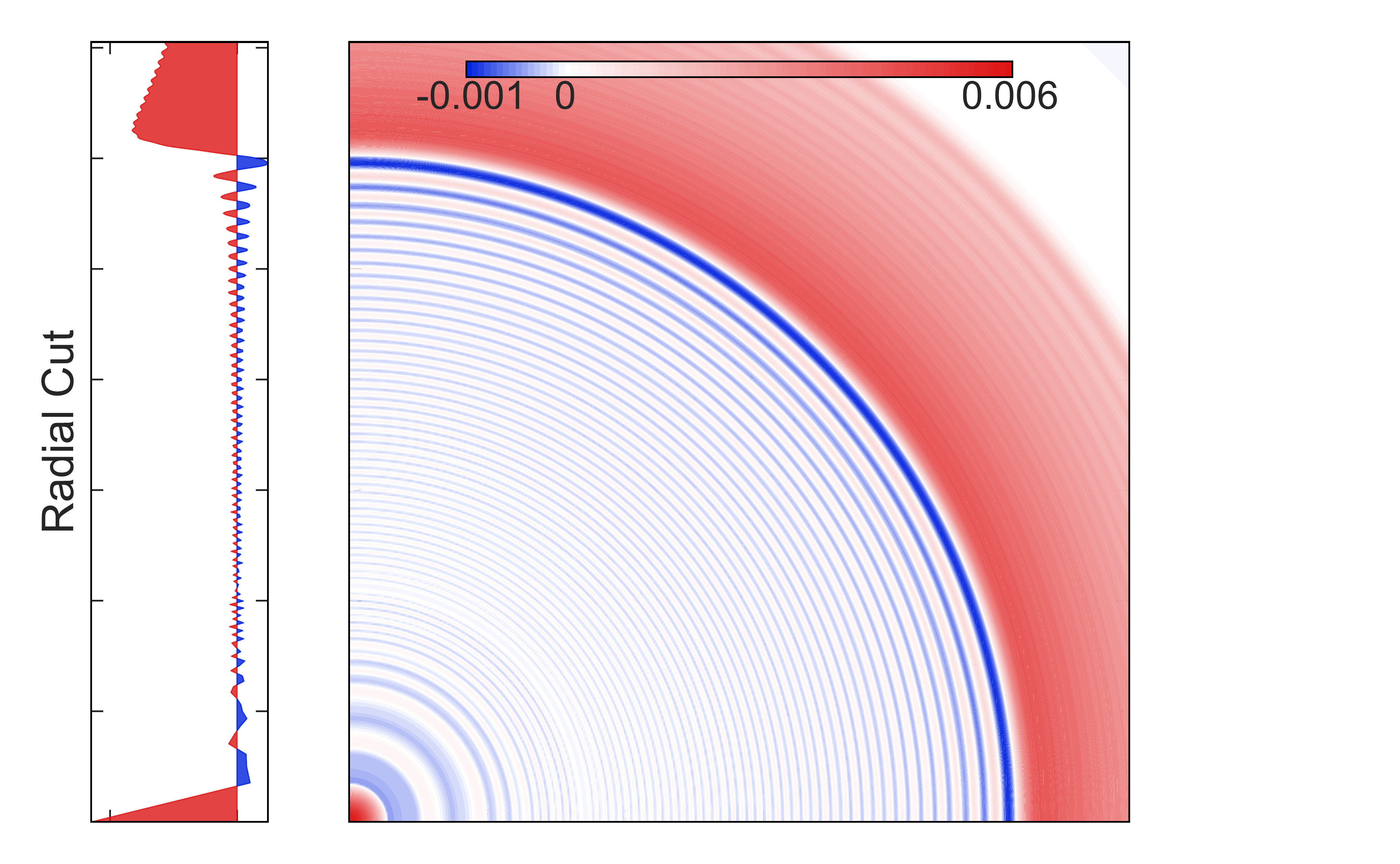}
    \vspace{-10pt}
    \caption{First quadrant of the rotationally symmetric Wigner quasi distribution of the upper BCS of Fig.~\ref{fig:Setup}(d,e,f). On the left, a radial cut emphasizes the relative size between the positive and negative parts of the distribution. The latter are a direct consequence of the non-classical nature of this BCS.}
    \label{fig:Wigner}
\end{figure}

Unconditioned detection of photocounts in either single-mode branch would yield thermal distributions~\cite{Scully1997,Barnett1997,Gardiner2004}, with excess noise compared to coherent probes.
However, conditioning of one of the measurements on the values of the other now exploits the entanglement properties of the twin beams and can be used to carve desired features in the distribution of photons in the probe.
To exemplify this property, Fig.~\ref{fig:Wigner} depicts the calculated Wigner quasi-distribution of a probe, obtained with the band-conditioned rule $S_i>S_\textrm{threshold}$ [see Fig.~\ref{fig:Setup}(d)].
The resulting annular shape of the Wigner representation resembles the conditioning protocol, while also sporting distinct regions of negative values, -- a hallmark sign of the non-classicality of the probe's state~\cite{Scully1997,Bachor2019}.

As displayed in their Wigner representation, BCS probes are quantum in nature and offer distinct advantages that can be used in field-based metrology.
To bring this out, we compare side by side the $D$ measures of a coherent and a BCS probe when sampling two distinct cases of incoming quantum fields: a vacuum and a cat state with a size and coupling to the probe adjusted so that its measured variance is identical to that of the vacuum.
In the case of the input vacuum of Ref.~\cite{Riek2015}, the BCS probes (orange line) offer a six-fold increase in amplitude of $D(\textrm{vac.})$ [Figure~\ref{fig:Setup}(e)].
Even more visible is the advantage of BCS when applied to the analysis of a few-photon cat state~\cite{Scully1997,Barnett1997,Bachor2019}, shown in Fig.~\ref{fig:Setup}(f).
While the coherent probe (blue line) barely registers any qualitative change between $D(\textrm{vac.})$ and $D(cat)$, the BCS probe shows a remarkable difference between the two quantum signals. 
To better understand why it is so, we rewrite Eq.~\ref{eq:P} as
\begin{equation}
    \label{eq:Pbis}
    P(\Delta n;\psi)=\sum_{k=0}^{+\infty} \chi_k(\Delta n)\langle\psi\rvert\hat{\mathcal{E}}^k\lvert\psi\rangle,
\end{equation}
where $\chi_k(\Delta n)$ represents the susceptibility of the EOS probe to the $k^{\textrm{th}}$ moment of the electric field distribution.
\begin{figure}
    \centering
    \includegraphics[width=0.485\linewidth, trim=5 15 95 25, clip]{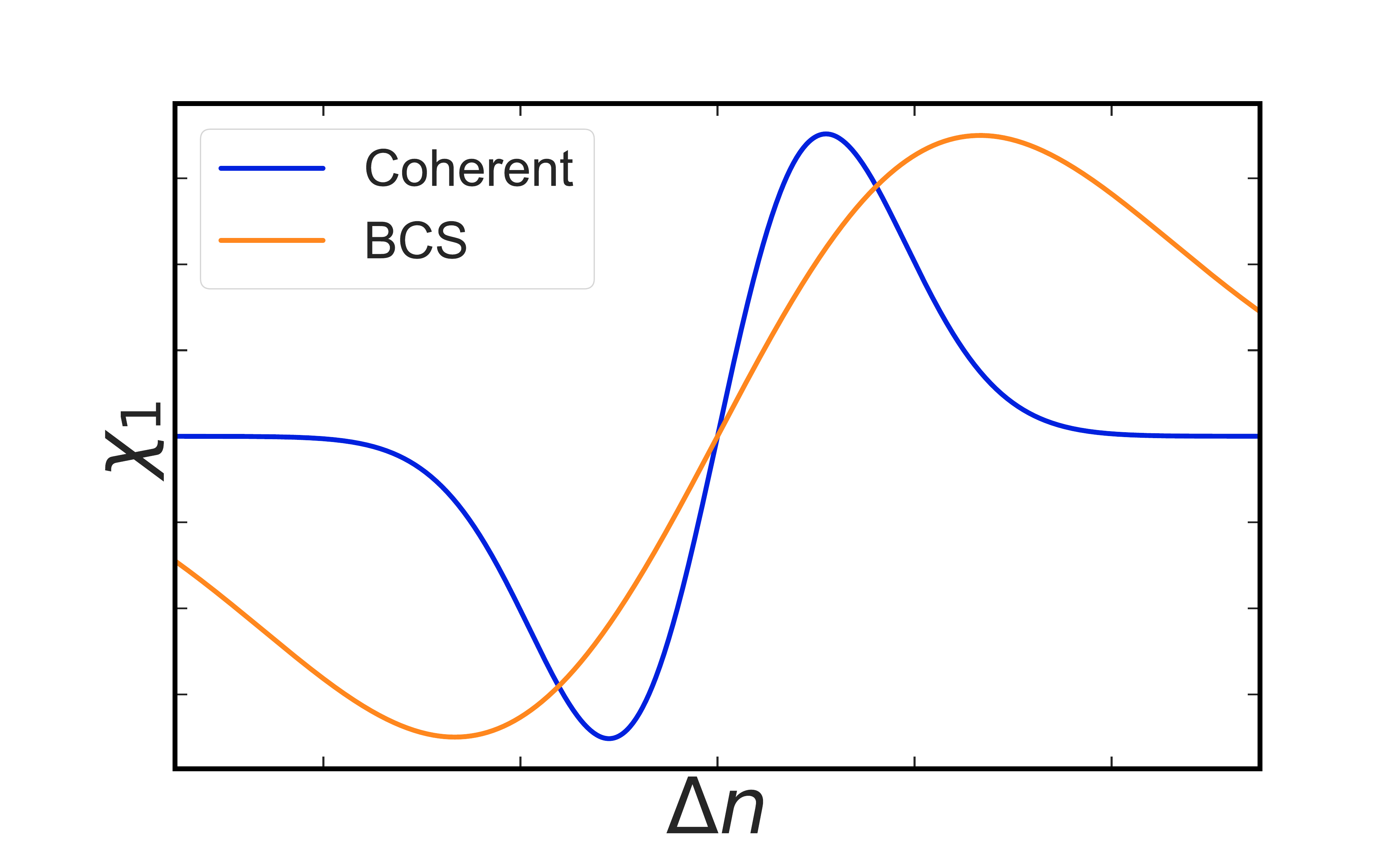}
    \hspace{0.005\linewidth}
    \includegraphics[width=0.485\linewidth, trim=5 15 95 25, clip]{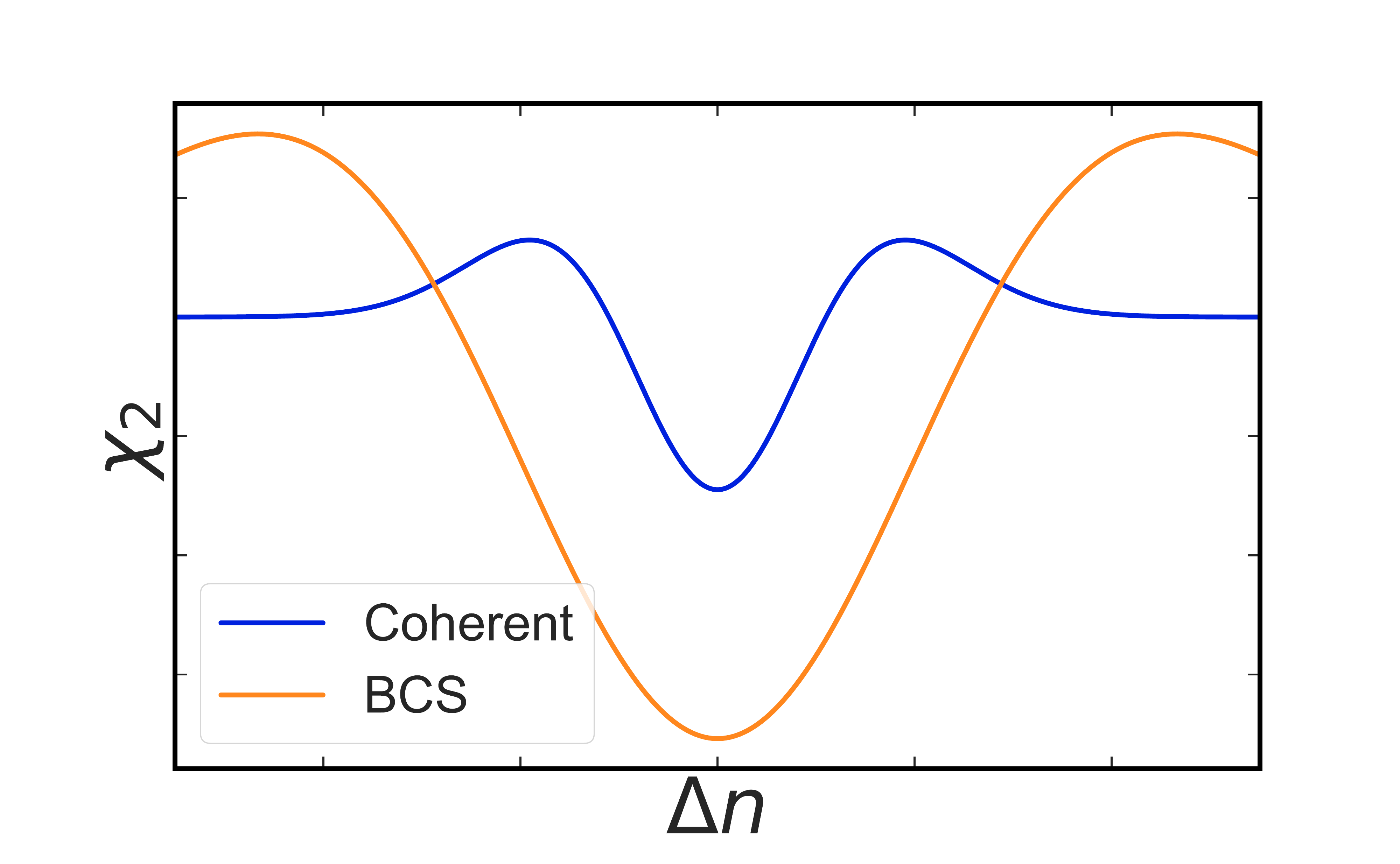}\\
    \includegraphics[width=0.485\linewidth, trim=5 15 95 25, clip]{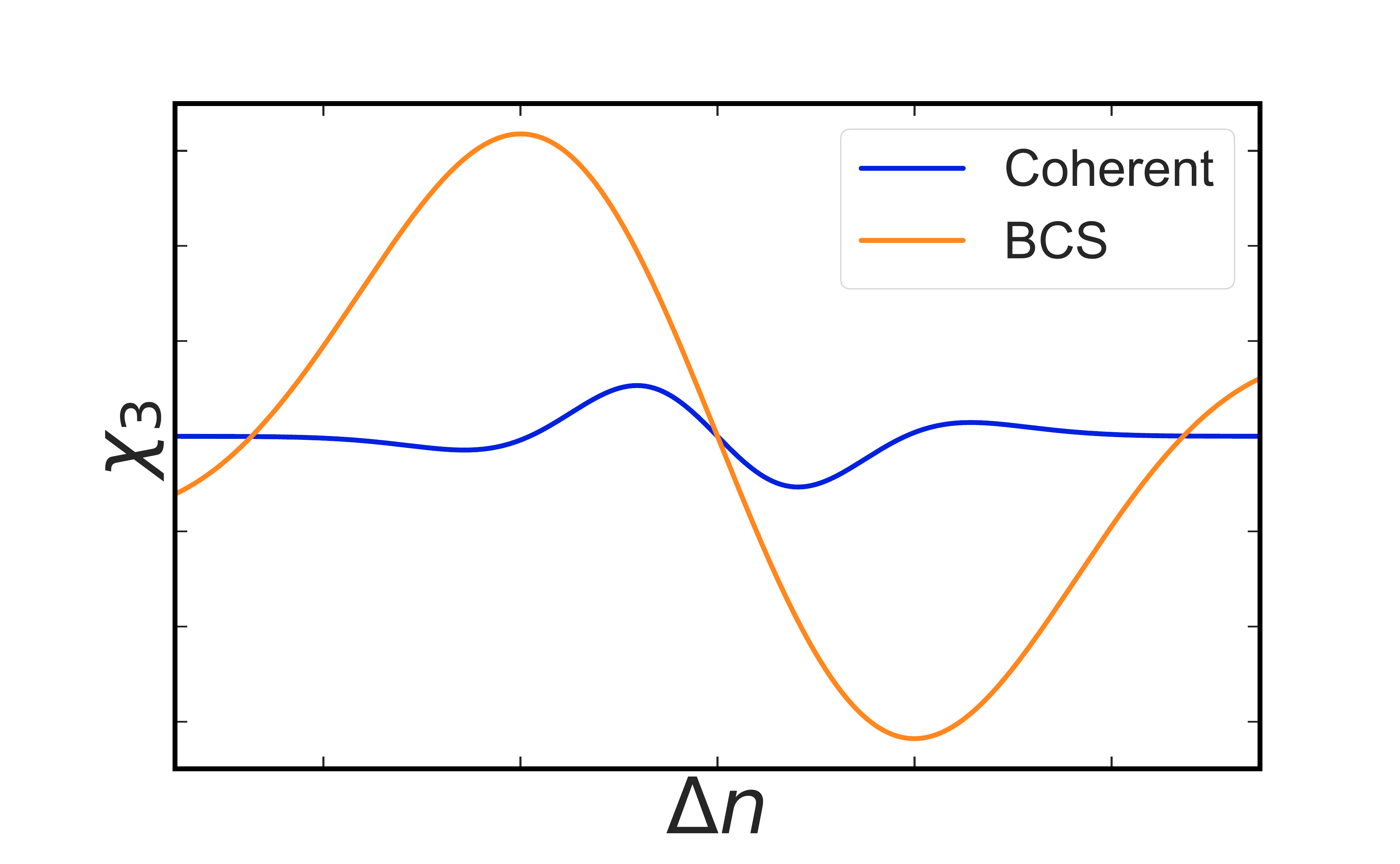}
    \hspace{0.005\linewidth}
    \includegraphics[width=0.485\linewidth, trim=5 15 95 25, clip]{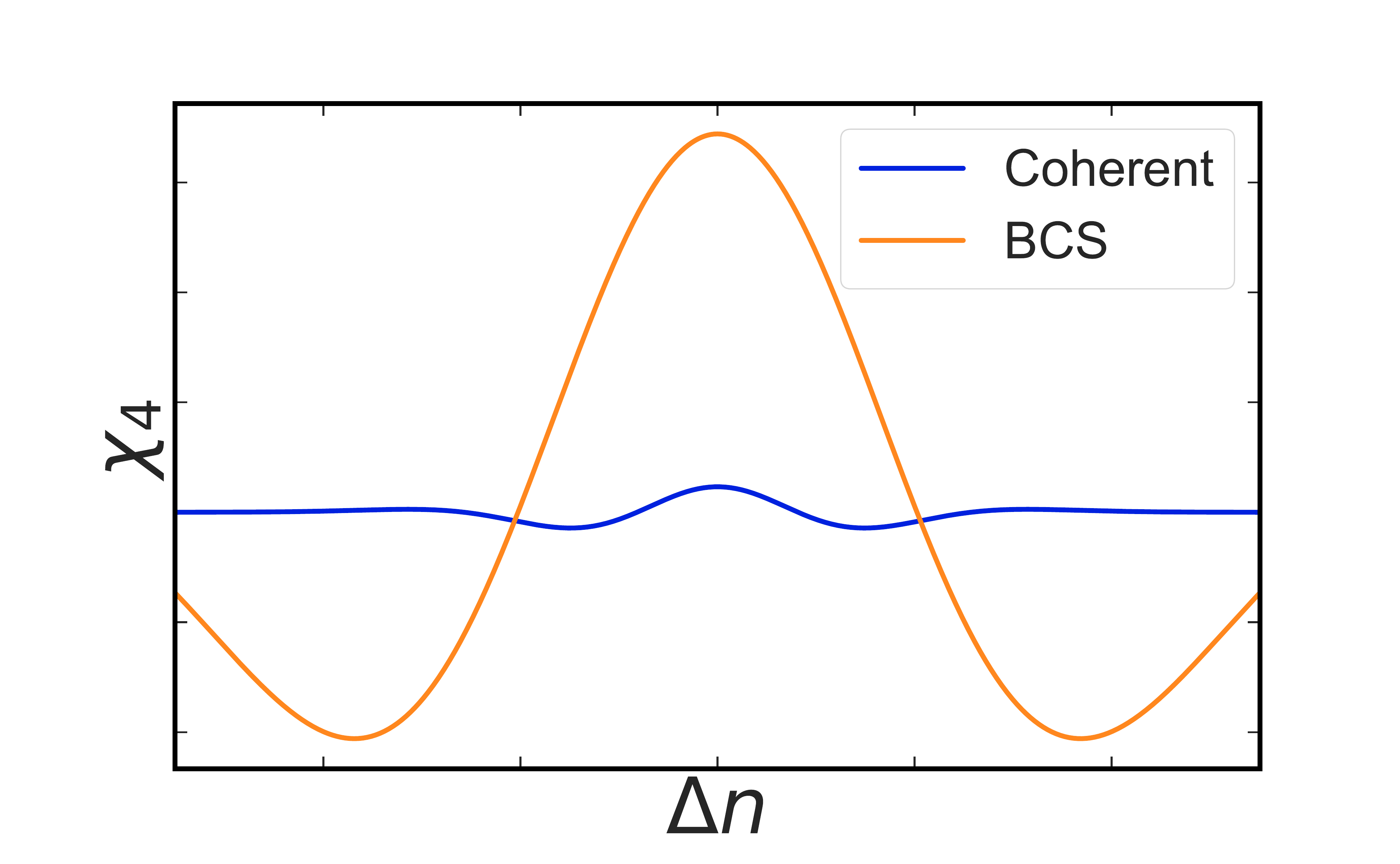}\\
     \includegraphics[width=0.485\linewidth, trim=5 15 95 25, clip]{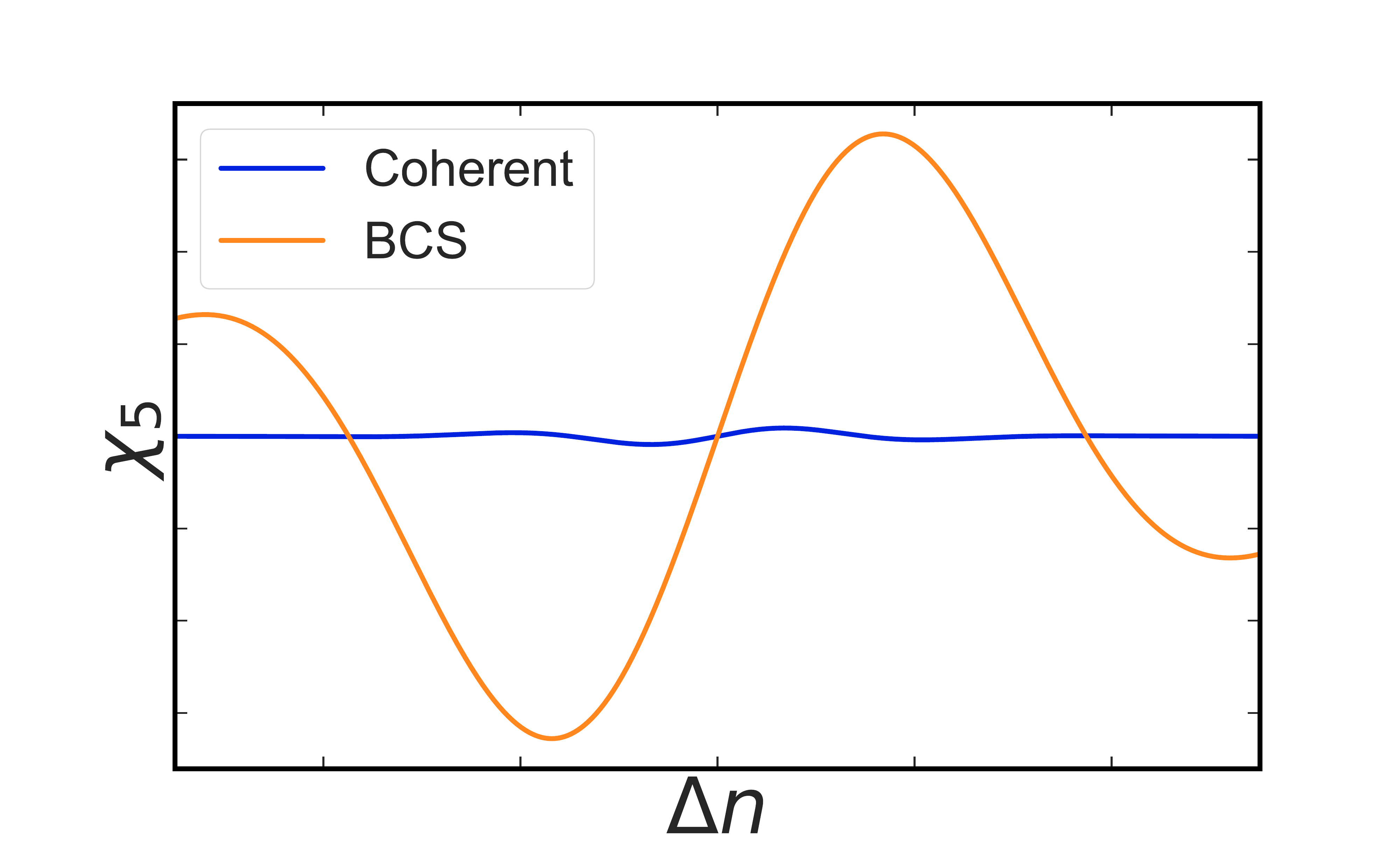}
    \hspace{0.005\linewidth}
    \includegraphics[width=0.485\linewidth, trim=5 15 95 25, clip]{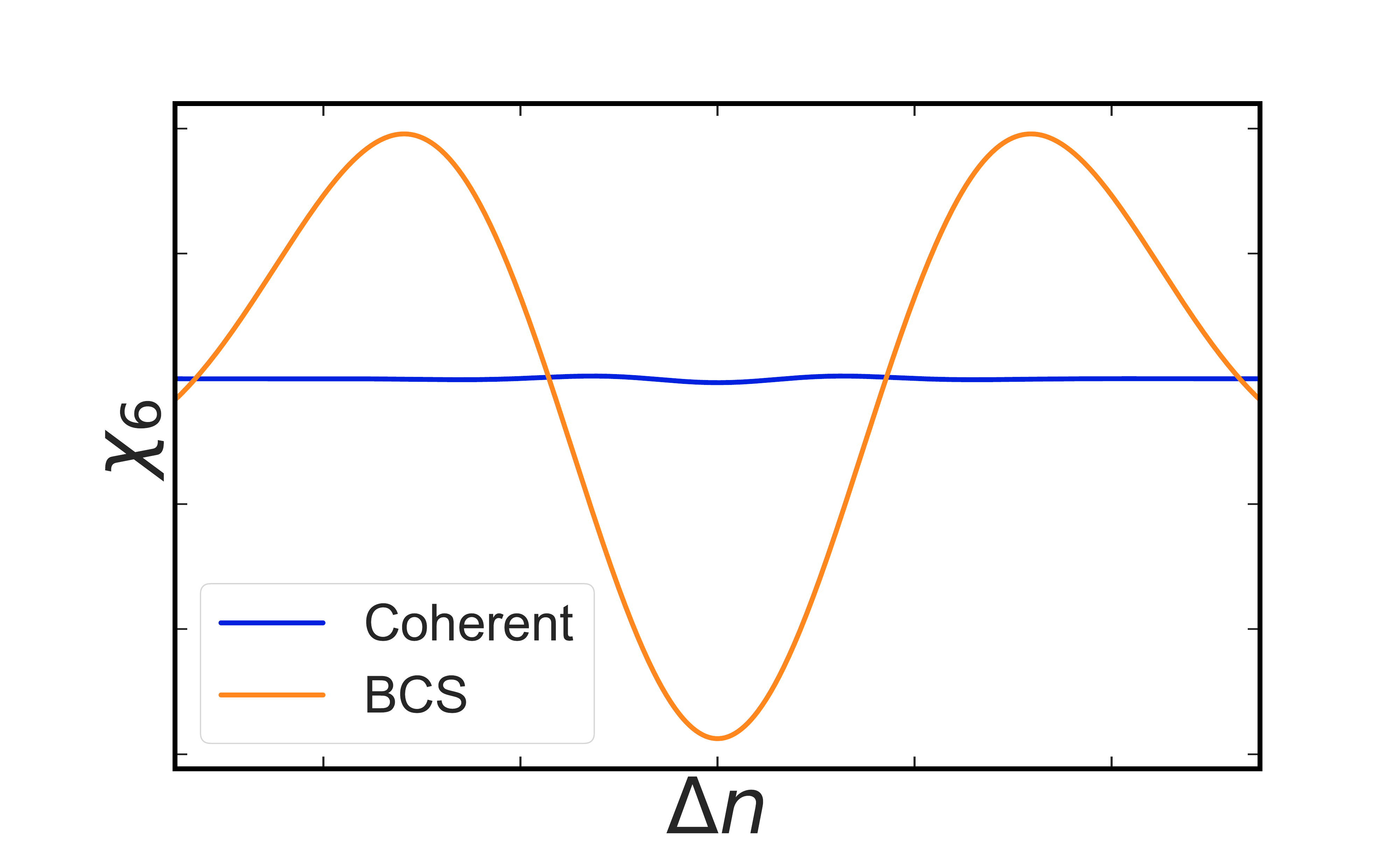}
    \vspace{-10pt}
    \caption{Susceptibilities to the first six moments of the distribution of the electric field, for a coherent probe (blue lines) and a BCS probe (orange lines). Although the higher moment susceptibilities quickly vanish for the coherent probe, they remain large for the BCS probe, which explains the very different outcomes of Fig.~\ref{fig:Setup}(e,f).}
    \label{fig:Chis}
\end{figure}
Fig.~\ref{fig:Chis} compares the susceptibilities of the signal to the first six moments of the electric field distribution for the coherent probe and the BCS probe of Fig.~\ref{fig:Setup}(d,e,f) and Fig.~\ref{fig:Wigner}.
Strikingly, the susceptibilities of the coherent probe vanish much quicker with increase in their order, relative to those of the BCS probe.
This explains the very different results for the two cases analyzed in Fig.~\ref{fig:Setup}(e,f): a pronounced distinction in the fourth-order moment between the few-photon cat state and the vacuum is clearly discerned by the BCS probe, while remaining almost imperceptible for the coherent case.
In fact, the improvement on the measurement of the second moment of the distribution is compounded onto each successive higher moment.
This sustained susceptibility to higher-order moments reveals the exciting potential of BCS probes toward robust sampling of strongly non-classical signals and high-precision quantum state tomography.   
An additional hallmark of BCS is its versatility. Conditioning protocols can be engineered to generate very distinct spectra of $\chi_k$ functions, as studied in the supplementary material for a variety of cases ~\cite{Note1}.
In particular, it is shown that BCS conditioned on multiple bands feature noticeable rapidly-varying features in their susceptibilities, further aiding in the robust discrimination and full tomography of non-trivial input quantum states. 

Finally, as expected for heralded photon sources~\cite{Virally2010}, their metrological advantage is sensitive to the imperfections of the conditioning branch.
To analyze this, we considered photon losses and the inability of square-law detectors to act as perfect photon-number resolving devices.
Our results~\cite{Note1} show that the advantage of BCS in EOS vanishes slowly and really comes into question only when losses reach the 50\% level.
Thankfully, high quantum efficiency detectors and low-loss optics are readily available across the visible and the near-infrared bands, posing no technological obstacle to the realization of quantum-enhanced EOS with BCS probes.

In conclusion, we introduced a new class of non-classical states of light, BCS, based on the band-conditioned detection of bright entangled twin beams derived from spontaneous parametric down conversion.
Through a detailed comparison with classical probing, we demonstrated the remarkable promise of these states toward subcycle metrology of quantum fields.
Two major improvements on the spectroscopy of quantum fields have been detailed: a six-fold improvement over current experiments, that adds up geometrically for each order of the statistical moments of the field distribution; and versatility in the post-selection scheme that greatly facilitates the process of recovering each successive moment.
It thus shatters some of the current limitations of the technique and opens the way to a full tomography of the quantum distribution of THz and mid-IR electric fields.
Remarkably, the conditioning protocols can be applied to the raw data sets in post-processing; it is thus expected that the full power of machine learning algorithms can be applied toward a variety of optimization tasks for robust extraction of quantum information.
In combination with quantum enhancement, the ability to probe the dynamics of quantum fields intrinsically in space-time promises unprecedented experimental access to relativistic quantum electrodynamics and time-domain quantum spectroscopy of matter.  

\vspace{.5\baselineskip}
\emph{Acknowledgements. }
We acknowledge useful discussions with A. Moskalenko.
This work was supported by Natural Sciences and Engineering Research Council of Canada (NSERC), via the Canada Research Chair program (CRC); Fonds de Recherche du Qu{\'e}bec – Nature et Technologies (FRQNT), via Institut Transdisciplinaire d'Information Quantique (INTRIQ).

\begin{appendix}
 
\section{Nonlinear beamsplitter}
\label{appendix}
\begin{figure}[b] 
    \centering
    \includegraphics[width=0.95\linewidth, trim=15 50 15 15, clip]{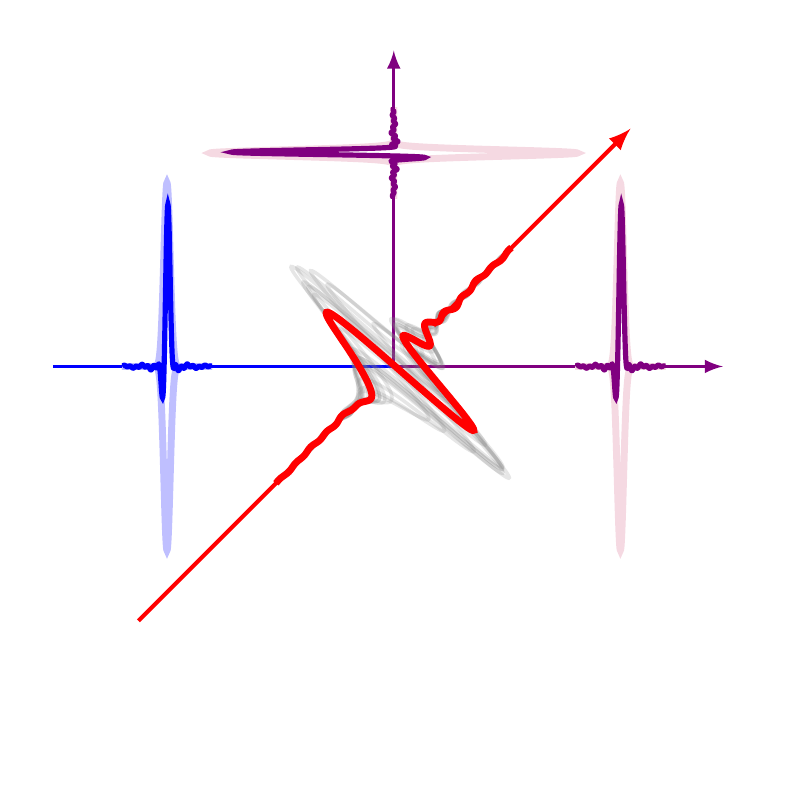}
    \caption{Nonlinear beamsplitter: an incoming probe pulse (in blue) is effectively split into two branches (in purple), by the nonlinear interaction with the terahertz (THz) or mid-infrared (mid-IR) signal (in red). The photocount difference $\Delta n$ between the output branches is proportional to the amplitude of the signal, as captured by a pair of balanced photodetectors (see also Fig. 1 of the main text).}
    \label{NLBS}
\end{figure}

As described in the main text, we model electro-optic sampling (EOS) as a ``nonlinear beamsplitter'', also sketched in Fig.~\ref{NLBS}.
The effect of the THz field inside the nonlinear crystal on the EOS probe is to generate photons in the orthogonal polarization. After the crystal, the amplitude in both branches is respectively $E_p\,\sqrt{1-\varepsilon^2}$ and $i\,E_p\,\varepsilon$, where $E_p$ is the original probe amplitude and $\varepsilon$ is proportional to the value of the THz electric field $\mathcal{E}$,
\begin{equation} 
    \varepsilon=\gamma\,\mathcal{E},
\end{equation}
where $\gamma$ is a nonlinear coupling coefficient. Following the quarter-plate and polarizer, the amplitudes in the two detection branches are respectively $E_p\,\frac{\sqrt{1-\varepsilon^2}+\varepsilon}{\sqrt{2}}$ and $-i\,E_p\,\frac{\sqrt{1-\varepsilon^2}-\varepsilon}{\sqrt{2}}$, that are balanced for $\mathcal{E}\rightarrow 0$. It is thus possible to consider the effect of EOS on the probe exactly like that of a beamsplitter, with transmission and reflection coefficients dependent on the signal $\varepsilon$. 

\section{Counting statistics}
\emph{Conditional and full probabilities. }
We start with the conditional probability $P$ of observing the result $\Delta n$, obtained when we subtract the photocounts on the balanced detectors considering a given number $n$ photons is incident on the beam splitter in each successive probe pulse. This results in 
\footnote[1]{For $\Delta n\ll n$, which we assume going forward, we use~\cite[p.~66]{Spencer2014}
\begin{equation}
    \label{epsilon1}
    \binom{n}{\frac{n+\Delta n}{2}}\simeq2^n\,\sqrt{\frac{2}{n\pi}}\;e^{-\frac{\Delta n^2}{2n}}.
\end{equation}}
\begin{multline}
    \label{binom}
    P(\Delta n;n,\varepsilon)=\binom{n}{\frac{n+\Delta n}{2}}\\\left(\frac{\sqrt{1-\varepsilon^2}+\varepsilon}{\sqrt{2}}\right)^{n+\Delta n}\left(\frac{\sqrt{1-\varepsilon^2}-\varepsilon}{\sqrt{2}}\right)^{n-\Delta n}.
\end{multline}

For small signal, we Taylor-expand this expression in $\varepsilon$,  
\begin{equation}
    \label{pmoments}
    P(\Delta n;n,\varepsilon)=\sum_{k=0}^{+\infty}\alpha_k(n,\Delta n)\,\varepsilon^k.
\end{equation}

For an incoming THz quantum state $\lvert\psi\rangle$ (which we assume is the case going forward) all powers $\mathcal{E}^k$ of the electric field are replaced by the moments $\langle\psi\rvert\hat{\mathcal{E}}^k\lvert\psi\rangle$ of the electric field operator $\hat{\mathcal{E}}$, and we have
\begin{equation}
    \label{EOSDeltan}
    P(\Delta n;n,\psi)=\sum_{k=0}^{+\infty}\alpha_k(n,\Delta n)\,\gamma^k\,\langle\psi\rvert\hat{\mathcal{E}}^k\lvert\psi\rangle.
\end{equation}
The full probability is ~\footnote[2]{As the average number of photons is large, all discrete sums are well approximated by their first-order Euler-Maclaurin expansions
\begin{equation}
    \sum_{n=a}^{b}f(n)\simeq\frac{f(a)+f(b)}{2}+\int_a^b f(x)\,dx,
\end{equation}
which we use in our numerical simulations.}
\begin{equation}
    \label{pdeltaepsilon}
    P(\Delta n;\psi)=\sum_{n=0}^{+\infty}P(\Delta n;n,\psi)\;P(n),
\end{equation}
where $P(n)$ is the probability of having $n$ photons in the probe.

\begin{figure}
    \centering
    \includegraphics[width=0.48\linewidth, trim=5 15 55 65, clip]{Coherent_SS.pdf}
    \put(-22,12){(a)}
    \hspace{0.025\linewidth}
    \includegraphics[width=0.48\linewidth, trim=5 15 55 65, clip]{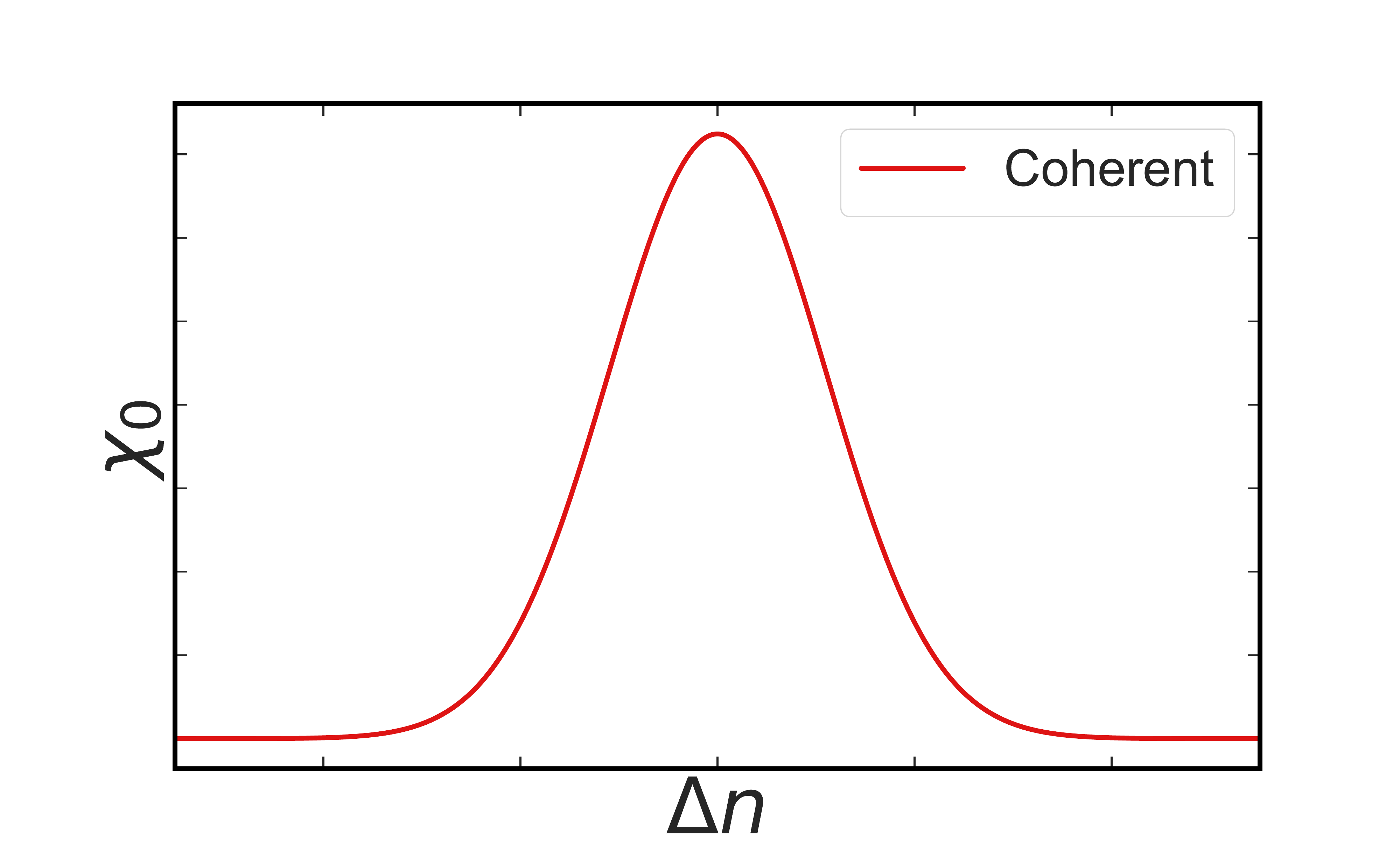}
    \put(-22,12){(b)}\\
    \includegraphics[width=0.48\linewidth, trim=5 15 55 65, clip]{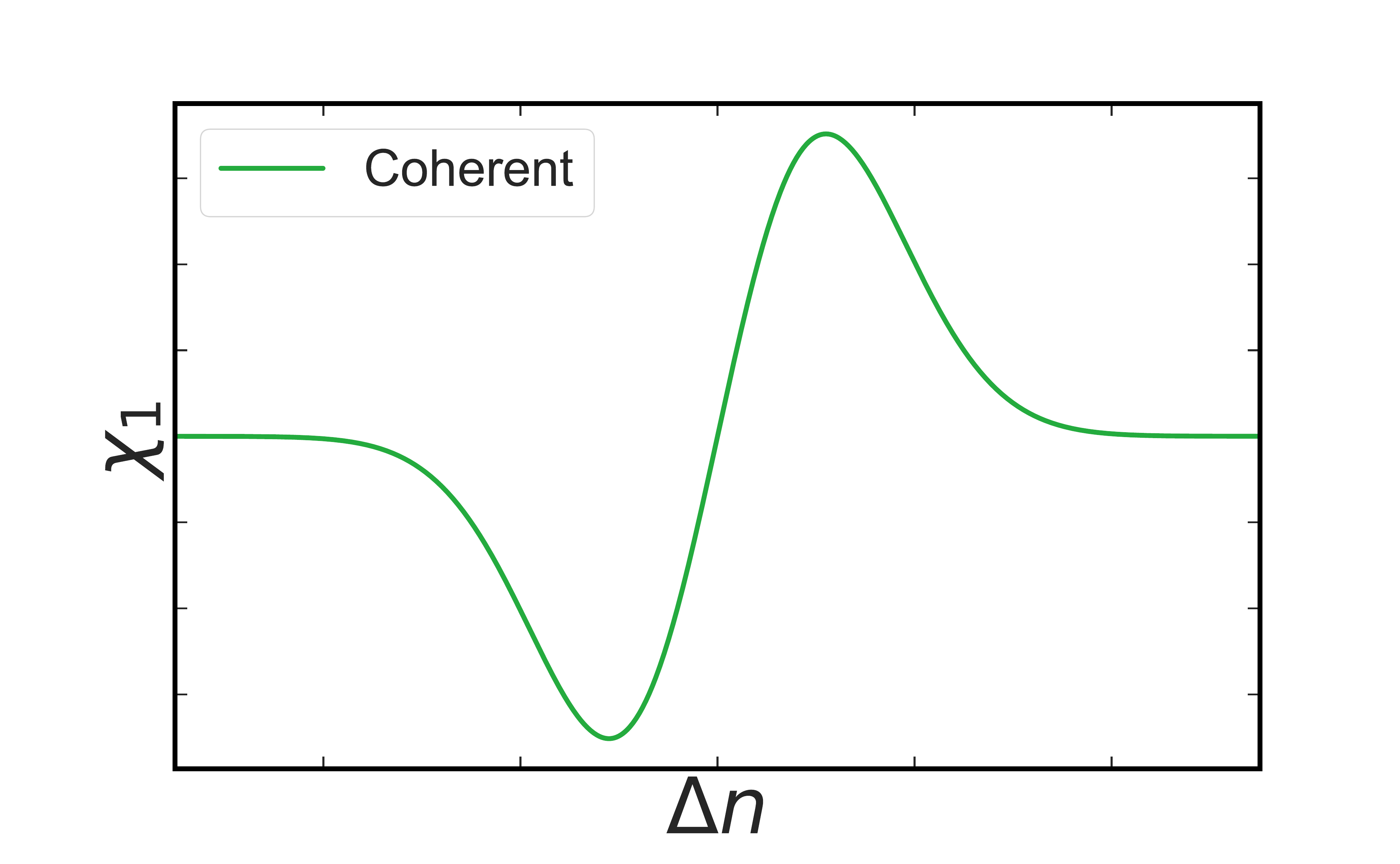}
    \put(-22,12){(c)}
    \hspace{0.025\linewidth}
    \includegraphics[width=0.48\linewidth, trim=5 15 55 65, clip]{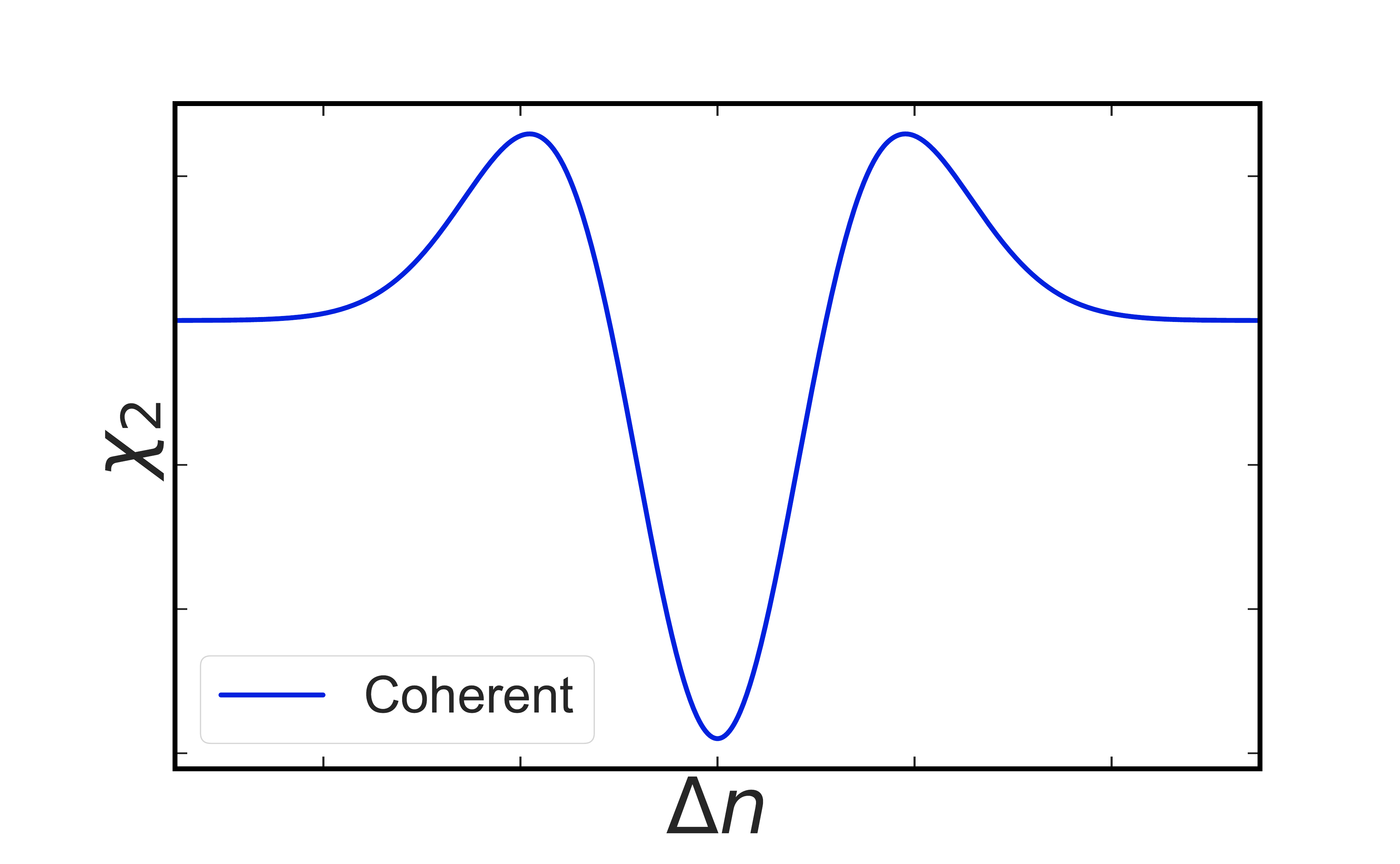}
    \put(-22,12){(d)}\\
    \includegraphics[width=0.48\linewidth, trim=5 15 55 65, clip]{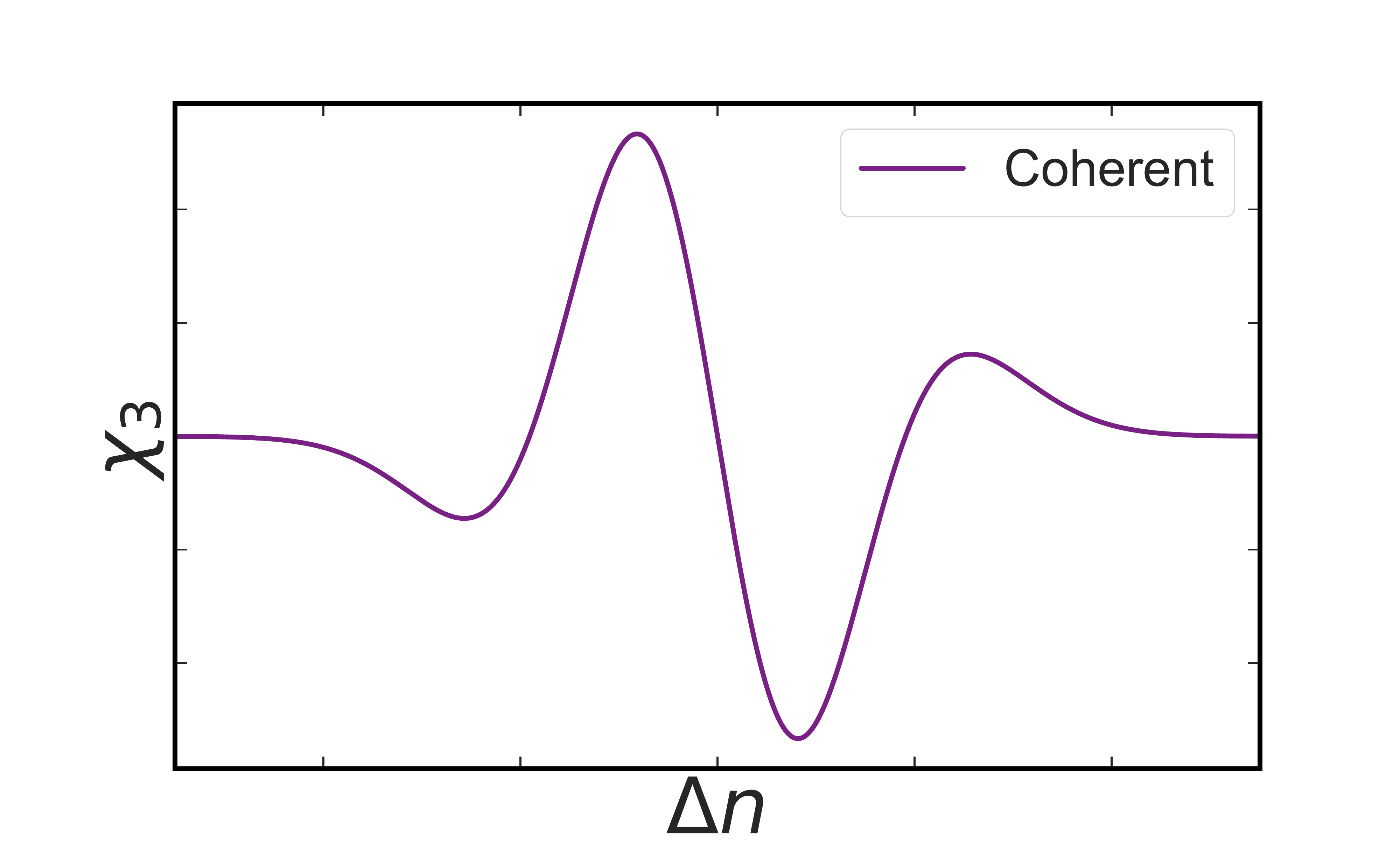}
    \put(-22,12){(e)}
    \hspace{0.025\linewidth}
    \includegraphics[width=0.48\linewidth, trim=5 15 55 65, clip]{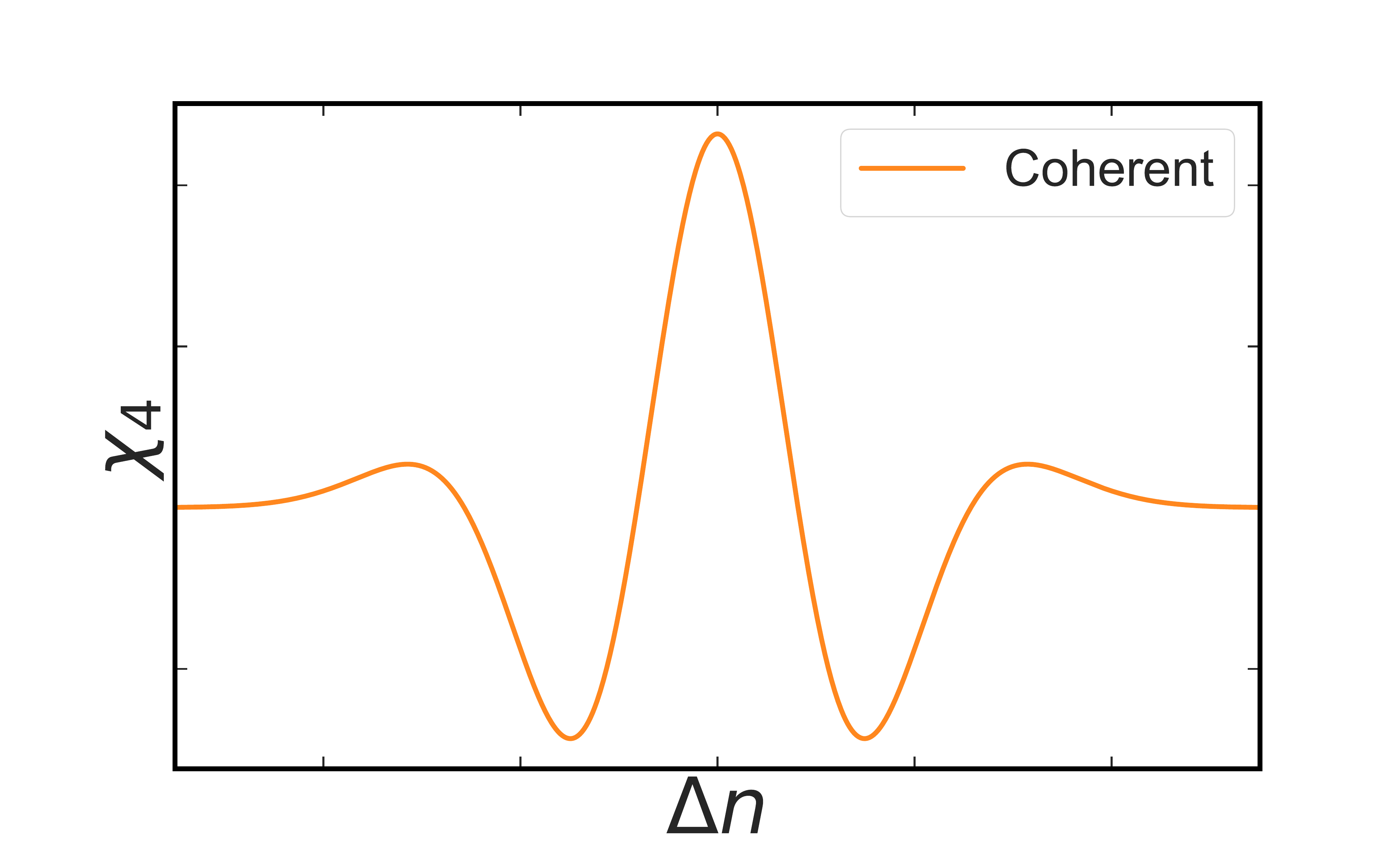}
    \put(-22,12){(f)}\\
    \includegraphics[width=0.48\linewidth, trim=5 15 55 65, clip]{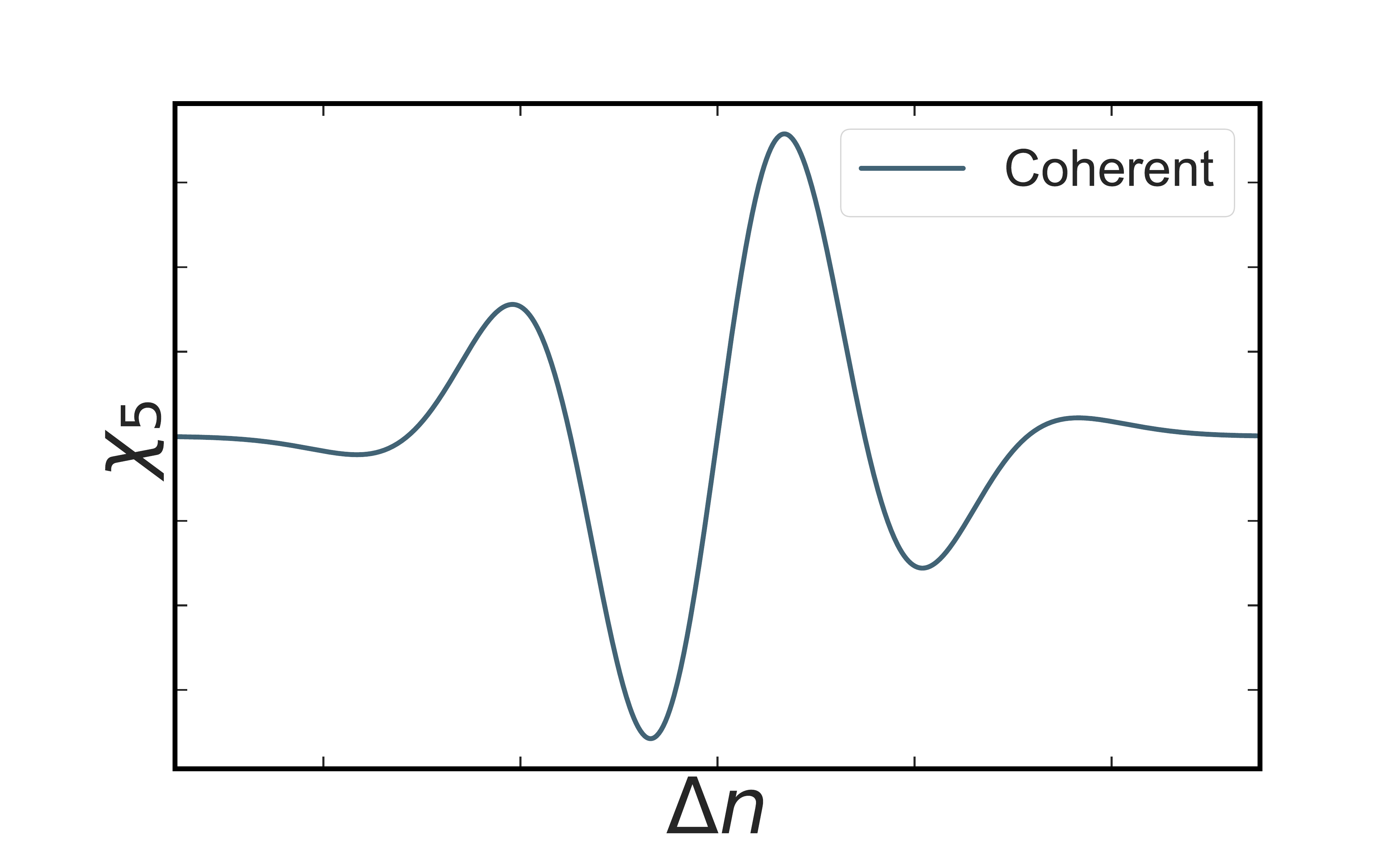}
    \put(-22,12){(g)}
    \hspace{0.025\linewidth}
    \includegraphics[width=0.48\linewidth, trim=5 15 55 65, clip]{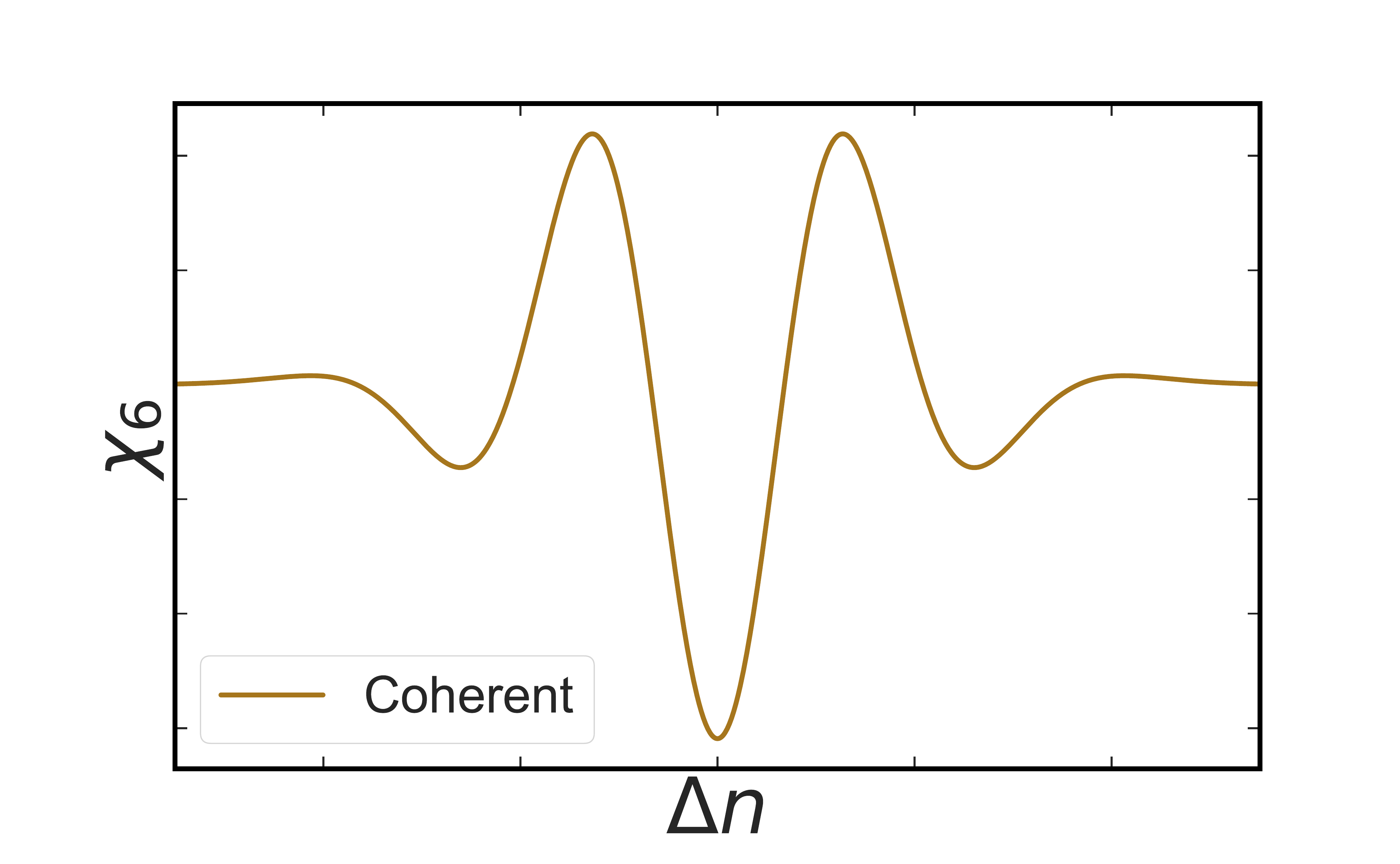}
    \put(-22,12){(h)}\\
    \includegraphics[width=0.48\linewidth, trim=5 15 55 65, clip]{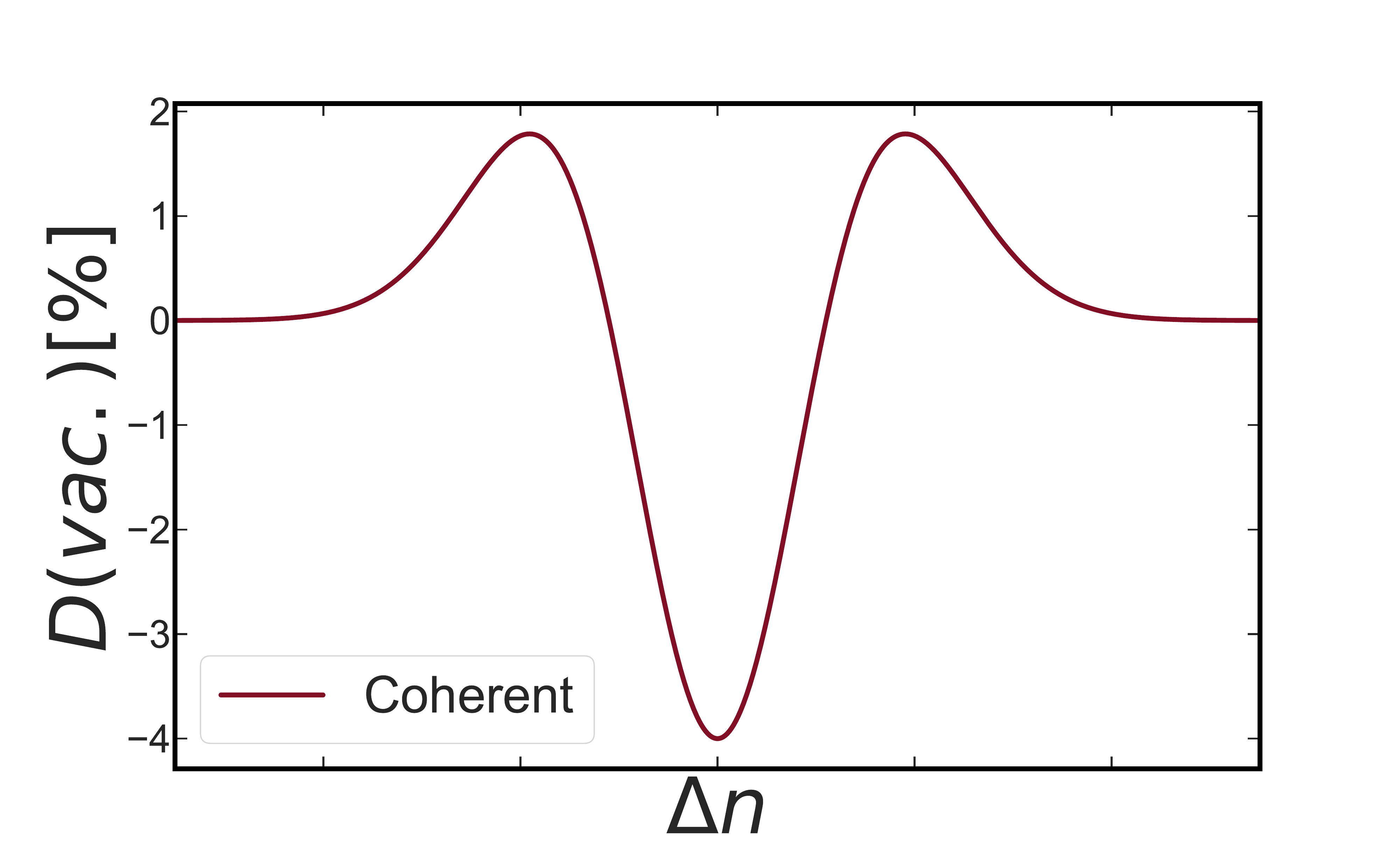}
    \put(-22,12){(i)}
    \hspace{0.025\linewidth}
    \includegraphics[width=0.48\linewidth, trim=5 15 55 65, clip]{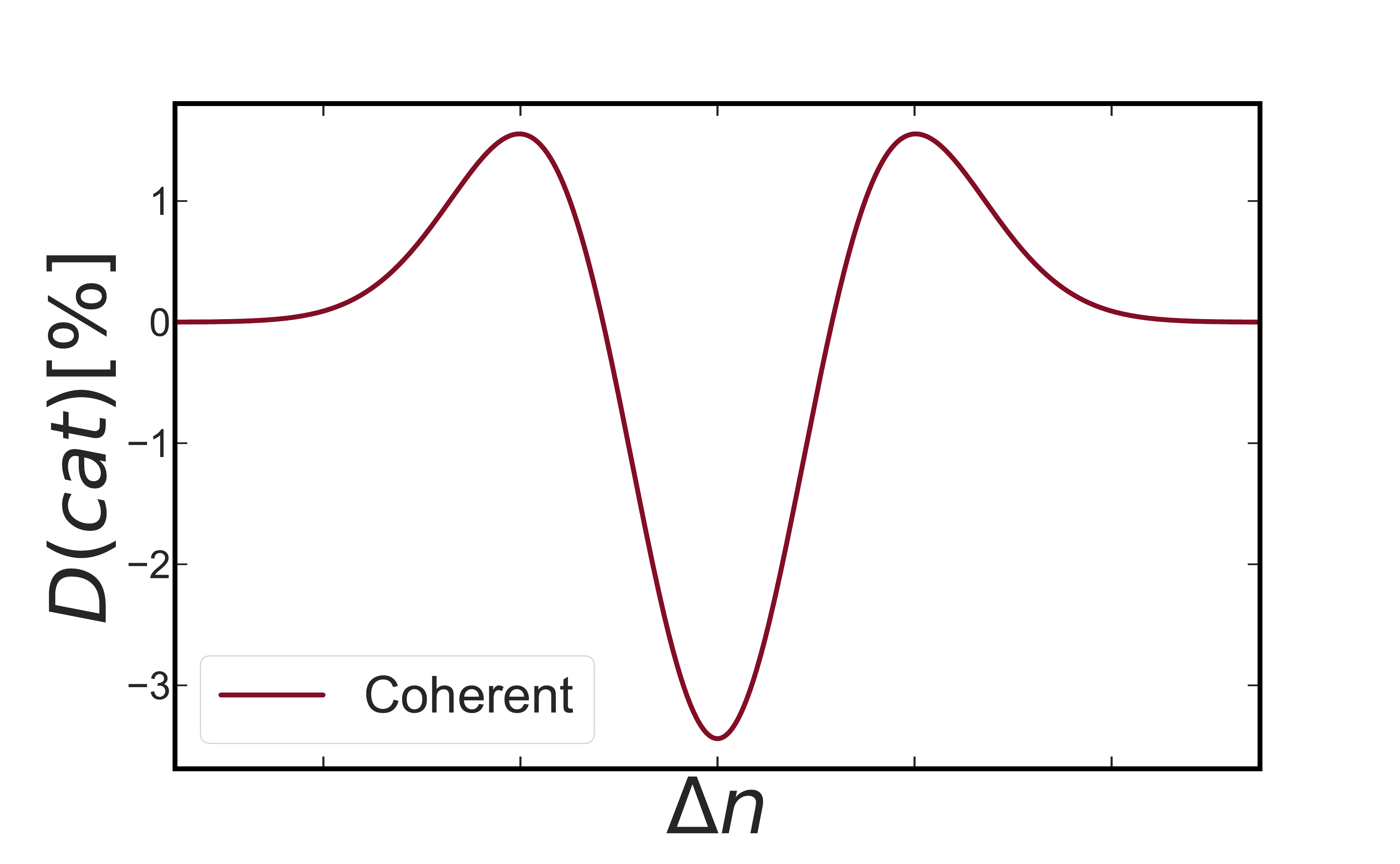}
    \put(-22,12){(j)}\\
    \includegraphics[width=0.99\linewidth, trim=20 10 200 25, clip]{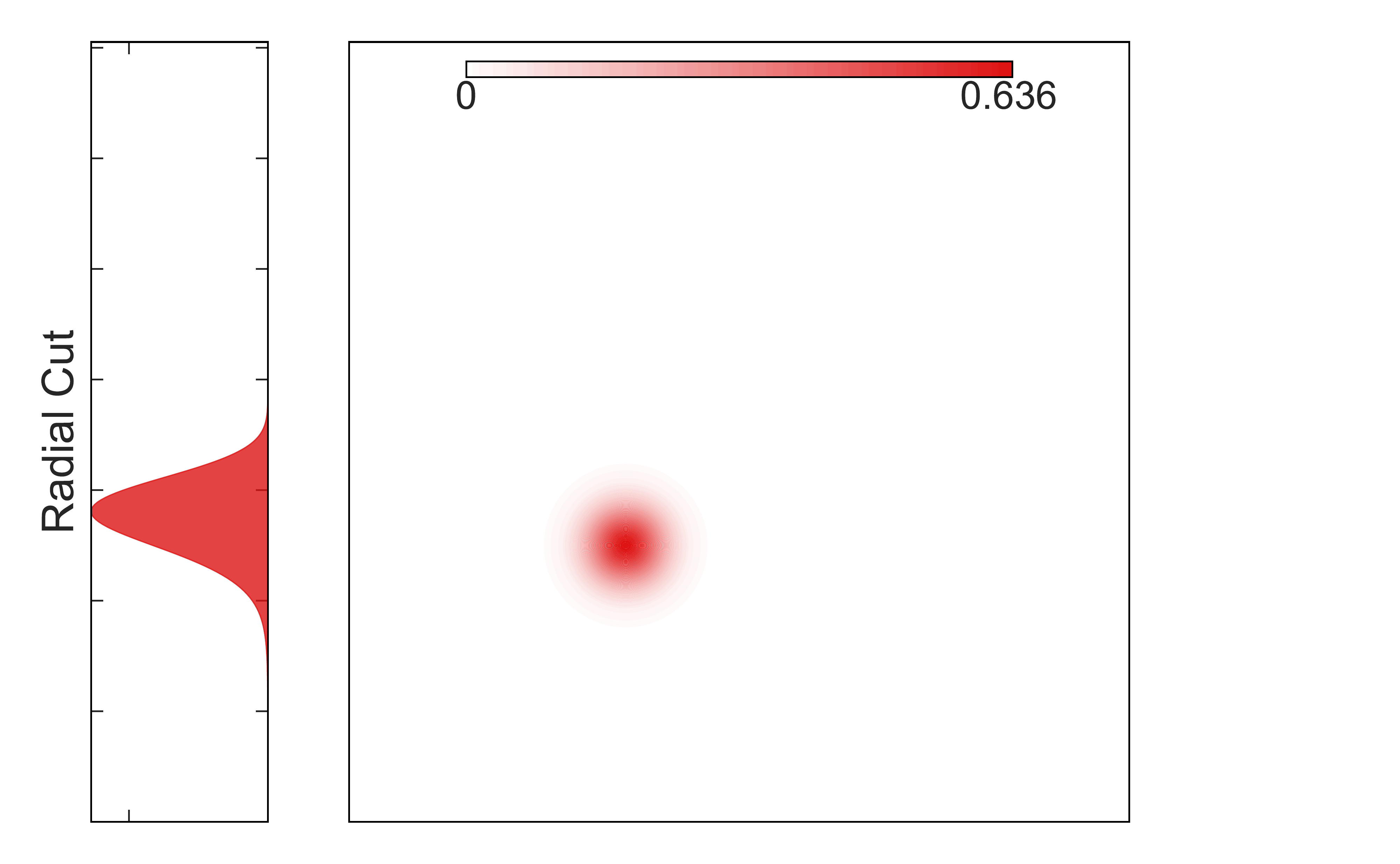}
    \put(-35,12){(k)}
    \vspace{-10pt}
    \caption{Coherent probe: (a) photon number distribution; (b) EOS in absence of THz signal; (c-h) normalized susceptibilities to statistical moments of the THz signal [Eq.~\eqref{susc}]; (i,j) relative differential noise amplitudes~\eqref{signaltonoise} for vacuum and cat signals; (k) first quadrant of the Wigner quasi-distribution.}
    \label{CoherentStats}
\end{figure}

\emph{Susceptibilities to statistical moments. }
We rewrite Eq.~\eqref{pdeltaepsilon} as
\begin{equation}
    P(\Delta n;\psi)=\sum_{k=0}^{+\infty}\chi_k(\Delta n)\,\langle\psi\rvert\hat{\mathcal{E}}^k\lvert\psi\rangle,
\end{equation}
with
\begin{equation}
    \label{susc}
    \chi_k(\Delta n)=\gamma^k\sum_{n=0}^{+\infty}\alpha_k(n,\Delta n)\;P(n)
\end{equation}
being the susceptibility of the EOS measurement to the $k^\textrm{th}$ moment of the signal distribution. $\chi_0(\Delta n)=P(\Delta n;0)$, the EOS distribution in absence of signal. Note that these are solely dependent on the EOS setup (including the statistics of the probe), and not on the signal itself.

\vspace{.5\baselineskip}
\emph{Relative differential noise amplitude. }
In order to compare EOS results with different probes, we define the relative differential noise amplitude
\begin{equation}
    \label{signaltonoise}
    D(\psi)=\frac{P(\Delta n;\psi)-P(\Delta n;0)}{\max[P(\Delta n;0)]}.
\end{equation}
This is a good metric for comparing probes, as it quantifies an effective signal-to-noise ratio.

\begin{figure}
    \centering
    \includegraphics[width=0.48\linewidth, trim=5 15 55 65, clip]{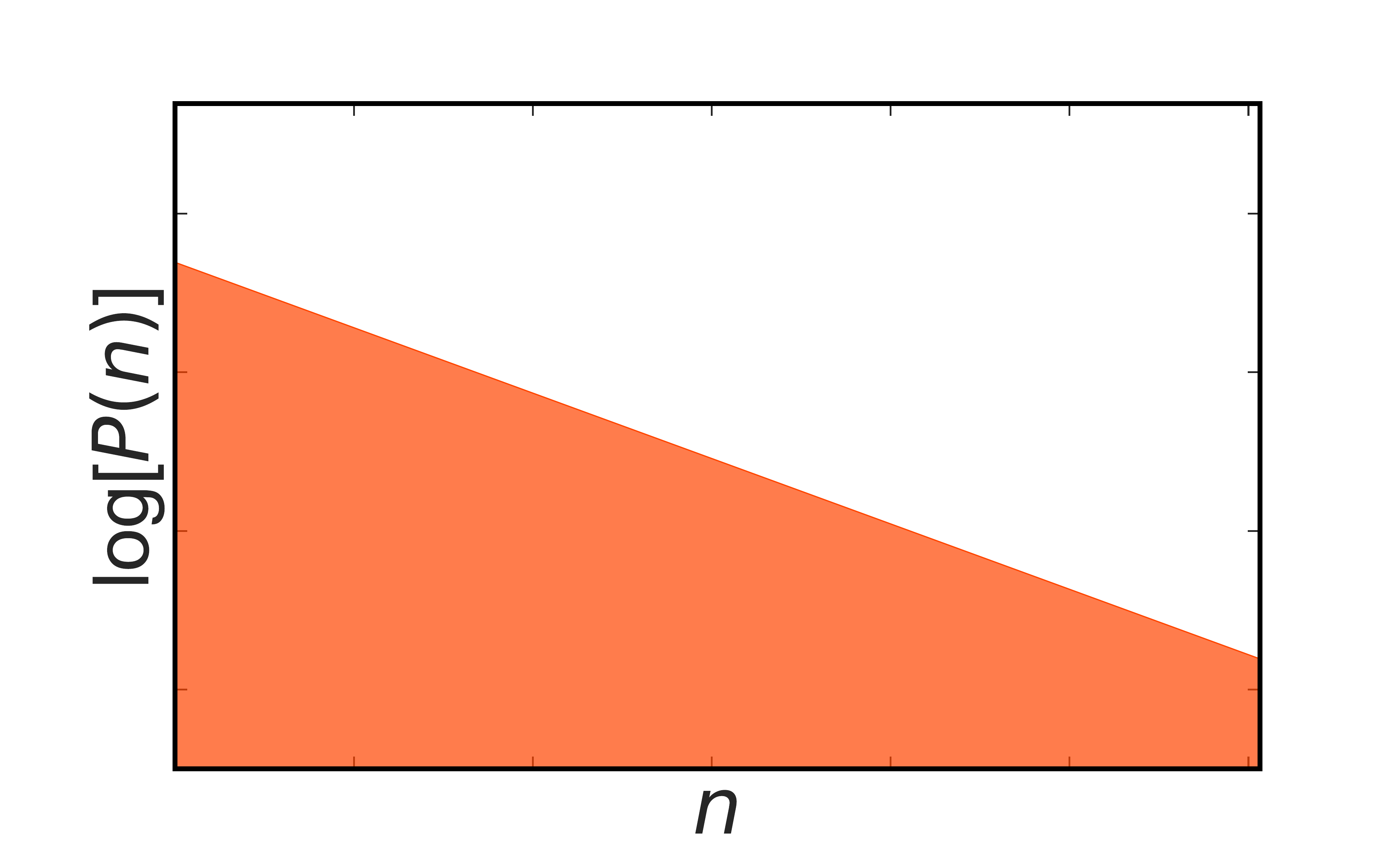}
    \put(-22,12){(a)}
    \hspace{0.025\linewidth}
    \includegraphics[width=0.48\linewidth, trim=5 15 55 65, clip]{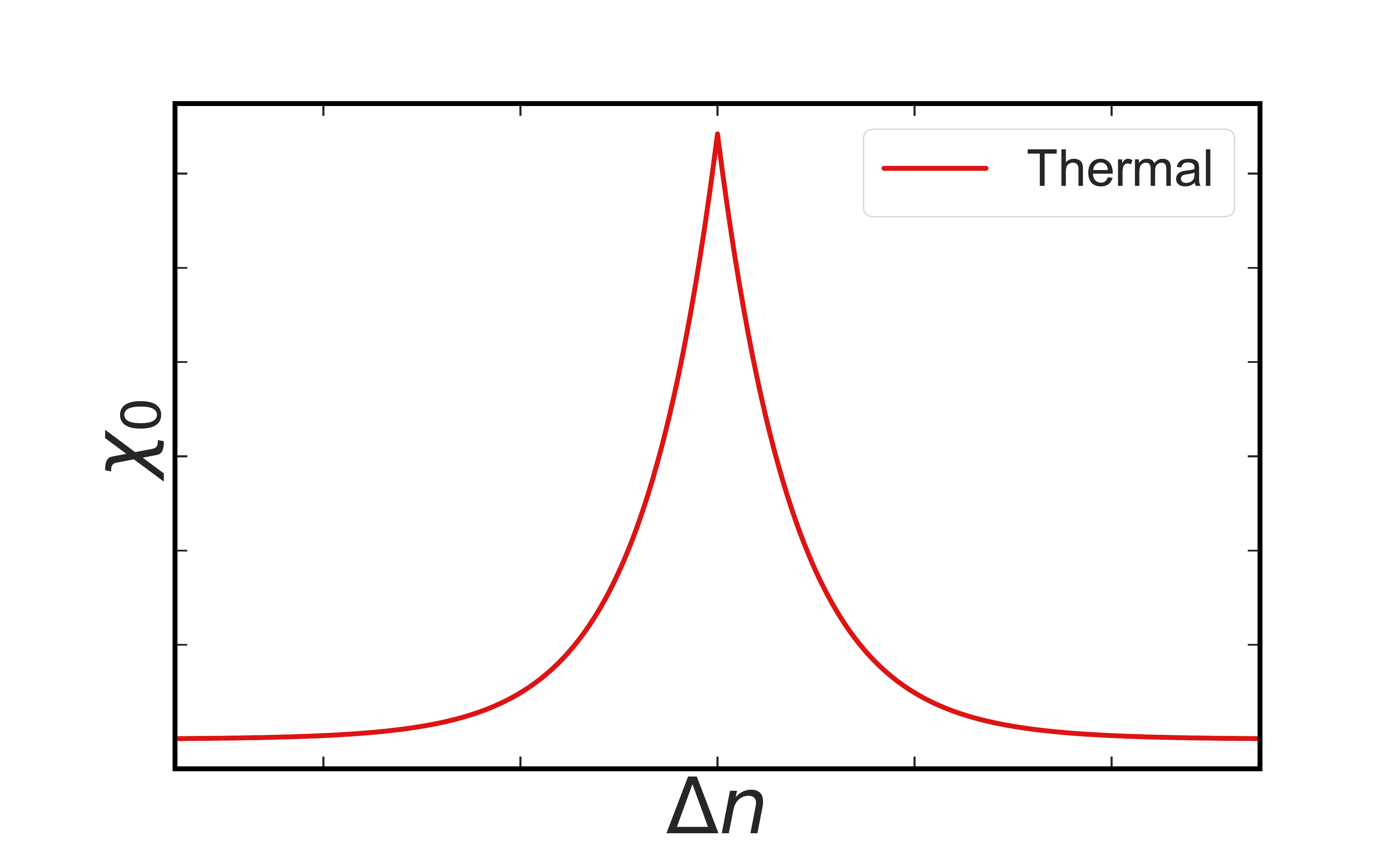}
    \put(-22,12){(b)}\\
    \includegraphics[width=0.48\linewidth, trim=5 15 55 65, clip]{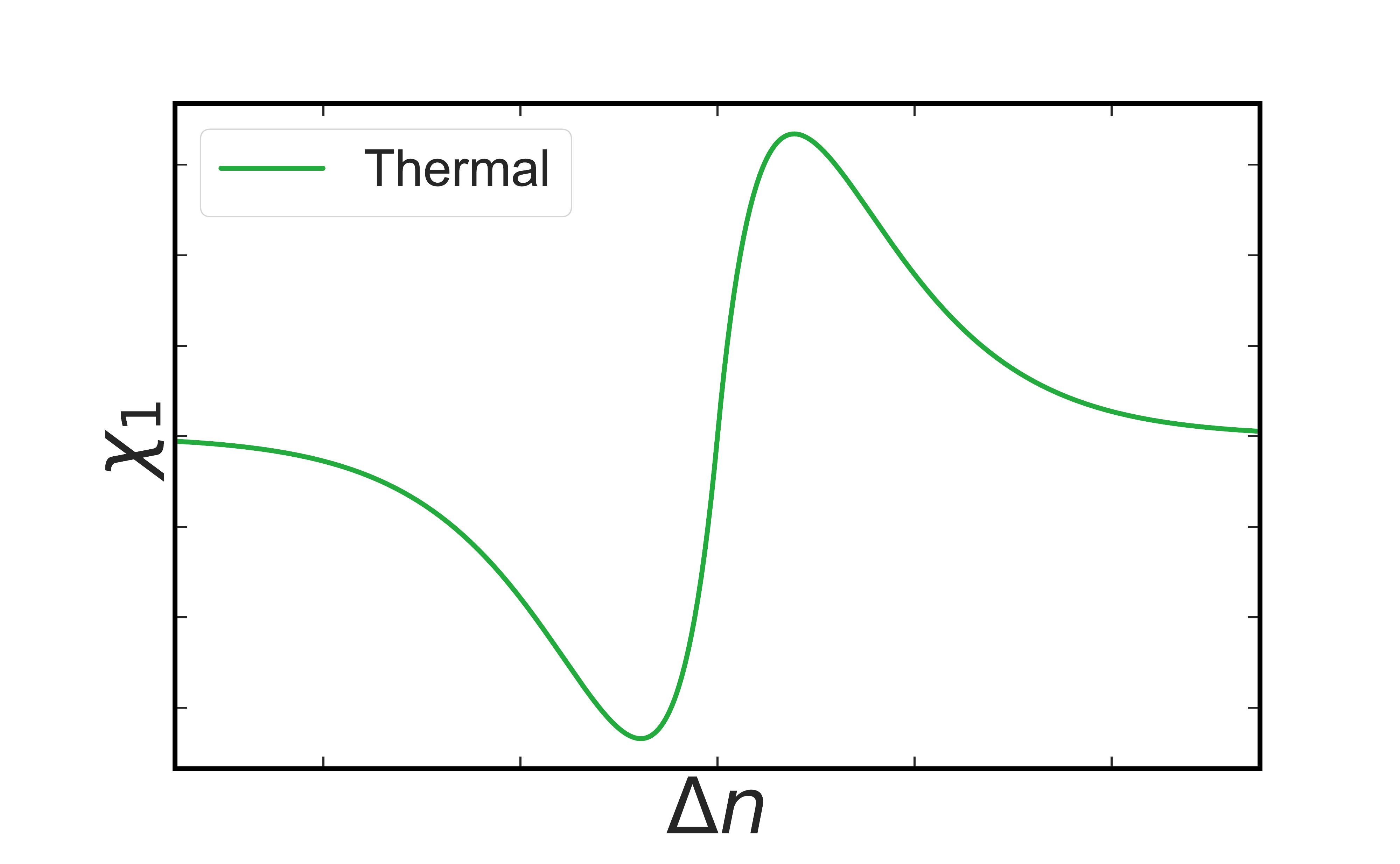}
    \put(-22,12){(c)}
    \hspace{0.025\linewidth}
    \includegraphics[width=0.48\linewidth, trim=5 15 55 65, clip]{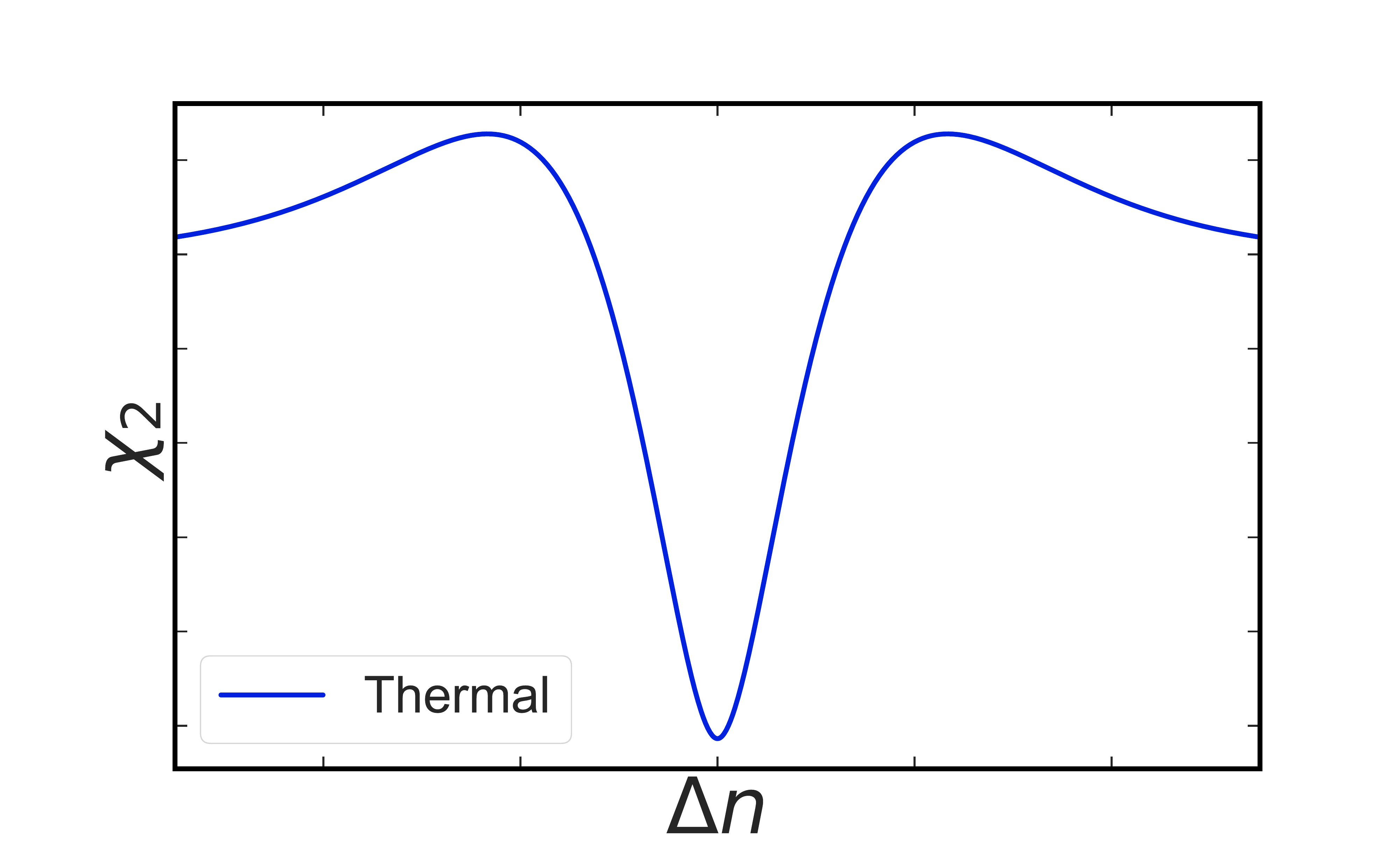}
    \put(-22,12){(d)}\\
    \includegraphics[width=0.48\linewidth, trim=5 15 55 65, clip]{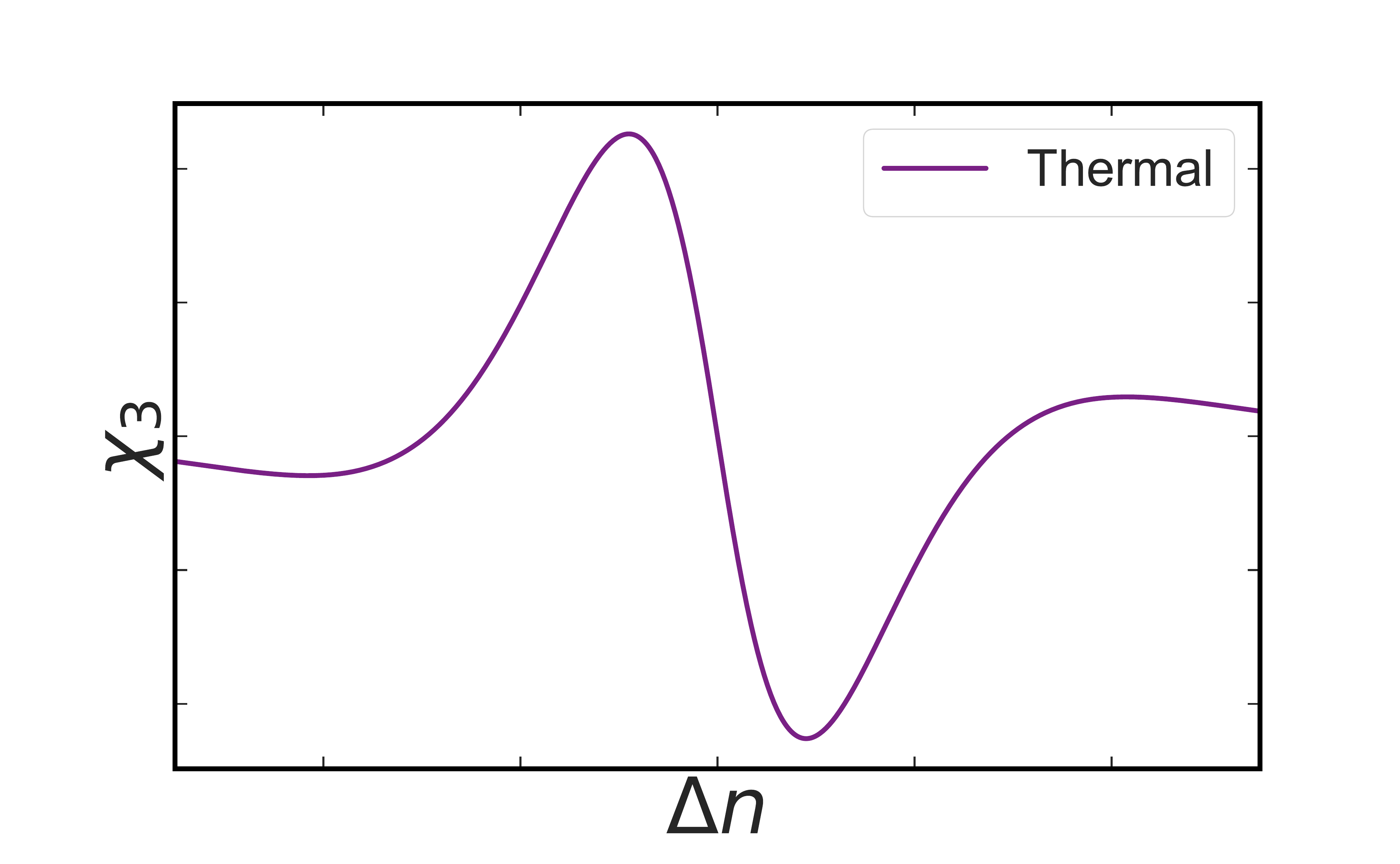}
    \put(-22,12){(e)}
    \hspace{0.025\linewidth}
    \includegraphics[width=0.48\linewidth, trim=5 15 55 65, clip]{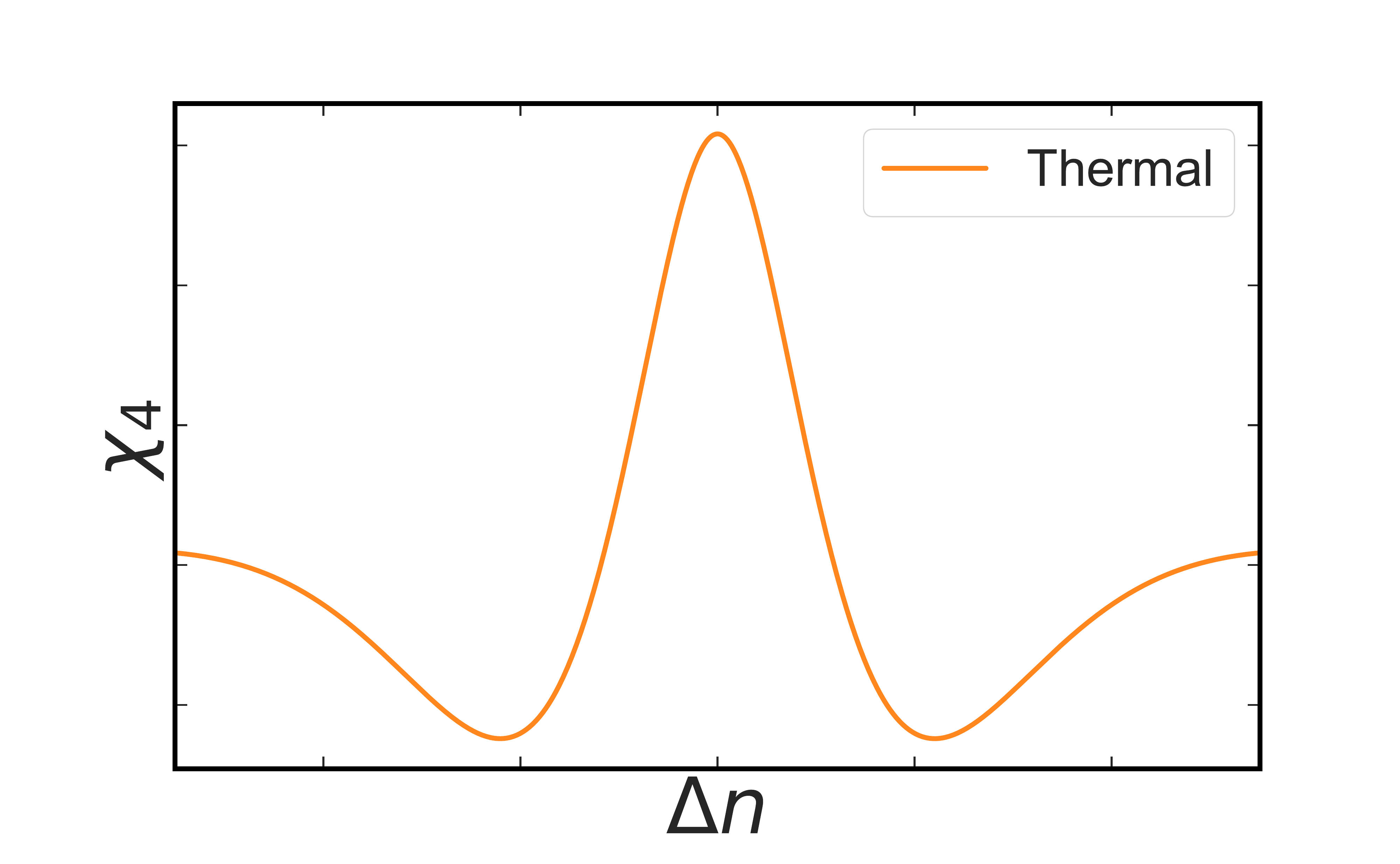}
    \put(-22,12){(f)}\\
    \includegraphics[width=0.48\linewidth, trim=5 15 55 65, clip]{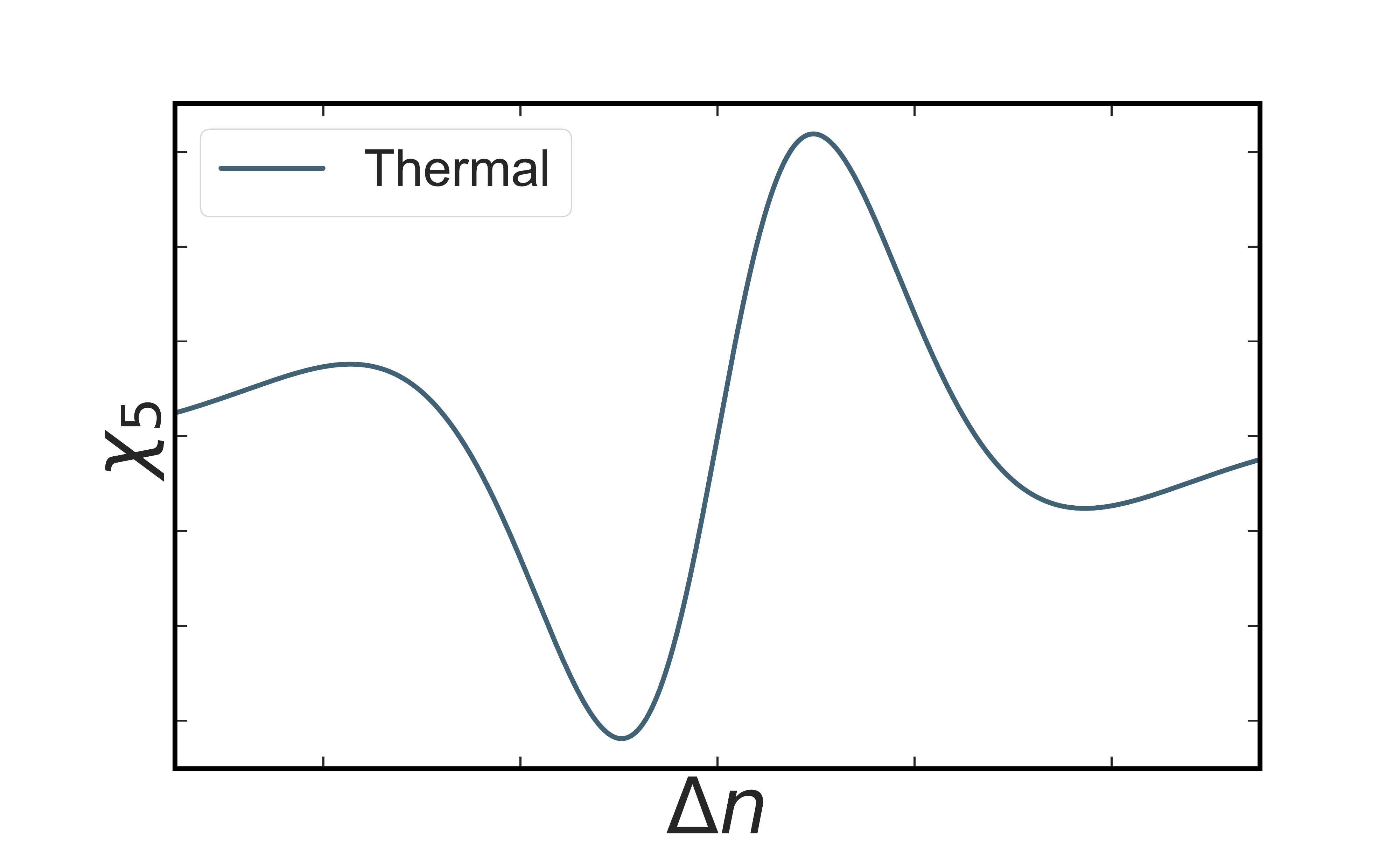}
    \put(-22,12){(g)}
    \hspace{0.025\linewidth}
    \includegraphics[width=0.48\linewidth, trim=5 15 55 65, clip]{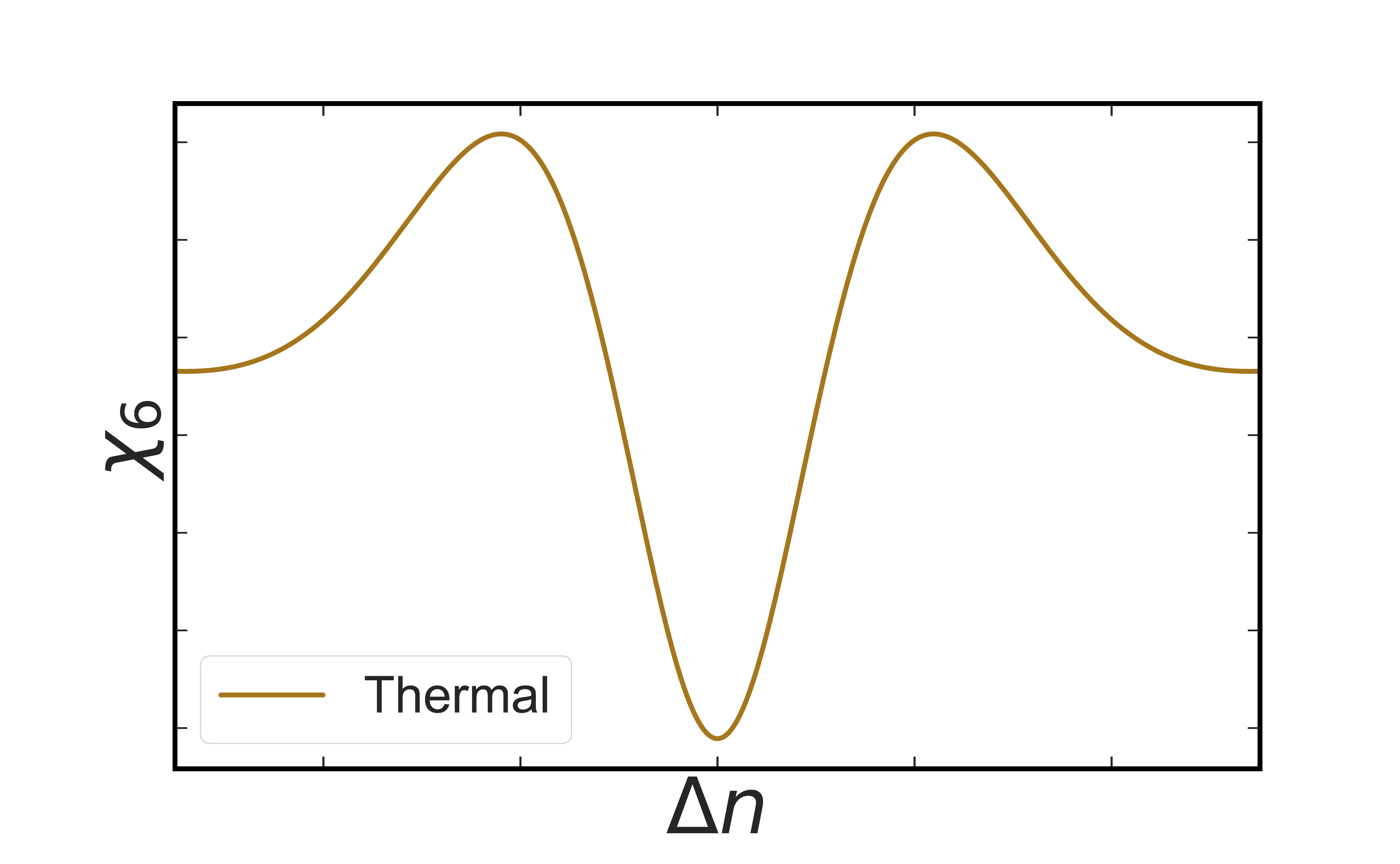}
    \put(-22,12){(h)}\\
    \includegraphics[width=0.48\linewidth, trim=5 15 55 65, clip]{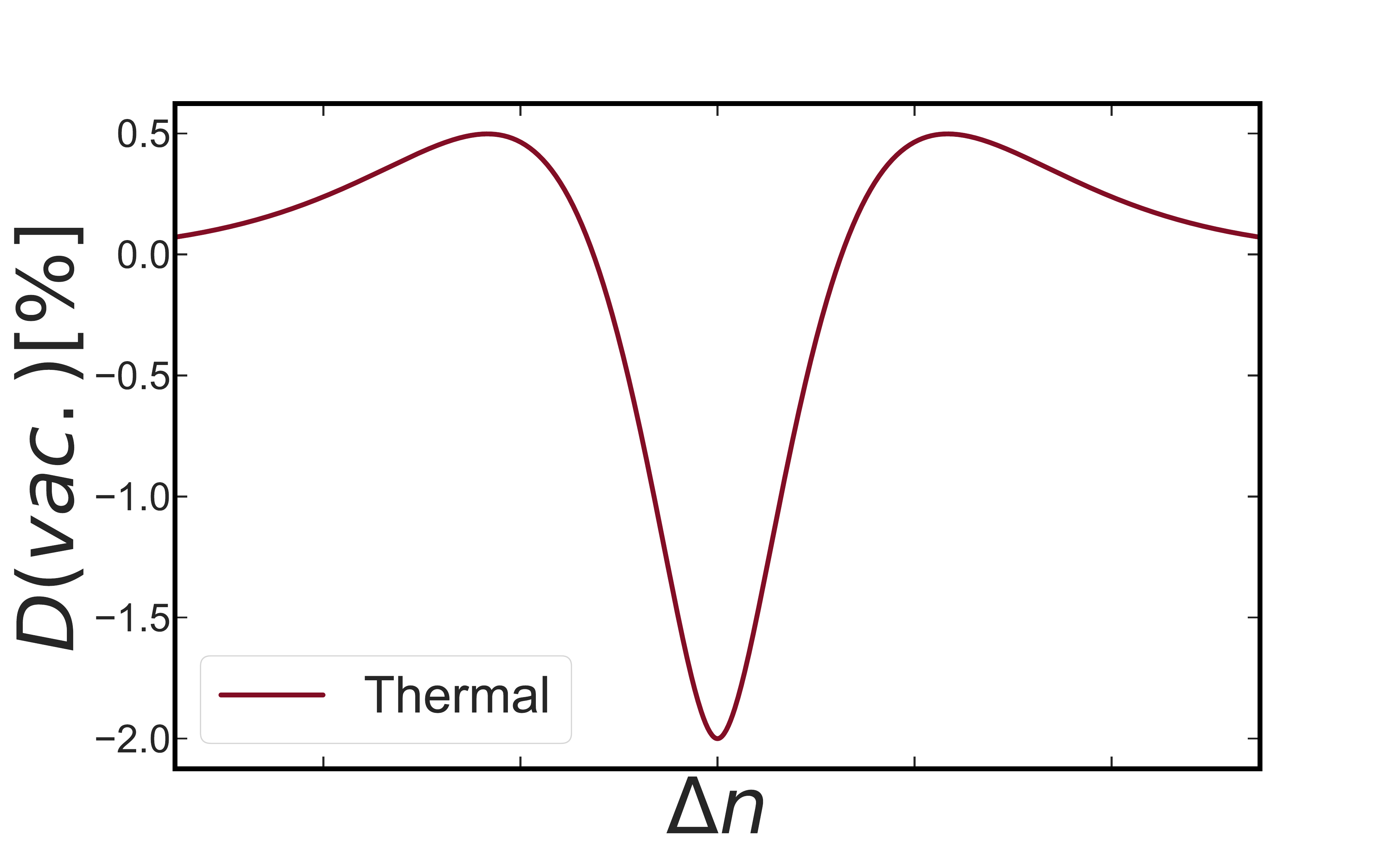}
    \put(-22,12){(i)}
    \hspace{0.025\linewidth}
    \includegraphics[width=0.48\linewidth, trim=5 15 55 65, clip]{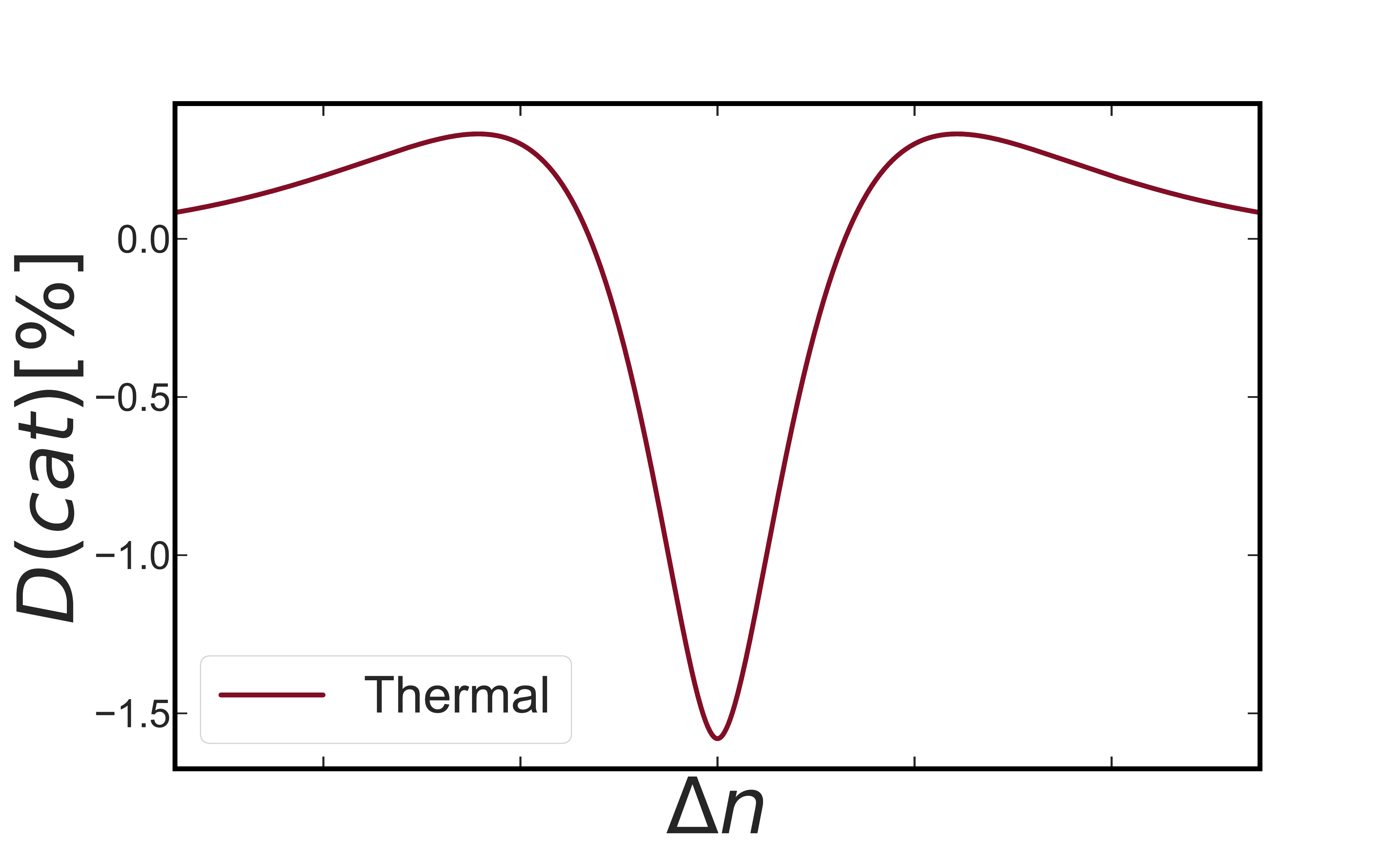}
    \put(-22,12){(j)}\\
    \includegraphics[width=0.99\linewidth, trim=20 10 200 25, clip]{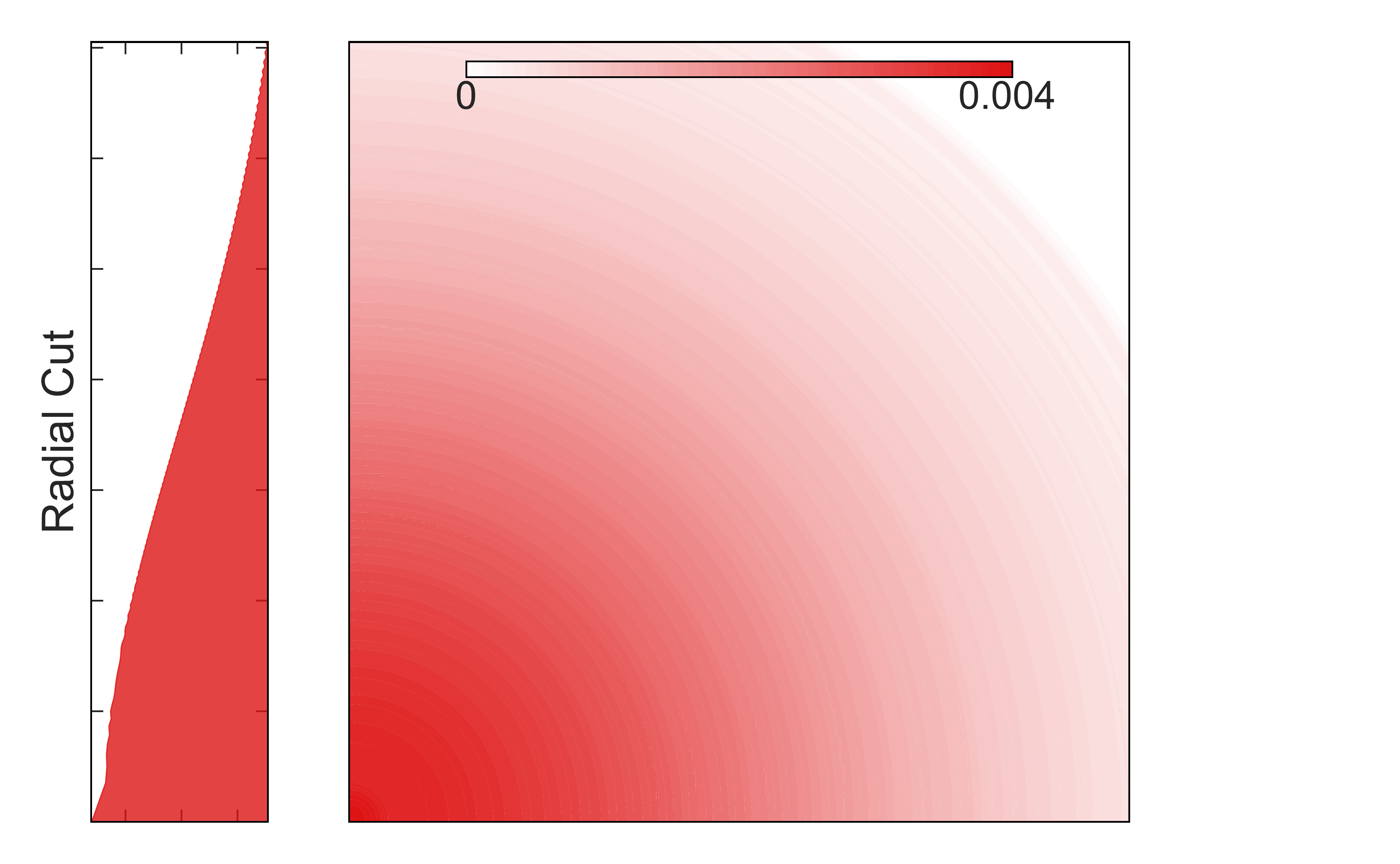}
    \put(-35,12){(k)}
    \vspace{-10pt}
    \caption{Thermal probe: see caption of Fig.~\ref{CoherentStats}.}
    \label{ThermalStats}
\end{figure}

\section{Statistical distributions of the probes}
\label{Appendix2}
As shown in Eq.~\eqref{pdeltaepsilon}, the photon statistics $P(n)$ of the probe play a major role in the detection. Here we explore some of the potential statistical distributions and their properties.

In all the following subsections, $\nu$ is the average number of photons in the probe.

\begin{figure}
    \centering
    \includegraphics[width=0.48\linewidth, trim=5 15 55 65, clip]{Coherent_SS.pdf}
    \put(-22,12){(a)}
    \hspace{0.025\linewidth}
    \includegraphics[width=0.48\linewidth, trim=5 15 55 65, clip]{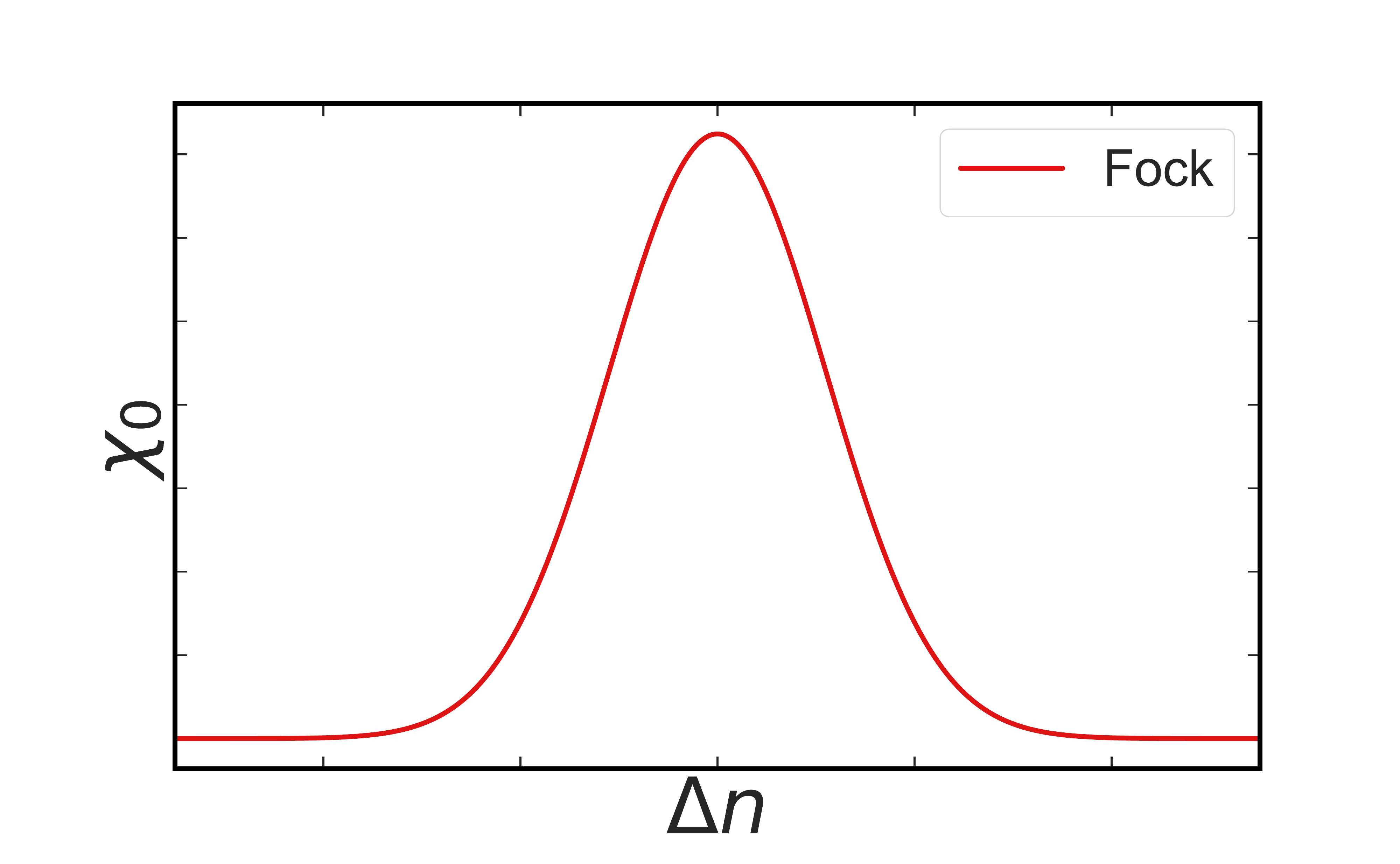}
    \put(-22,12){(b)}\\
    \includegraphics[width=0.48\linewidth, trim=5 15 55 65, clip]{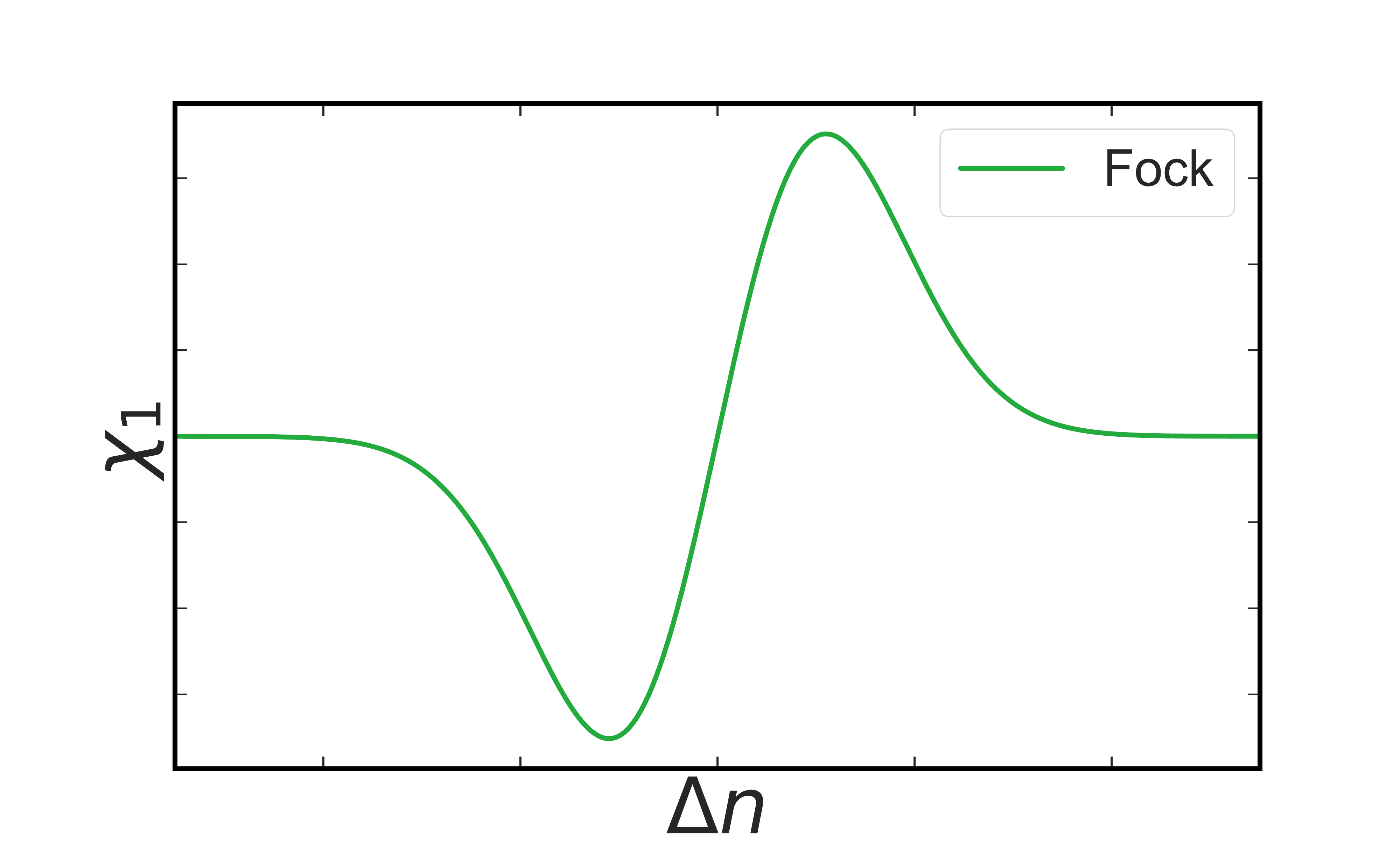}
    \put(-22,12){(c)}
    \hspace{0.025\linewidth}
    \includegraphics[width=0.48\linewidth, trim=5 15 55 65, clip]{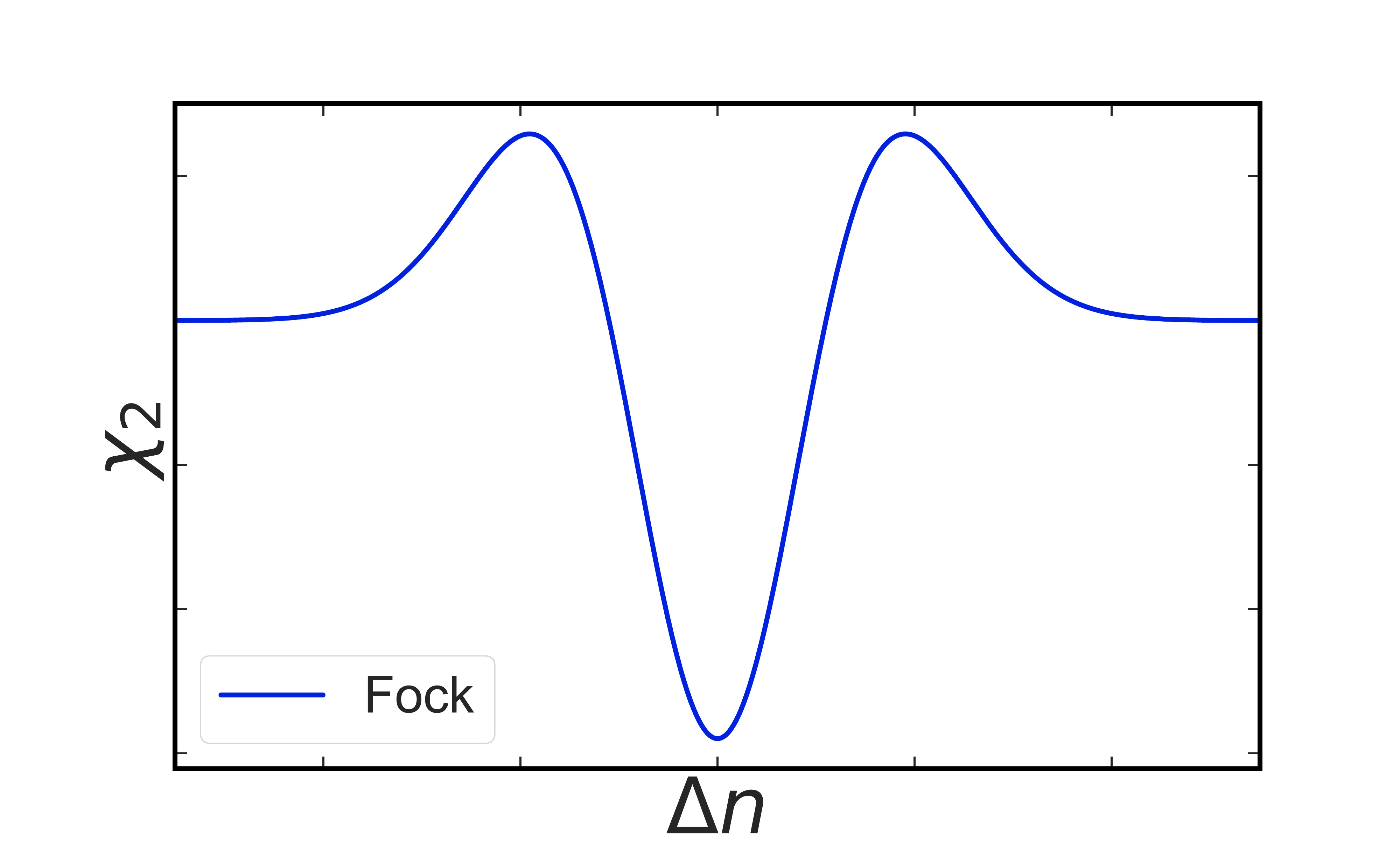}
    \put(-22,12){(d)}\\
    \includegraphics[width=0.48\linewidth, trim=5 15 55 65, clip]{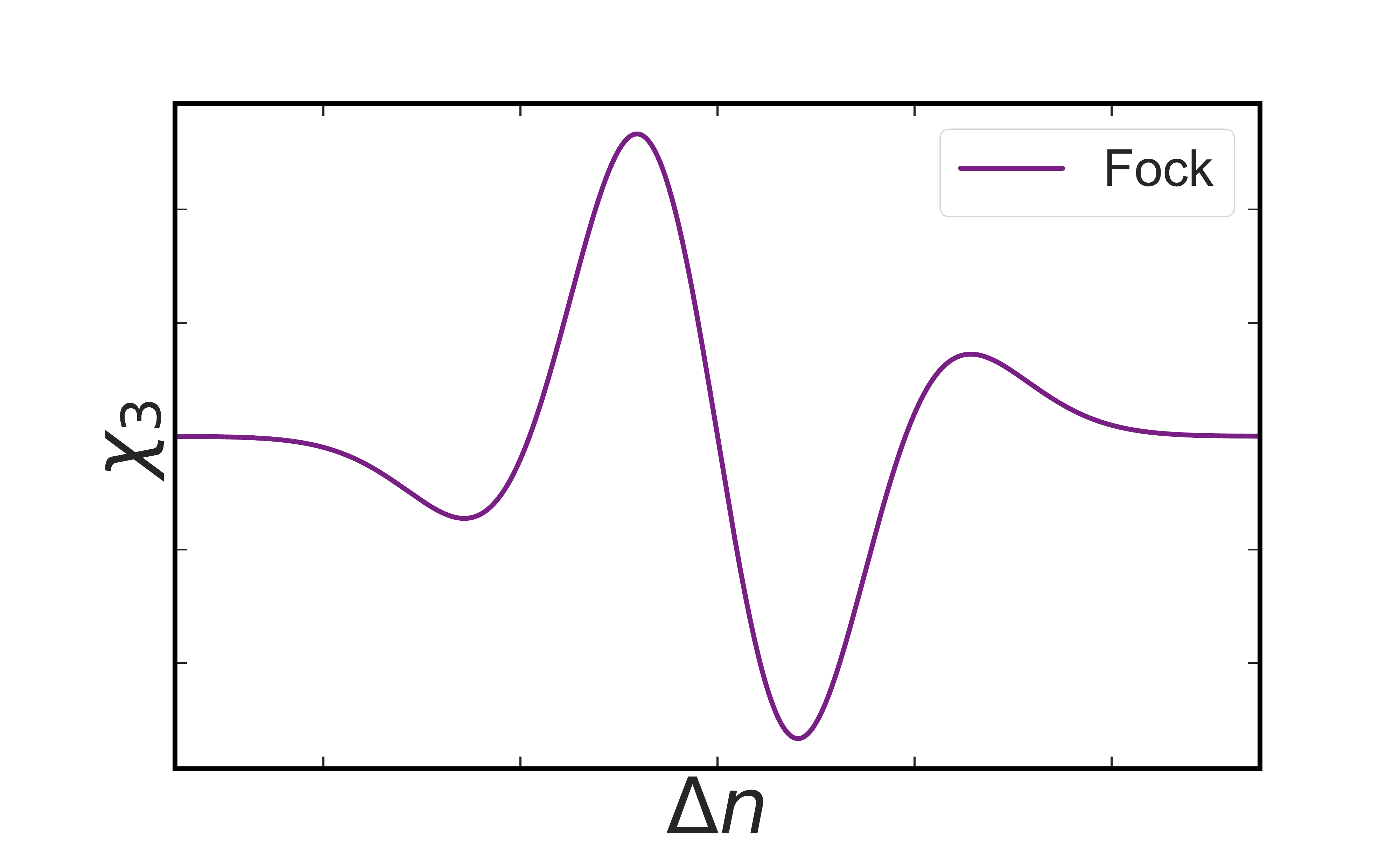}
    \put(-22,12){(e)}
    \hspace{0.025\linewidth}
    \includegraphics[width=0.48\linewidth, trim=5 15 55 65, clip]{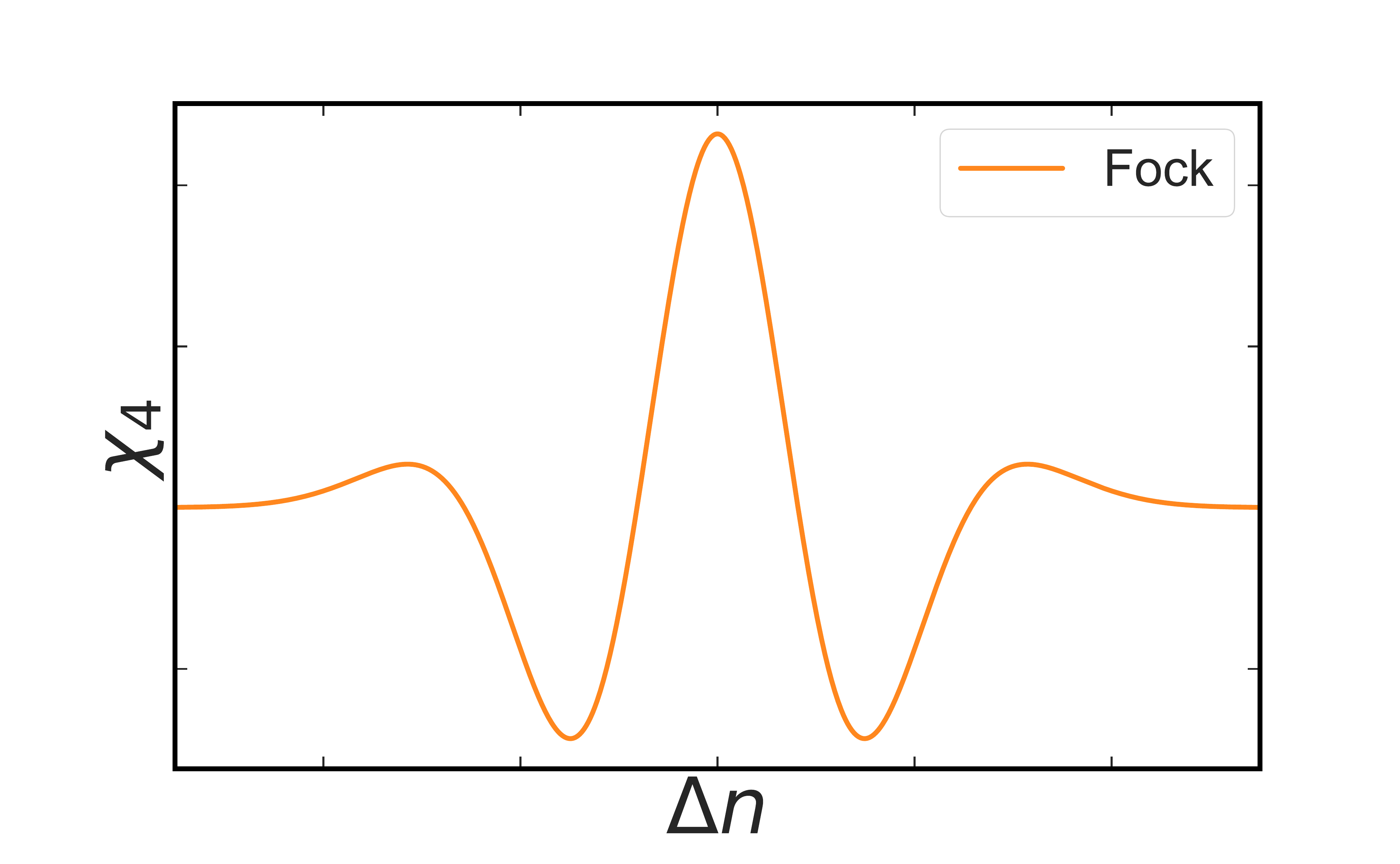}
    \put(-22,12){(f)}\\
    \includegraphics[width=0.48\linewidth, trim=5 15 55 65, clip]{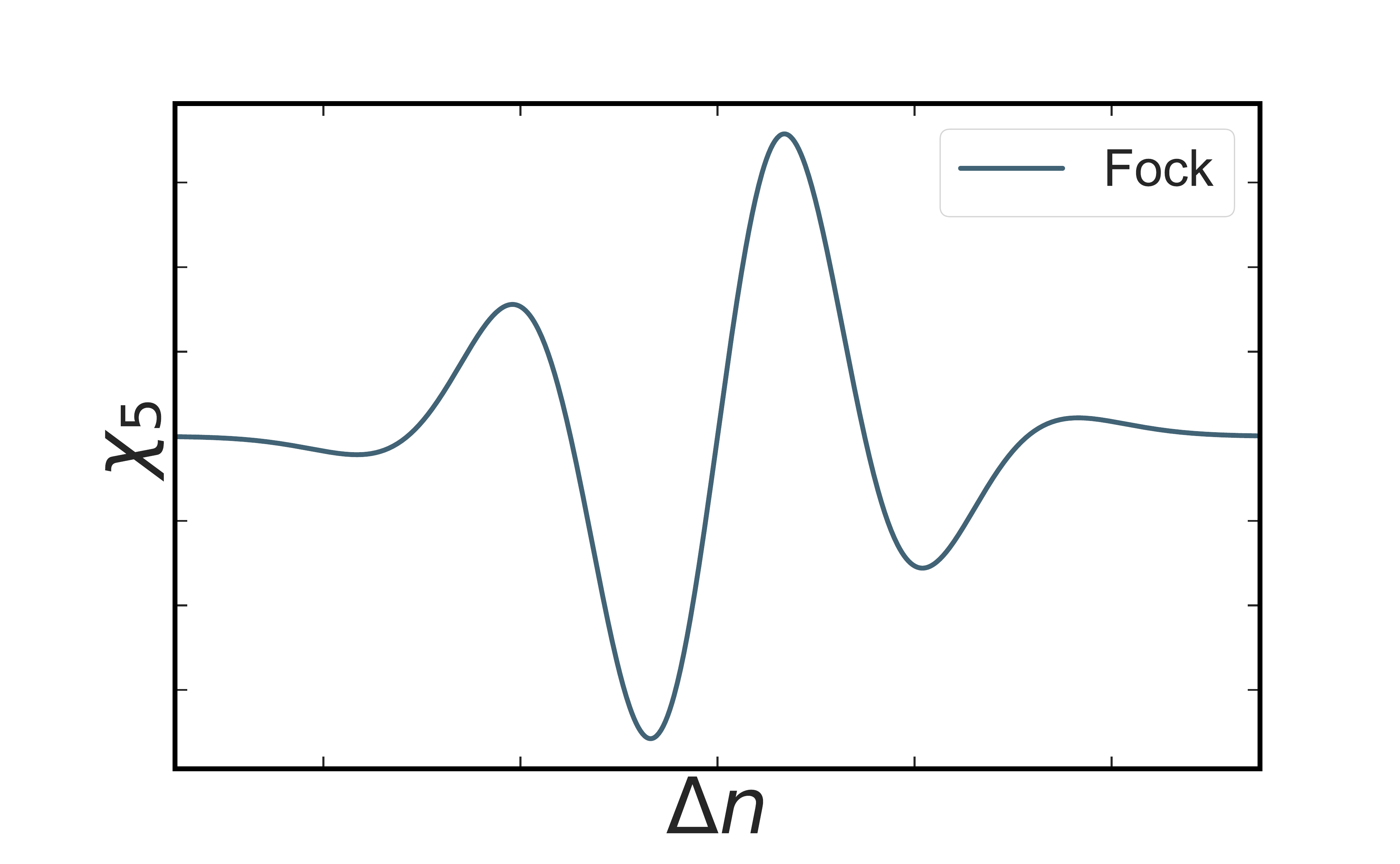}
    \put(-22,12){(g)}
    \hspace{0.025\linewidth}
    \includegraphics[width=0.48\linewidth, trim=5 15 55 65, clip]{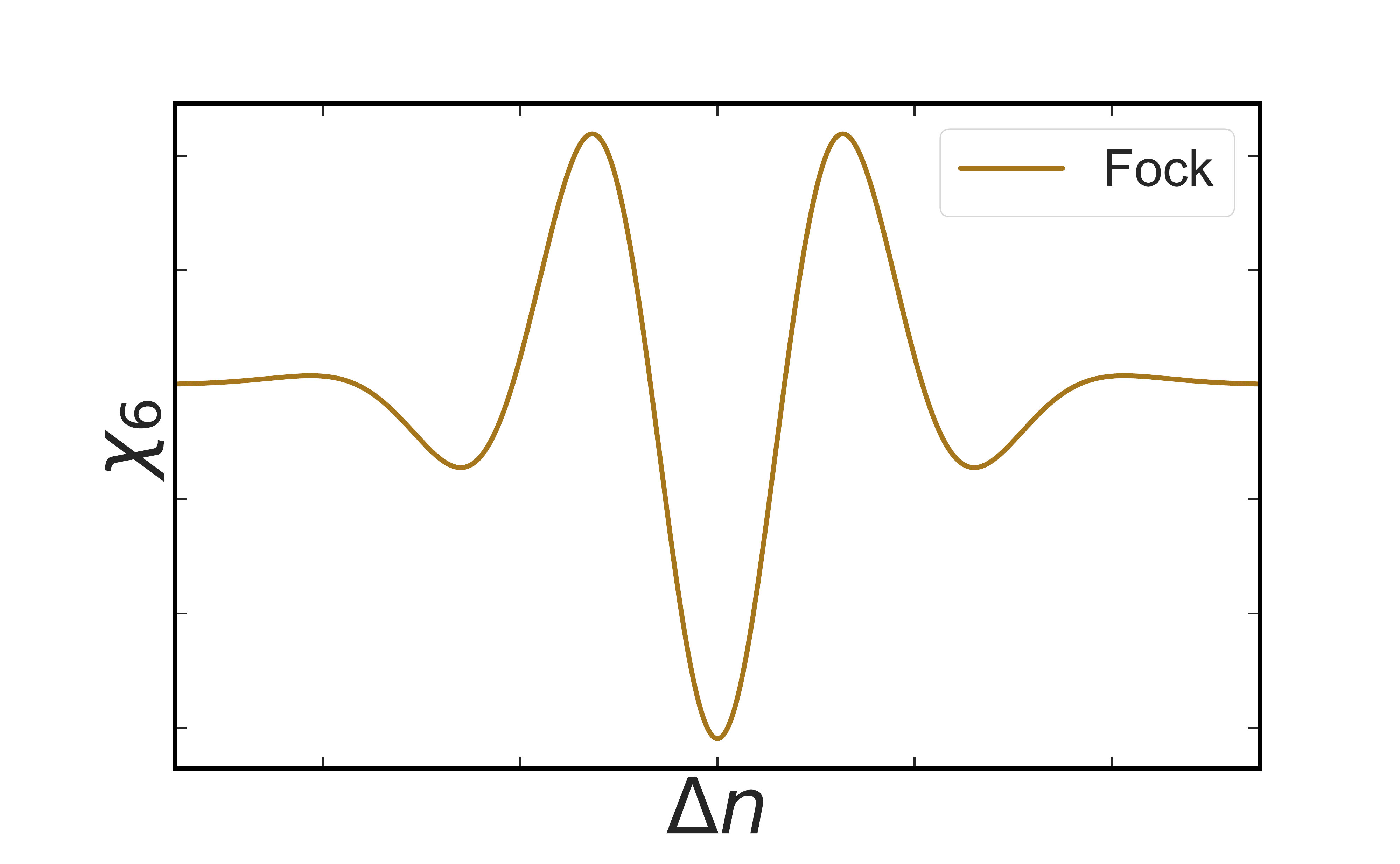}
    \put(-22,12){(h)}\\
    \includegraphics[width=0.48\linewidth, trim=5 15 55 65, clip]{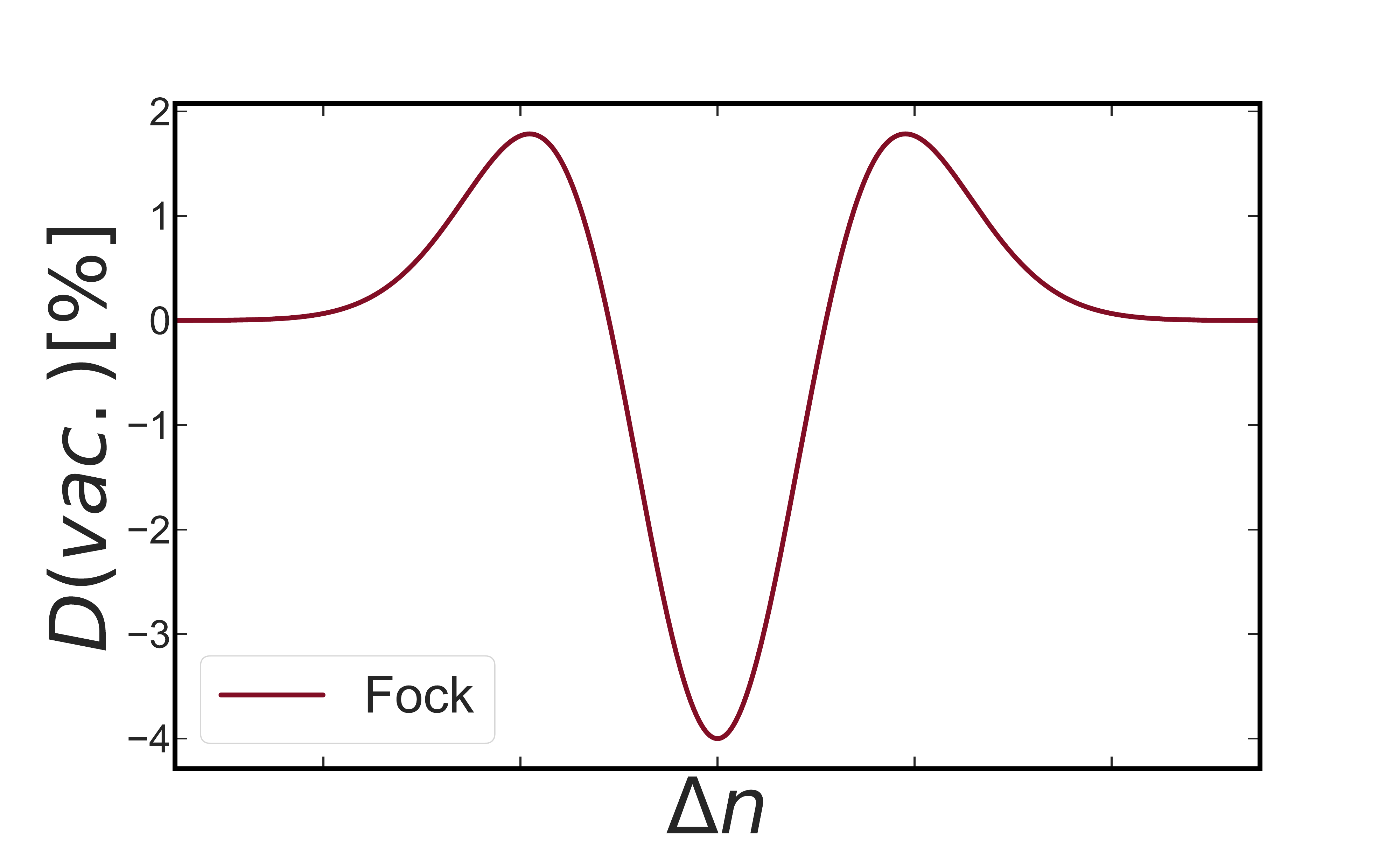}
    \put(-22,12){(i)}
    \hspace{0.025\linewidth}
    \includegraphics[width=0.48\linewidth, trim=5 15 55 65, clip]{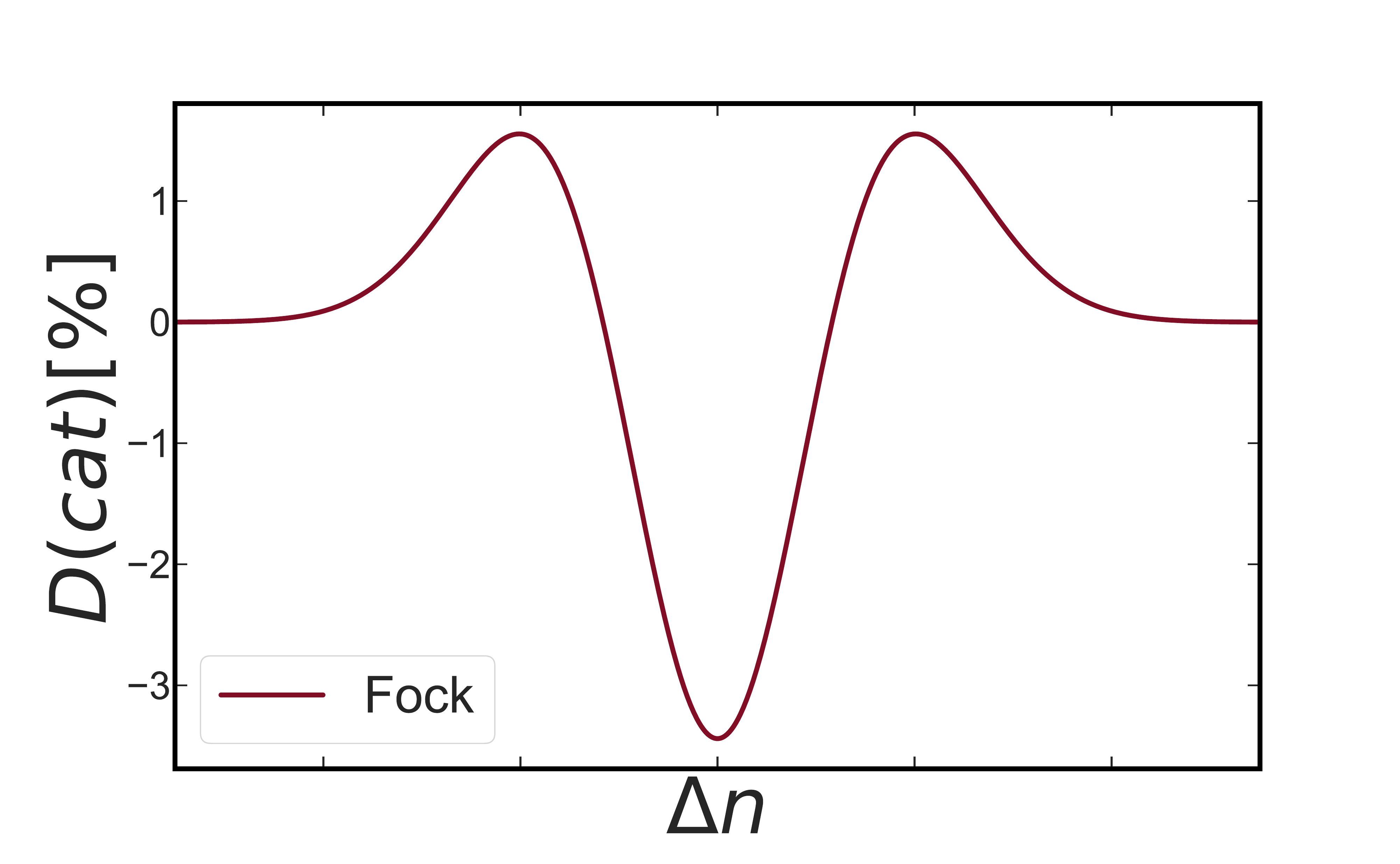}
    \put(-22,12){(j)}\\
    \includegraphics[width=0.99\linewidth, trim=20 10 200 25, clip]{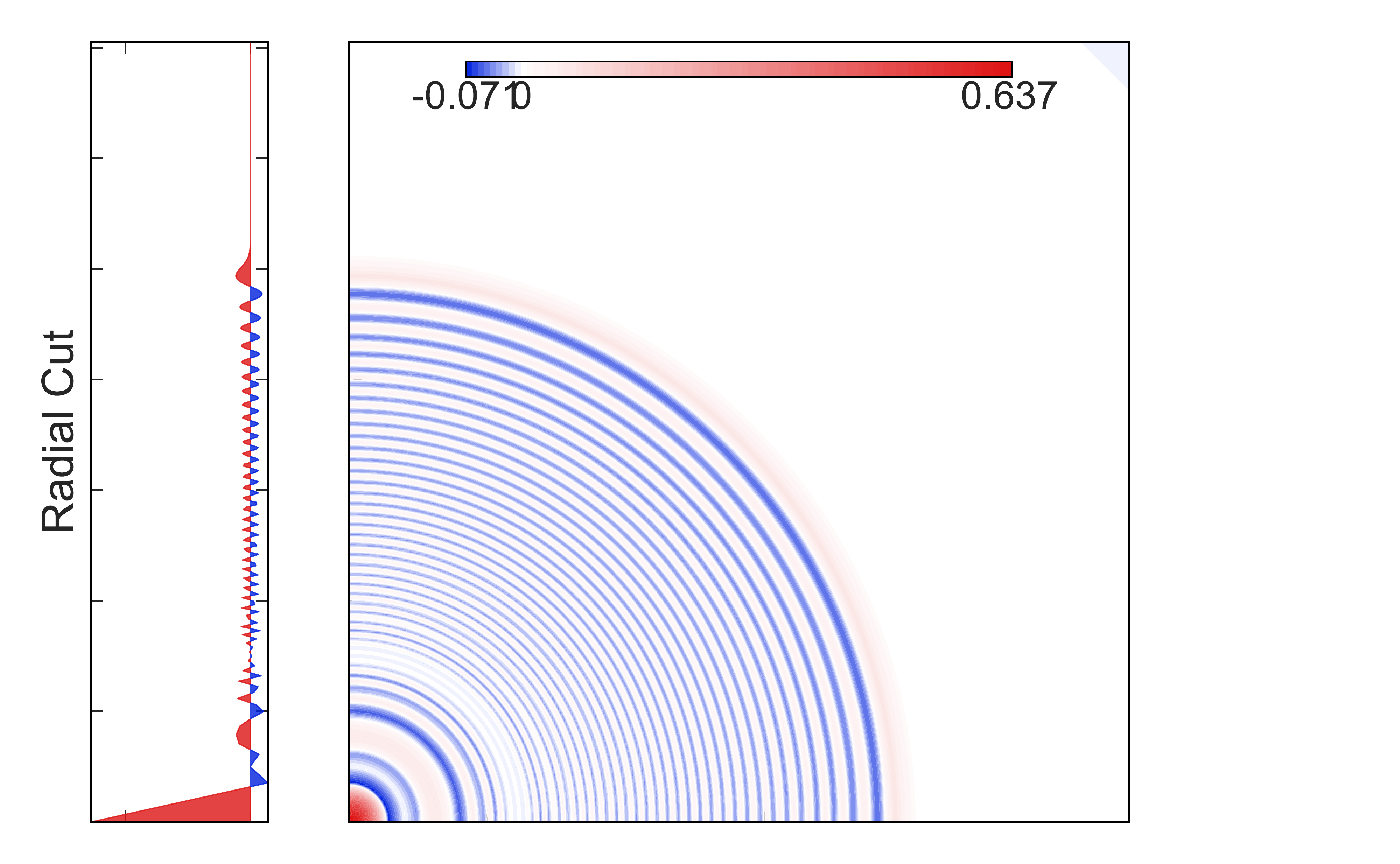}
    \put(-35,12){(k)}
    \vspace{-10pt}
    \caption{Fock probe: see caption of Fig.~\ref{CoherentStats}.}
    \label{FockStats}
\end{figure}

\begin{figure}
    \centering
    \includegraphics[width=0.48\linewidth, trim=5 15 55 65, clip]{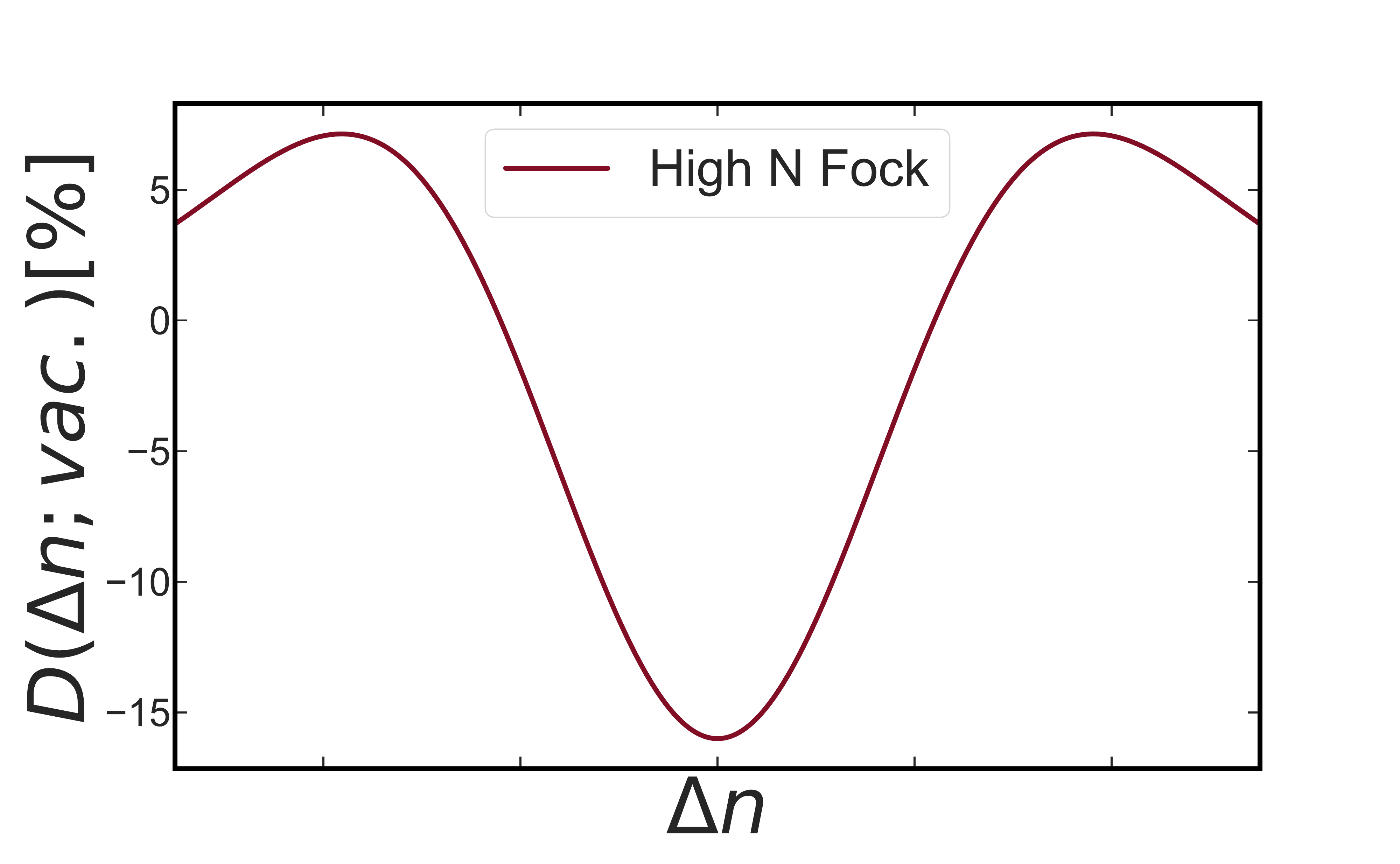}
    \put(-22,12){(a)}
    \hspace{0.025\linewidth}
    \includegraphics[width=0.48\linewidth, trim=5 15 55 65, clip]{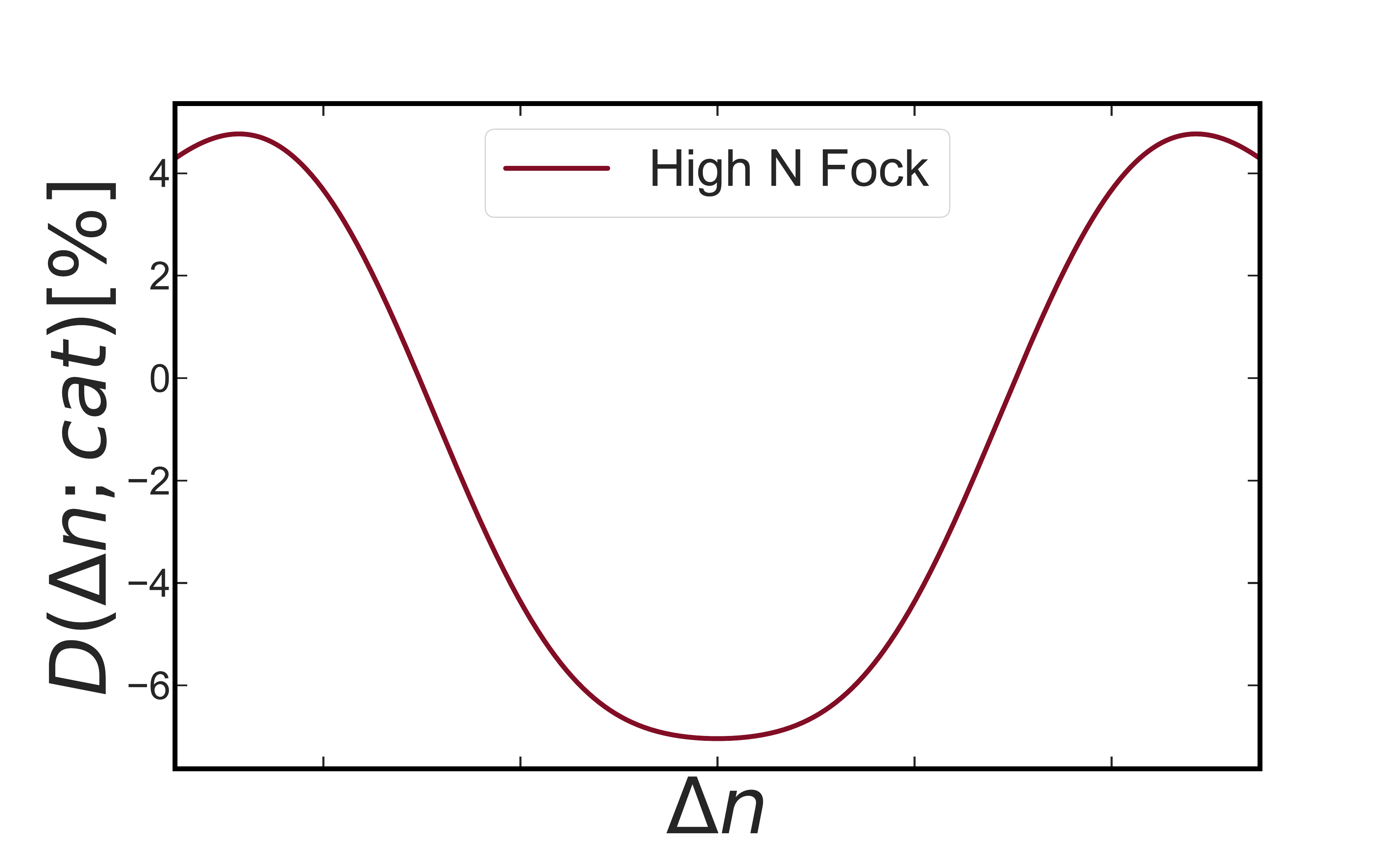}
    \put(-22,12){(b)}
    \vspace{-10pt}
    \caption{High-photon number Fock probe: relative differential noise amplitudes~\eqref{signaltonoise} for vacuum (a) and cat (b) signals.}
    \label{HighNFockStats}
\end{figure}

\vspace{.5\baselineskip}
\emph{Coherent state. }
Traditional EOS schemes have been based on coherent probes. A probe $\ket{\alpha}$ features a Poisson probability distribution with parameter $\nu=\abs{\alpha}^2$,
\begin{equation}
    P(n)\equiv P_{\ket{\alpha}}(n)=e^{-\nu}\frac{\nu^n}{n!}.
\end{equation}

The ratio of the standard deviation to the mean number of photons varies as the inverse of the square root of $\nu$.
Hence, the signal to noise ratio of the EOS measurement scales with the brightness of the probe. Fig.~\ref{CoherentStats} displays the properties of the probe for EOS.
In particular, Fig.~\ref{CoherentStats}(i) provides the expected results for the vacuum signal studied in Ref.~\cite{Riek2015}. Fig.~\ref{CoherentStats}(j) provides expected results for a few-photon cat state~\cite{Asavanant2017} with the same coupling to the probe as that of the vacuum state to which it is compared.

\vspace{.5\baselineskip}
\emph{Thermal state. }
Another possible classical probe is the thermal state. It is not a pure state but a full statistical mixture expressible as a diagonal density matrix. It is fully characterized by the parameter $\xi=\nu/(\nu+1)$, as 
\begin{equation}
    \label{thermal}
    P(n)\equiv P_\xi(n)=(1-\xi)\;\xi^n.
\end{equation}

The statistics is very different from that of the coherent state, being exponentially heavier on the lower photon counts. With the same average number of photons, the probability of getting a very large photon counts is much greater than for the coherent case. In general, the thermal distribution is a great starting point for discrimination schemes, as it features non-vanishing probabilities over a very large spectrum of photon counts, however yielding lower $D$ measures, e.~g.~ as shown for vacuum and cat states in comparison to the coherent state probe [Fig.~\ref{ThermalStats}(i,j)].

\vspace{.5\baselineskip}
\emph{Fock state. }
For the Fock state  $\ket{\nu}$, we have
\begin{equation}
    P(n)\equiv P_{\ket{\nu}}(n)=\delta_{n,\nu}.
\end{equation}

\emph{A priori}, it might seem that Fock states would be ideal as an EOS probe. However, a Fock state with the same number of photons as the mean of the coherent state of Fig.~\ref{CoherentStats} provides essentially the same EOS signal, as seen in Fig.~\ref{FockStats}. A Fock state with a 4 times higher number of photons would provide a factor of 4x higher signal-to-noise ratio, as displayed in Fig.~\ref{HighNFockStats}. The advantage is only due to the fact that the binomial distribution of Eq.~\eqref{binom} has a standard deviation that varies as $\sqrt{n}$. Hence, a coherent state with the same increase in the number of photons would lead to an essentially identical result.

The discriminating schemes that are presented next are meant to offer substantial statistical advantages over classical probes, as detailed in the main text.

\begin{figure}
    \centering
    \includegraphics[width=0.48\linewidth, trim=5 15 55 65, clip]{StatLog_Upper.pdf}
    \put(-22,12){(a)}
    \hspace{0.025\linewidth}
    \includegraphics[width=0.48\linewidth, trim=5 15 55 65, clip]{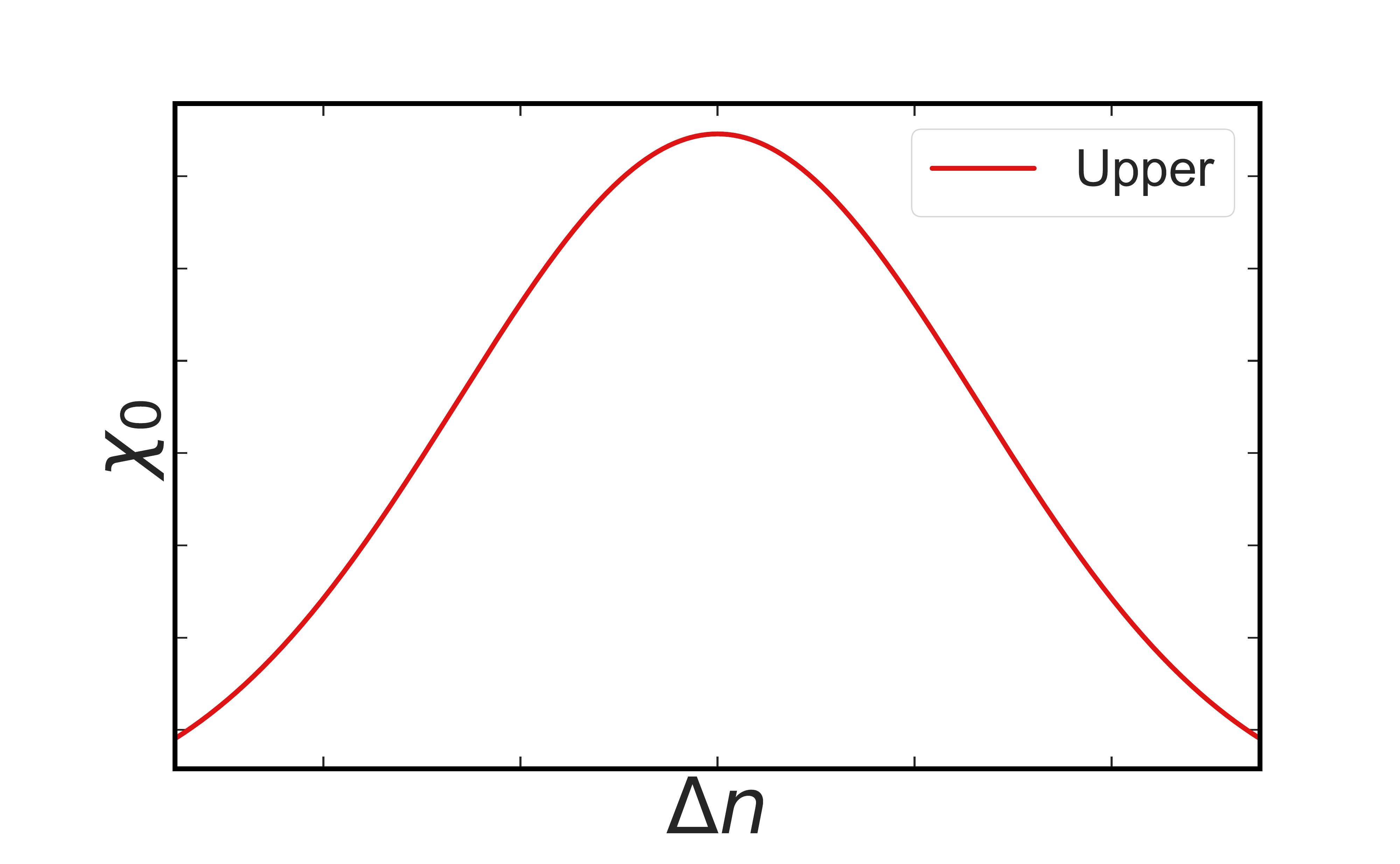}
    \put(-22,12){(b)}\\
    \includegraphics[width=0.48\linewidth, trim=5 15 55 65, clip]{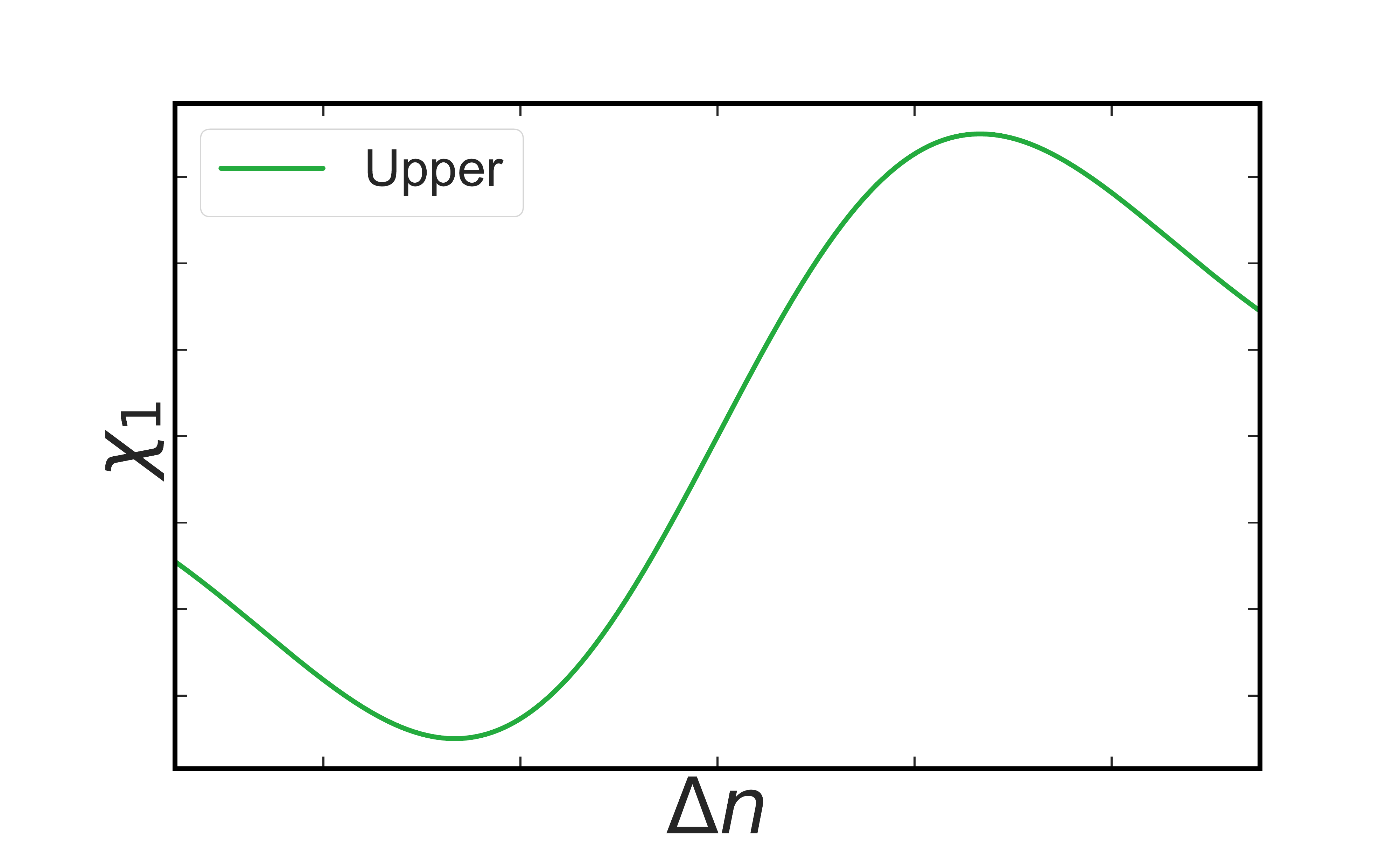}
    \put(-22,12){(c)}
    \hspace{0.025\linewidth}
    \includegraphics[width=0.48\linewidth, trim=5 15 55 65, clip]{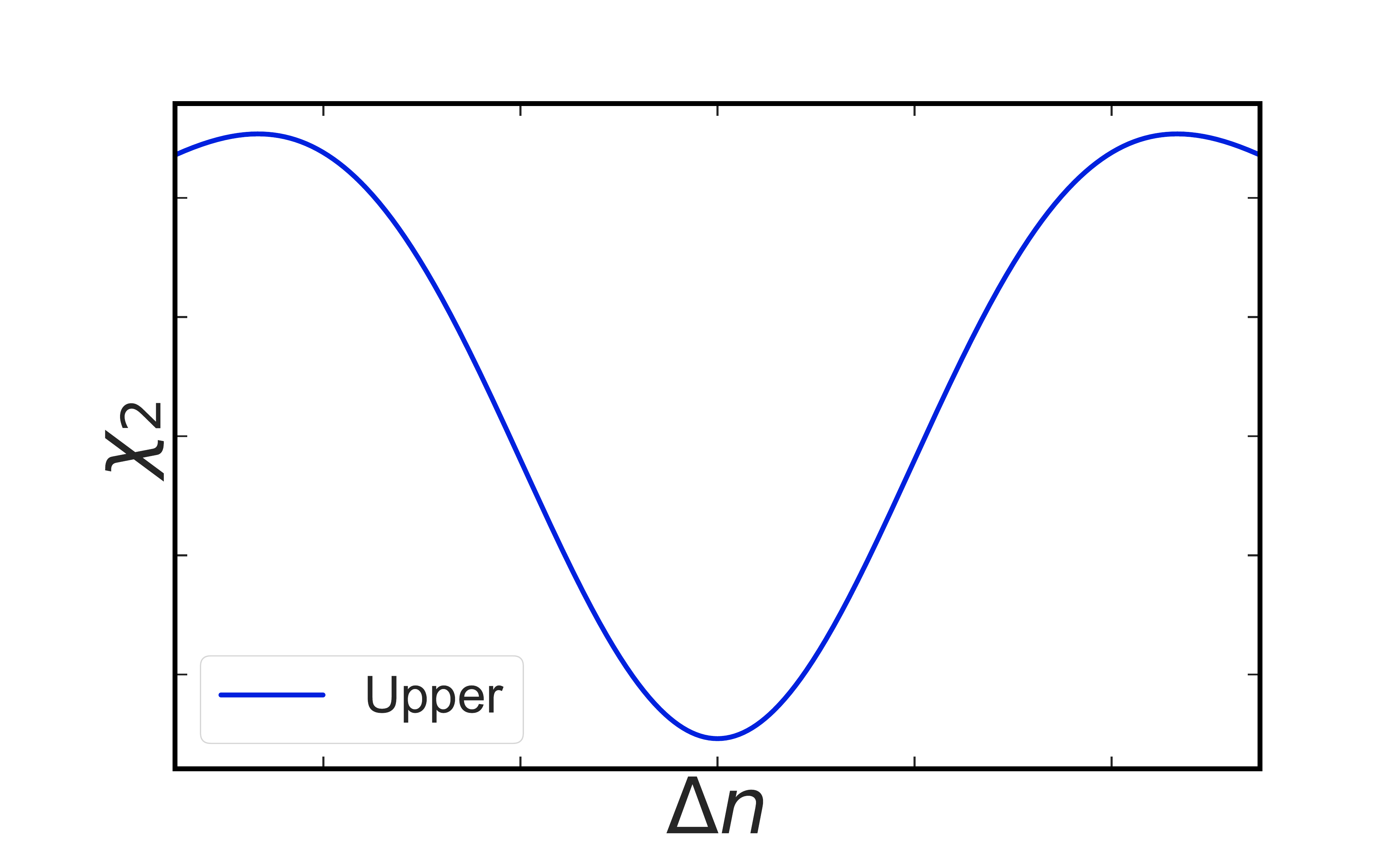}
    \put(-22,12){(d)}\\
    \includegraphics[width=0.48\linewidth, trim=5 15 55 65, clip]{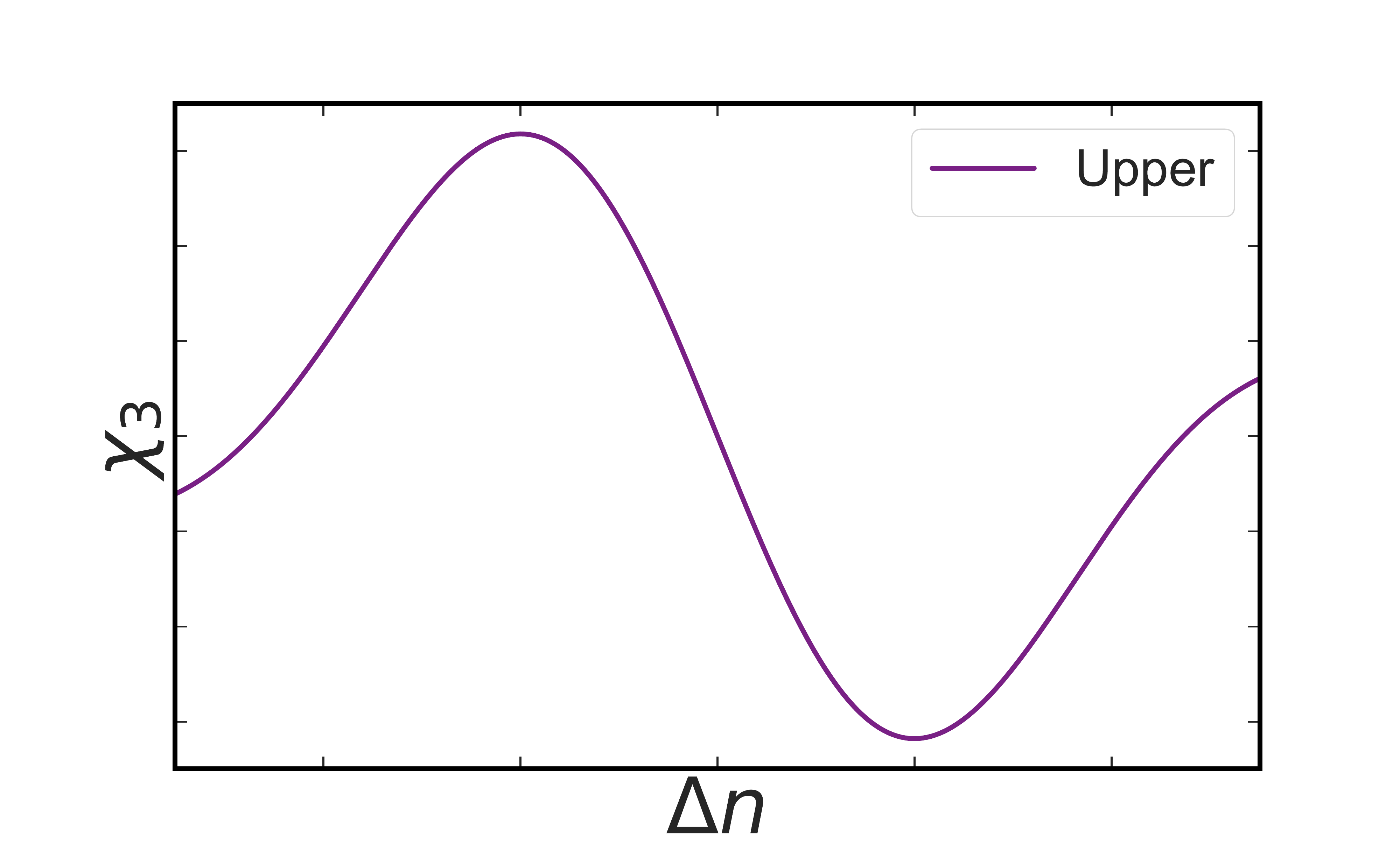}
    \put(-22,12){(e)}
    \hspace{0.025\linewidth}
    \includegraphics[width=0.48\linewidth, trim=5 15 55 65, clip]{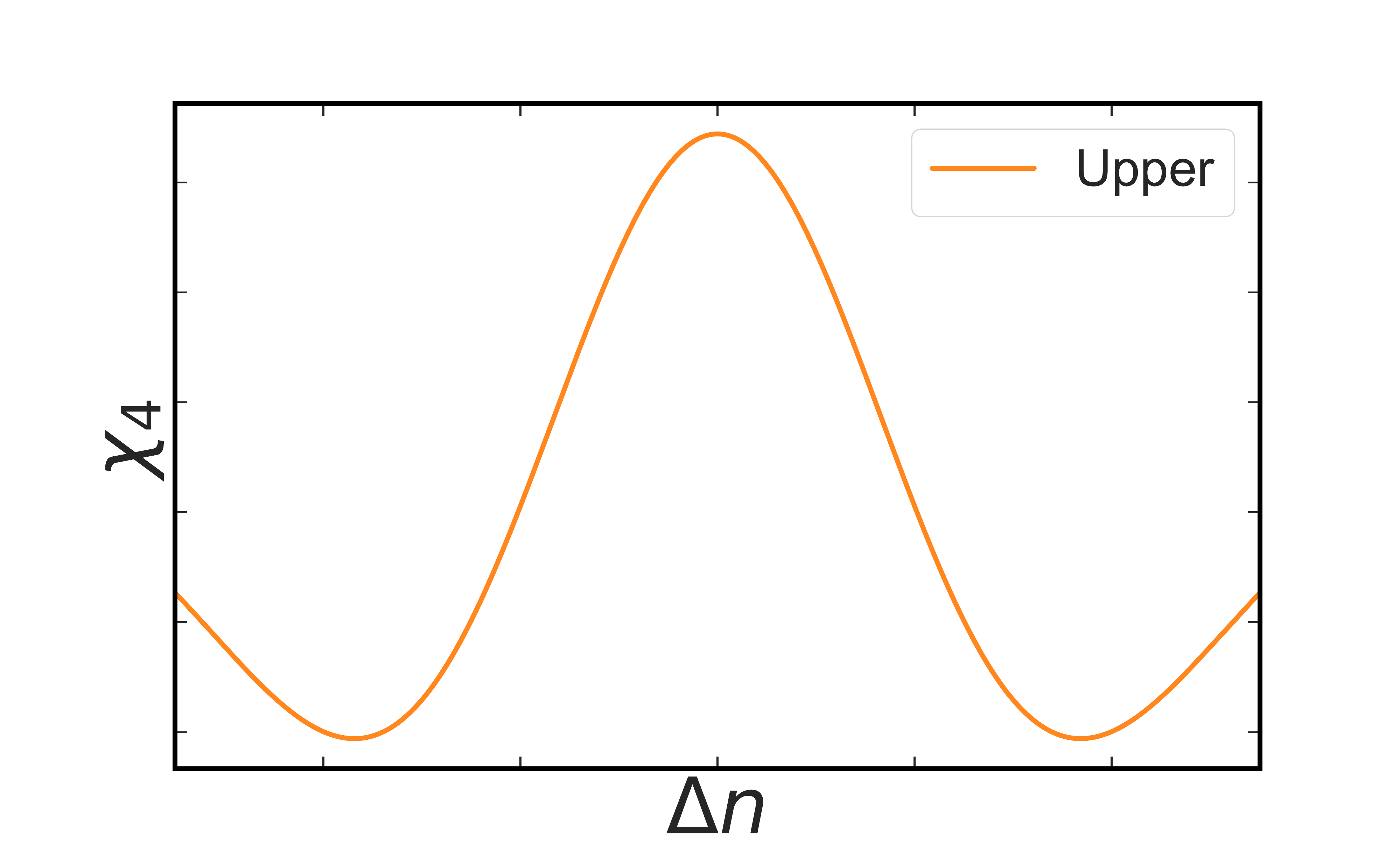}
    \put(-22,12){(f)}\\
    \includegraphics[width=0.48\linewidth, trim=5 15 55 65, clip]{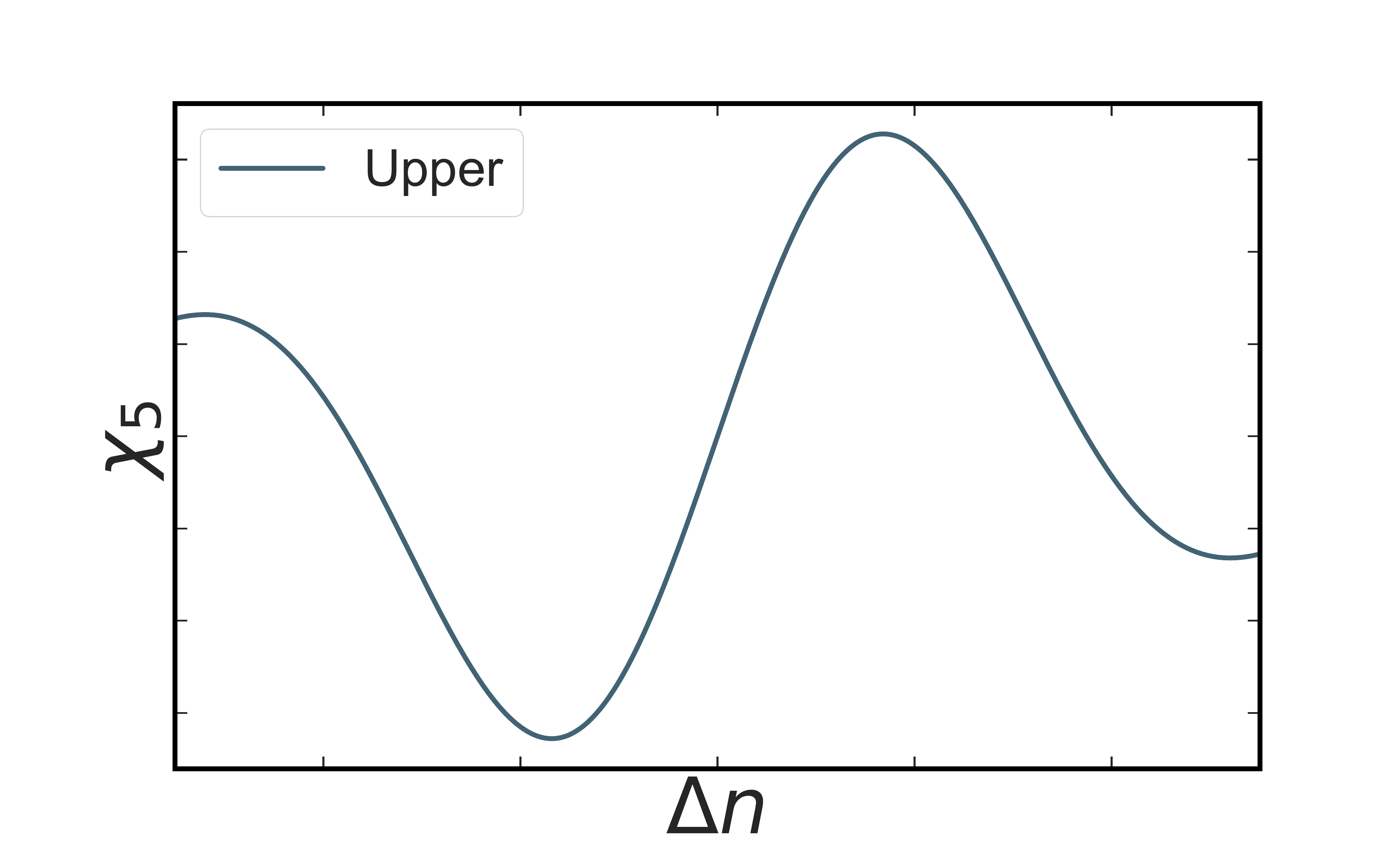}
    \put(-22,12){(g)}
    \hspace{0.025\linewidth}
    \includegraphics[width=0.48\linewidth, trim=5 15 55 65, clip]{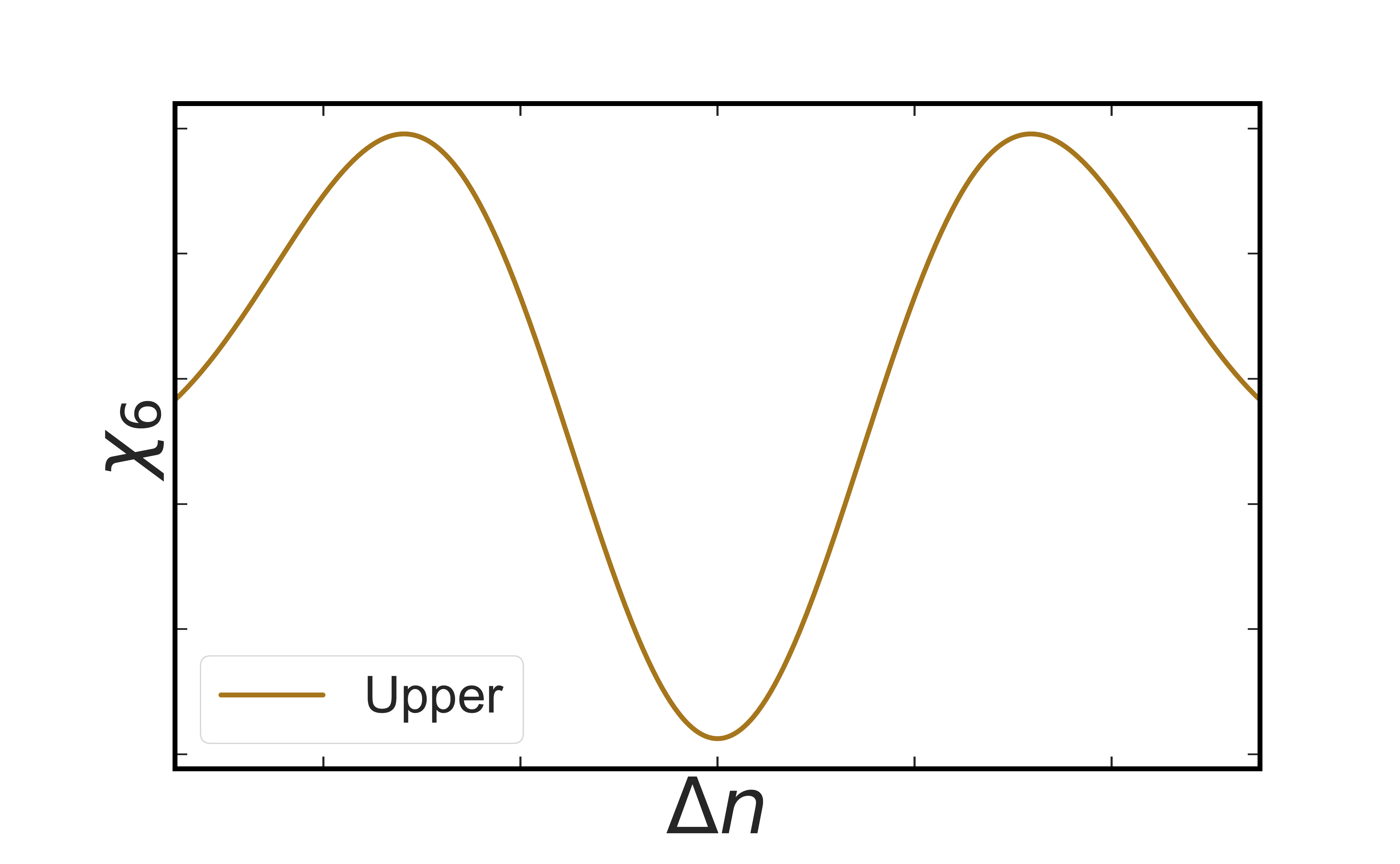}
    \put(-22,12){(h)}\\
    \includegraphics[width=0.48\linewidth, trim=5 15 55 65, clip]{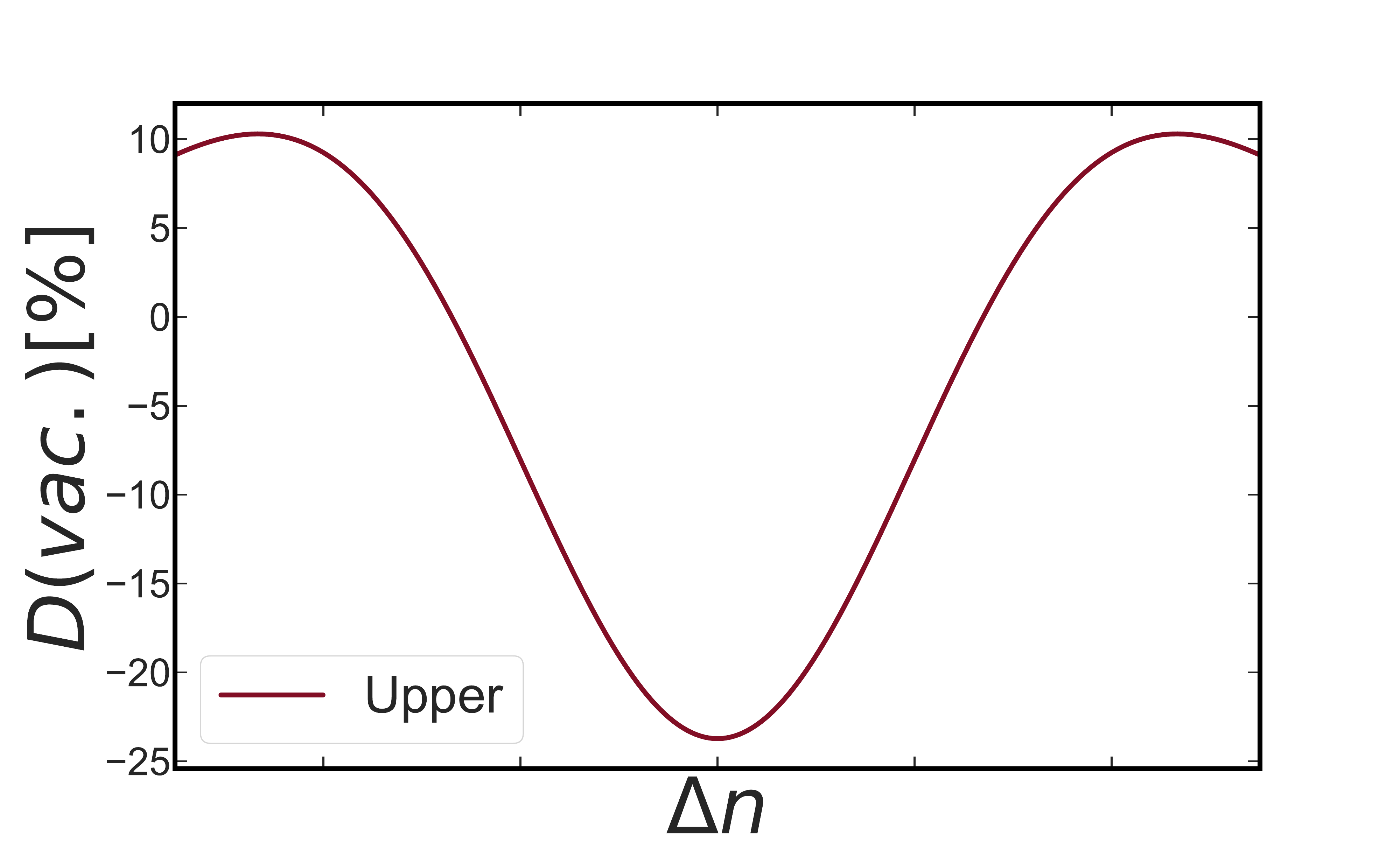}
    \put(-22,12){(i)}
    \hspace{0.025\linewidth}
    \includegraphics[width=0.48\linewidth, trim=5 15 55 65, clip]{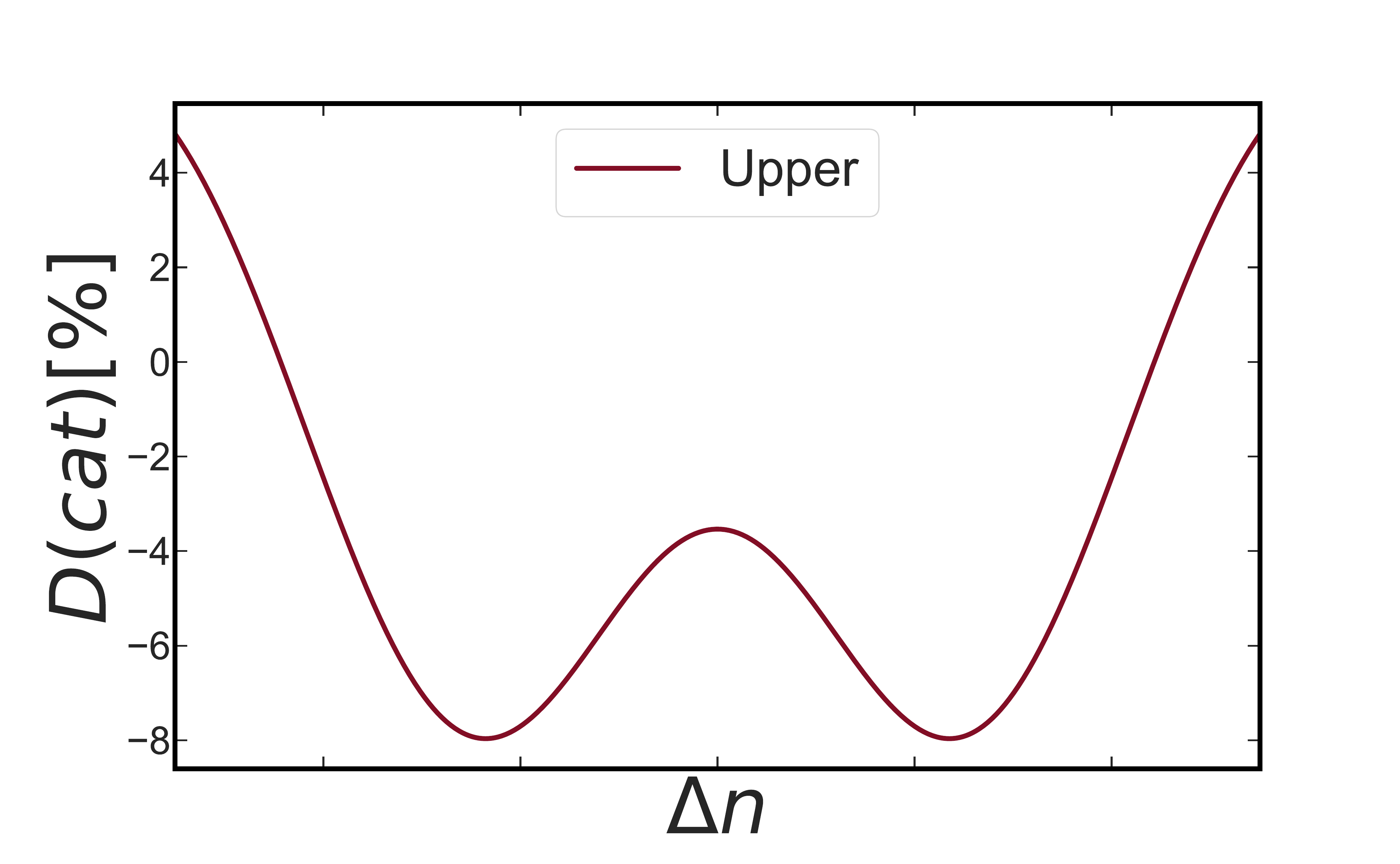}
    \put(-22,12){(j)}\\
    \includegraphics[width=0.99\linewidth, trim=20 10 200 25, clip]{Wigner_Upper.pdf}
    \put(-35,12){(k)}
    \vspace{-10pt}
    \caption{Upper BCS probe: see caption of Fig.~\ref{CoherentStats}.}
    \label{UpperStats}
\end{figure}

\begin{figure}
    \centering
    \includegraphics[width=0.48\linewidth, trim=5 15 55 65, clip]{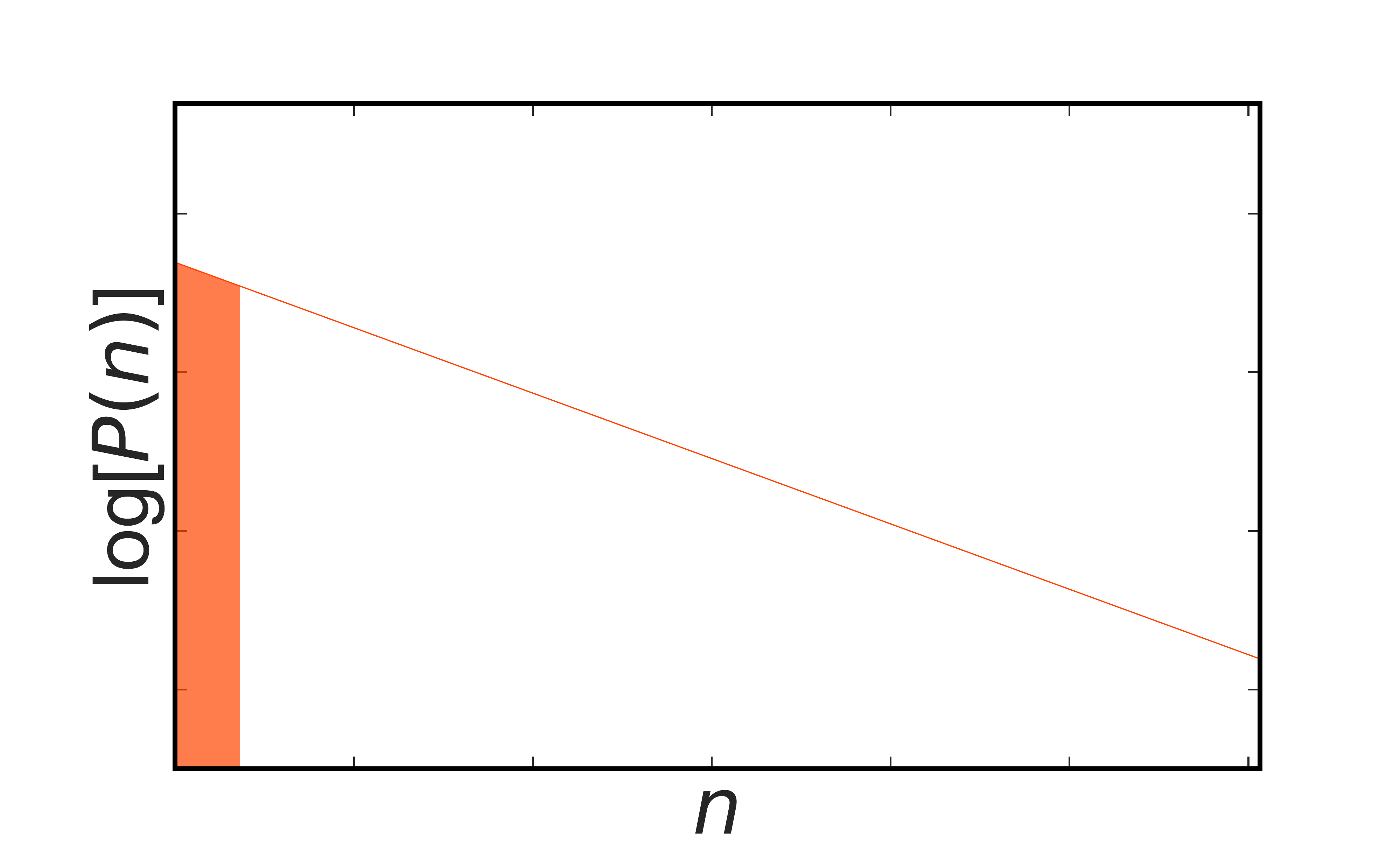}
    \put(-22,12){(a)}
    \hspace{0.025\linewidth}
    \includegraphics[width=0.48\linewidth, trim=5 15 55 65, clip]{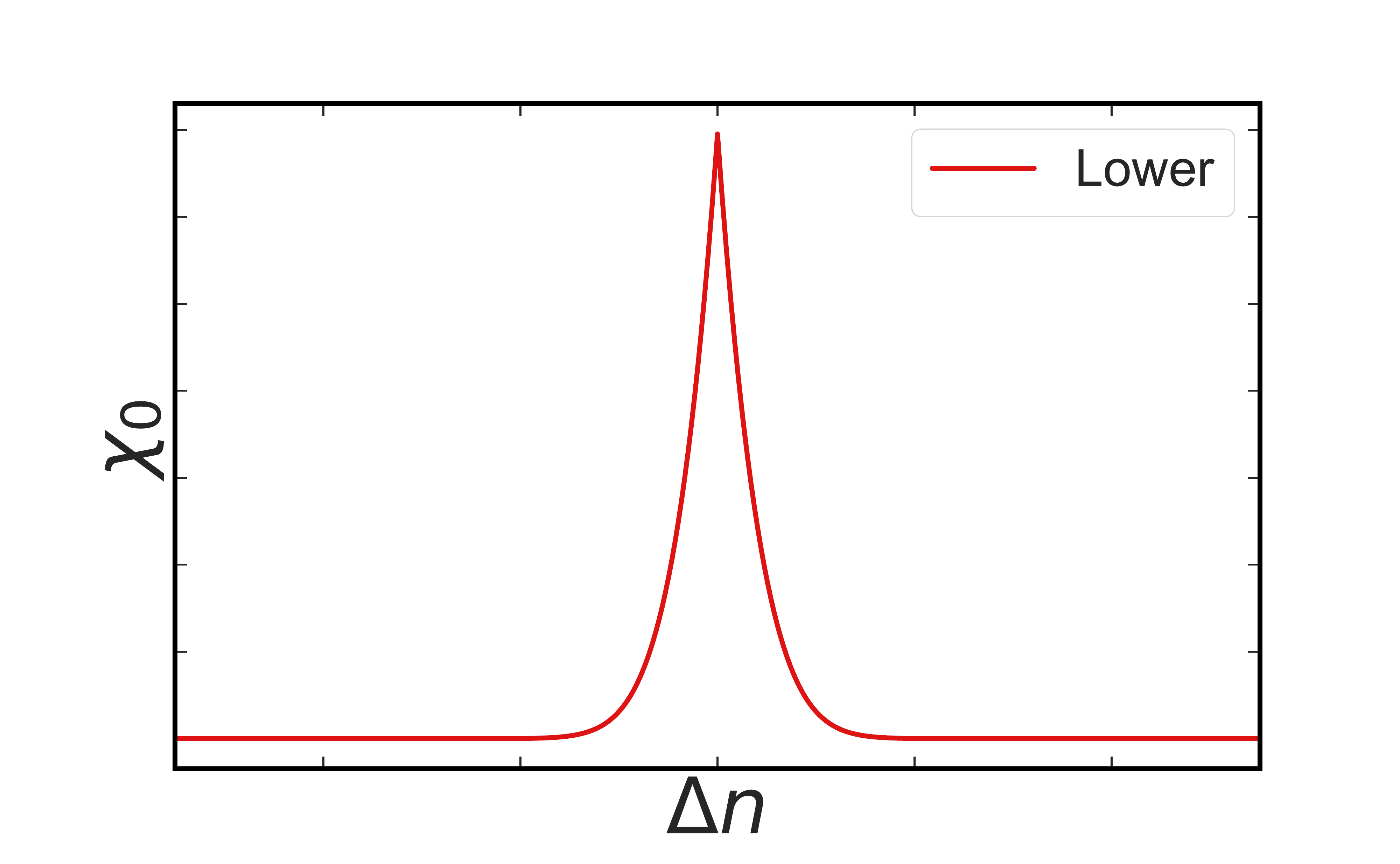}
    \put(-22,12){(b)}\\
    \includegraphics[width=0.48\linewidth, trim=5 15 55 65, clip]{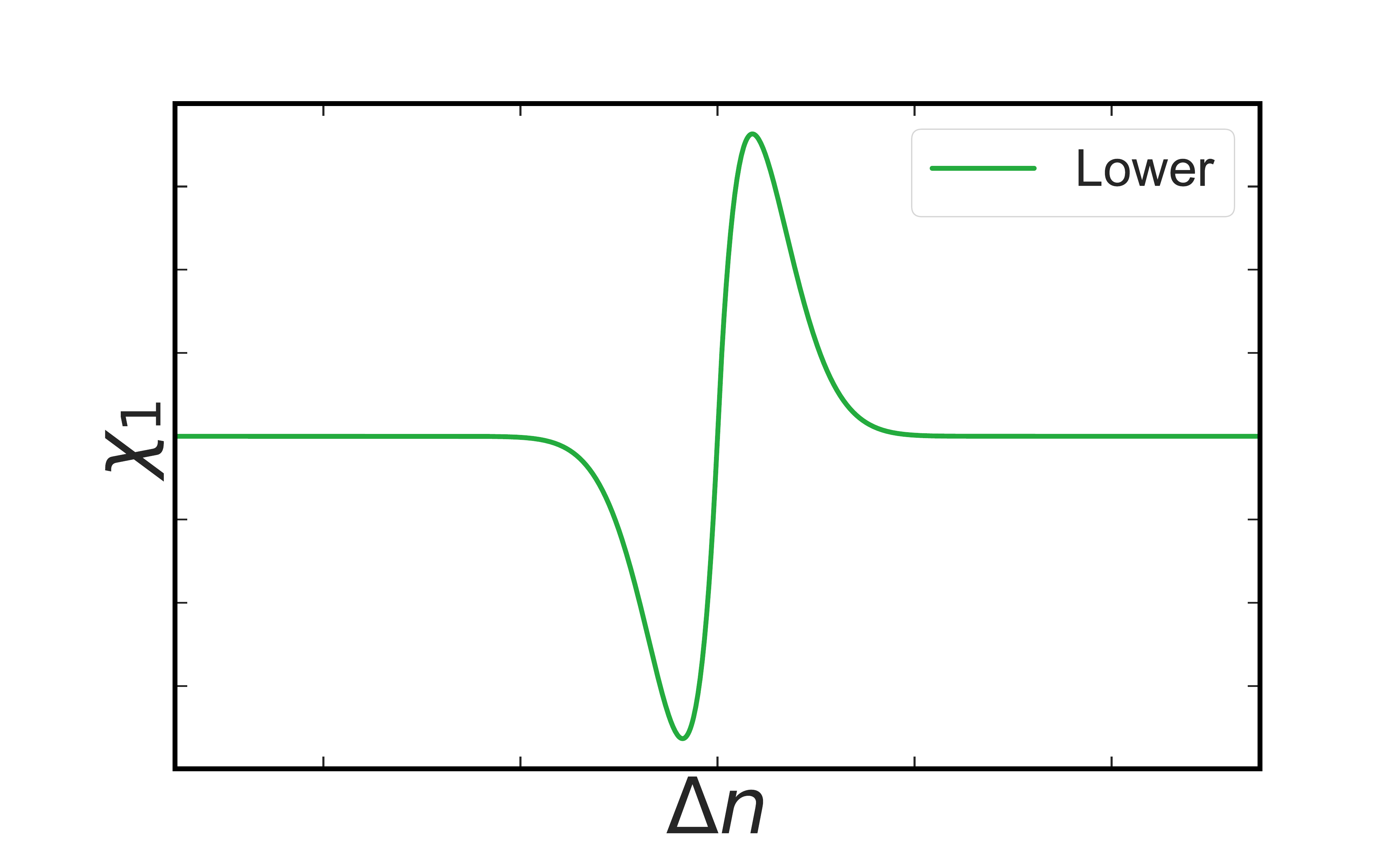}
    \put(-22,12){(c)}
    \hspace{0.025\linewidth}
    \includegraphics[width=0.48\linewidth, trim=5 15 55 65, clip]{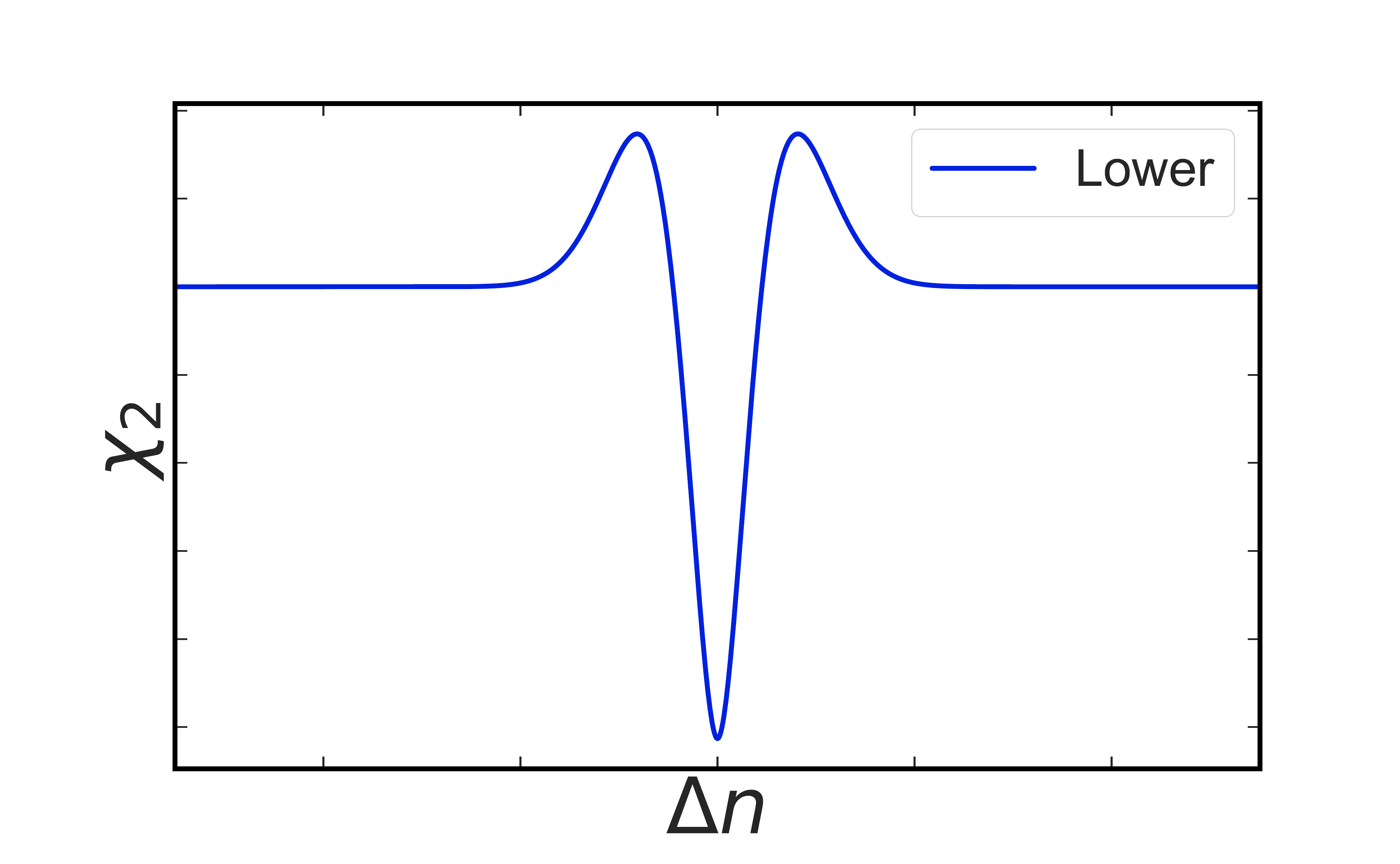}
    \put(-22,12){(d)}\\
    \includegraphics[width=0.48\linewidth, trim=5 15 55 65, clip]{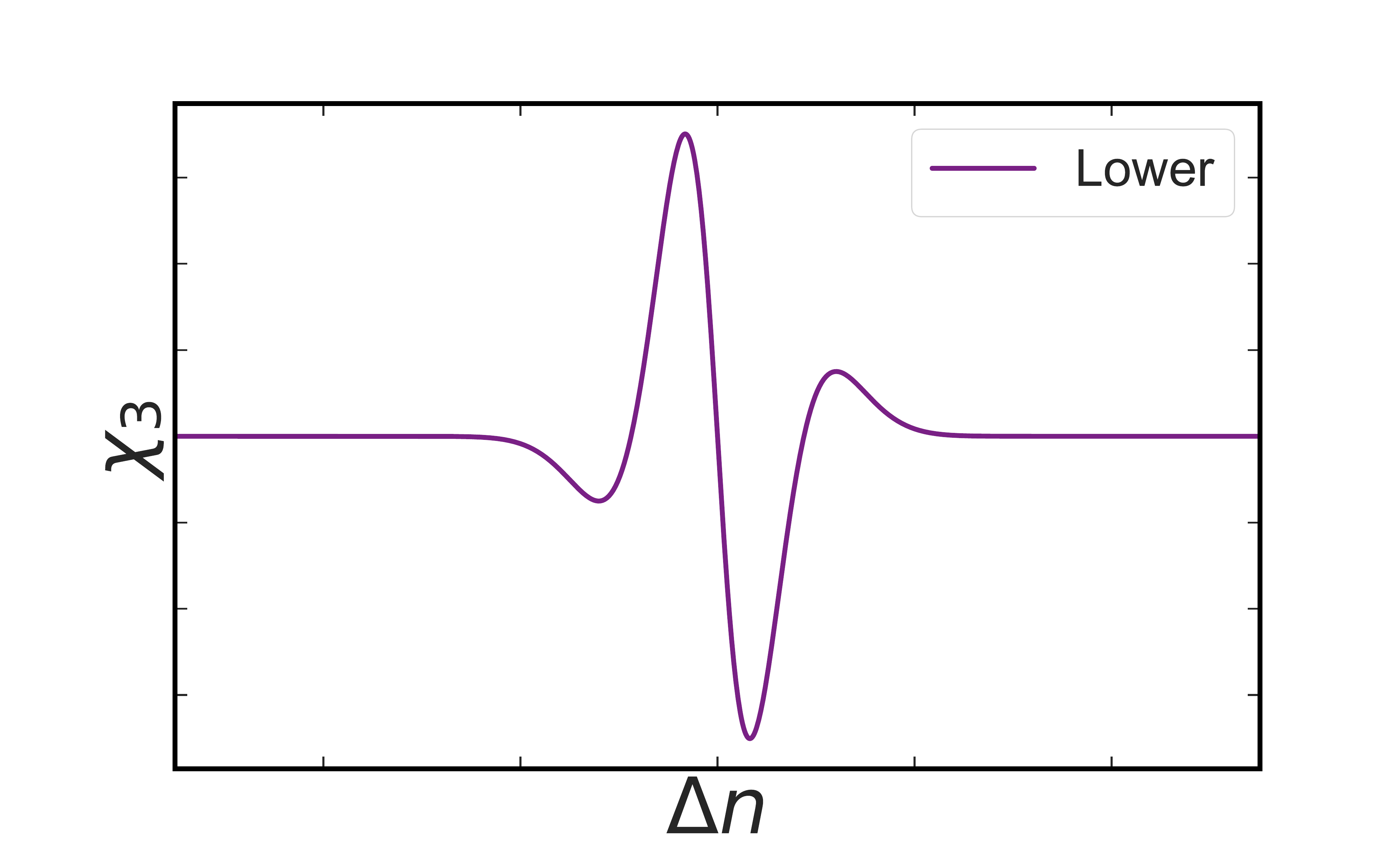}
    \put(-22,12){(e)}
    \hspace{0.025\linewidth}
    \includegraphics[width=0.48\linewidth, trim=5 15 55 65, clip]{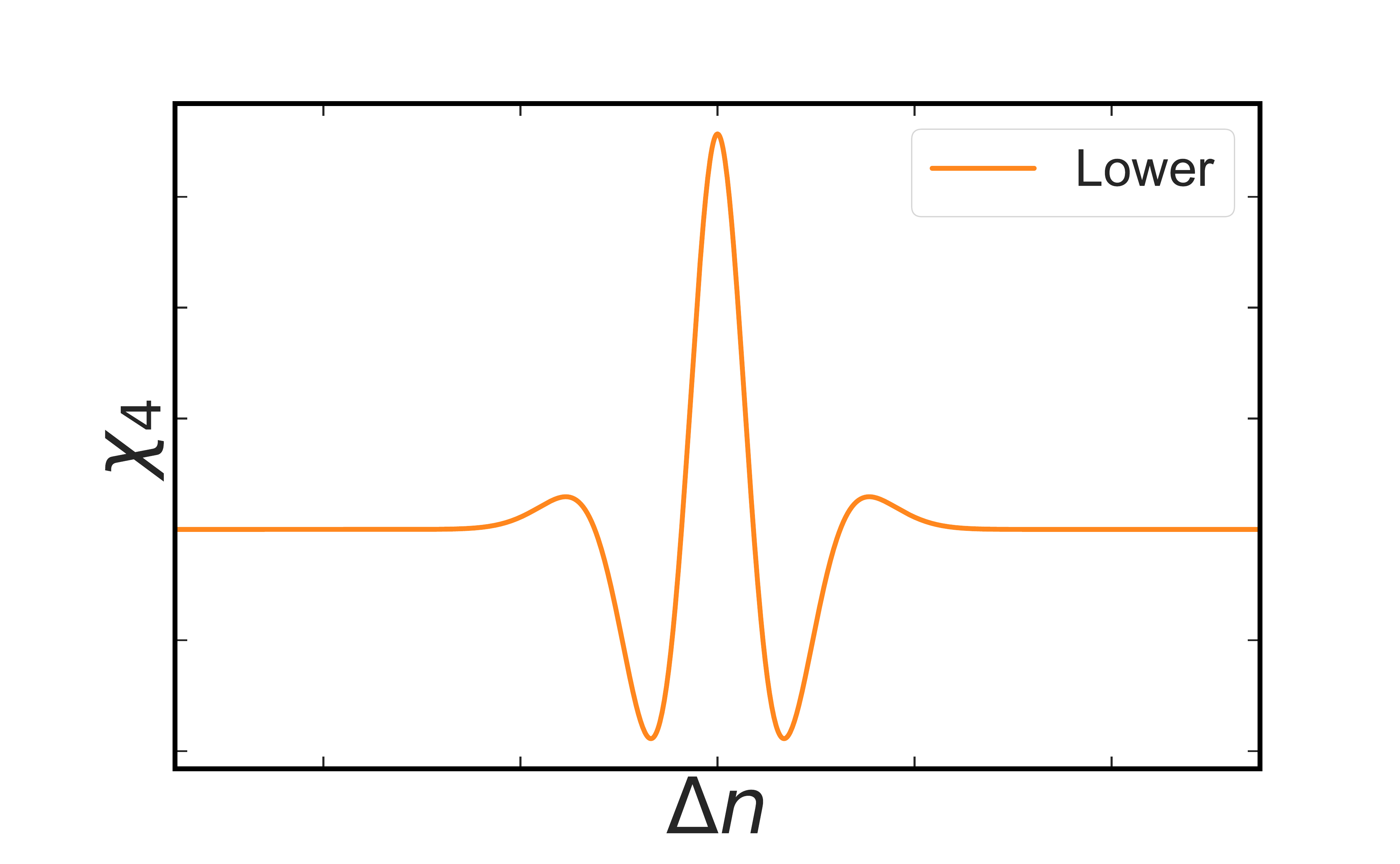}
    \put(-22,12){(f)}\\
    \includegraphics[width=0.48\linewidth, trim=5 15 55 65, clip]{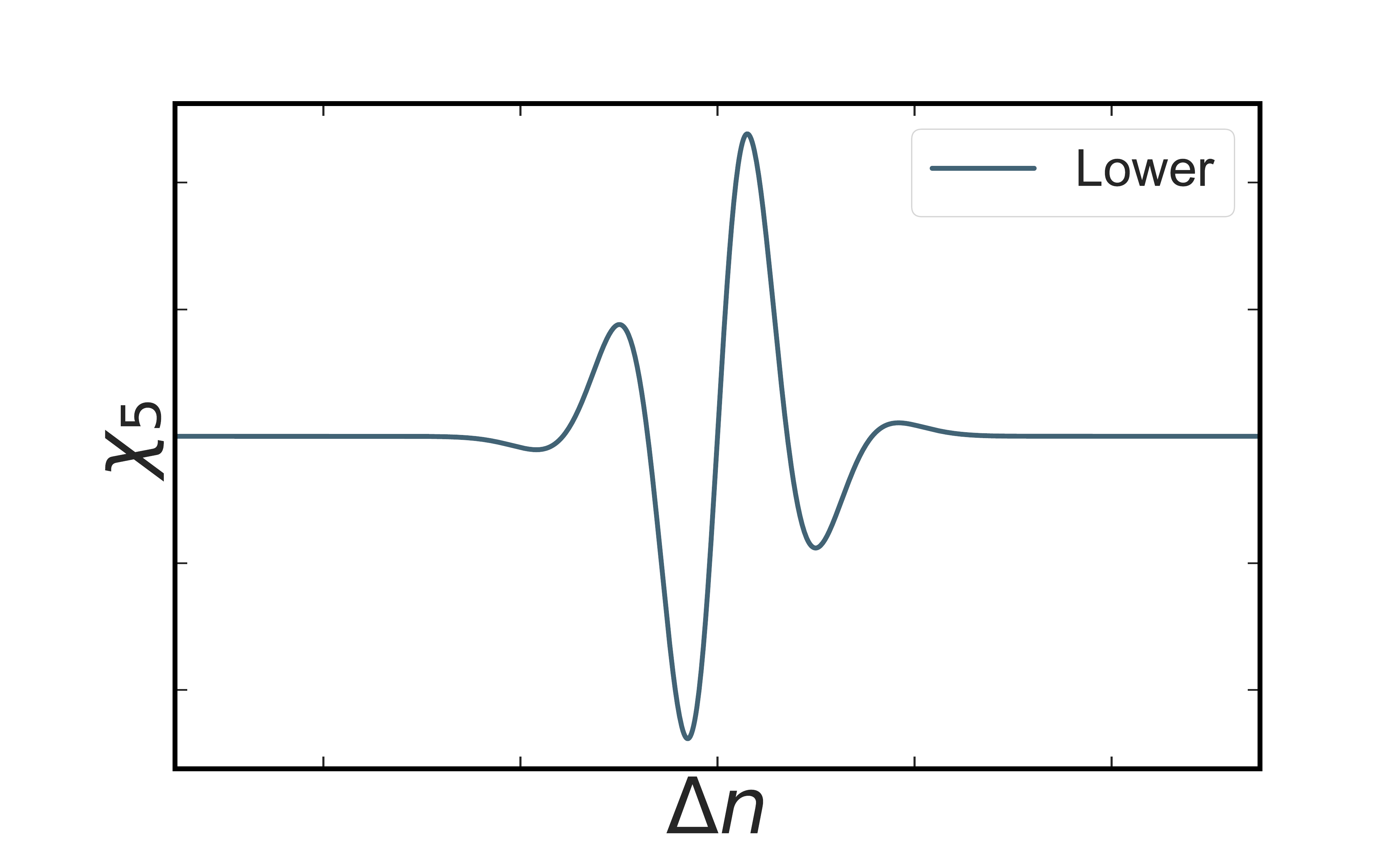}
    \put(-22,12){(g)}
    \hspace{0.025\linewidth}
    \includegraphics[width=0.48\linewidth, trim=5 15 55 65, clip]{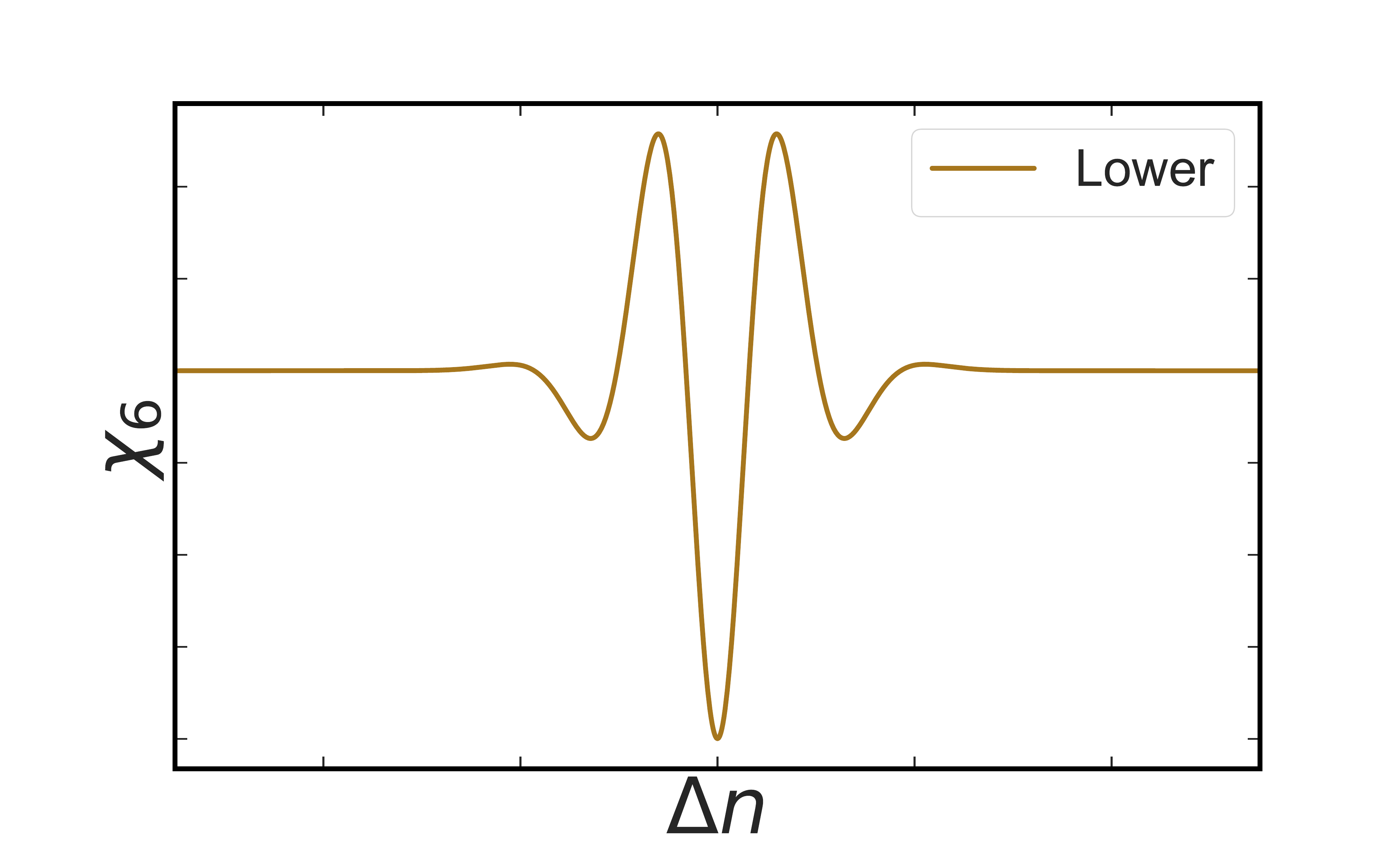}
    \put(-22,12){(h)}\\
    \includegraphics[width=0.48\linewidth, trim=5 15 55 65, clip]{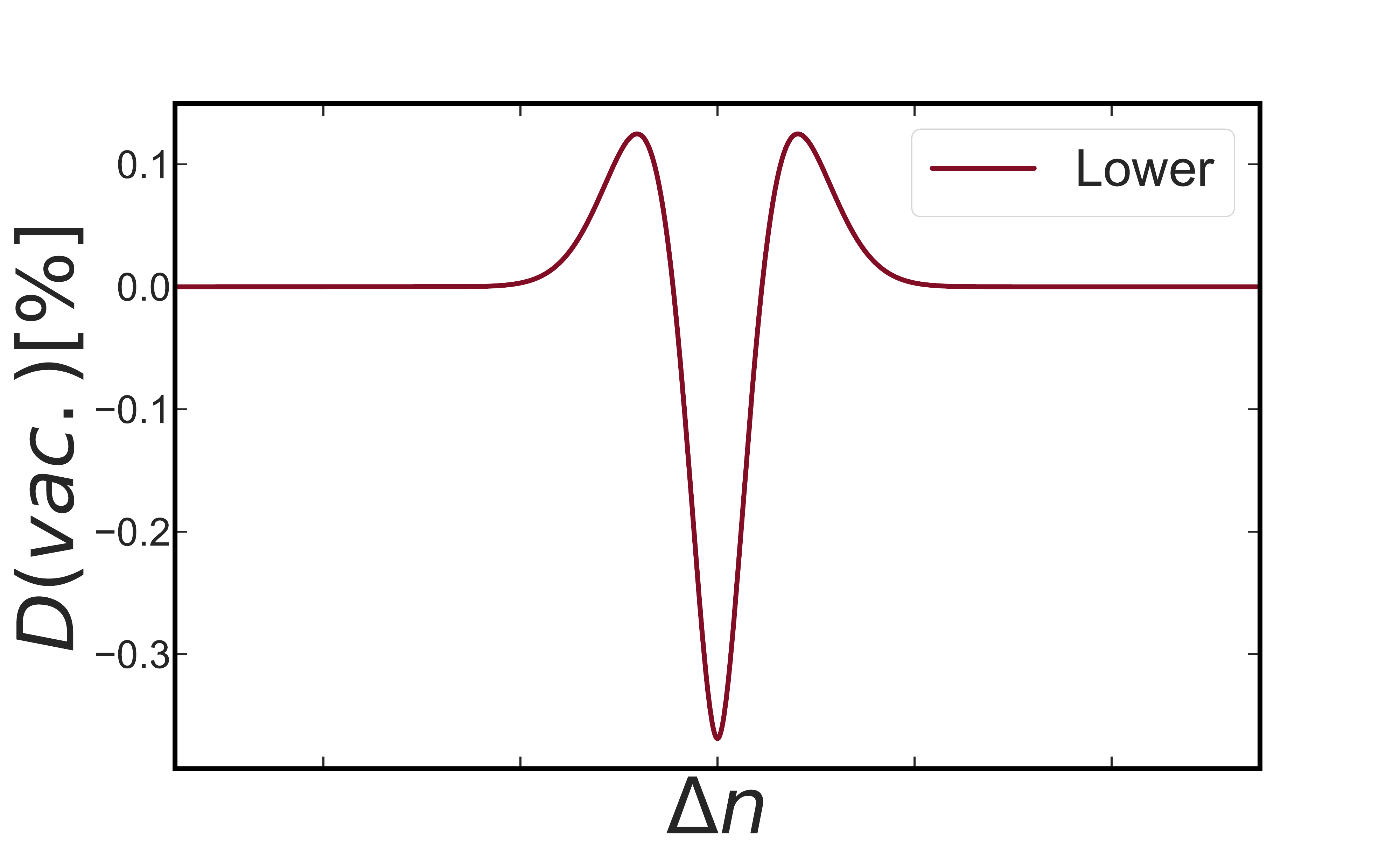}
    \put(-22,12){(i)}
    \hspace{0.025\linewidth}
    \includegraphics[width=0.48\linewidth, trim=5 15 55 65, clip]{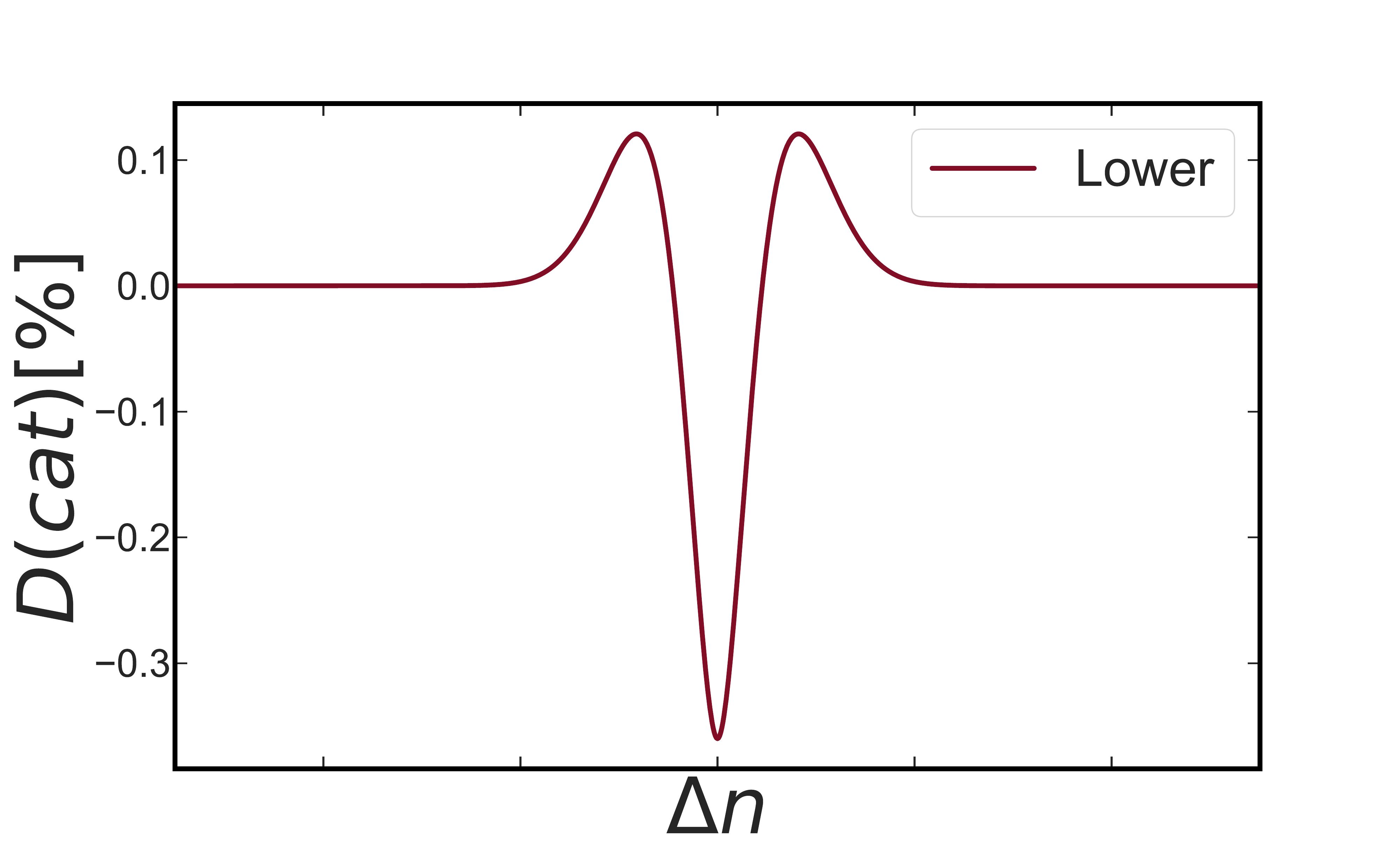}
    \put(-22,12){(j)}\\
    \includegraphics[width=0.99\linewidth, trim=20 10 200 25, clip]{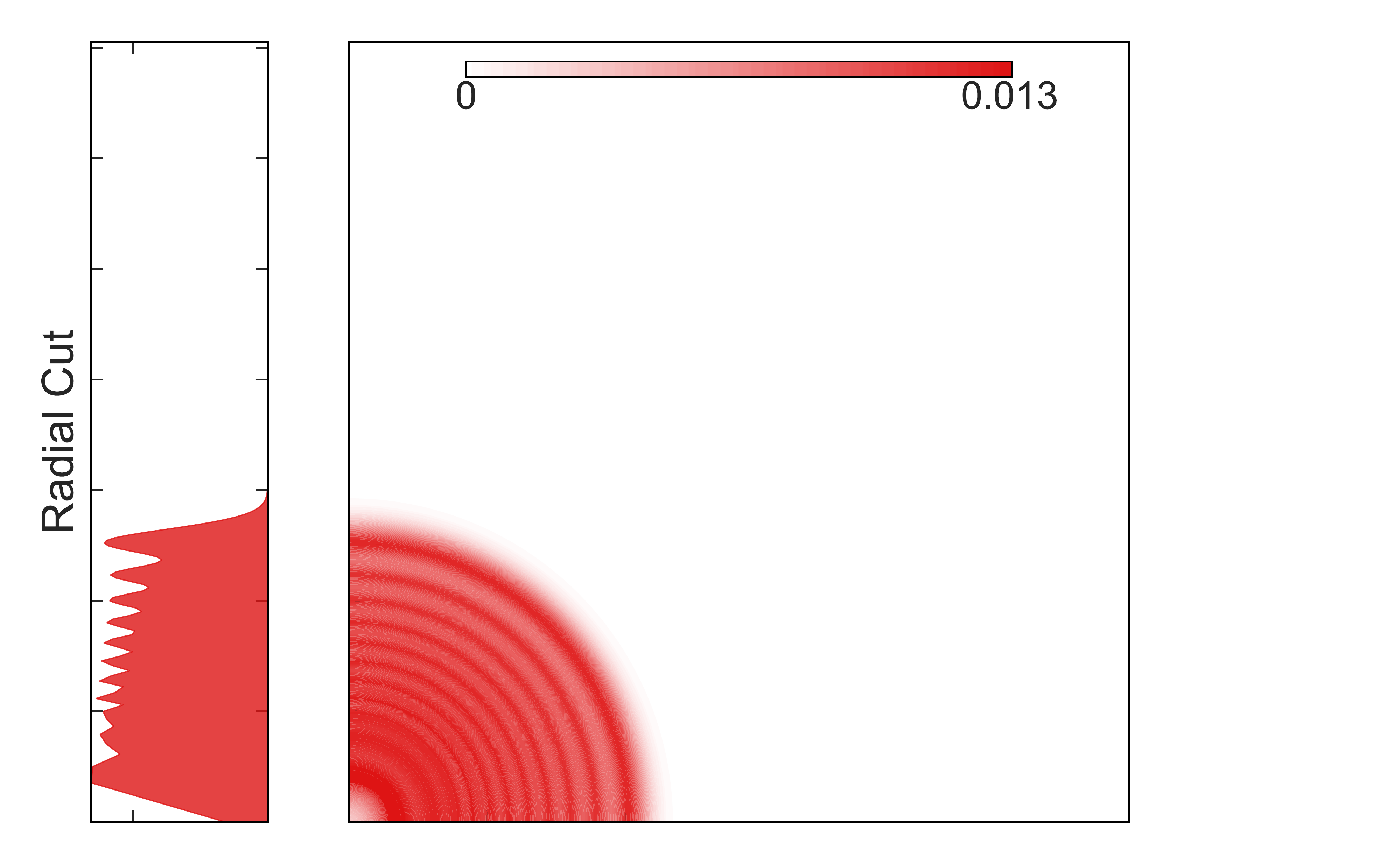}
    \put(-35,12){(k)}
    \vspace{-10pt}
    \caption{Lower BCS probe: see caption of Fig.~\ref{CoherentStats}.}
    \label{LowerStats}
\end{figure}

\begin{figure}
    \centering
    \includegraphics[width=0.48\linewidth, trim=5 15 55 65, clip]{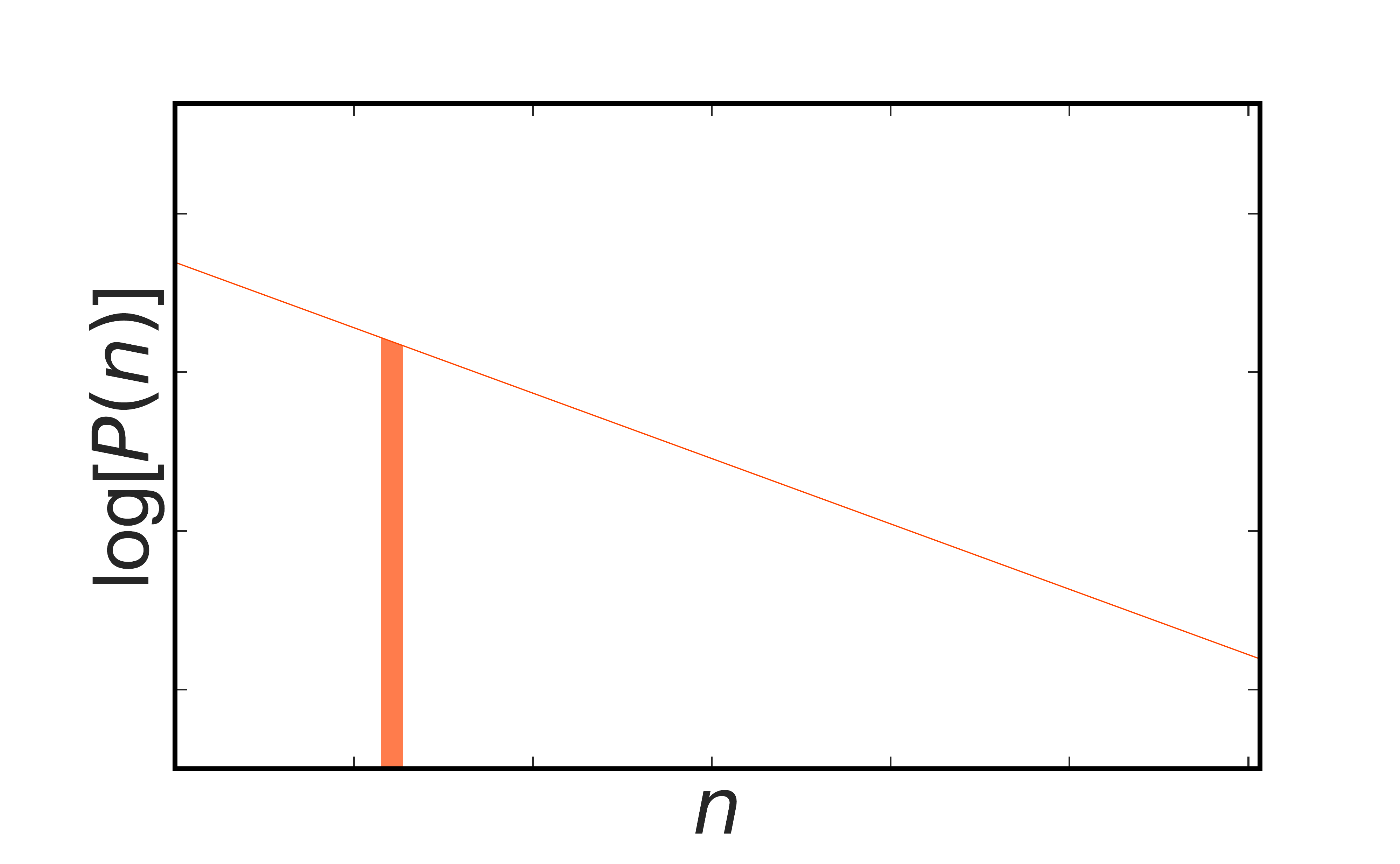}
    \put(-22,12){(a)}
    \hspace{0.025\linewidth}
    \includegraphics[width=0.48\linewidth, trim=5 15 55 65, clip]{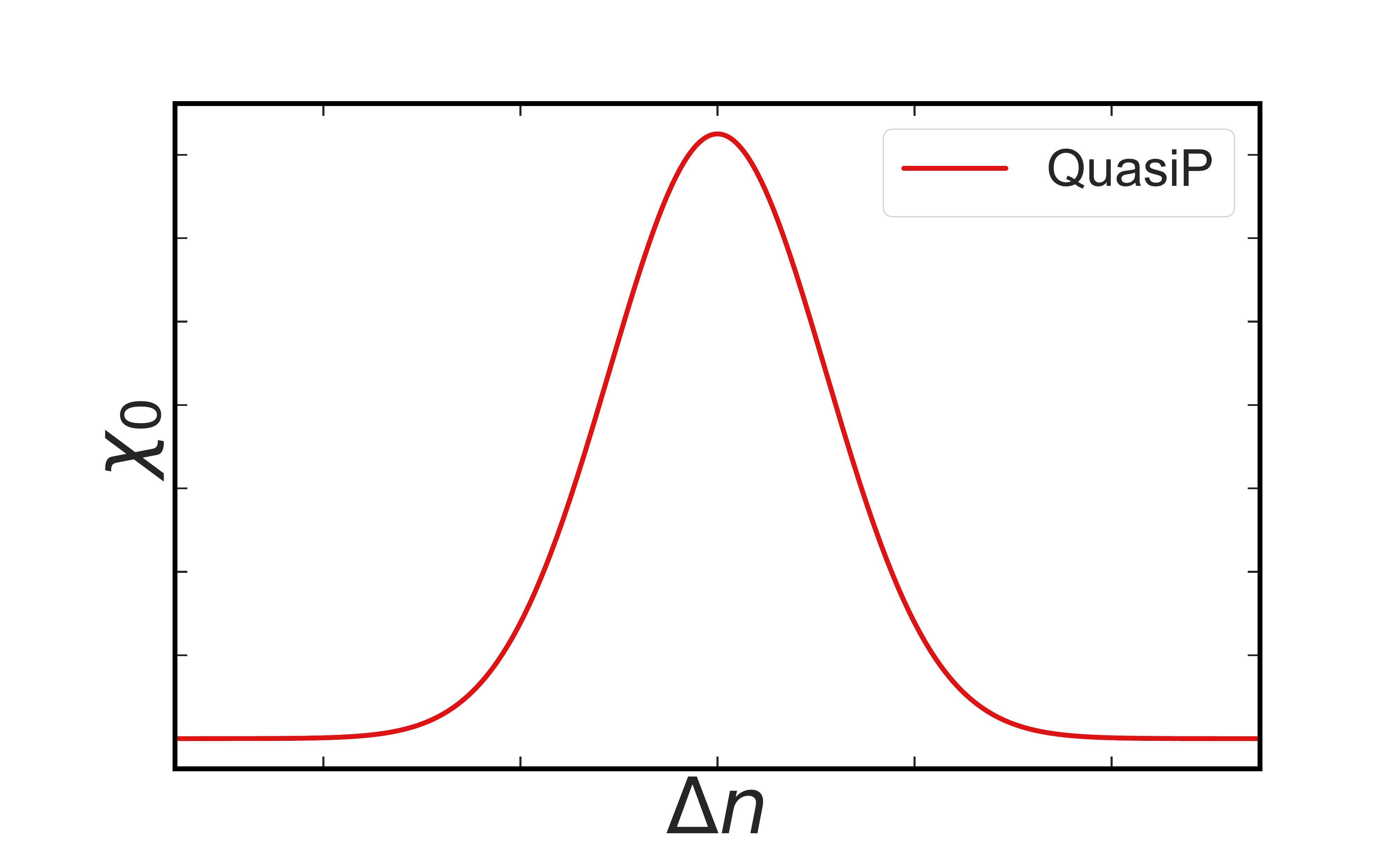}
    \put(-22,12){(b)}\\
    \includegraphics[width=0.48\linewidth, trim=5 15 55 65, clip]{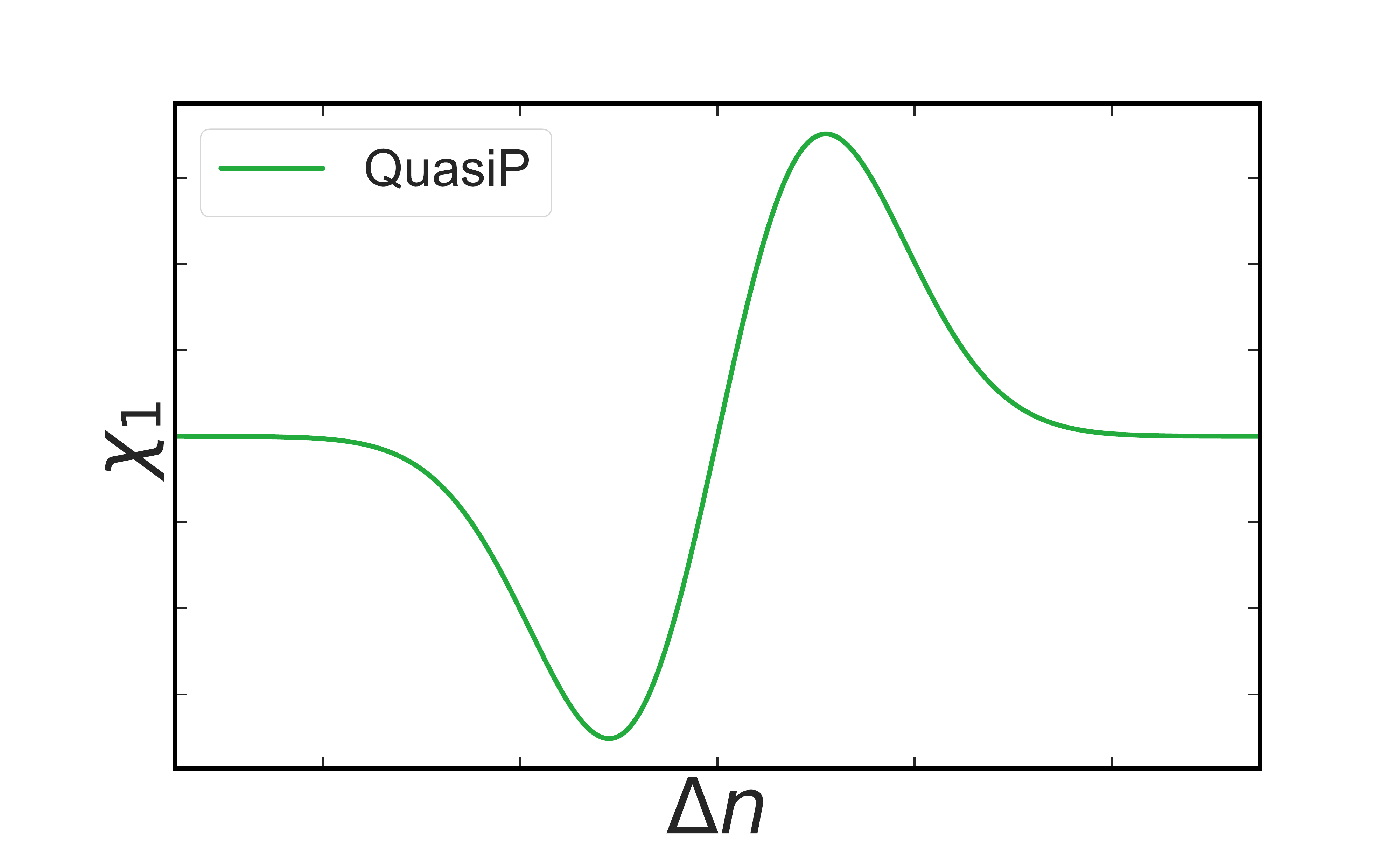}
    \put(-22,12){(c)}
    \hspace{0.025\linewidth}
    \includegraphics[width=0.48\linewidth, trim=5 15 55 65, clip]{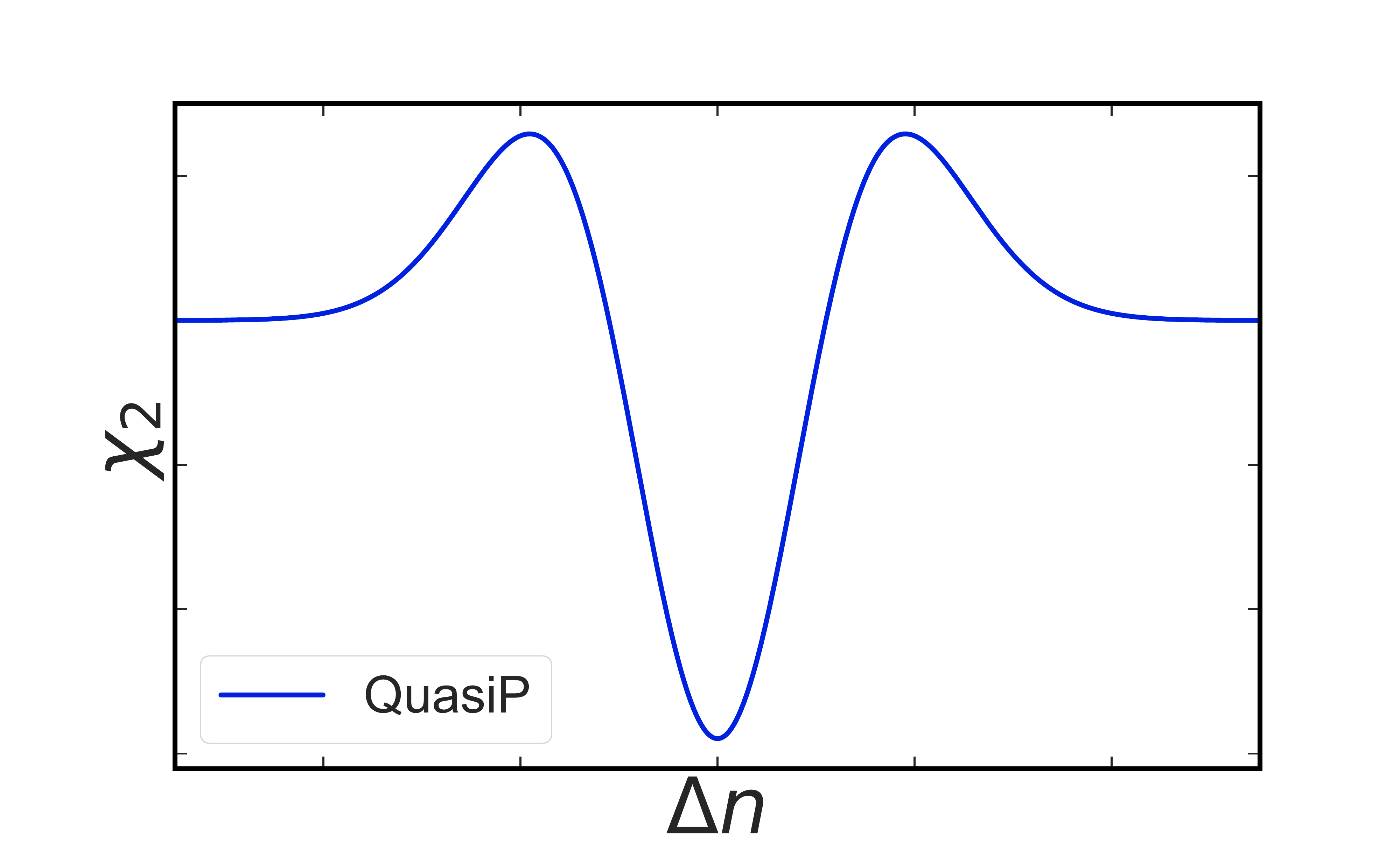}
    \put(-22,12){(d)}\\
    \includegraphics[width=0.48\linewidth, trim=5 15 55 65, clip]{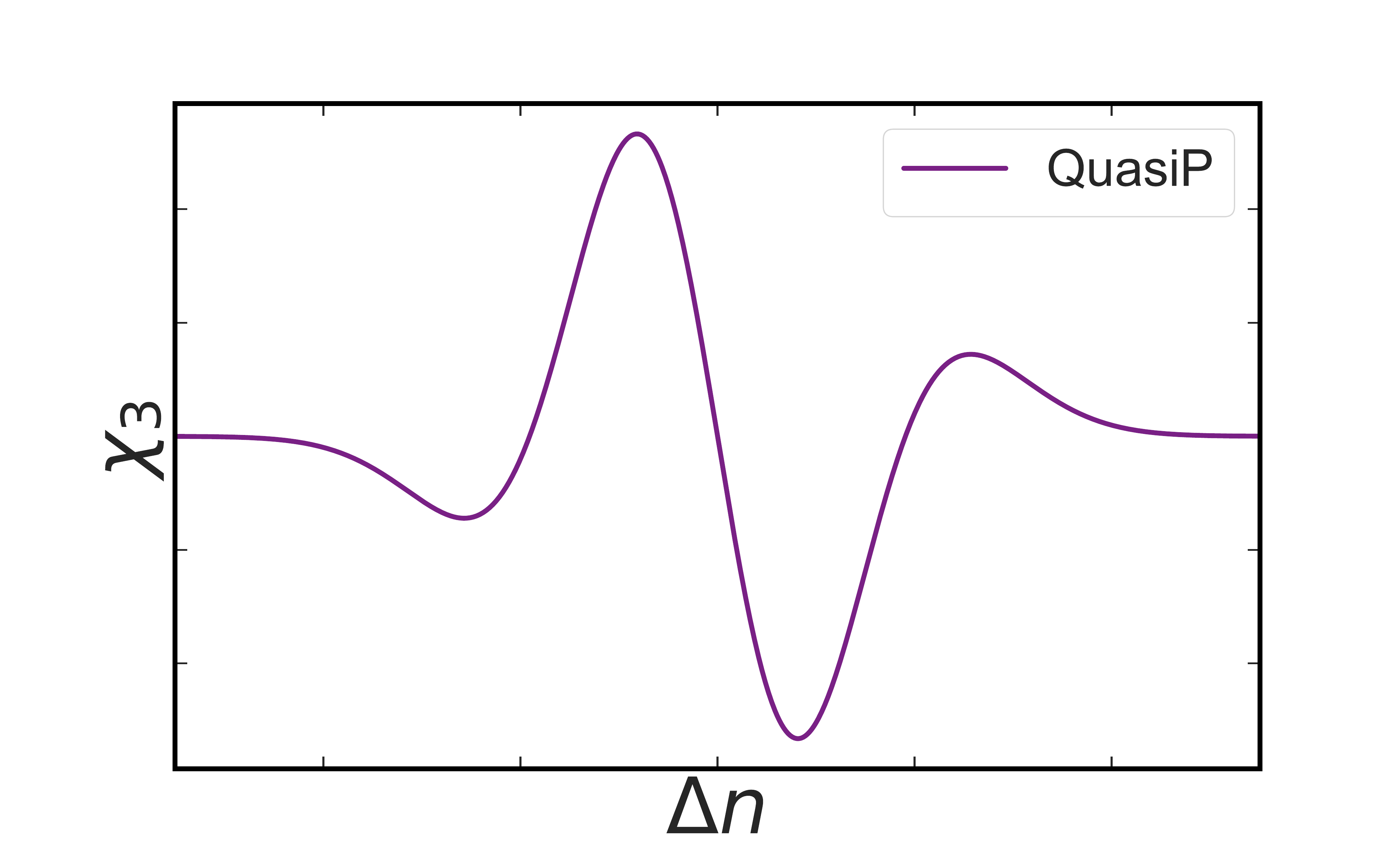}
    \put(-22,12){(e)}
    \hspace{0.025\linewidth}
    \includegraphics[width=0.48\linewidth, trim=5 15 55 65, clip]{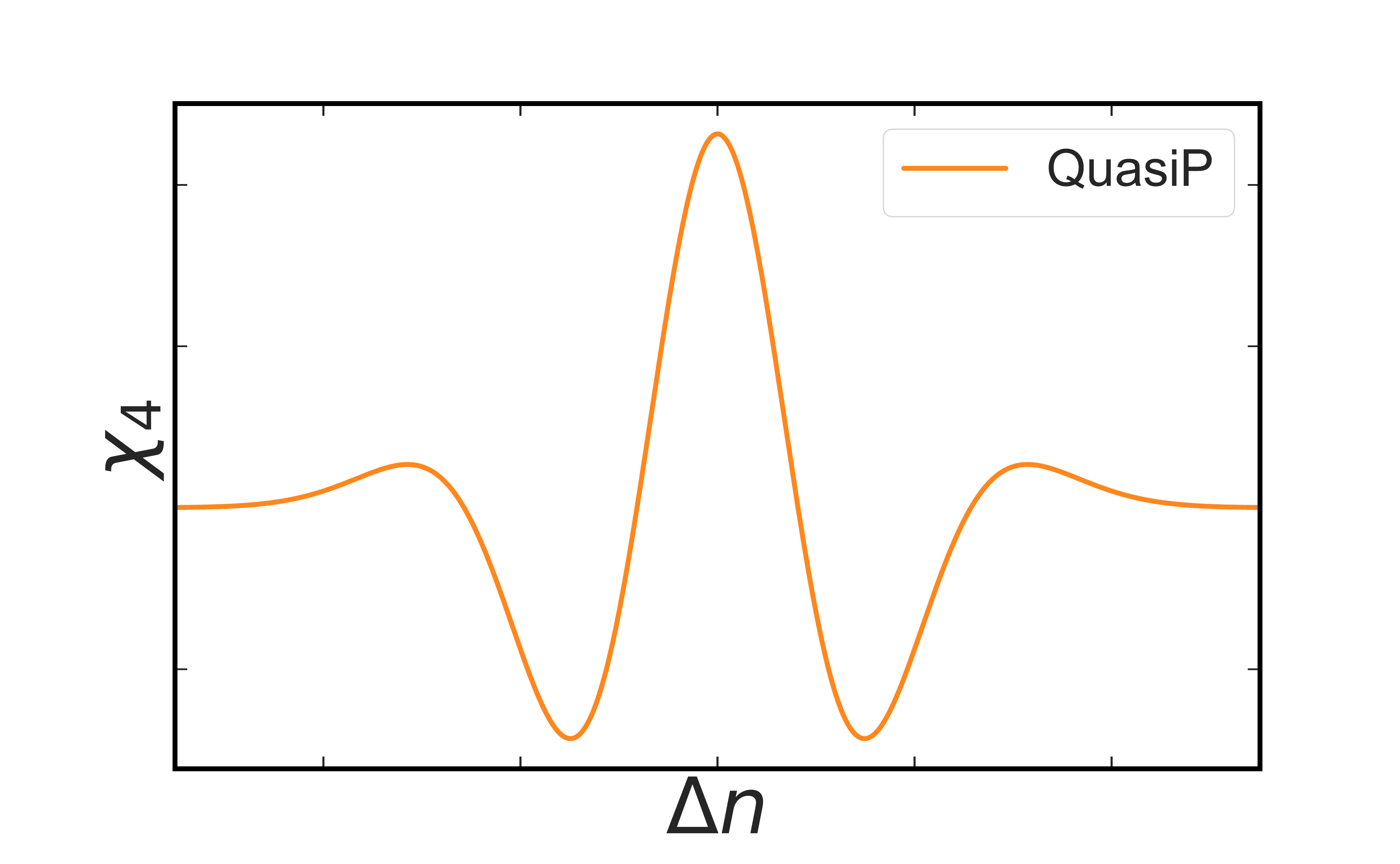}
    \put(-22,12){(f)}\\
    \includegraphics[width=0.48\linewidth, trim=5 15 55 65, clip]{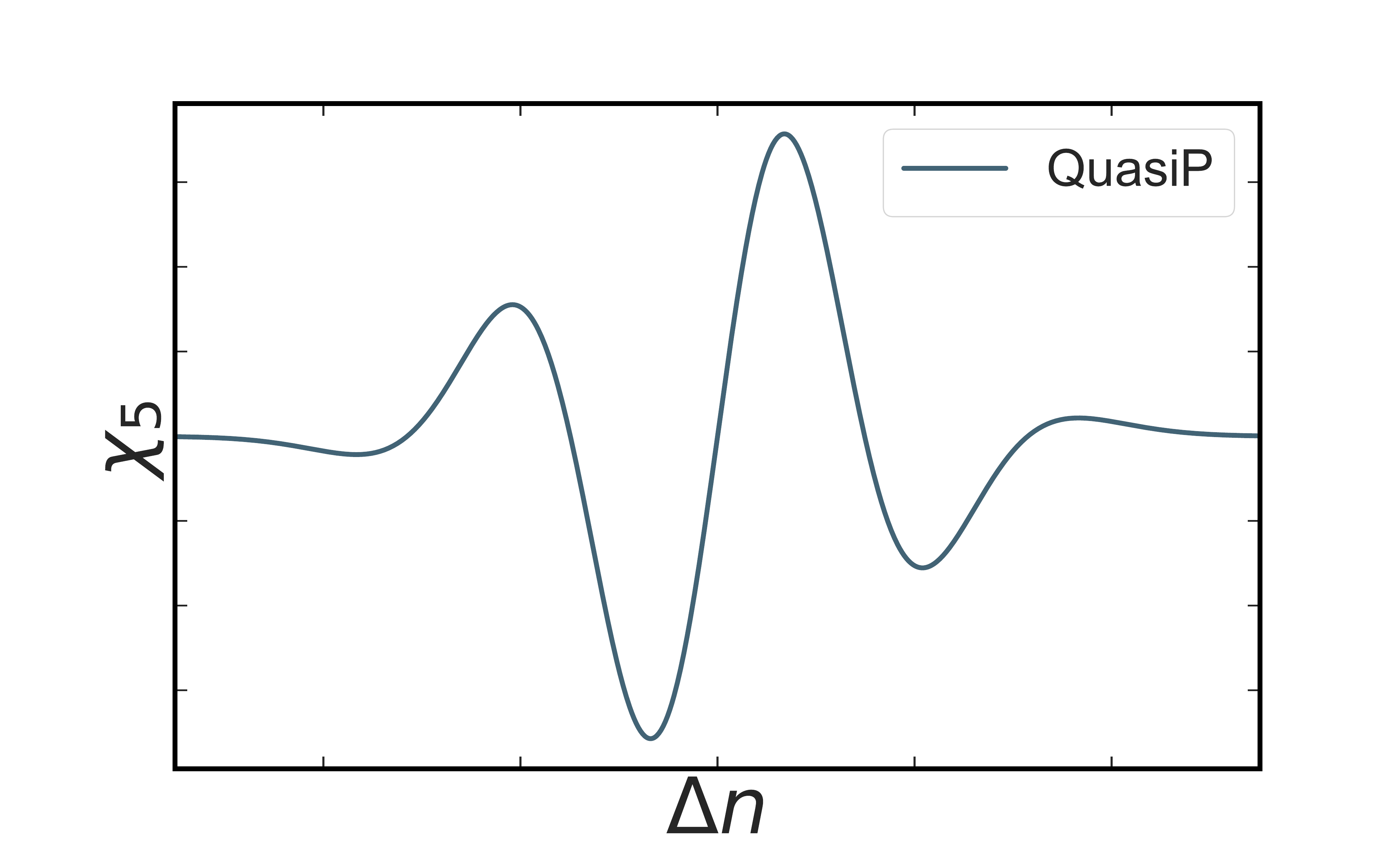}
    \put(-22,12){(g)}
    \hspace{0.025\linewidth}
    \includegraphics[width=0.48\linewidth, trim=5 15 55 65, clip]{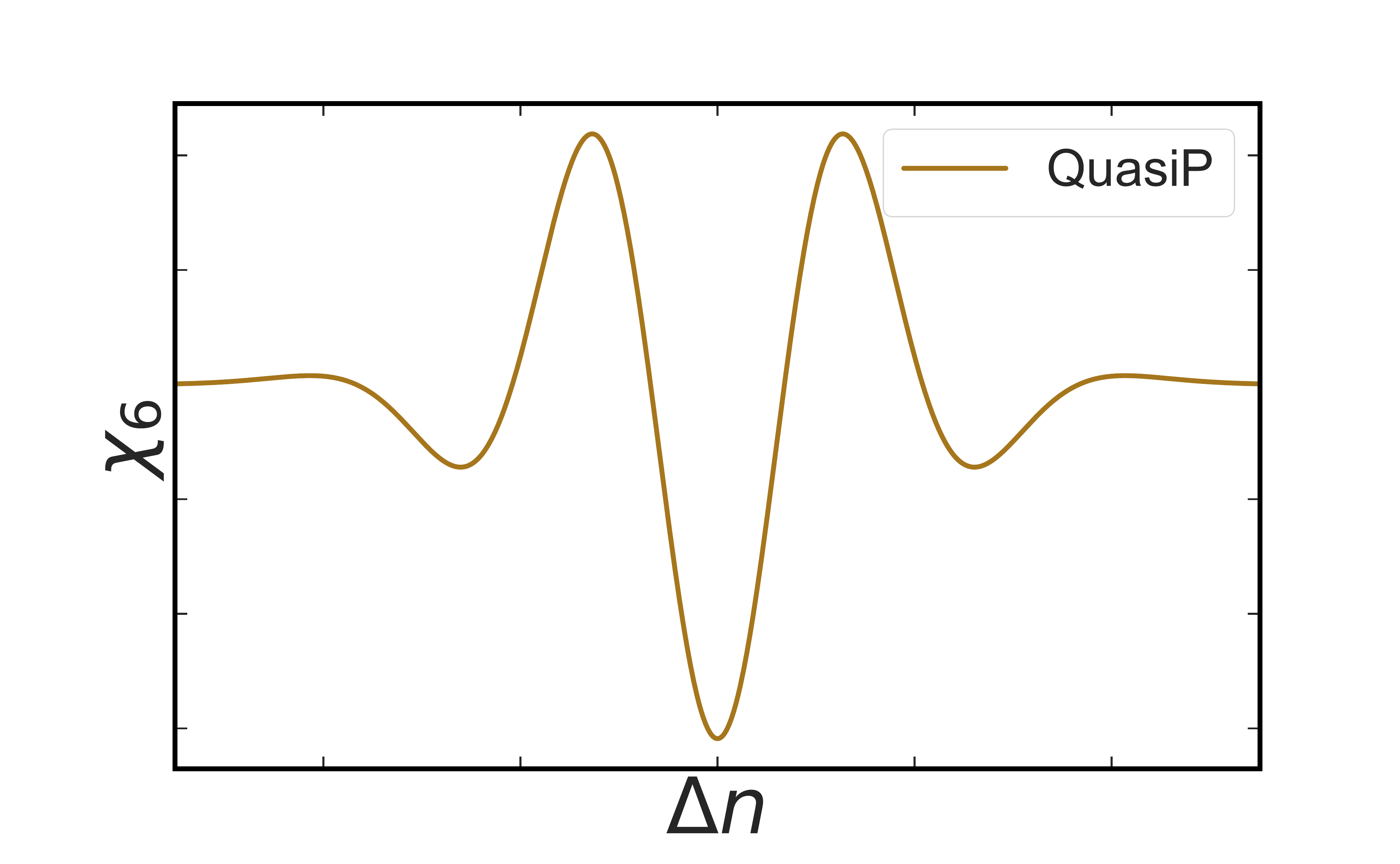}
    \put(-22,12){(h)}\\
    \includegraphics[width=0.48\linewidth, trim=5 15 55 65, clip]{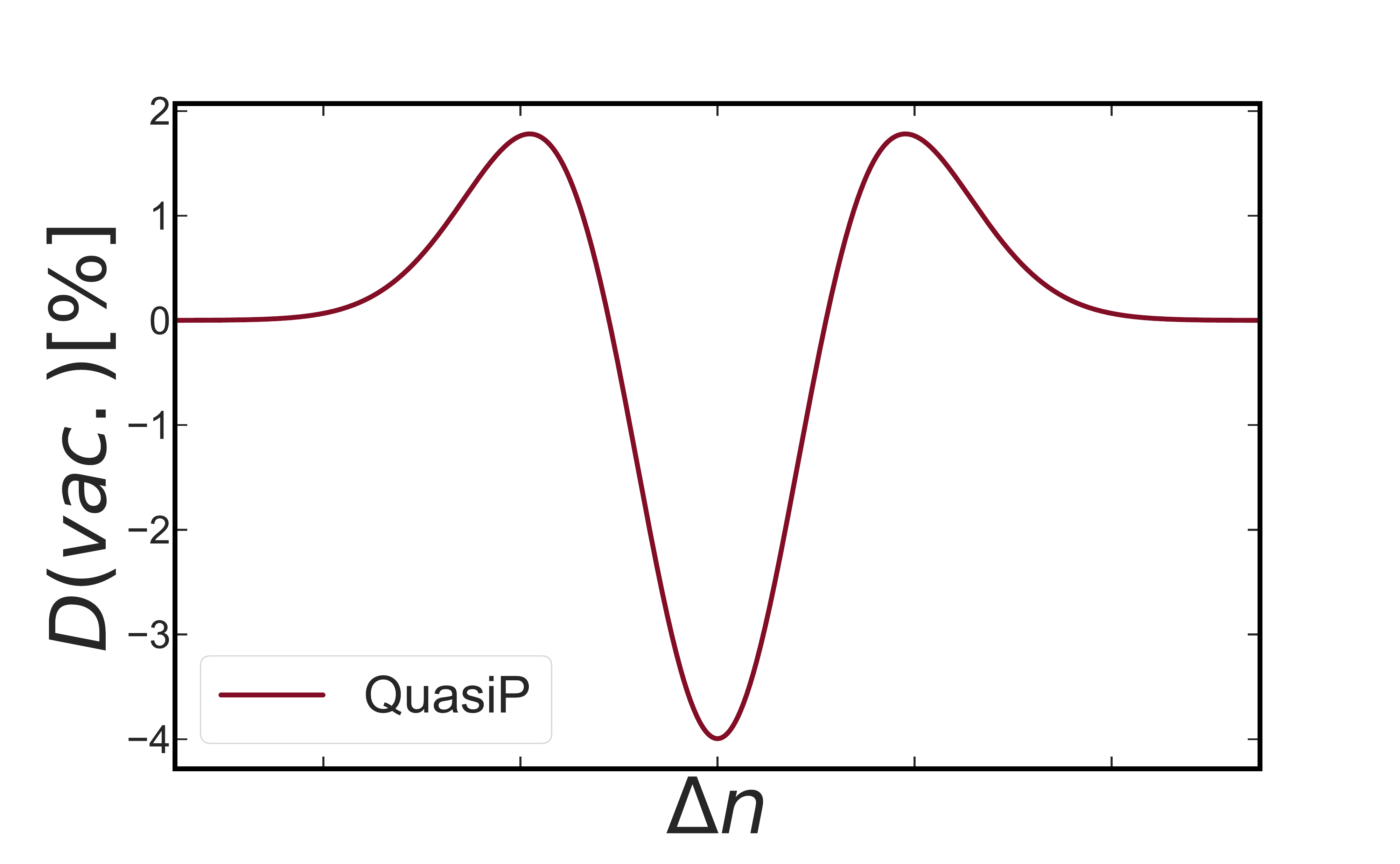}
    \put(-22,12){(i)}
    \hspace{0.025\linewidth}
    \includegraphics[width=0.48\linewidth, trim=5 15 55 65, clip]{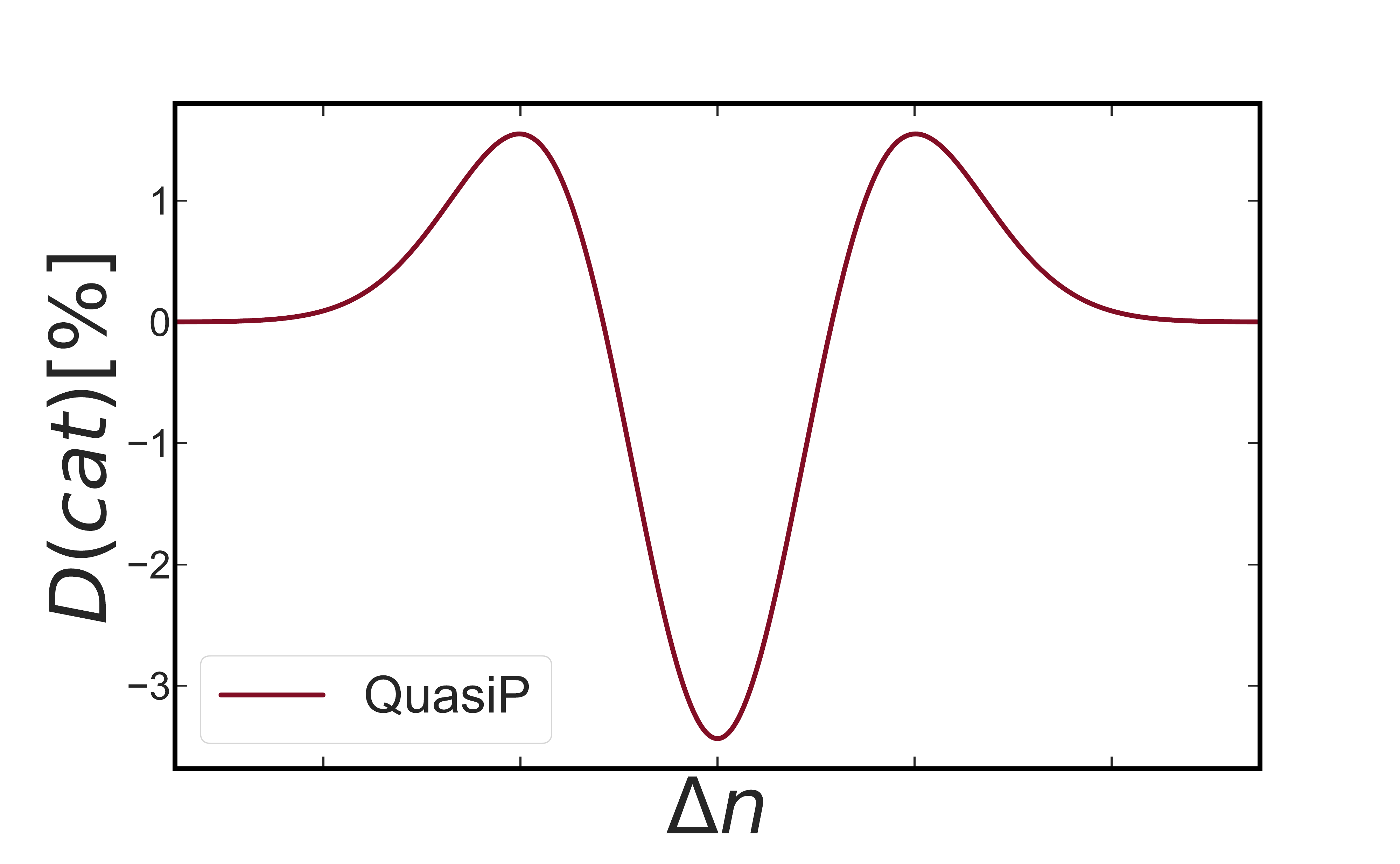}
    \put(-22,12){(j)}\\
    \includegraphics[width=0.99\linewidth, trim=20 10 200 25, clip]{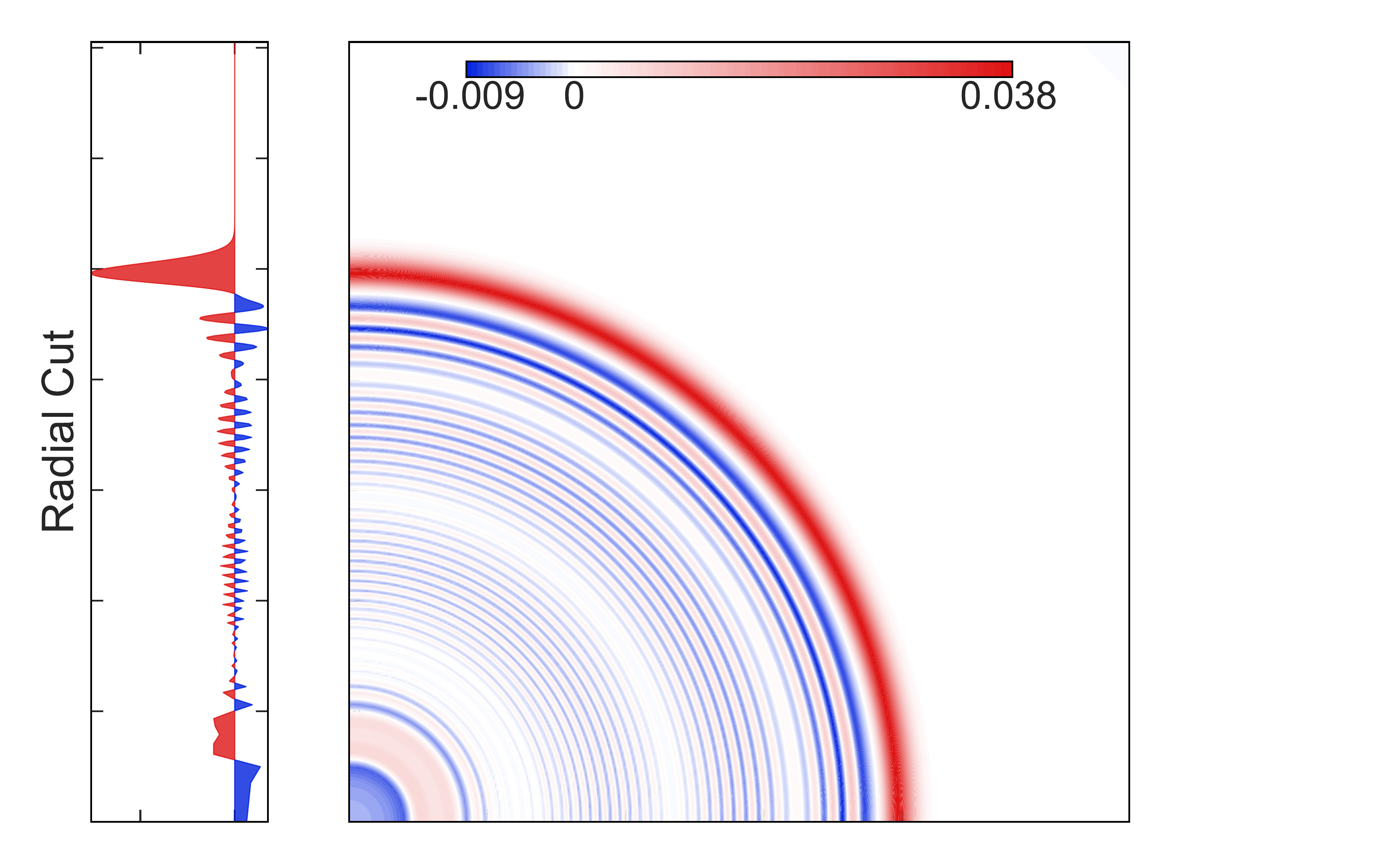}
    \put(-35,12){(k)}
    \vspace{-10pt}
    \caption{Quasi-Poissonian BCS probe: see caption of Fig.~\ref{CoherentStats}.}
    \label{QuasiPStats}
\end{figure}

\begin{figure}
    \centering
    \includegraphics[width=0.48\linewidth, trim=5 15 55 65, clip]{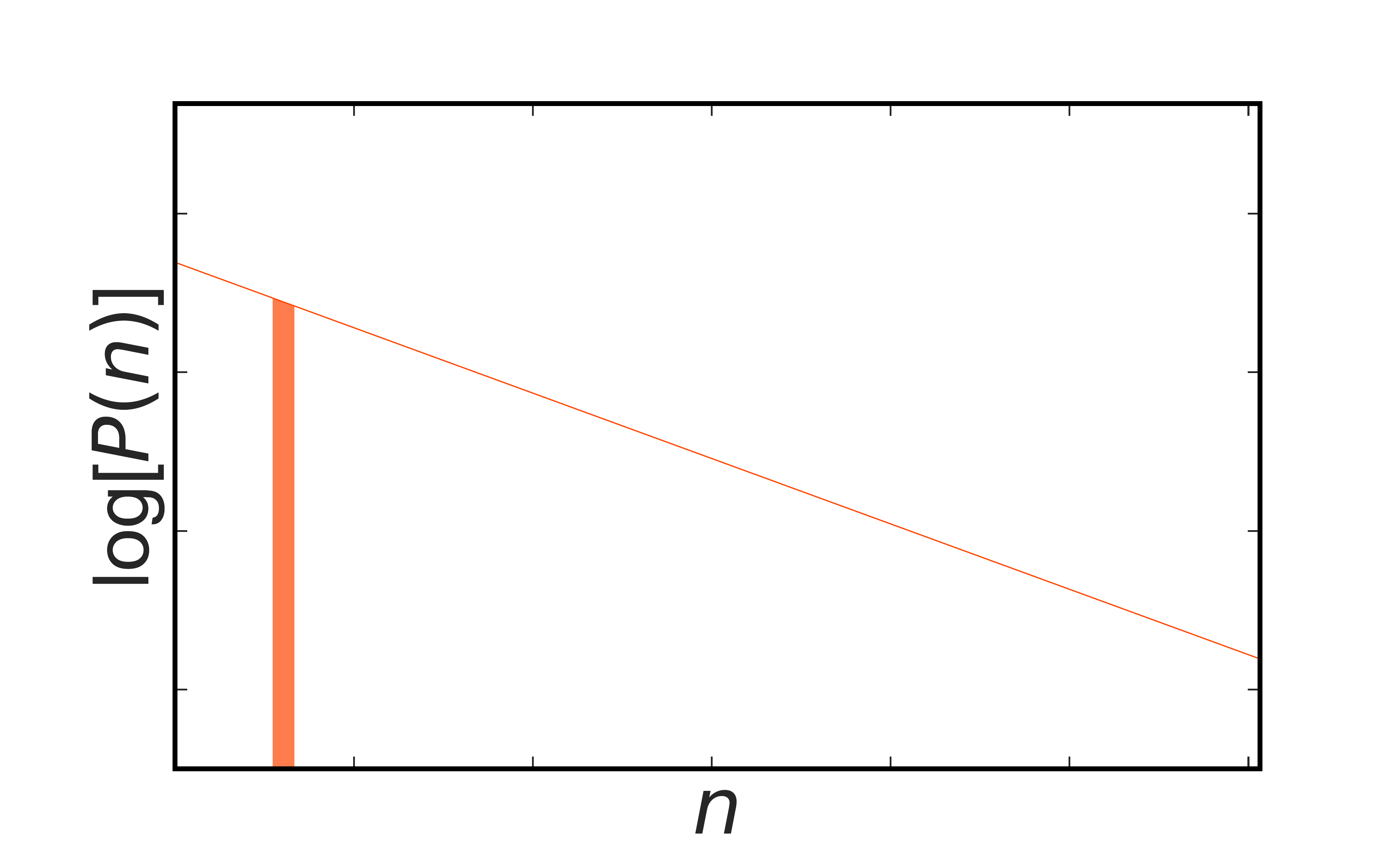}
    \put(-22,12){(a)}
    \hspace{0.025\linewidth}
    \includegraphics[width=0.48\linewidth, trim=5 15 55 65, clip]{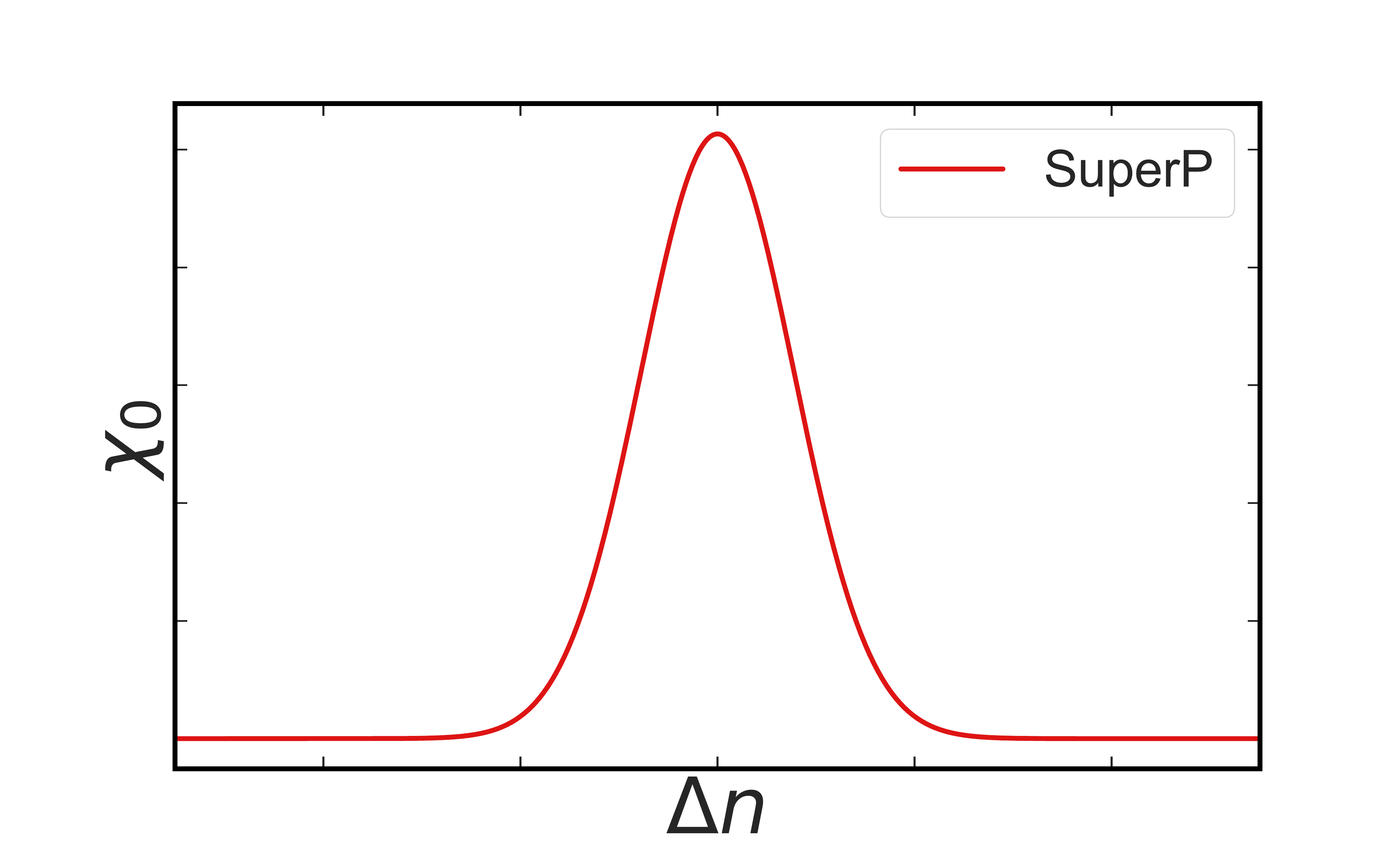}
    \put(-22,12){(b)}\\
    \includegraphics[width=0.48\linewidth, trim=5 15 55 65, clip]{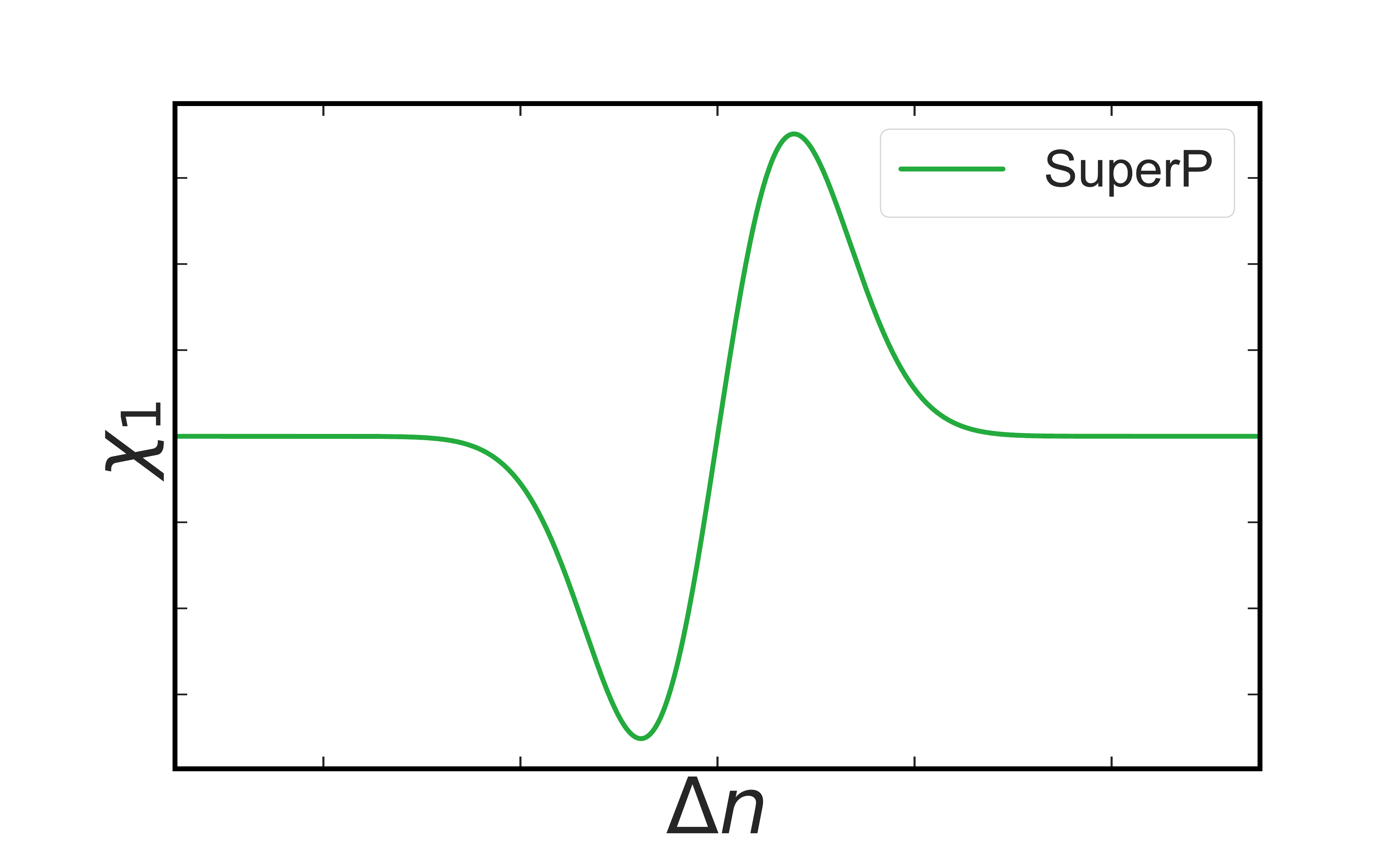}
    \put(-22,12){(c)}
    \hspace{0.025\linewidth}
    \includegraphics[width=0.48\linewidth, trim=5 15 55 65, clip]{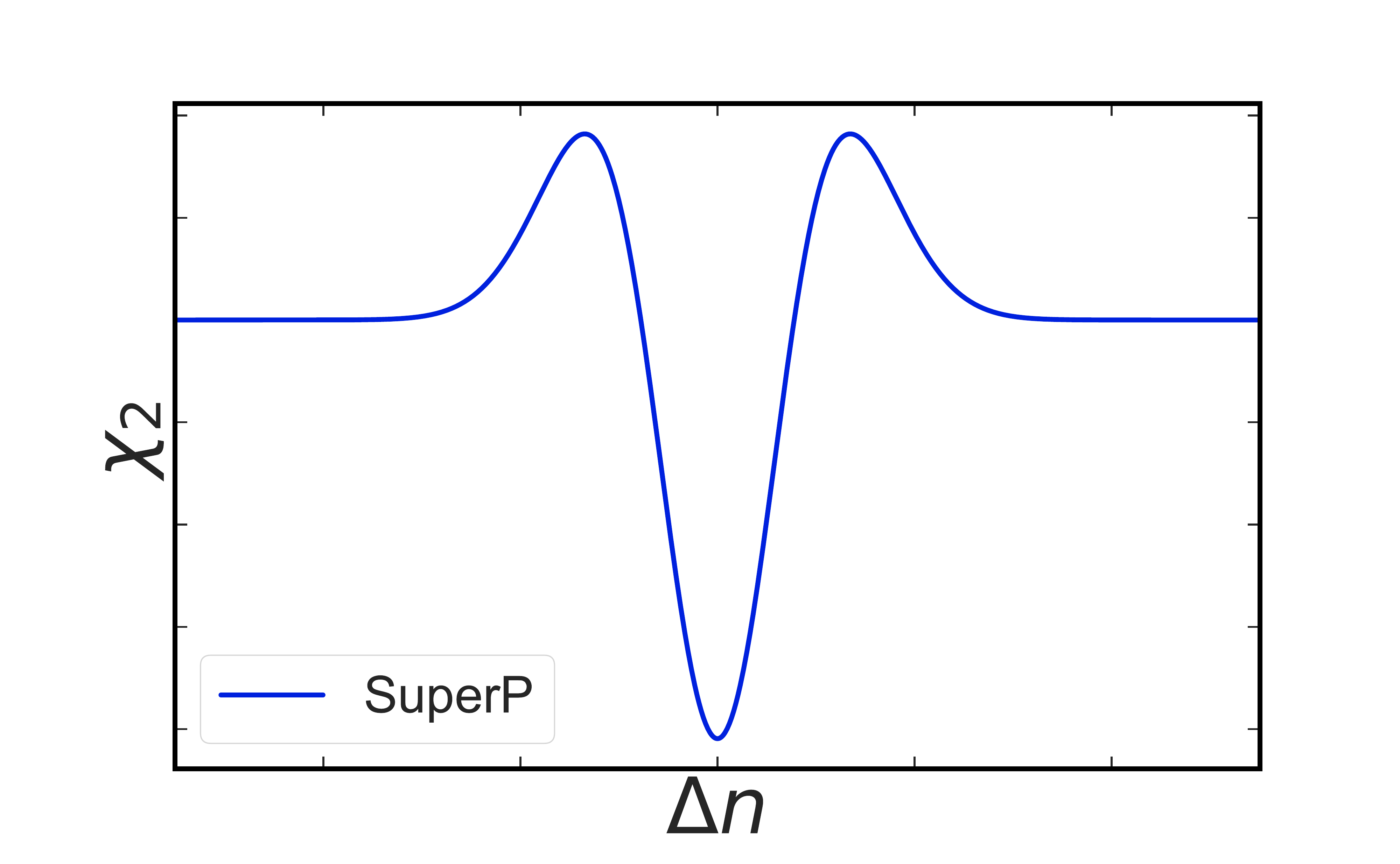}
    \put(-22,12){(d)}\\
    \includegraphics[width=0.48\linewidth, trim=5 15 55 65, clip]{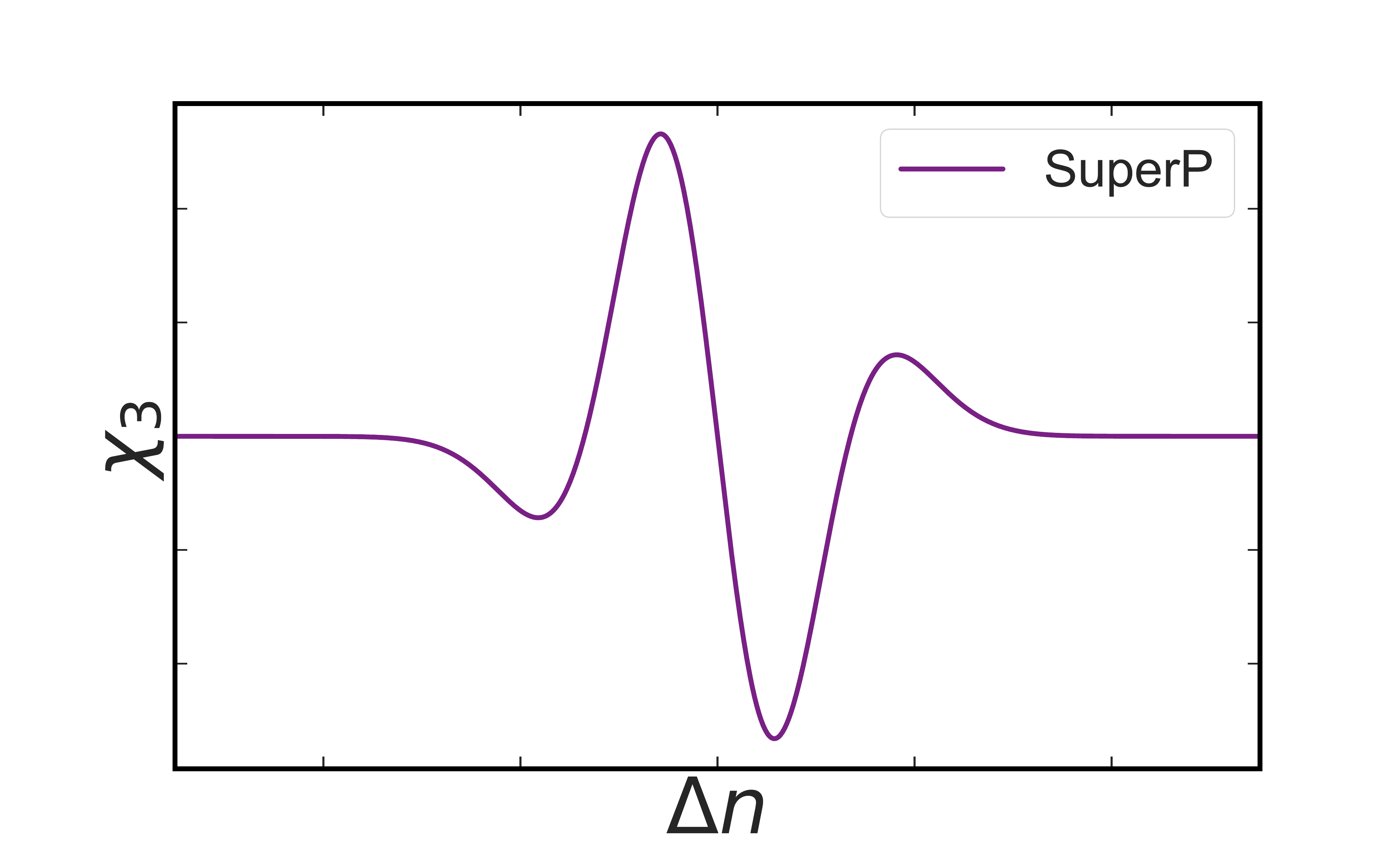}
    \put(-22,12){(e)}
    \hspace{0.025\linewidth}
    \includegraphics[width=0.48\linewidth, trim=5 15 55 65, clip]{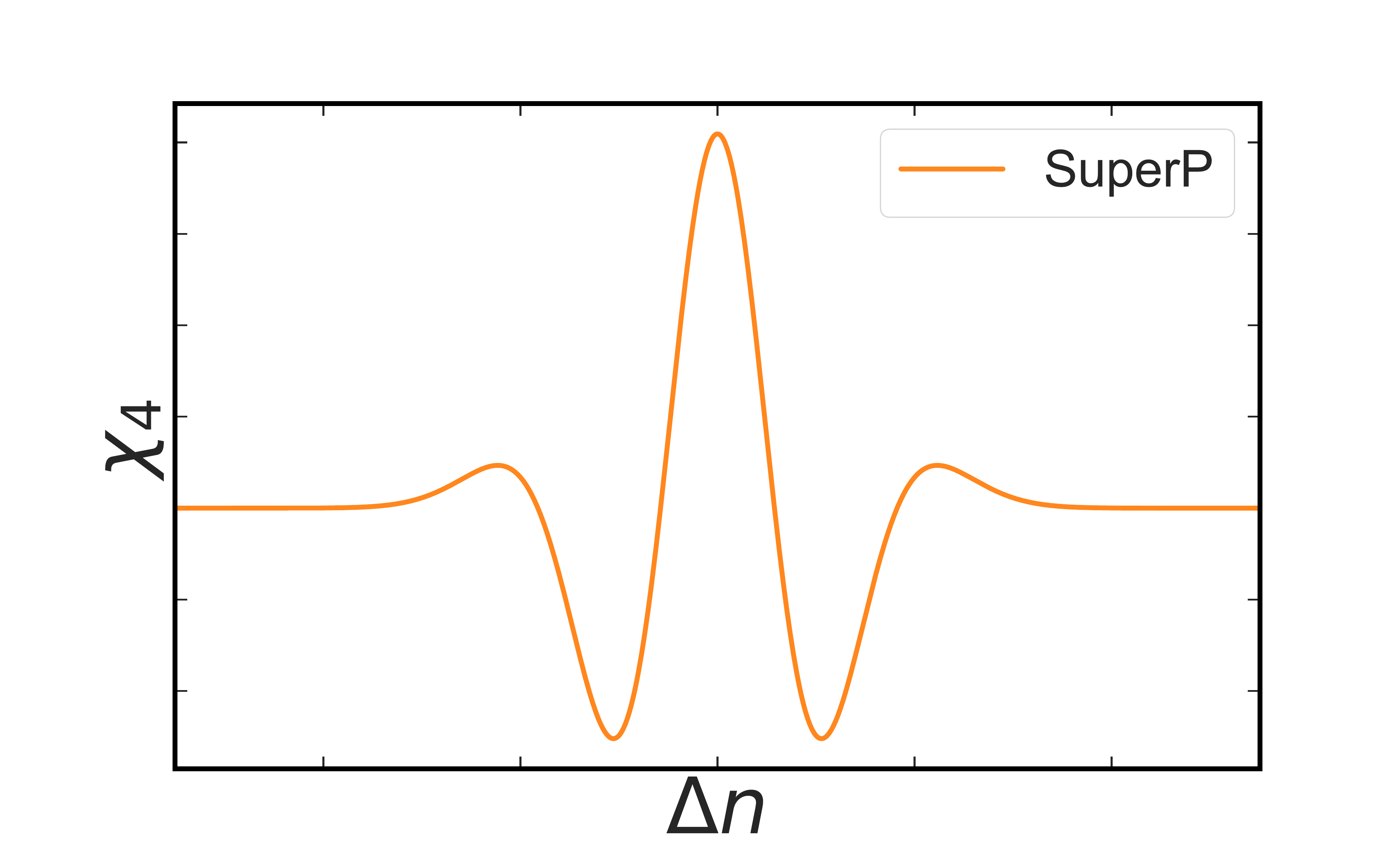}
    \put(-22,12){(f)}\\
    \includegraphics[width=0.48\linewidth, trim=5 15 55 65, clip]{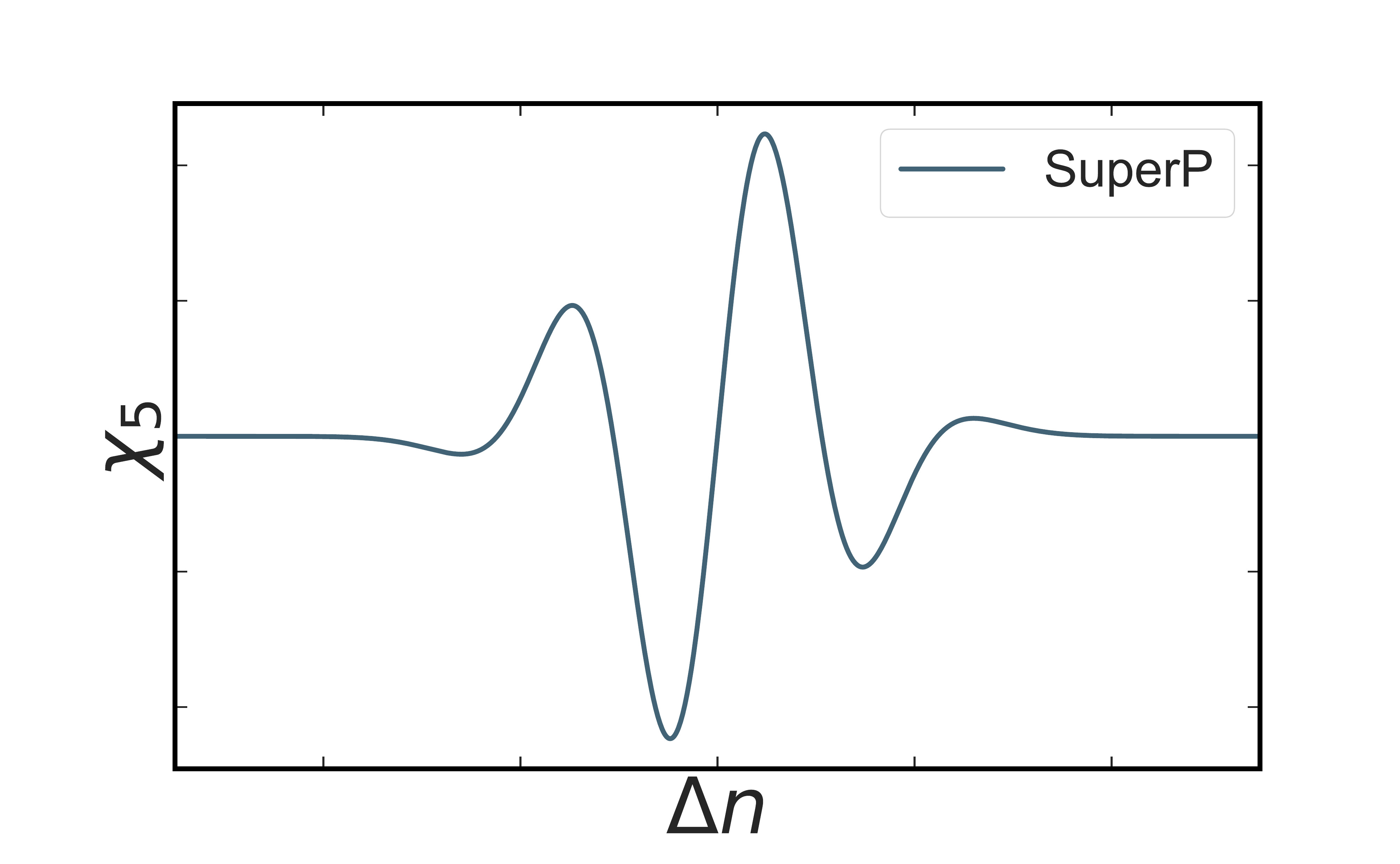}
    \put(-22,12){(g)}
    \hspace{0.025\linewidth}
    \includegraphics[width=0.48\linewidth, trim=5 15 55 65, clip]{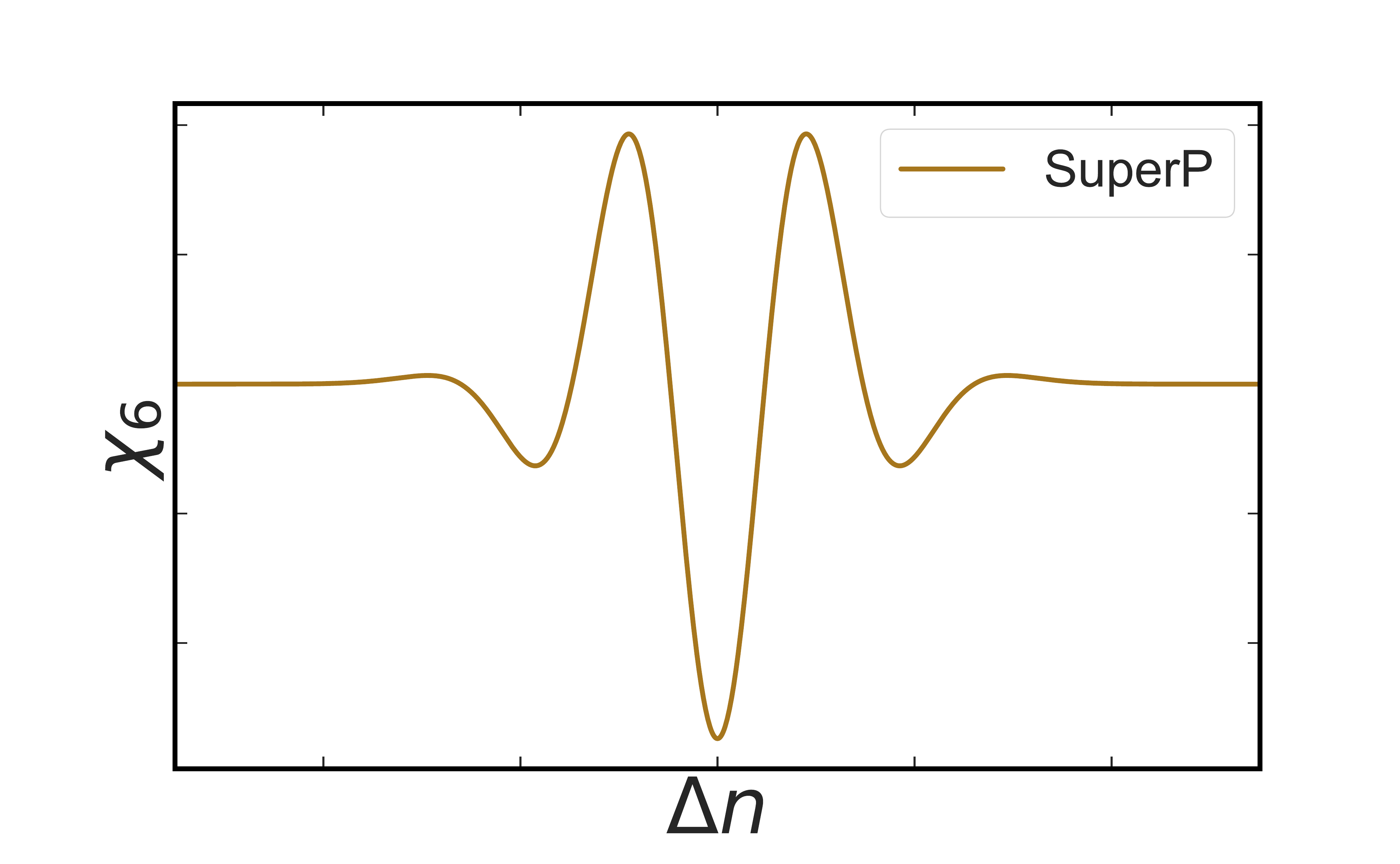}
    \put(-22,12){(h)}\\
    \includegraphics[width=0.48\linewidth, trim=5 15 55 65, clip]{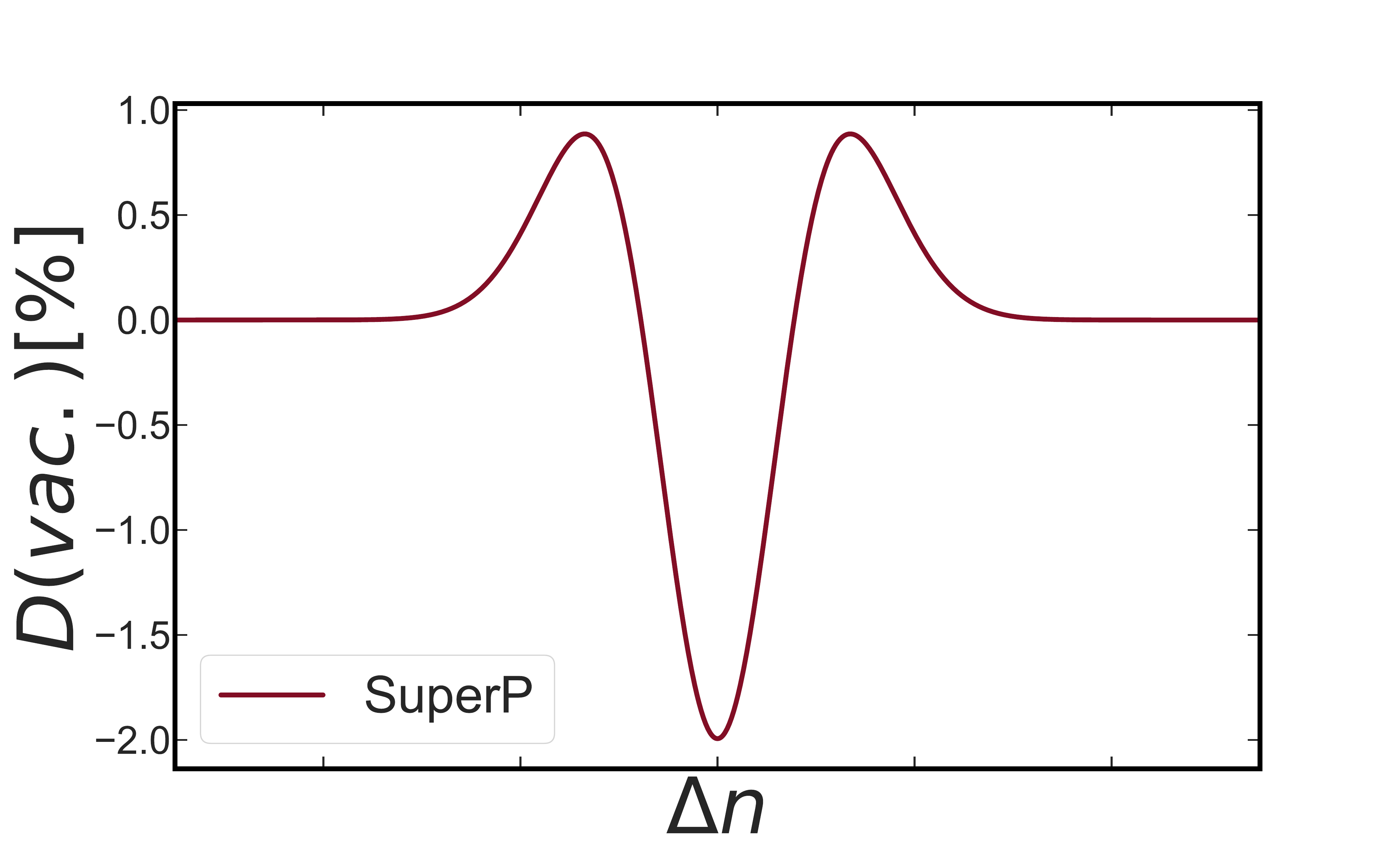}
    \put(-22,12){(i)}
    \hspace{0.025\linewidth}
    \includegraphics[width=0.48\linewidth, trim=5 15 55 65, clip]{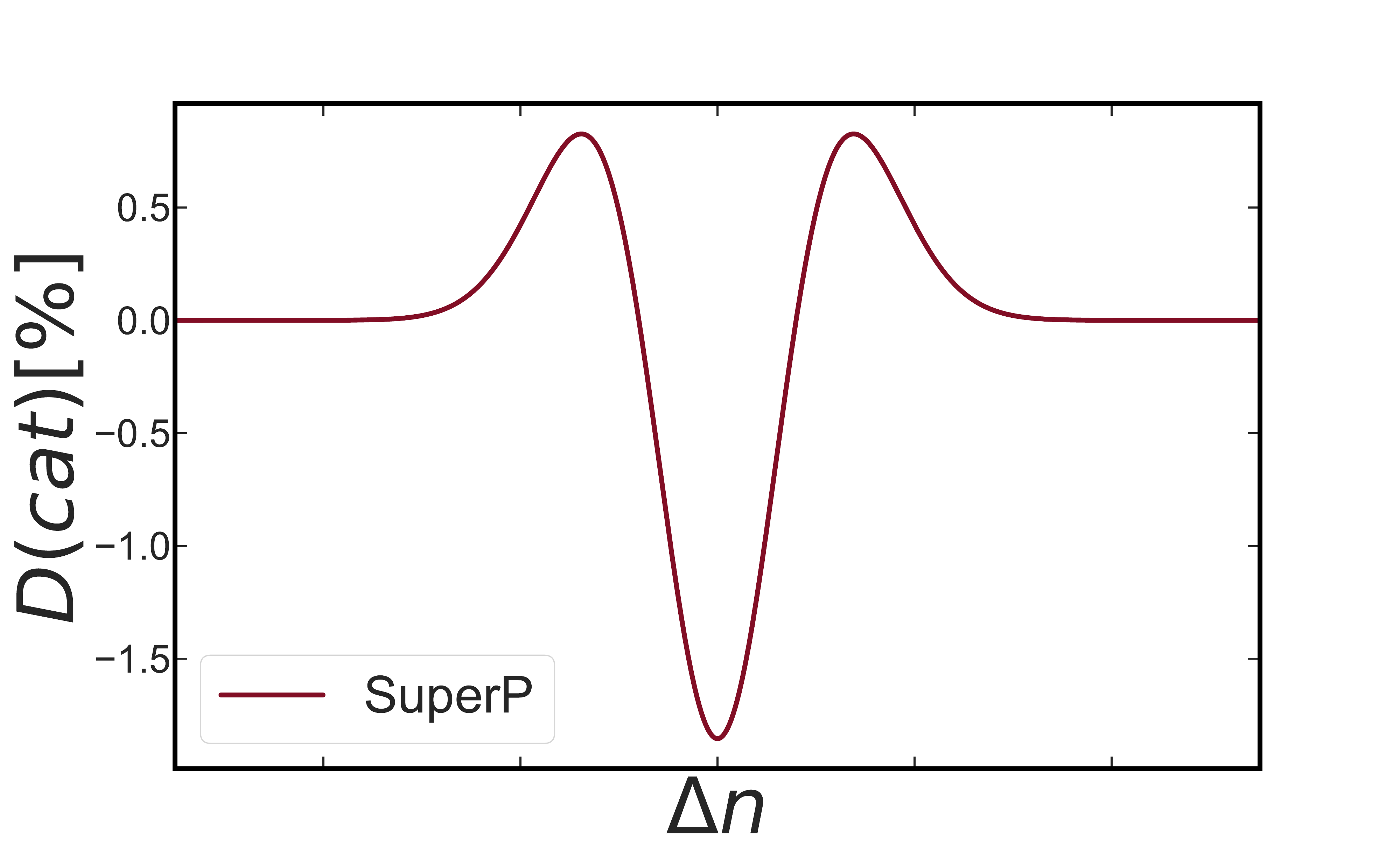}
    \put(-22,12){(j)}\\
    \includegraphics[width=0.99\linewidth, trim=20 10 200 25, clip]{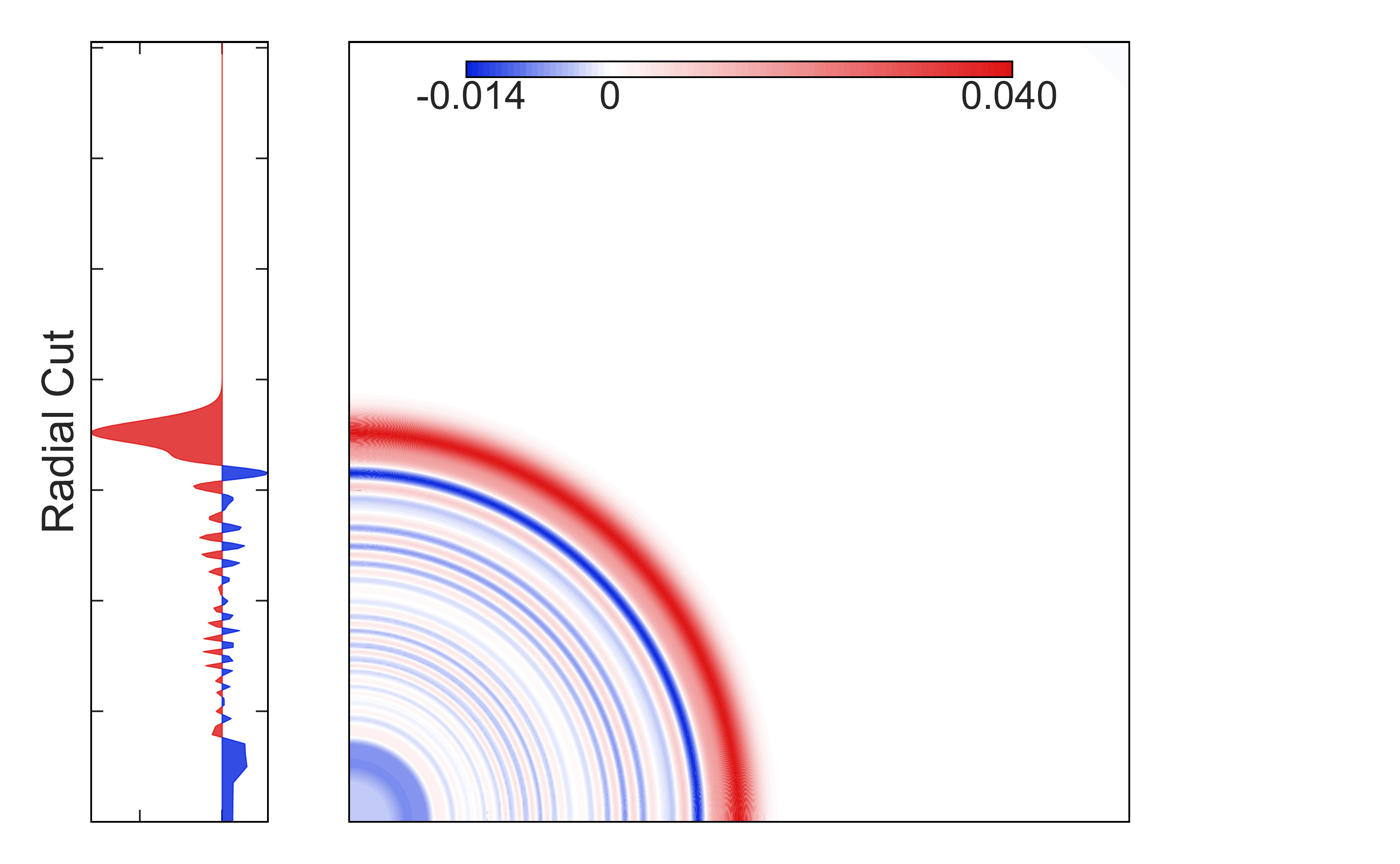}
    \put(-35,12){(k)}
    \vspace{-10pt}
    \caption{Super-Poissonian BCS probe: see caption of Fig.~\ref{CoherentStats}.}
    \label{SuperPStats}
\end{figure}

\begin{figure}
    \centering
    \includegraphics[width=0.48\linewidth, trim=5 15 55 65, clip]{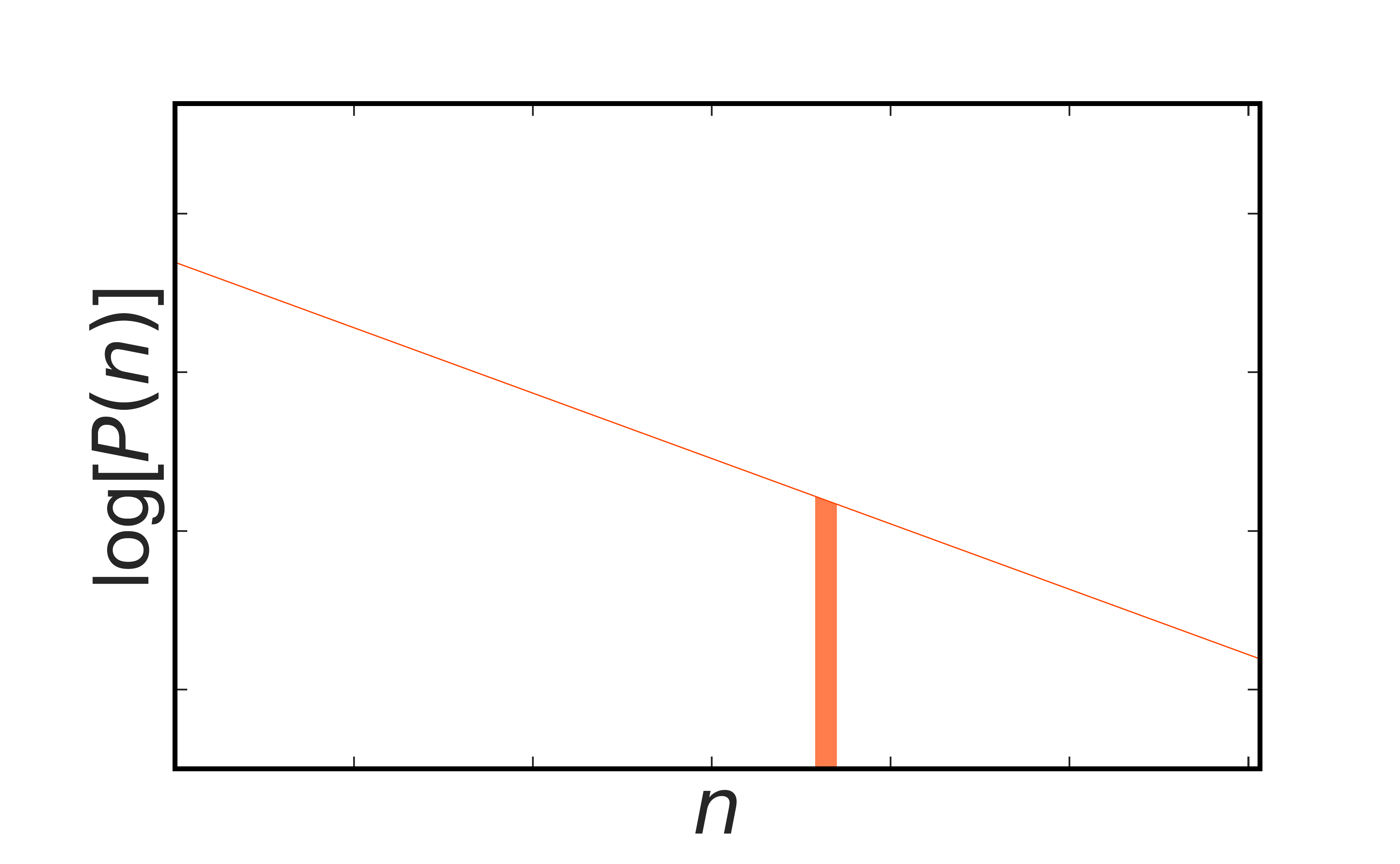}
    \put(-22,12){(a)}
    \hspace{0.025\linewidth}
    \includegraphics[width=0.48\linewidth, trim=5 15 55 65, clip]{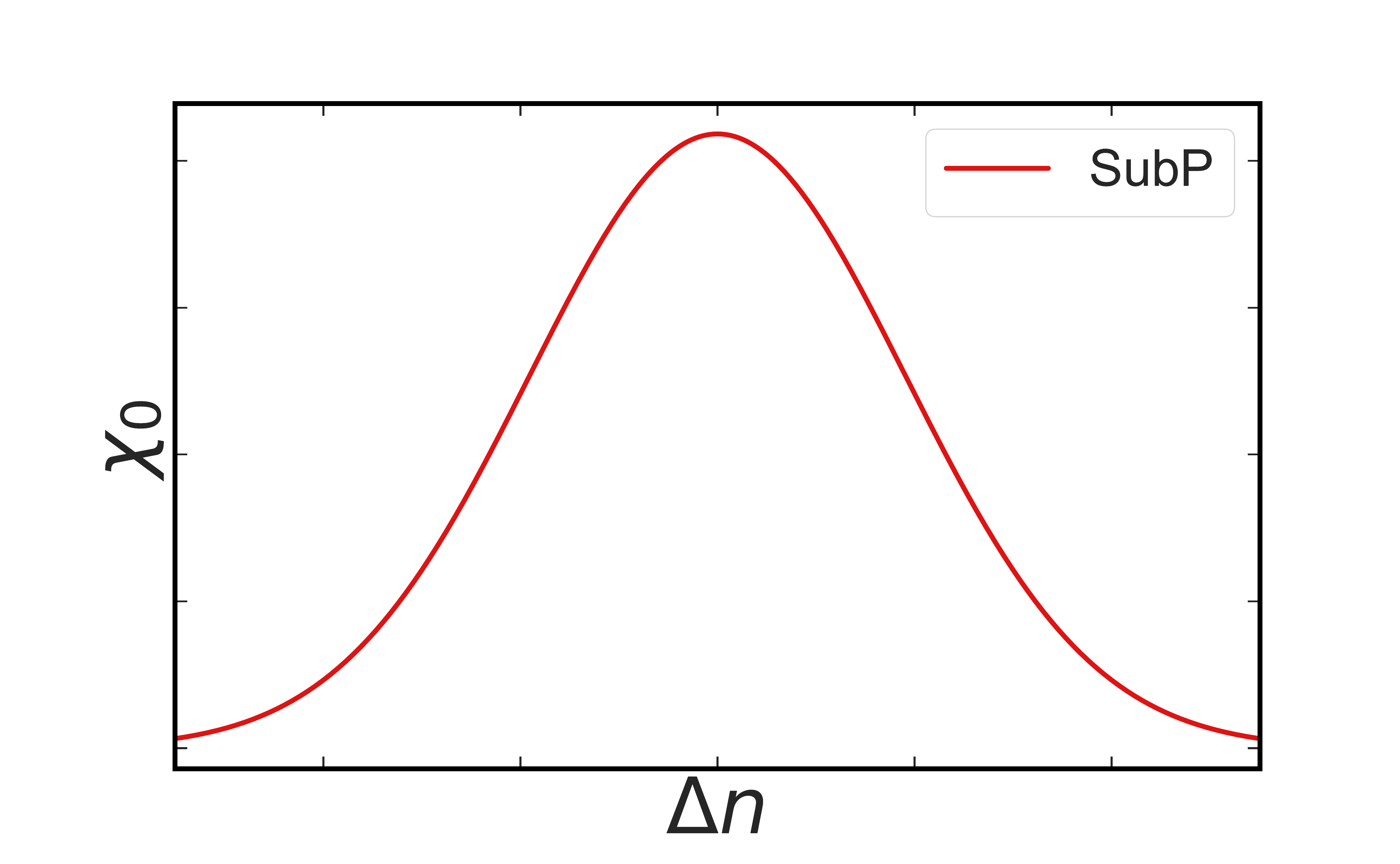}
    \put(-22,12){(b)}\\
    \includegraphics[width=0.48\linewidth, trim=5 15 55 65, clip]{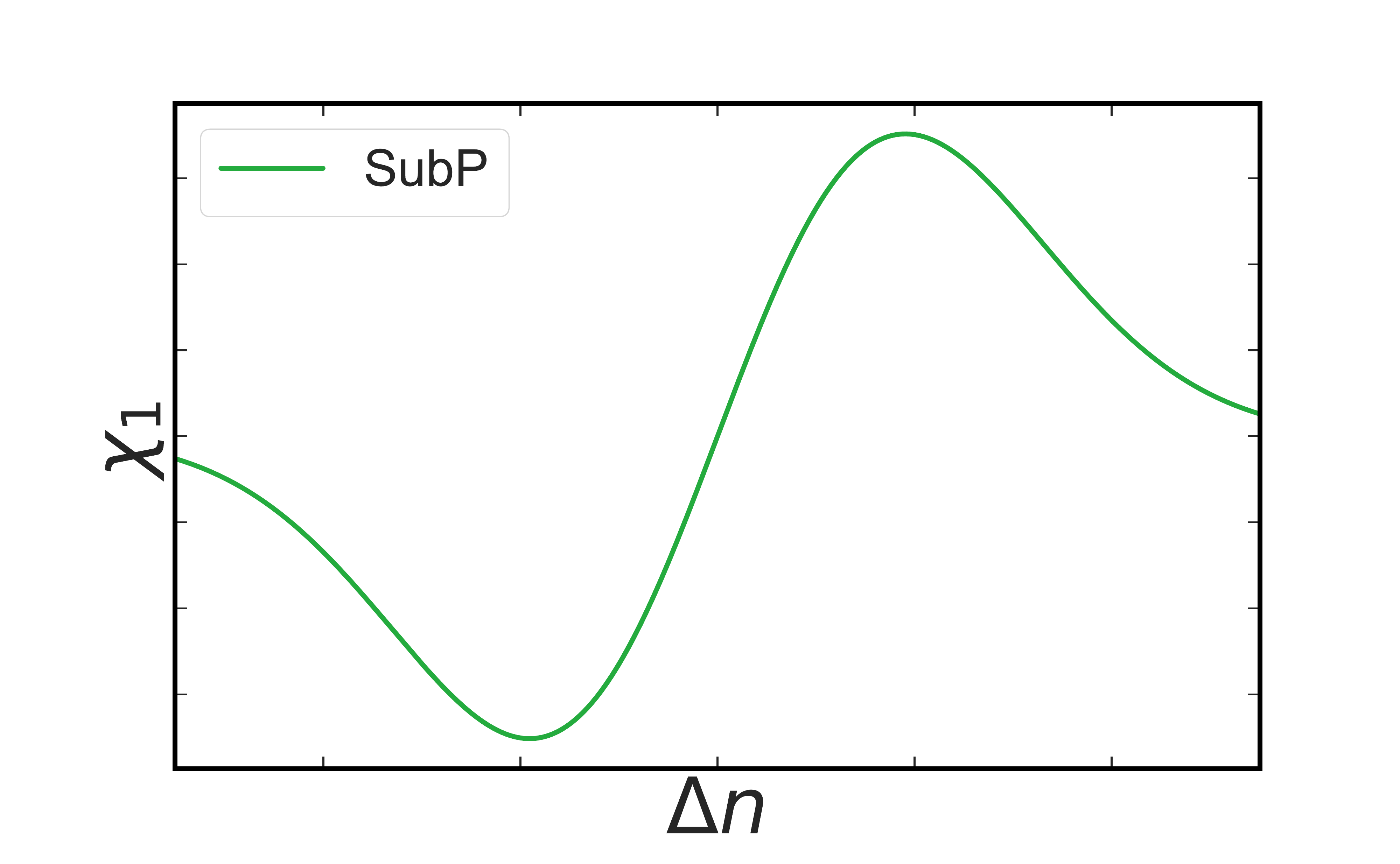}
    \put(-22,12){(c)}
    \hspace{0.025\linewidth}
    \includegraphics[width=0.48\linewidth, trim=5 15 55 65, clip]{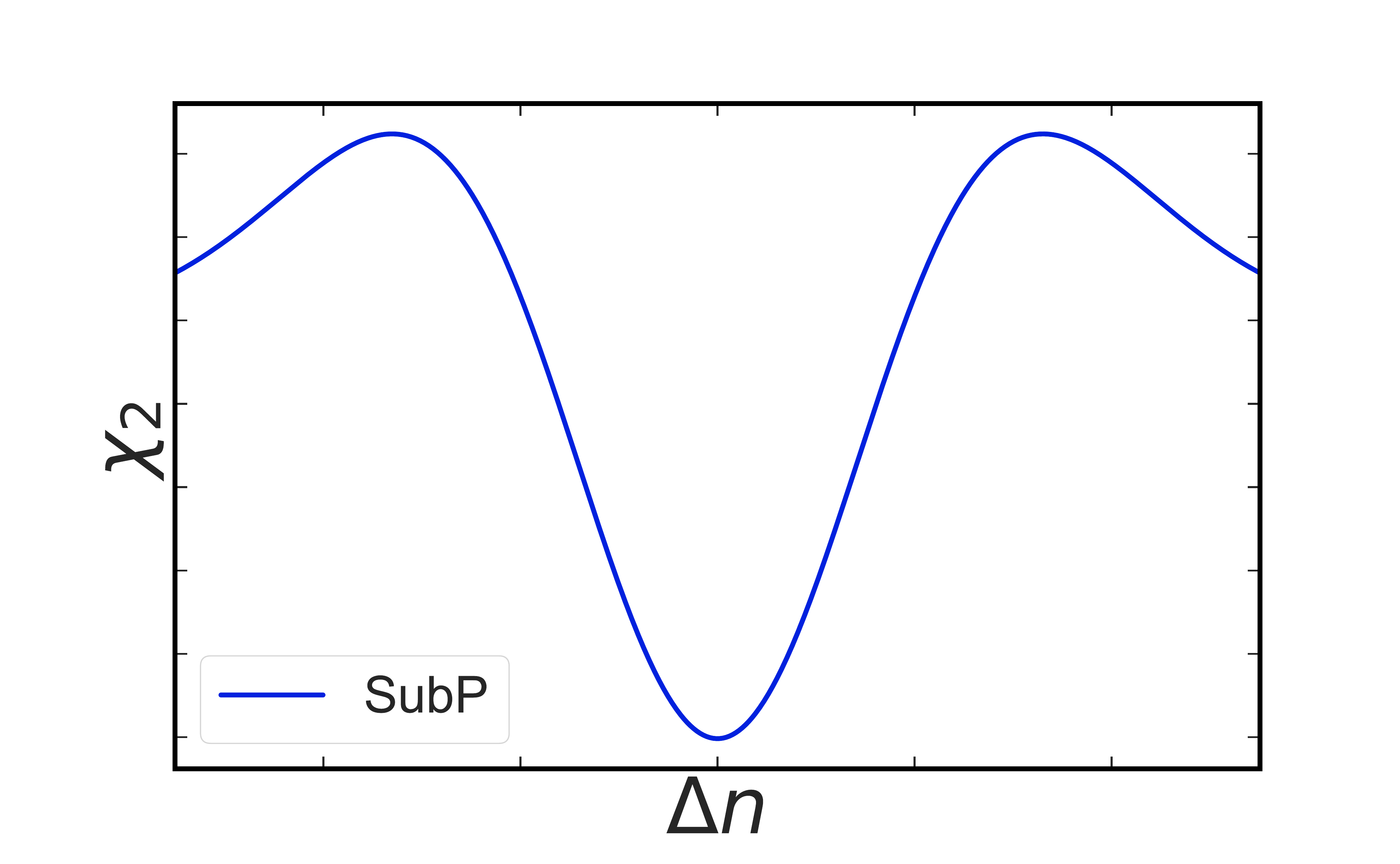}
    \put(-22,12){(d)}\\
    \includegraphics[width=0.48\linewidth, trim=5 15 55 65, clip]{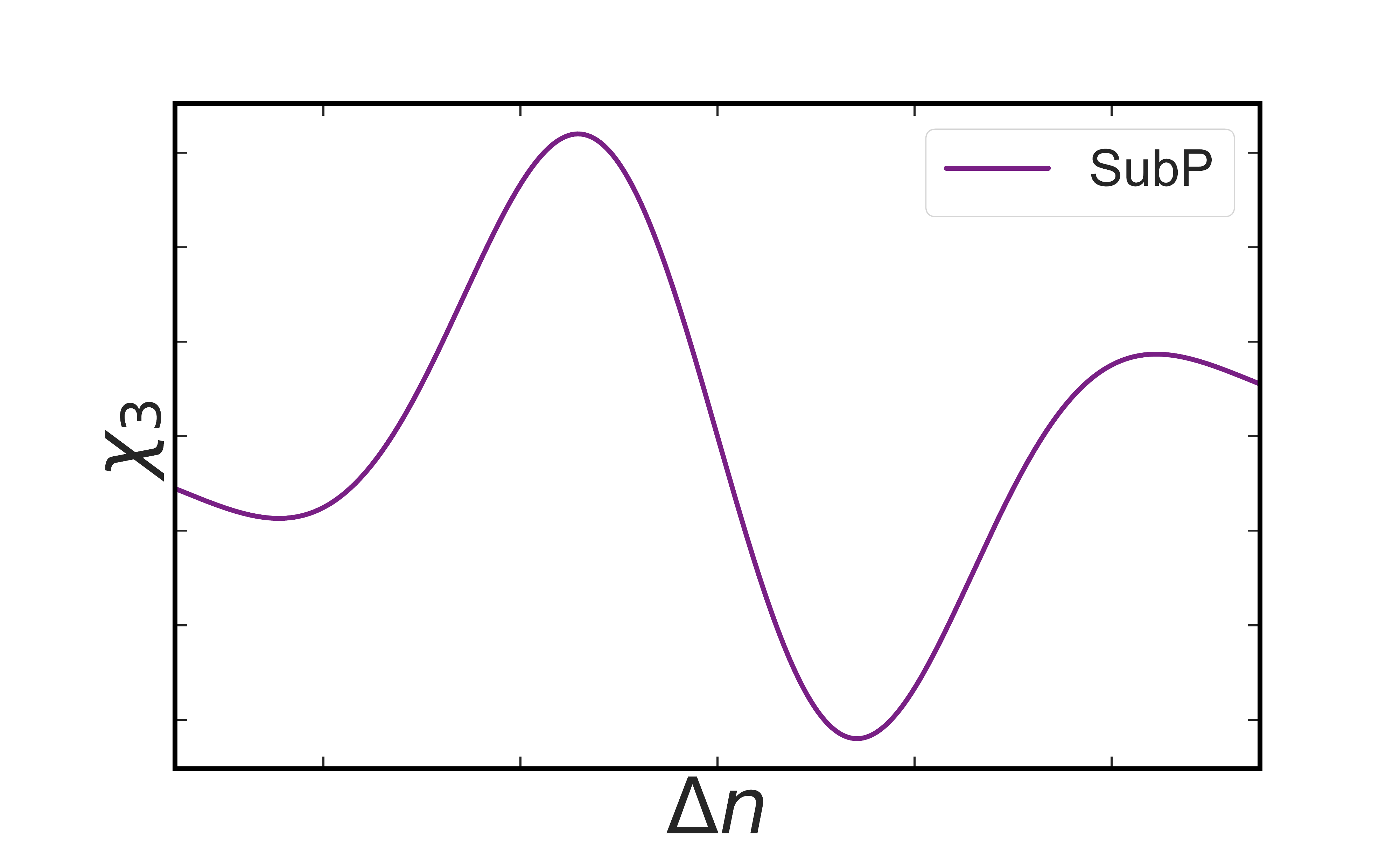}
    \put(-22,12){(e)}
    \hspace{0.025\linewidth}
    \includegraphics[width=0.48\linewidth, trim=5 15 55 65, clip]{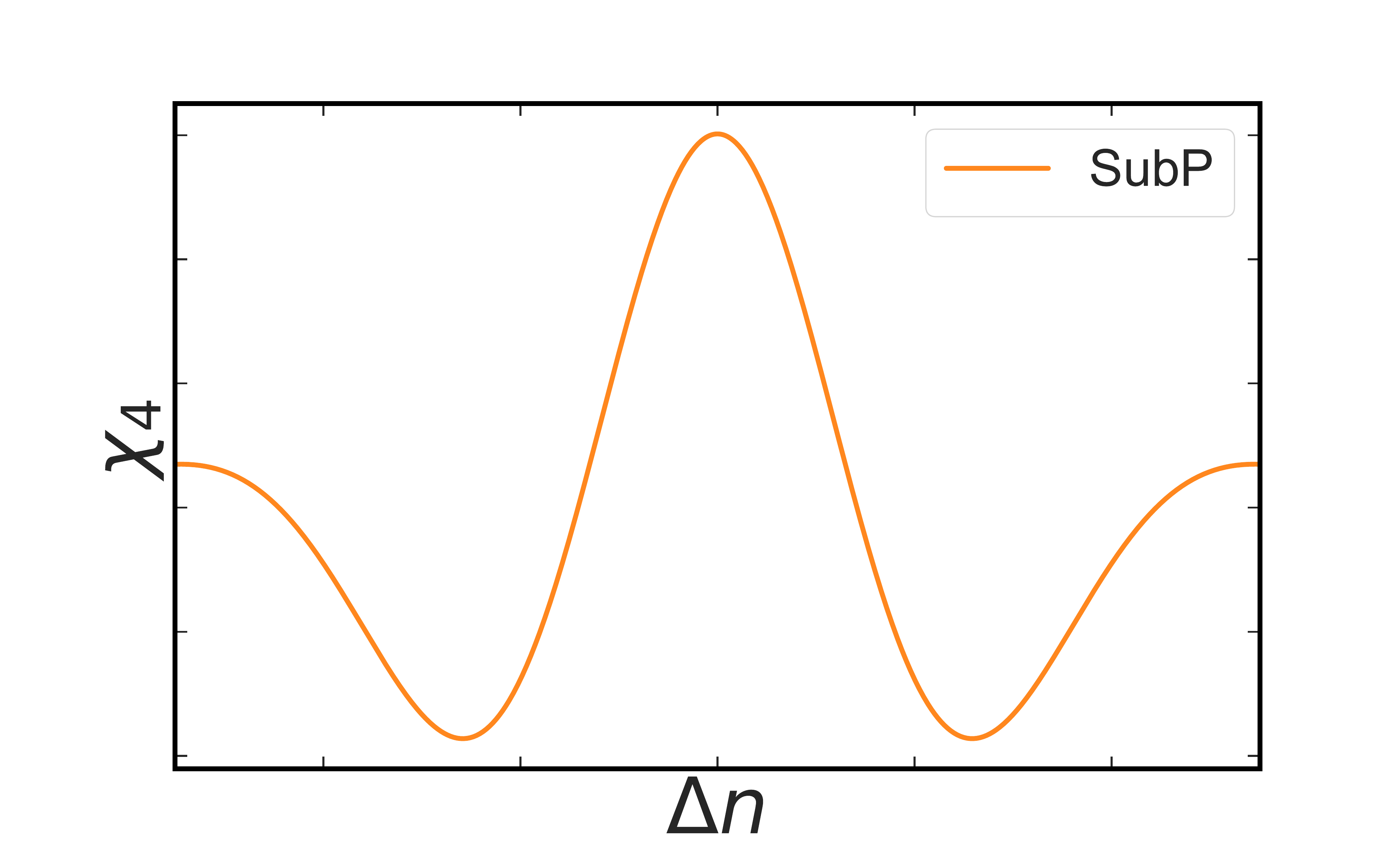}
    \put(-22,12){(f)}\\
    \includegraphics[width=0.48\linewidth, trim=5 15 55 65, clip]{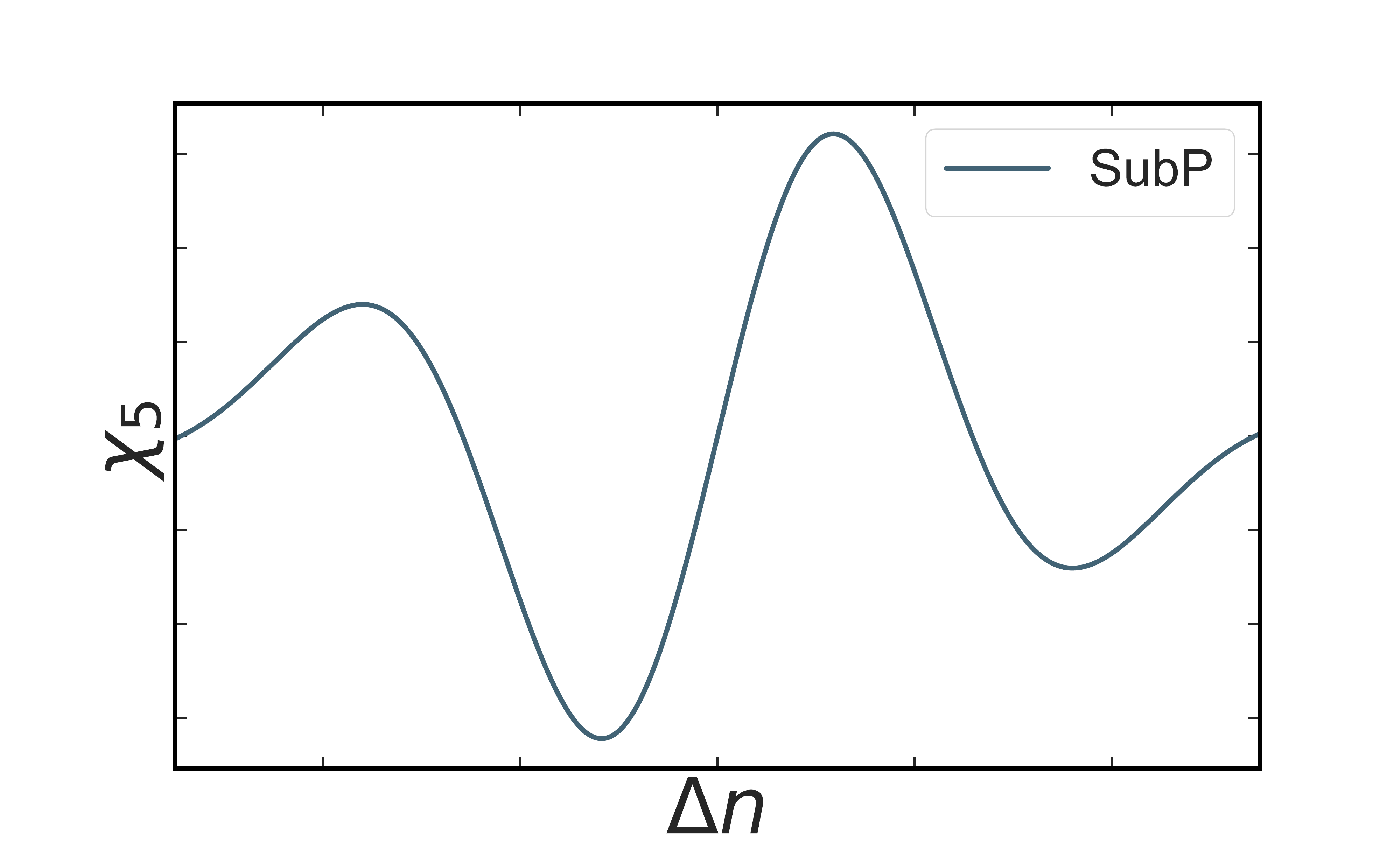}
    \put(-22,12){(g)}
    \hspace{0.025\linewidth}
    \includegraphics[width=0.48\linewidth, trim=5 15 55 65, clip]{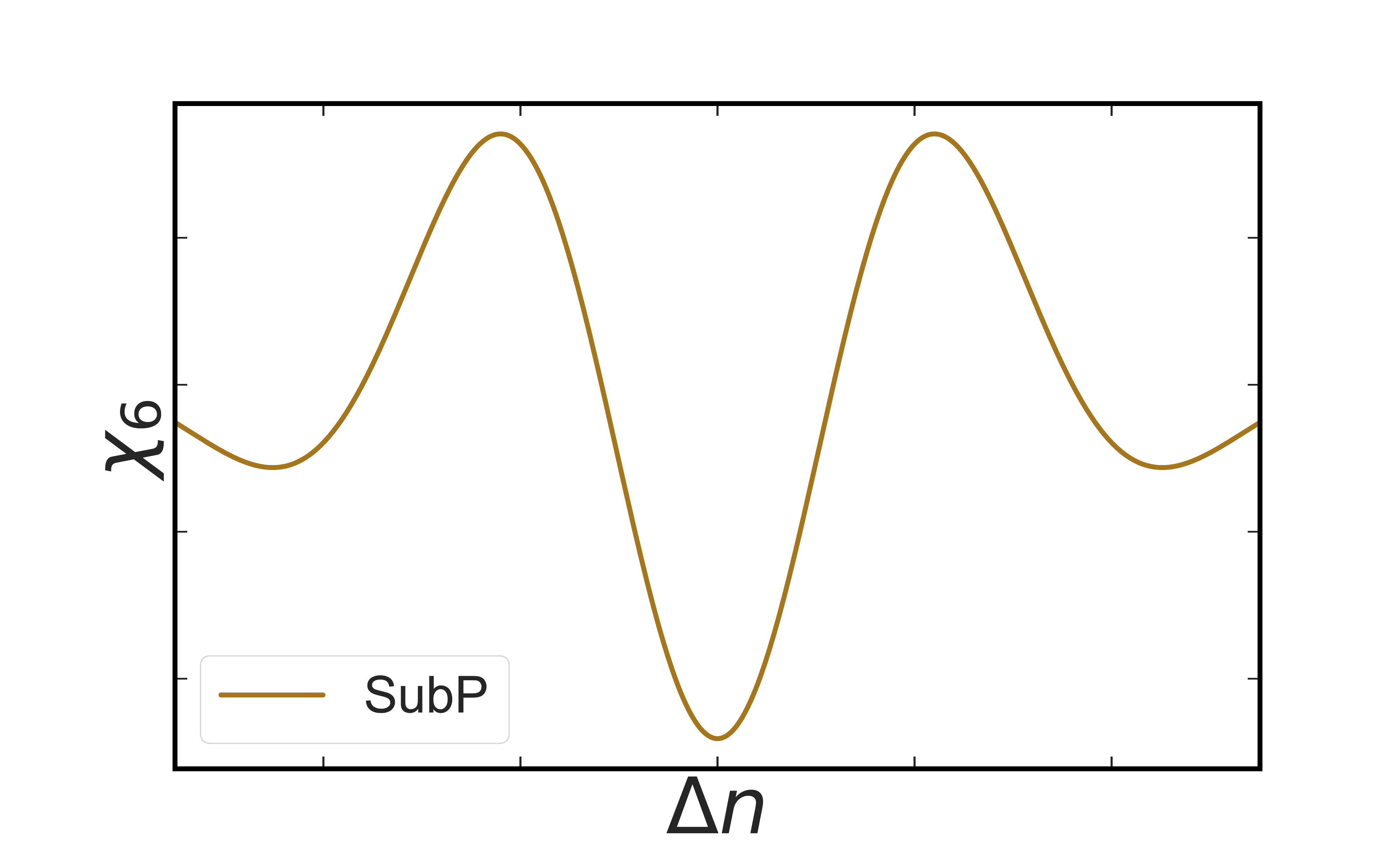}
    \put(-22,12){(h)}\\
    \includegraphics[width=0.48\linewidth, trim=5 15 55 65, clip]{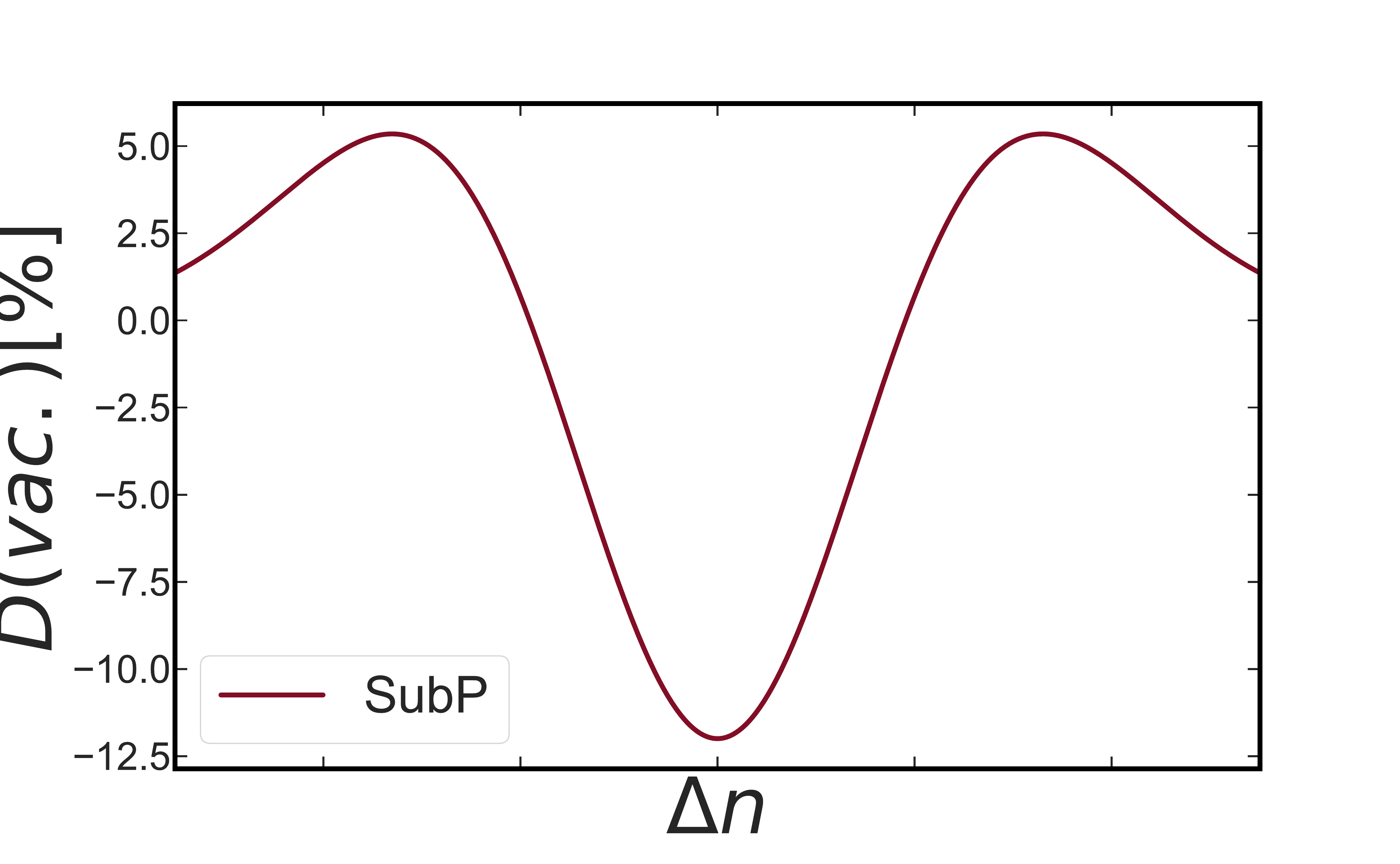}
    \put(-22,12){(i)}
    \hspace{0.025\linewidth}
    \includegraphics[width=0.48\linewidth, trim=5 15 55 65, clip]{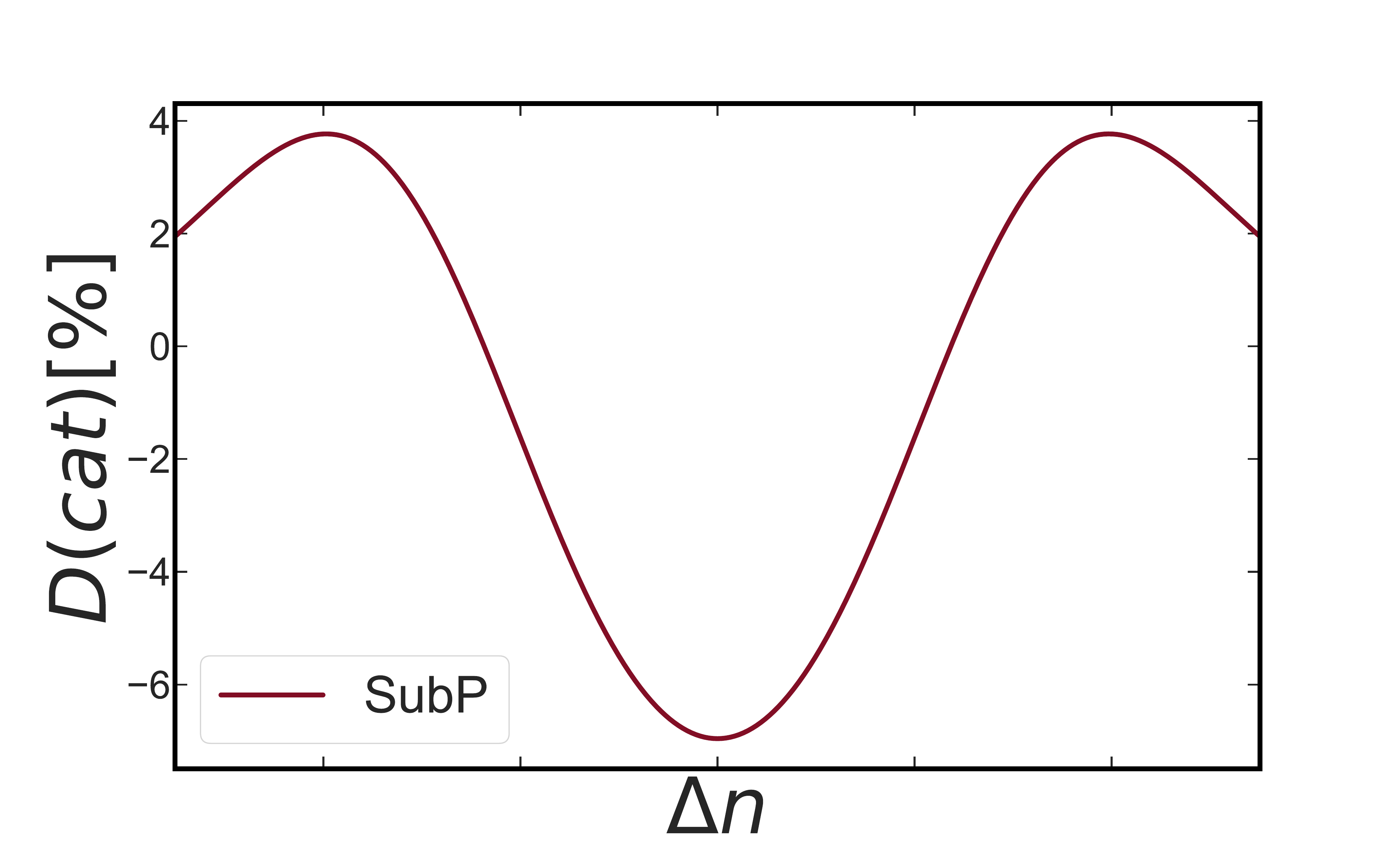}
    \put(-22,12){(j)}\\
    \includegraphics[width=0.99\linewidth, trim=20 10 200 25, clip]{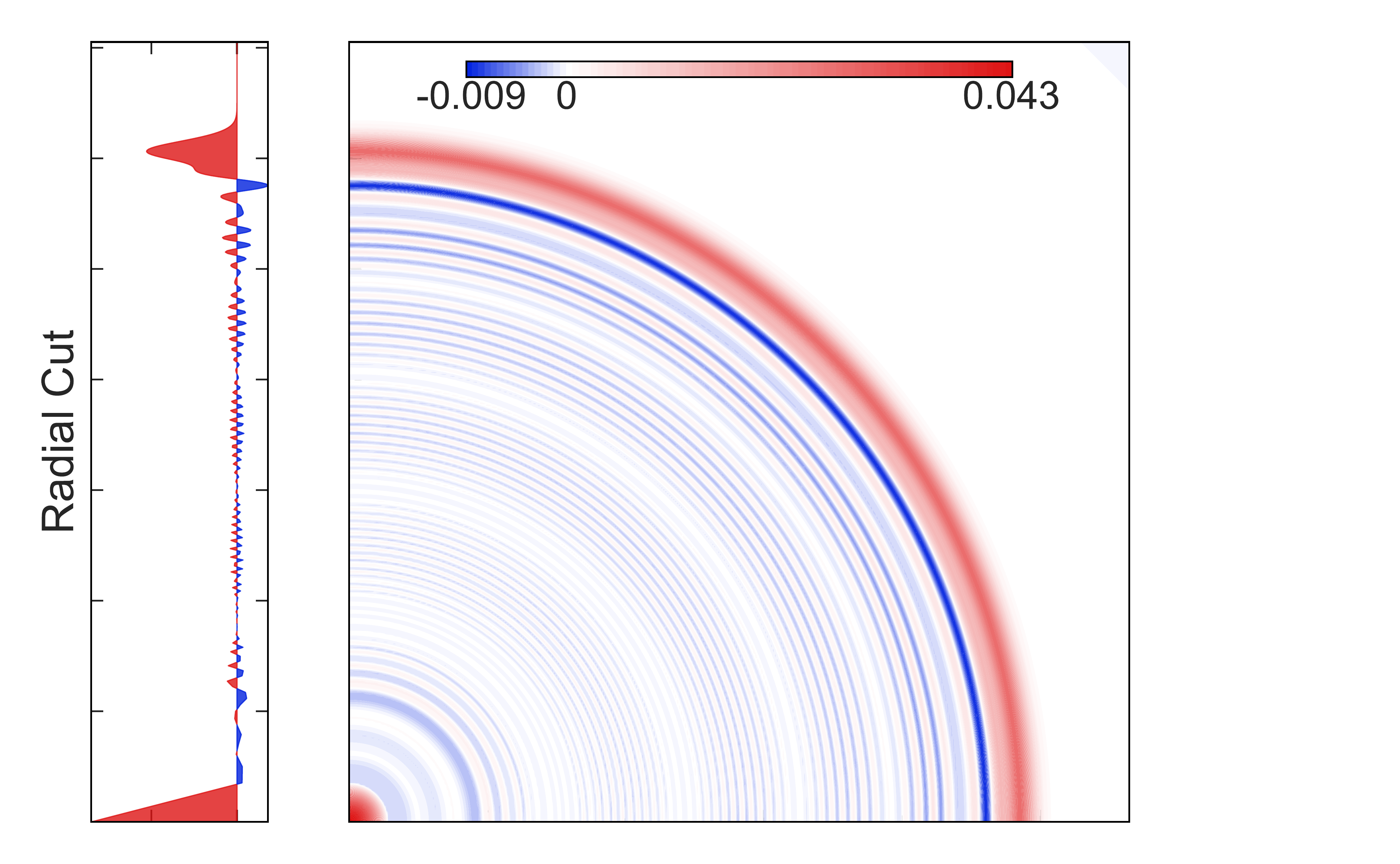}
    \put(-35,12){(k)}
    \vspace{-10pt}
    \caption{Sub-Poissonian BCS probe: see caption of Fig.~\ref{CoherentStats}.}
    \label{SubPStats}
\end{figure}

\begin{figure}
    \centering
    \includegraphics[width=0.48\linewidth, trim=5 15 55 65, clip]{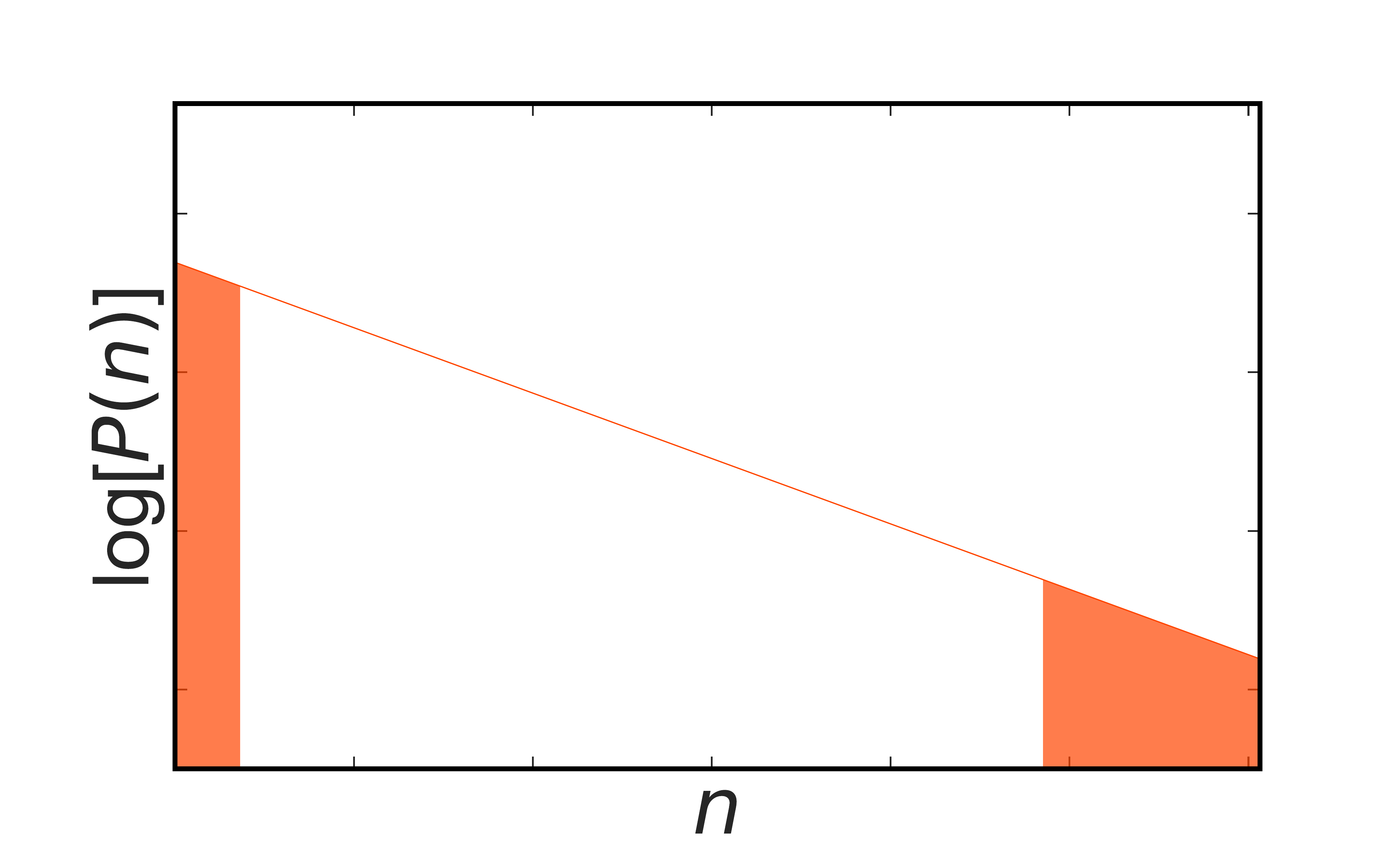}
    \put(-22,12){(a)}
    \hspace{0.025\linewidth}
    \includegraphics[width=0.48\linewidth, trim=5 15 55 65, clip]{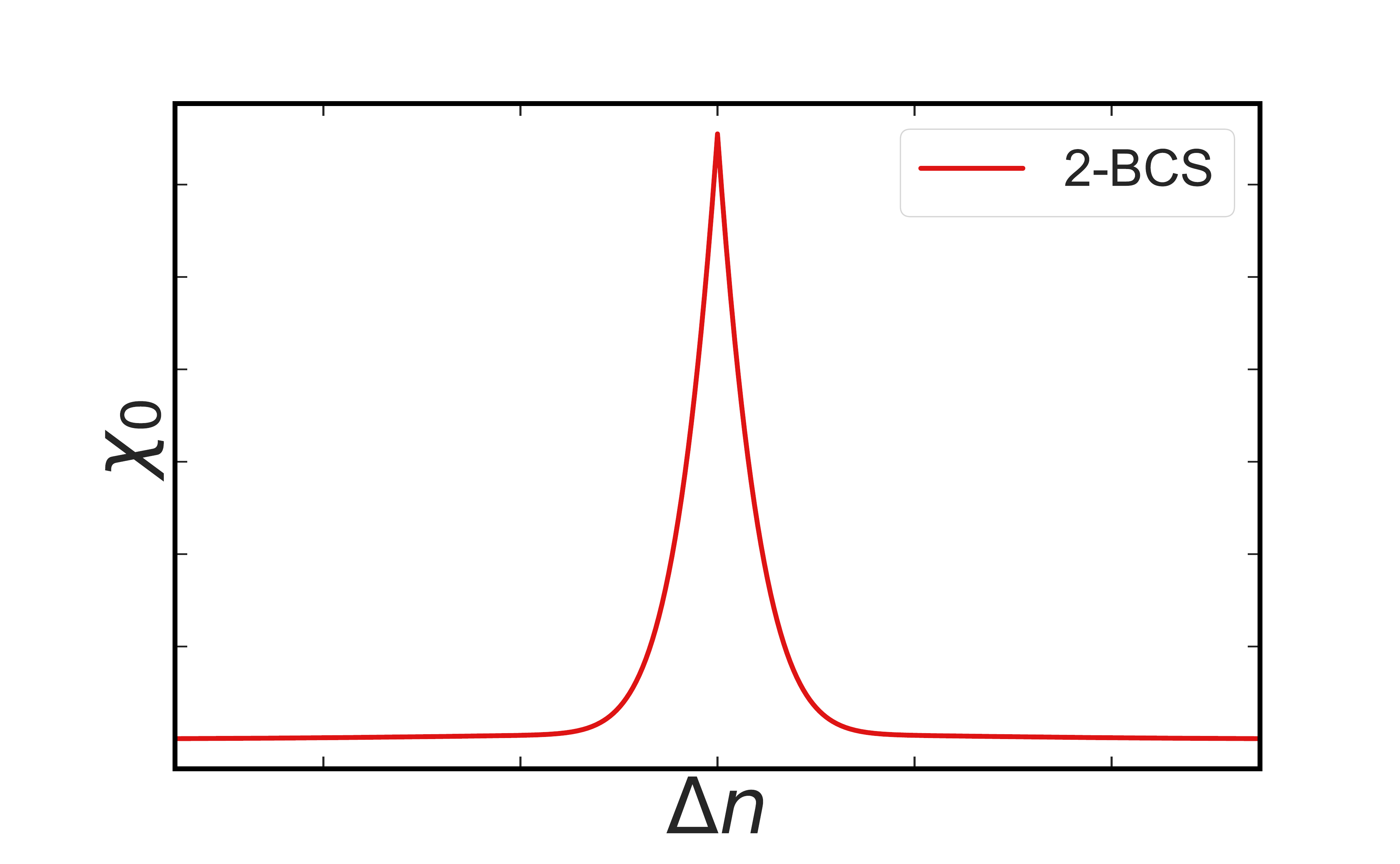}
    \put(-22,12){(b)}\\
    \includegraphics[width=0.48\linewidth, trim=5 15 55 65, clip]{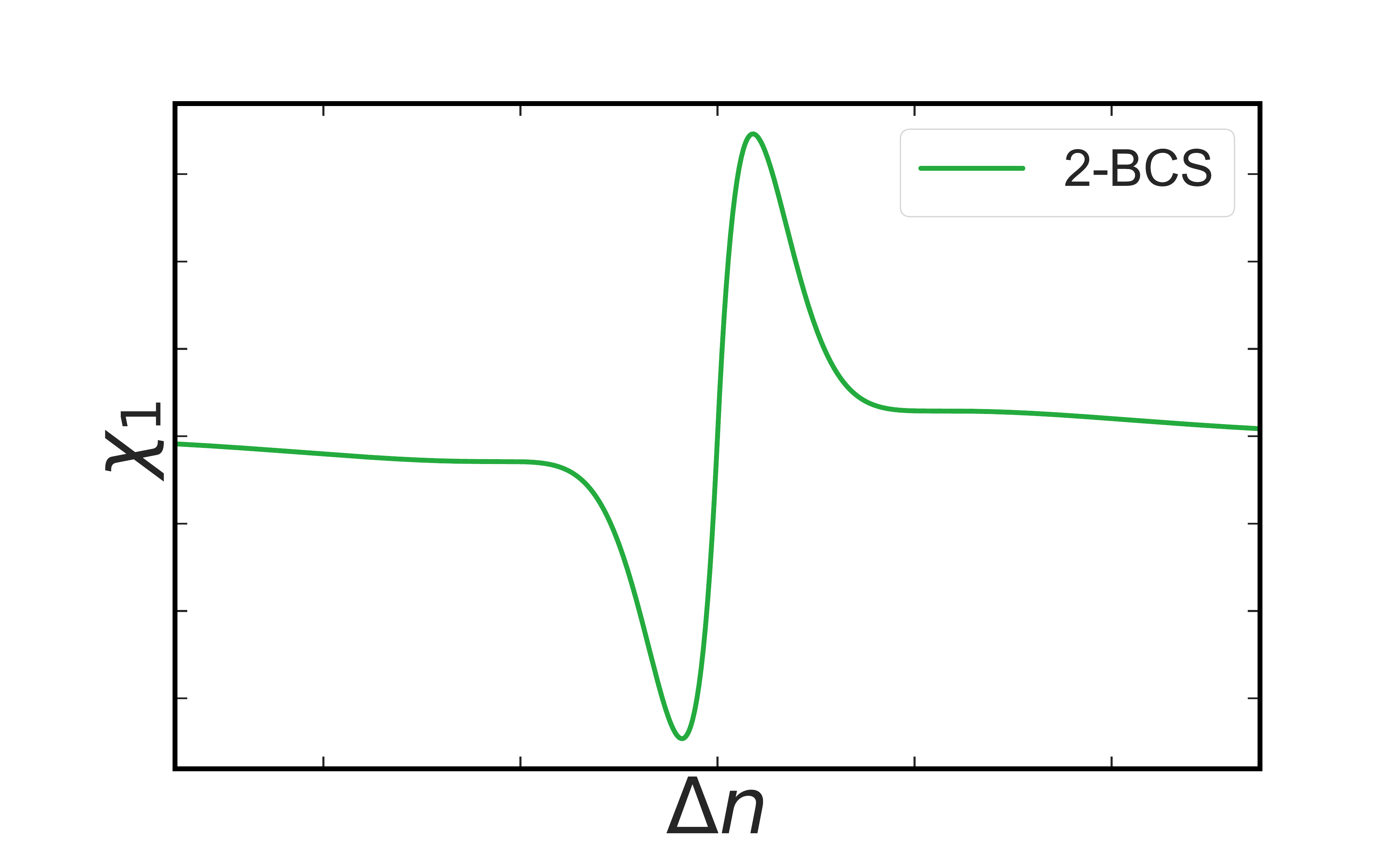}
    \put(-22,12){(c)}
    \hspace{0.025\linewidth}
    \includegraphics[width=0.48\linewidth, trim=5 15 55 65, clip]{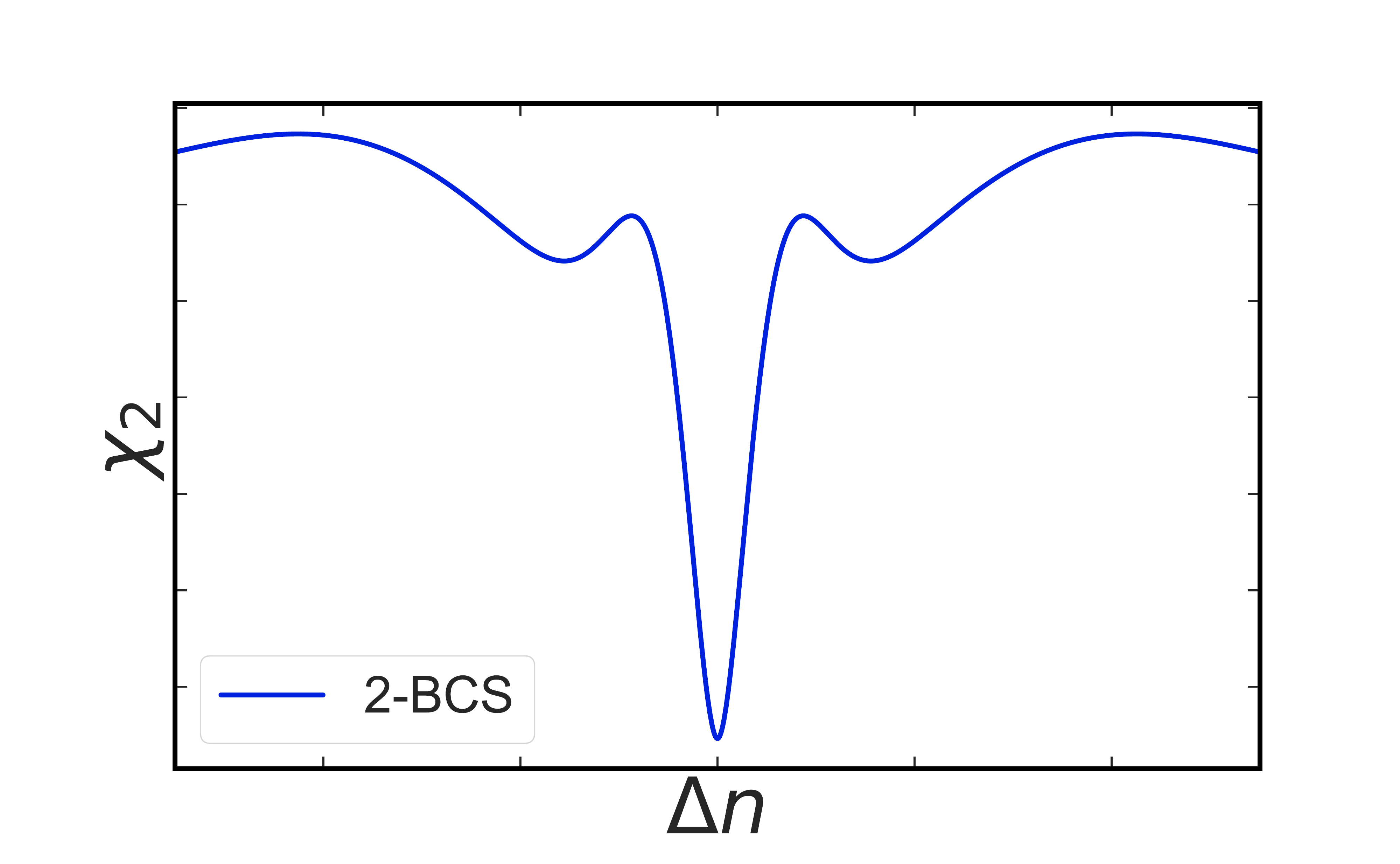}
    \put(-22,12){(d)}\\
    \includegraphics[width=0.48\linewidth, trim=5 15 55 65, clip]{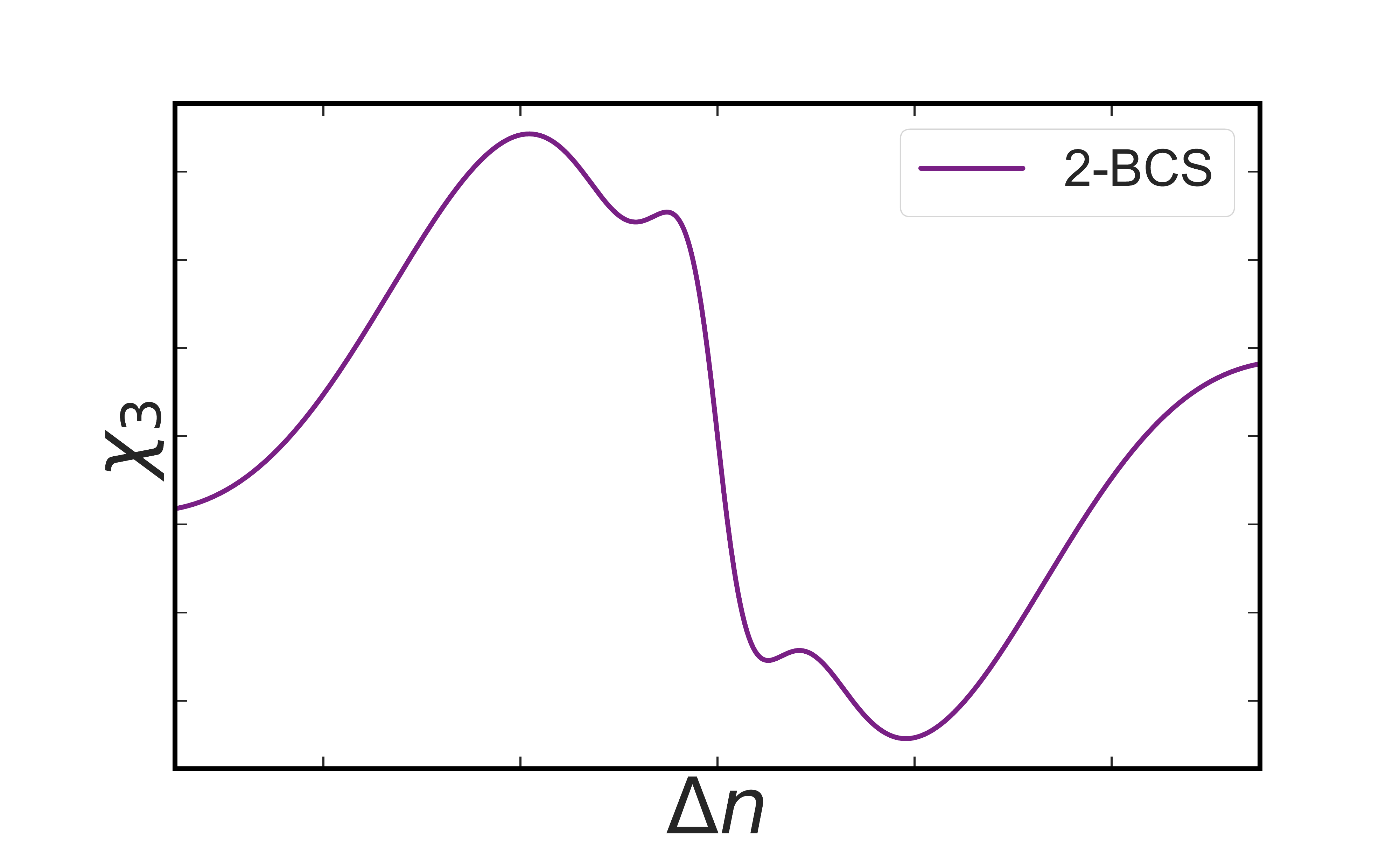}
    \put(-22,12){(e)}
    \hspace{0.025\linewidth}
    \includegraphics[width=0.48\linewidth, trim=5 15 55 65, clip]{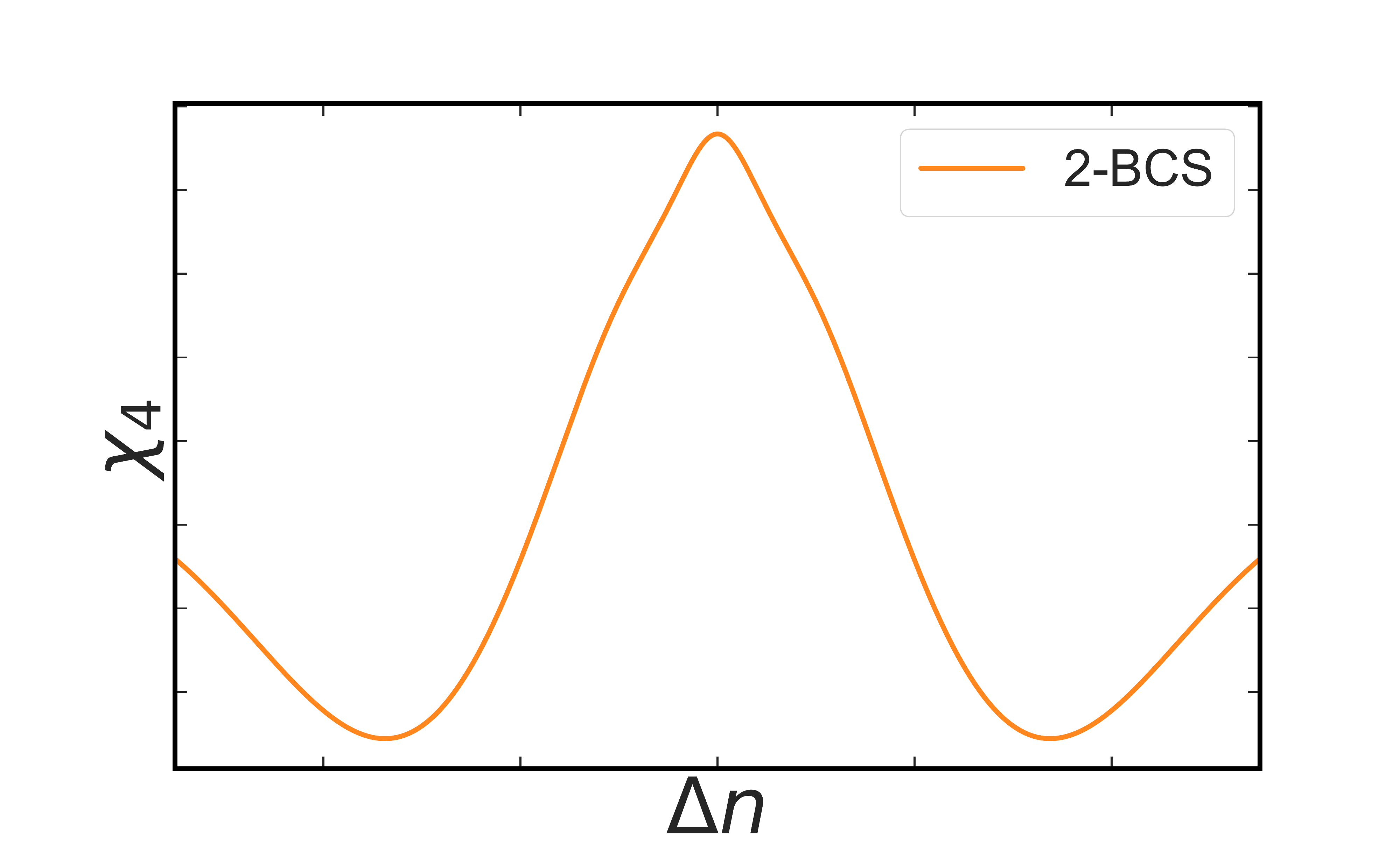}
    \put(-22,12){(f)}\\
    \includegraphics[width=0.48\linewidth, trim=5 15 55 65, clip]{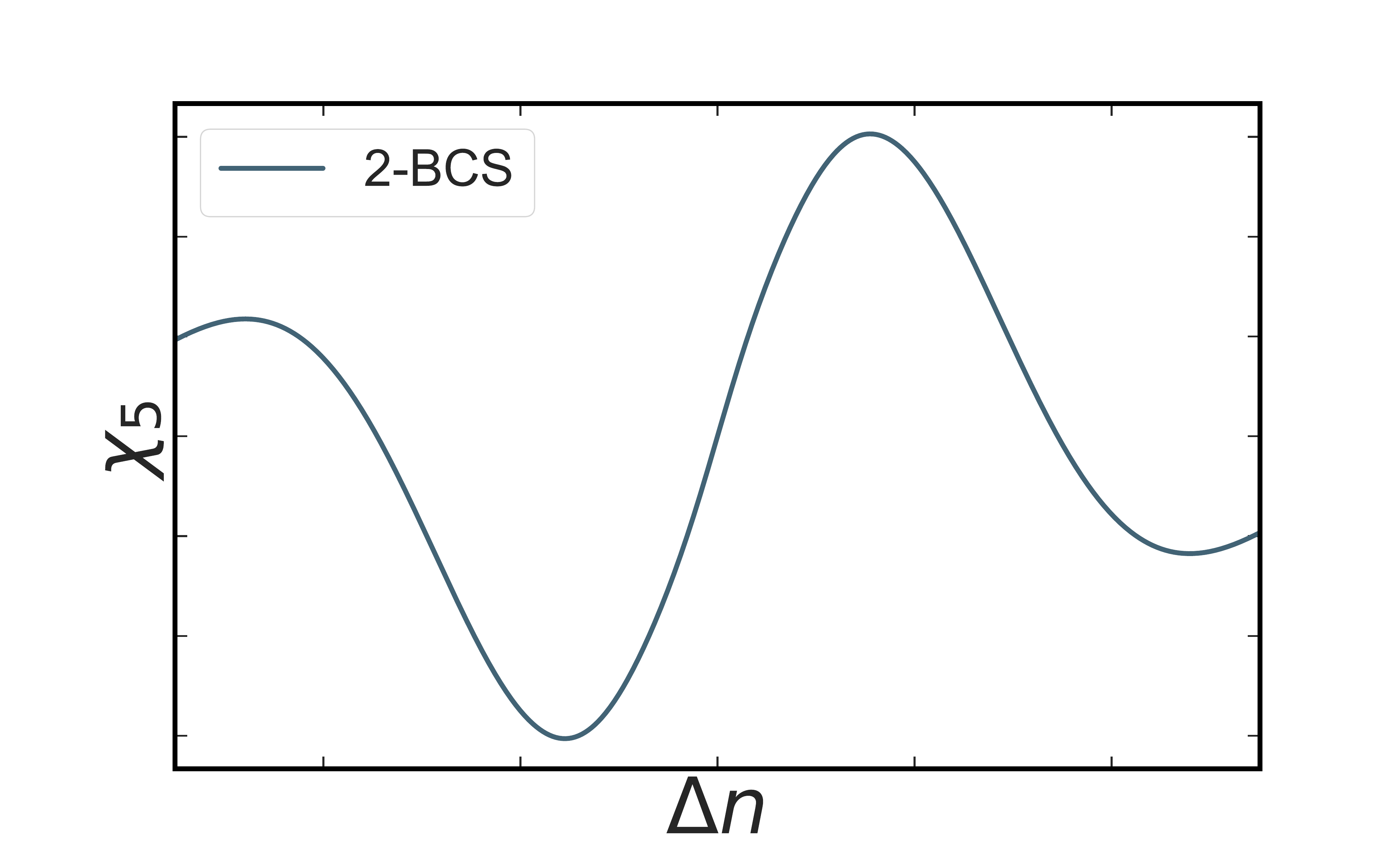}
    \put(-22,12){(g)}
    \hspace{0.025\linewidth}
    \includegraphics[width=0.48\linewidth, trim=5 15 55 65, clip]{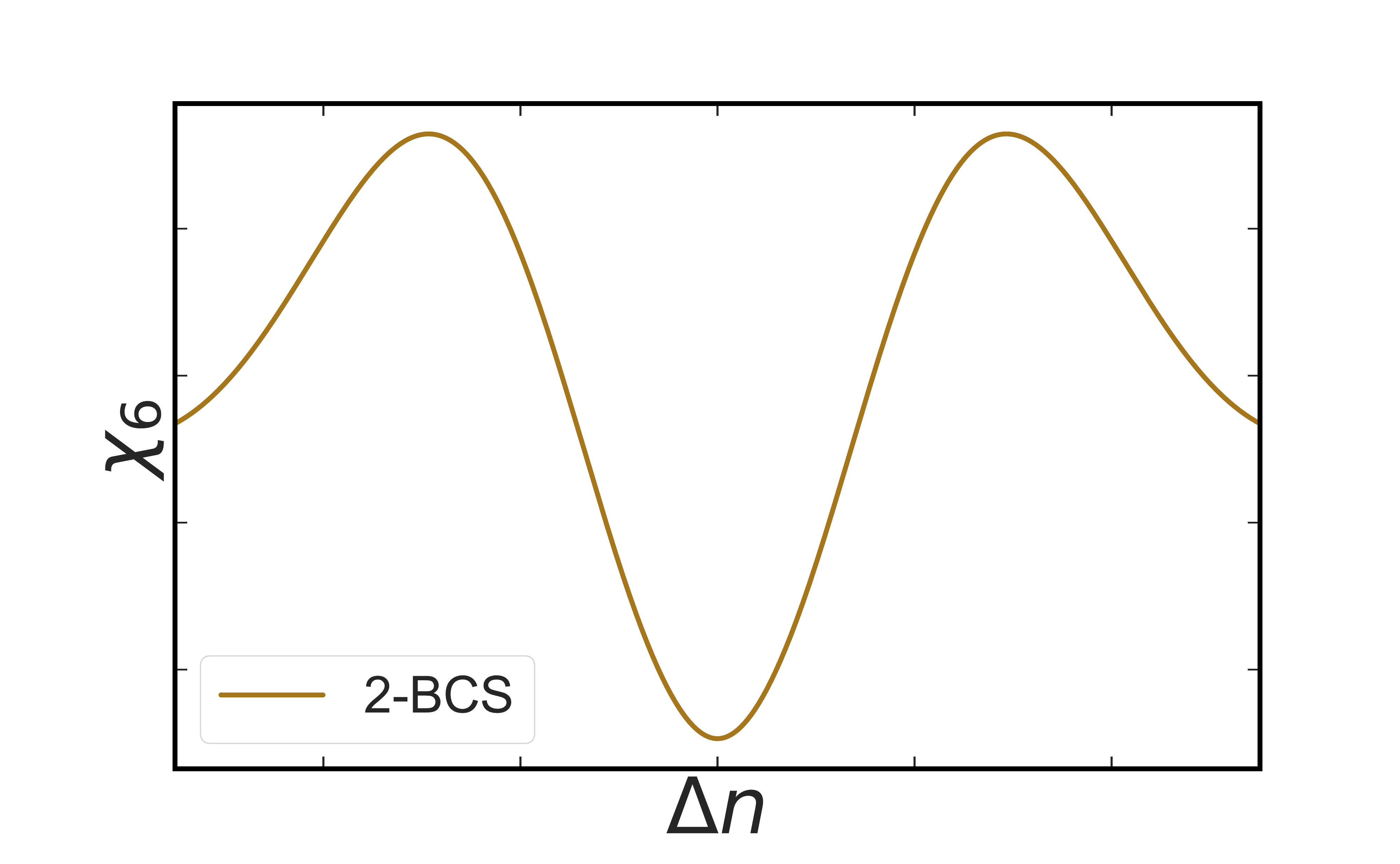}
    \put(-22,12){(h)}\\
    \includegraphics[width=0.48\linewidth, trim=5 15 55 65, clip]{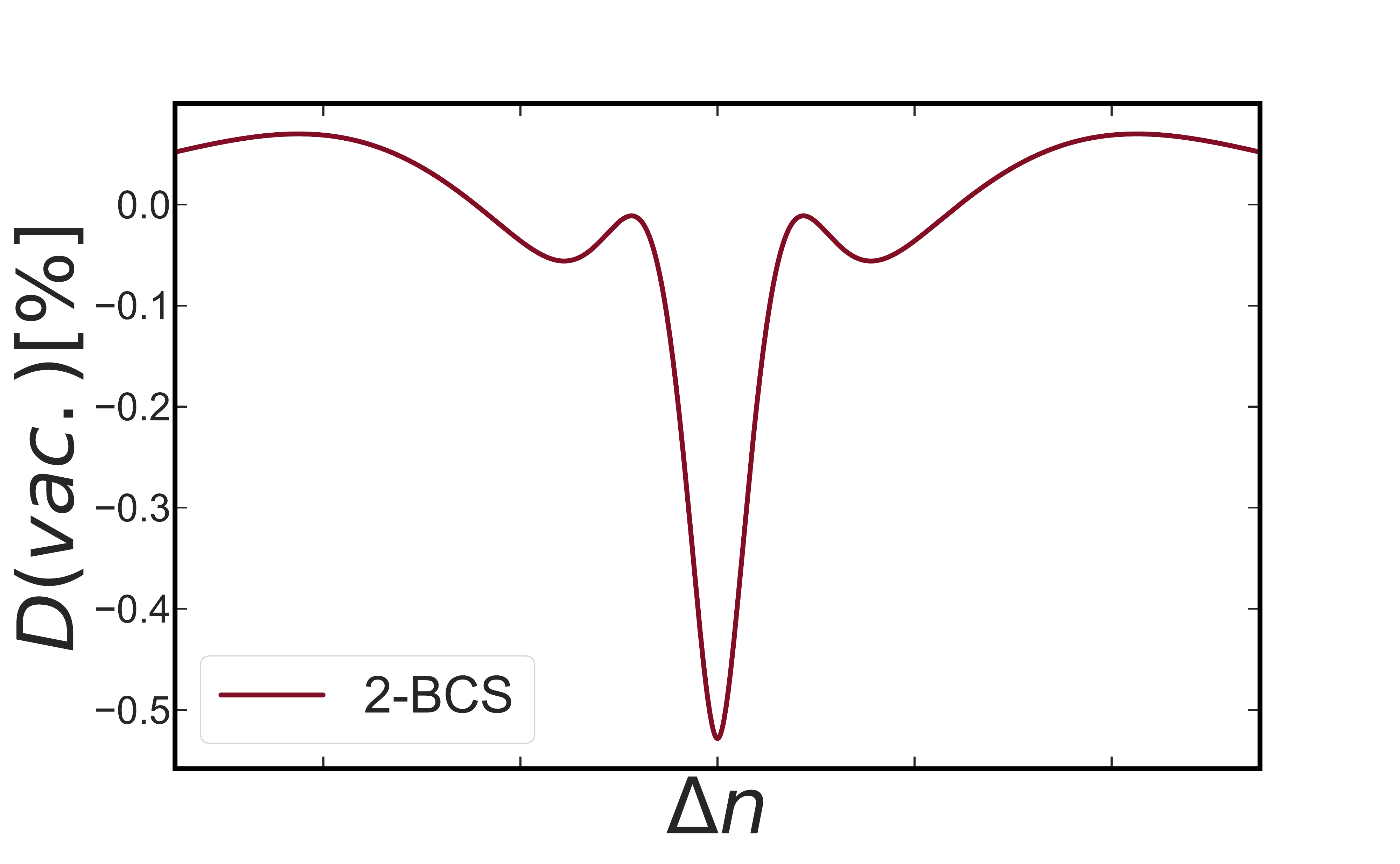}
    \put(-22,12){(i)}
    \hspace{0.025\linewidth}
    \includegraphics[width=0.48\linewidth, trim=5 15 55 65, clip]{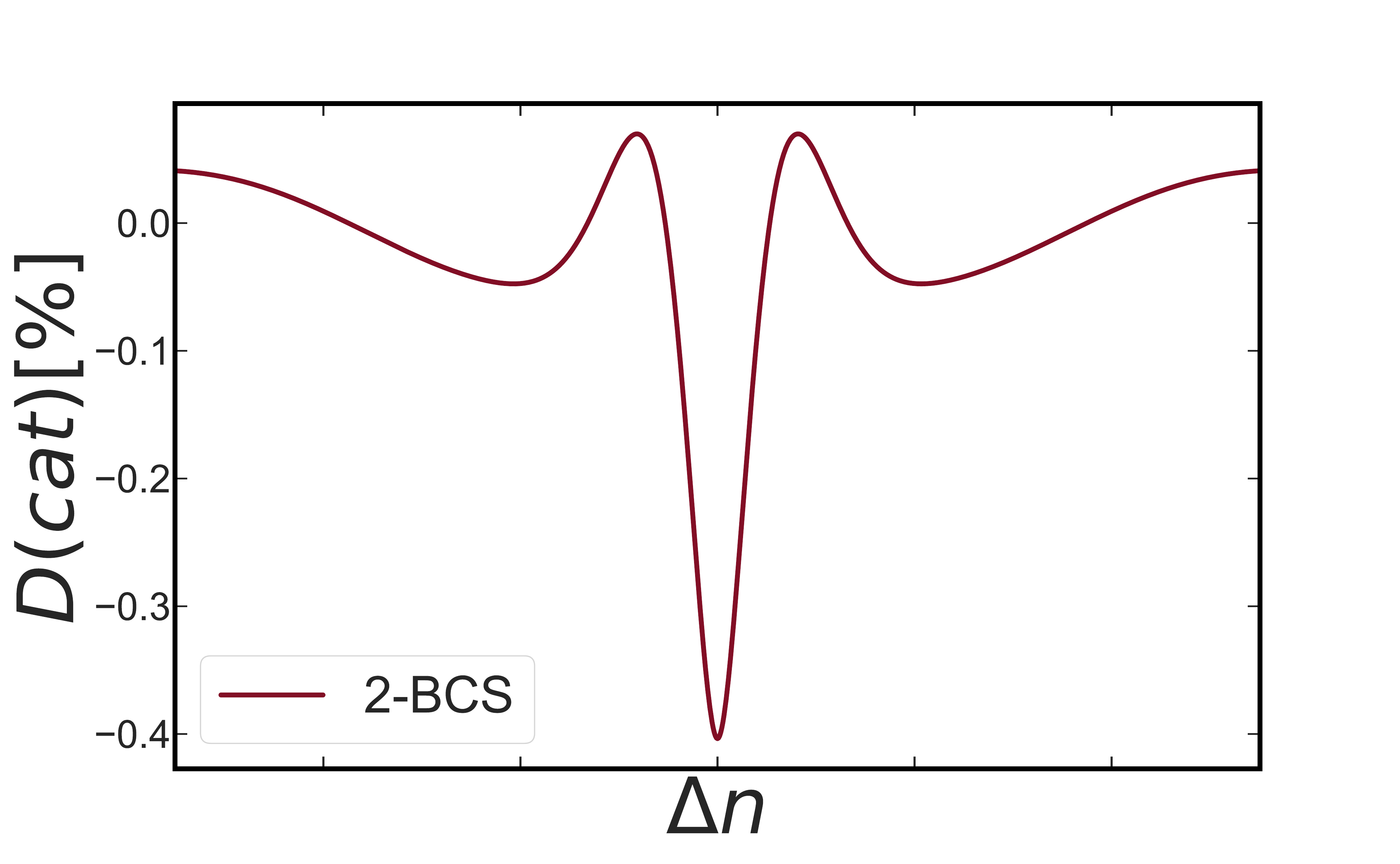}
    \put(-22,12){(j)}\\
    \includegraphics[width=0.99\linewidth, trim=20 10 200 25, clip]{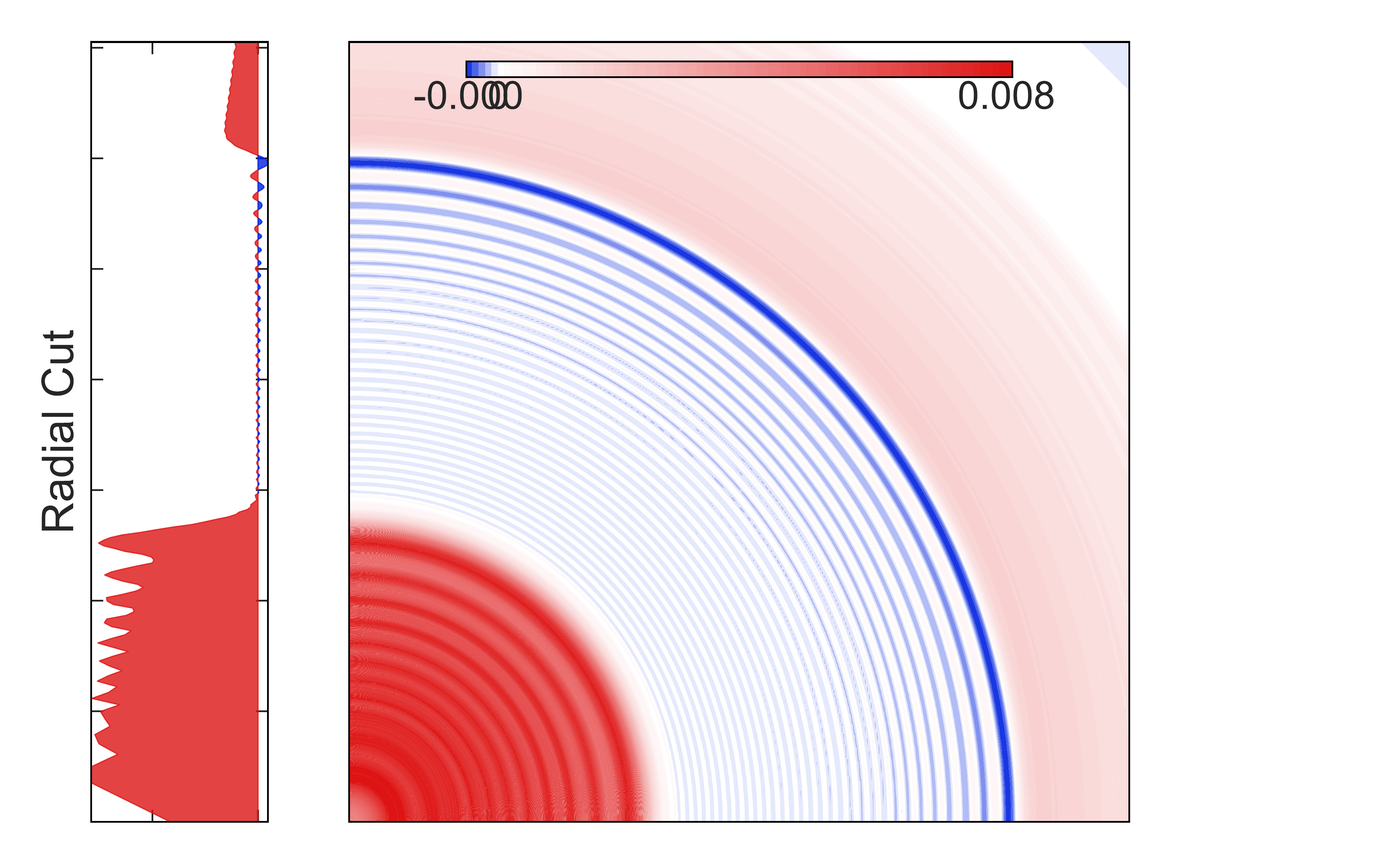}
    \put(-35,12){(k)}
    \vspace{-10pt}
    \caption{2-BCS probe: see caption of Fig.~\ref{CoherentStats}.}
    \label{2BCSStats}
\end{figure}

\vspace{.5\baselineskip}
\emph{Band-conditioned states. }
As described in the main text, we propose to use entanglement produced by spontaneous parametric down-conversion to derive new quantum-enabled EOS probe states. At the output of the SPDC process, we obtain a two-mode squeezed vacuum~\cite{Lvovsky2014}
\begin{equation}
    \label{TMSV}
    \ket{\Psi}=\sqrt{1-\xi}\,\sum_{n=0}^{+\infty}\,\xi^{n/2}\ket{nn},
\end{equation}
with $\nu=\xi/(1-\xi)$ the average photon number in each mode. If the modes are spatially distinct, we can use one as the probe and condition it on the number of photons in the probe by using a photodetector on the twin mode.

Without any conditioning, each mode independently (tracing the twin) exhibits thermal noise with parameter $\xi$, as in Eq.~\eqref{thermal}.

We condition the results by keeping only events for which the number of photons $m$ belongs in the set formed by a union  $\bm{\sigma}=\bigcup_{\ell}\;[\![A_\ell;B_\ell]\!]$ of $\ell$ segments (bands). The conditional probability is then
\begin{equation}
\label{DiscP}
P_{\bm\sigma,\xi}(n)=\frac{\sum_{m\in\bm\sigma} \Pi_\xi(m)\, \pi(n|m)}{\sum_{m\in\bm\sigma} \Pi_\xi(m)},
\end{equation}
where $\pi(n|m)$ is the probability of having $n$ photons in the probe mode, having detected $m$ photons in the heralding mode. For a perfect photodetector and without losses in the heralding branch, $\pi(n|m)=\delta_{n,m}$, and  $\Pi_\xi(m)=P_\xi(m)$, as in Eq.~(\ref{thermal}). This is not the case in practice, but we will use it as a good first approximation. Imperfections in the setup will be treated separately.

We call the resulting states band-conditioned states (BCS) as seen in the main text. They are nonclassical in nature, as witnessed by the negative values in their phase-independent Wigner quasi-distributions. 
\begin{equation}
    W_{\bm\sigma,\xi}(\beta)=\frac{2}{\pi}e^{-2\abs{\beta}^2}\sum_{m\in\bm\sigma}(-1)^m\mathcal{L}_m\!\left(4\abs{\beta}^2\right)\;P_{\bm\sigma,\xi}(m),
\end{equation}
where $\mathcal{L}_m$ are Laguerre polynomials.

Their nonclassical nature is one of the reasons for the advantages of the discriminated probes. The resulting statistical distributions offer better resources for EOS metrology than those obtained with classical (coherent and thermal) probes.

\vspace{.5\baselineskip}
\emph{Upper BCS. }
Because EOS is more efficient for a higher number of photons, a natural discrimination scheme is to keep only events where the number of photons exceeds a specific threshold, i.e. only the upper band of detected intensities. We call this an `` upper BCS'', and it is the band scheme that provides the highest values for $D(\textrm{vac.})$ of Eq.~\eqref{signaltonoise} among all the examples presented here. 
It is also a highly nonclassical state, with large swaths of negative values in the Wigner quasi-probability. Fig.~\ref{UpperStats} shows the advantage of this type of state over the classical ones.
In particular, it is observed that the relative differential noise amplitude $D(\textrm{vac.})$  is six times larger with this BCS state than with the coherent state. In addition, it clearly discriminates between a vacuum and a few-photon cat state much better than the coherent probe.

\vspace{.5\baselineskip}
\emph{Lower BCS. }
In contrast to the upper BCS scheme, a discrimination scheme that keeps only low photocounts actually lowers the amplitude of $D(\textrm{vac.})$, as seen in Fig.~\ref{LowerStats}.

\vspace{.5\baselineskip}
\emph{Quasi-Poissonian BCS. }
It is also possible to mimic the coherent state with a discriminator that keeps only photocounts near the average. The distribution is quasi-poissoninan and, as can be expected, the result for EOS is almost indistinguishable from the coherent case, as seen in Fig.~\ref{QuasiPStats}. Contrary to the coherent case however, the phase of the probe photons is completely unknown.

\vspace{.5\baselineskip}
\emph{Sub- and super-Poissonian BCS. }
It is interesting to compare cases that feature narrow discrimination bands both below and above the average number of photons. We should expect that the states remain nonclassical and feature negative Wigner values. However, we expect, and indeed find, that the super-Poissonian distribution of the BCS with lower counts (Fig.~\ref{SuperPStats}) is detrimental to EOS, whereas the sub-Poissonian distribution of the BCS with higher counts (Fig.~\ref{SubPStats}) is beneficial. 

\vspace{.5\baselineskip}
\emph{Multiple BCS. }
Schemes that feature multiple bands can be used to engineer specific features in the $D$ distributions (e.g. narrow peaks, wide tails...) that can be tailored to the input quantum signals. An example is the 2-BCS probe of Fig.~\ref{2BCSStats}. Already, in Fig.~\ref{2BCSStats} we observe non-trivial features in the relative susceptibilities. Such features can be exploited to efficiently evaluate the various moments of the quantum distribution. Many more types of probes can of course be envisioned, with specific features at specific positions, greatly enhancing the efficiency of the quantum statistics recovery process. Overall, the presented approach to quantum-enabled EOS probes (see also Fig. 1 of the main manuscript) is quite amendable to machine learning techniques for additional harvesting of metrological advantage.

\begin{figure}
    \centering
    \includegraphics[width=0.48\linewidth, trim=5 15 55 65, clip]{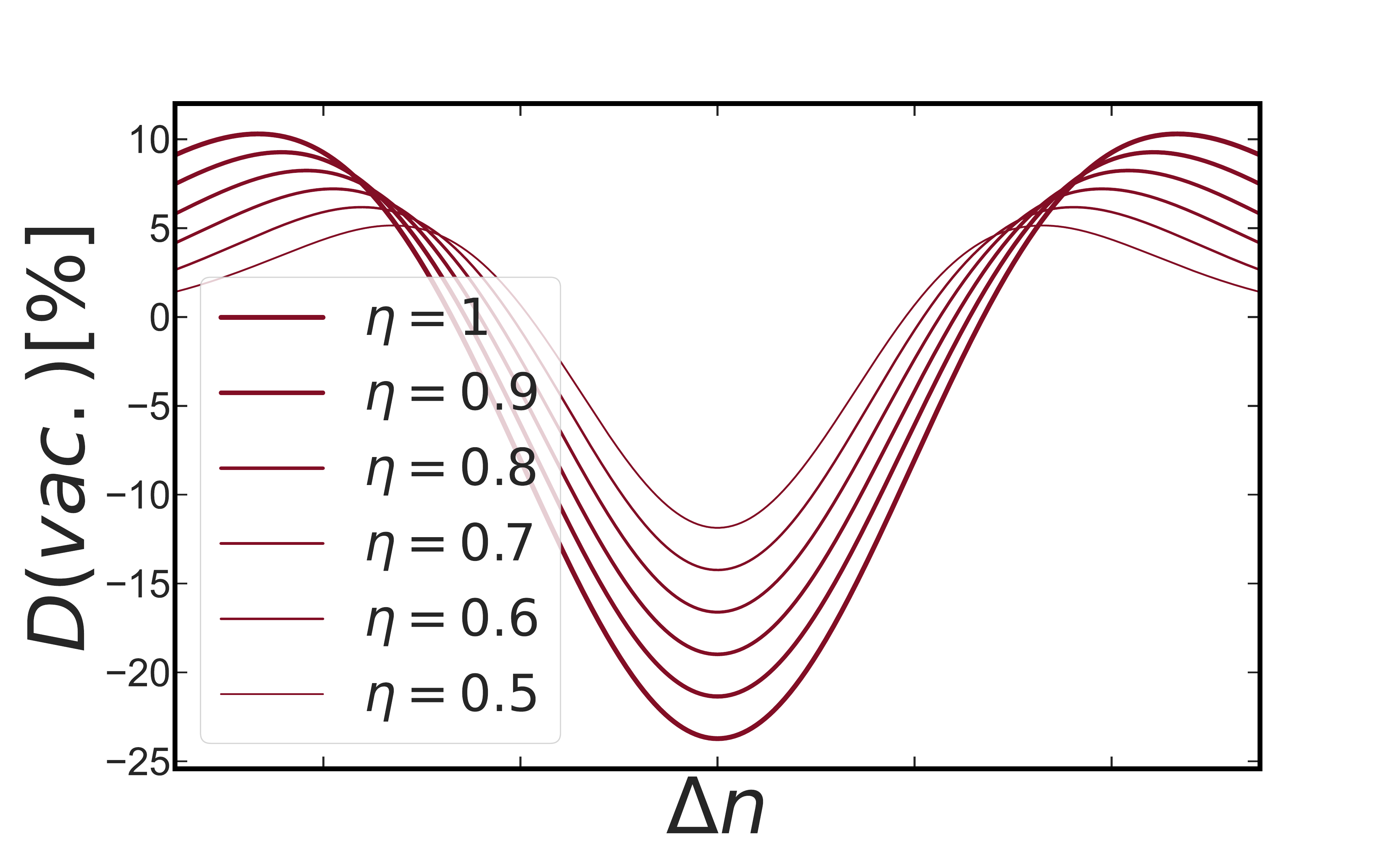}
    \put(-22,12){(a)}
    \hspace{0.025\linewidth}
    \includegraphics[width=0.48\linewidth, trim=5 15 55 65, clip]{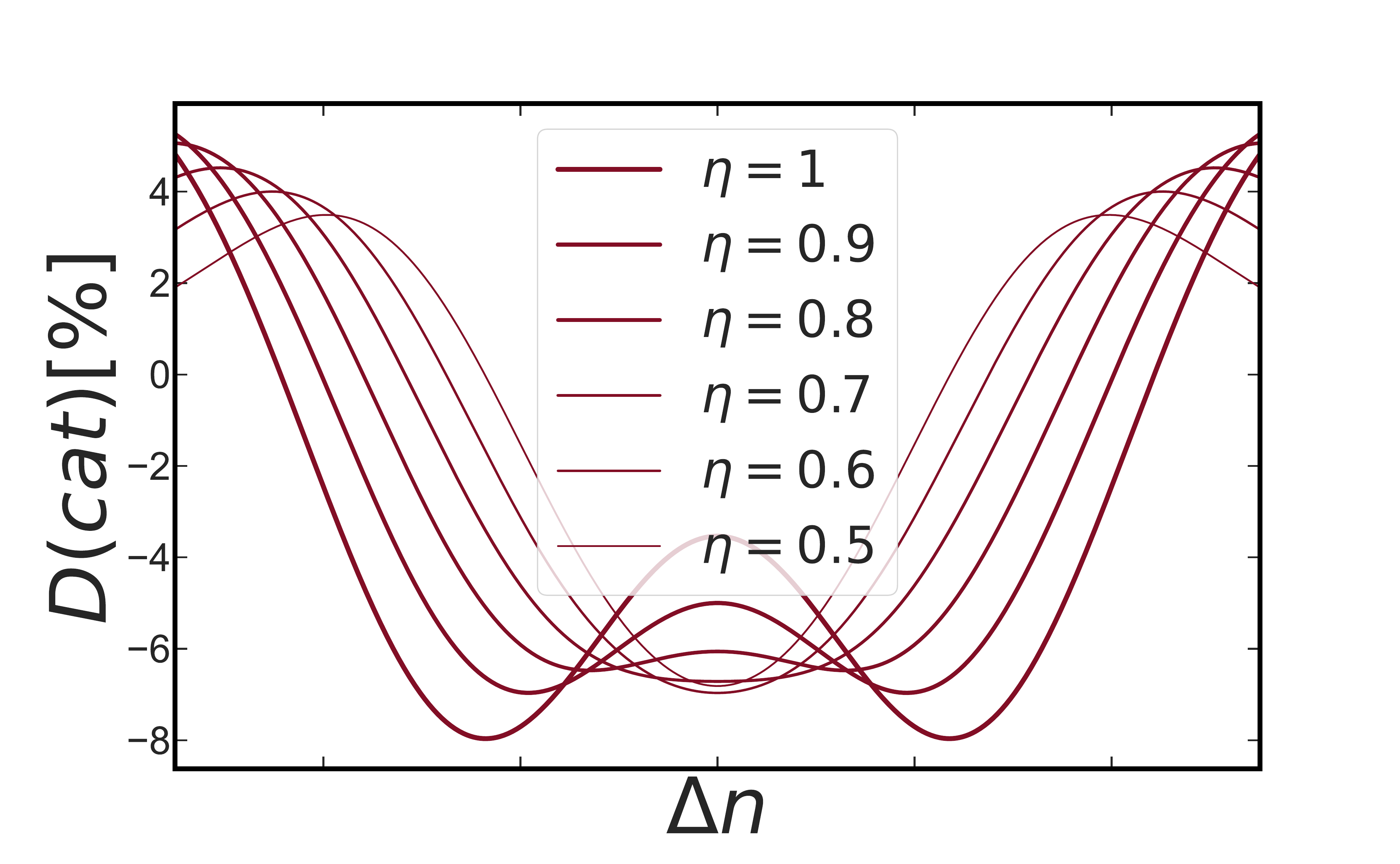}
    \put(-22,12){(b)}
    \caption{Effect of quantum efficiency in the discrimination branch on the relative differential noise function, for the upper BCS scheme of Fig.~\ref{UpperStats}. The advantage over classical states remains high for easily achievable quantum efficiencies ($\eta\simeq 0.8$), although it vanishes rapidly.}
    \label{eta}
\end{figure}

\section{Non-ideal photodetectors}
Eq.~\eqref{DiscP} takes into account the possibility of having a non-ideal photodetector. Discrimination on a large number of photons can be viewed as a natural extension of the historical heralded-single-photon experiments~\cite{Hong1986}. These sources are known to be primarily limited by the quantum efficiency of and noise in the heralding branch; in contrast, the probe branch, which is in any case the same as in the classical case, is much less affected by losses and noise~\cite{Virally2010}. Hence, it is important to understand what happens with a non-ideal detector in the conditioning branch.

We model the non-ideal conditioning process with two parameters: a non-unity effective quantum efficiency $\eta$ and a gain $\gamma$. The quantum efficiency is treated as usual like a beam-splitter with branching ratios $\eta$ and $1-\eta$. The gain is treated as follows: each photon generates a number of electrons that follows a Poisson distribution with parameter $\gamma$. In that case, the probability of detecting $m$ electrons in the heralding detector remains quasi-thermal,
\begin{equation}
    \Pi_\xi(m)\simeq P_\zeta(m)\text{, with }\zeta=\frac{\gamma\eta\xi}{1+\xi(\gamma\eta-1)}.
\end{equation}
At the same time, the conditional probability $\pi(n|m)$ of having $n$ photons in the probe when $m$ electrons have been detected in the conditioning branch is no longer a Dirac delta but becomes quasi-Gaussian with mean $m/\gamma\eta$ and variance $m[1+\gamma(1-\eta)]/\gamma^2\eta^2$.

It is thus possible to simulate the effect of all types of gains and losses on the conditioning branch. As expected~\cite{Virally2010}, the most important parameter is $\eta$. For large $m$ and $\eta\ge0.5$, the quasi-Gaussian distribution is so narrow that it is well approximated as a delta, leading to
\begin{equation}
    \label{pNetaapprox}
    P_{\bm\sigma,\zeta,\eta,\gamma}(n)\simeq\frac{P_\zeta(\gamma\eta n)}{\sum_{m\in\bm\sigma} P_\zeta(m)}.
\end{equation}

Fig.~\ref{eta} shows that the advantages of the BCS over the classical states vanishes rapidly as the quantum efficiency of the detecting branch decreases. However, in the range of real-life photodetectors ($\eta\simeq 0.8$), the advantage remains important. Unique double-peaked feature of $D(cat)$ sported by BCS vanishes for $\eta < 0.75$, again dictating stringent but realistic experimental constraints on the realization of the quantum-enhanced EOS.   

\end{appendix}

\bibliography{BCS}

\end{document}